\documentclass[smallextended]{svjour3}       % onecolumn (second format)
\smartqed  % flush right qed marks, e.g. at end of proof
\usepackage{graphicx}
\begin{document}

\title{Our astrochemical heritage}
%\subtitle{Do you have a subtitle?\\ If so, write it here}

%\titlerunning{Short form of title}        % if too long for running head

\author{Paola Caselli         \and
        Cecilia Ceccarelli %etc.
}

\authorrunning{Caselli \& Ceccarelli} % if too long for running head

\institute{P. Caselli \at
              School of Physics and Astronomy, University of Leeds, Leeds LS2 9JT, UK \\
              Tel.: +44-(0)113-3434065\\
              Fax: +44-(0)113-3433900\\
              \email{P.Caselli@leeds.ac.uk}           %  \\
%             \emph{Present address:} of F. Author  %  if needed
           \and
           C. Ceccarelli \at
              UJF-Grenoble 1/CNRS-INSU, Institut de Plan\'etologie et d'Astrophysique de Grenoble (IPAG) UMR 5274, Grenoble, F-38041, France \\
             Tel.: +33-(0)476-514201\\
              Fax: +33-(0)476-448821\\
              \email{Cecilia.Ceccarelli@obs.ujf-grenoble.fr}
}

\date{Received: 3 September 2012 / Accepted: 7 October 2012}
% The correct dates will be entered by the editor

\maketitle

\begin{abstract}
%  Please provide an abstract of 150 to 250 words. The abstract should
%  not contain any undefined abbreviations or unspecified references.

  Our Sun and planetary system were born about 4.5 billion years
  ago. How did this happen and what is our heritage from these 
  early times?  This review tries to address these questions from an astrochemical point of view. On the one hand, we have some crucial information from meteorites, comets and other
  small bodies of the Solar System. On the other hand, we have the
  results of studies on the formation process of Sun-like stars in
  our Galaxy.  These results tell us that Sun-like stars form in dense
  regions of molecular clouds and that three major steps are involved
  before the planet formation period.  They are represented by the
  pre-stellar core, protostellar envelope and protoplanetary disk
  phases. Simultaneously with the evolution from one phase to the
  other, the chemical composition gains increasing complexity. 
 
 In this review, we first present the information on the chemical
 composition of meteorites, comets and other small bodies of the
 Solar System, which is potentially linked to the first phases of the
 Solar System's formation. Then we describe the observed chemical
 composition in the pre-stellar core, protostellar envelope and protoplanetary disk phases,
including the processes that lead to them. Finally, we draw together pieces from
 the different objects and phases to understand whether and how much
 we inherited chemically from the time of the Sun's birth. 
 
\end{abstract}

%%%%%%%%%%%%%%%%%%% Cecilia %%%%%%%%%%%%%%%%%%%%%%%%%%%%%%%%%%%%%%%%%%%%%
%%%%%%%%%%%%%%%%%%% Cecilia %%%%%%%%%%%%%%%%%%%%%%%%%%%%%%%%%%
% just to facilitate the counting of the pages
\newpage
\section{Introduction}   % Cecilia, about 5 pages
\label{intro}
%It should point out (among other things) what we know about our Solar
%System bodies and our astrochemical heritage (eg isotopic
%fractionation (D, 15N, ...); organic heritage; Al, Ca, Cl --
%irradiation).

Once upon a time, there was a small cold cloud of gas and dust 
in an interstellar medium broken into several clumps and filaments of
different masses and dimensions. Then, about 4.5 billion years ago,
the small cloud became the Solar System. What happened to that
primordial cloud? When, why and how did it happen? Does the
Earth receive a heritage from those old eons?  Can this heritage help us to
understand our origins? 

The answers to these questions can only come from putting together
many pieces of a giant puzzle that covers different research fields:
from what the Earth is made of to its evolution, from what are the
most pristine meteorites from outer space that have fallen on
Earth to their present composition, from which other small bodies of
the Solar System, comets and asteroids, have the imprint of the first
composition of the solar nebula to their origin and evolution. Last
but not least, the study of other small clouds and young
Sun-like stars in our Galaxy gives us the wide range of possible 
outcomes of star and planet formation, from which we would like to 
understand why the Solar System and the Earth chose one of them. 

Each single piece of the puzzle brings precise and precious
information. The problem is that sometimes the information is
hidden in a scrambled code whose key is unknown. Take
meteorites as an example. As explained to us by an expert colleague, 
assessing the composition of the Solar Nebula from the study of
the meteorites is like trying to assess Napoleon's army structure
looking at the few survivors of the Russian war. How representative
are those survivors?  Although, evidently, they still provide very
precious information, extracting the whole information from them is far from
obvious. The example is to say that every single piece of the puzzle
is important, even the pieces that seem to be redundant. Actually, the
redundant ones are likely the most important, as they may allow to
distinguish and disentangle all the various intervening effects.  In
this context, the study of the objects similar to the Solar System
progenitor takes a particular relevance, because it can provide us with
plenty of pieces to compare with the other pieces from the
present Solar System. The hope is that they will provide us with the
keys of the scrambled codes.

In this review, we will focus on just a subset of these pieces, those
coming from the study of the chemical composition during the birth of
stars and planetary systems like our Solar System. In
\S\ref{sec:star-form-chem}, we will first give a very general
overview of how we think the Solar System and stars of similar mass
have formed and how this process influences the chemistry. This is
based on the ensemble of observations and studies on star-forming
regions and Solar System objects. Then, in
\S\ref{sec:pieces-puzzle}, we will describe in detail some
pieces of the puzzle which potentially connect what we observe in the
objects of the Solar System nowadays and what we know about star
formation in our Galaxy. The next sections will discuss 
star and planet formation studies.  We will describe how the evolution of the
matter from a cold cloud (\S \ref{sec:calm-before-storm}) to a
protostellar envelope (\S \ref{sec:violent-origin-sun}) and a protoplanetary disk
(\S \ref{sec:turb-assembl-plan}) corresponds to an increase of the
molecular complexity.  Section
\ref{sec:assembl-some-piec} will provide specific examples on the link
between the present Solar System small bodies with the pre- and
proto-stellar phase.  A final section will try to draw some
conclusions.

We emphasize that the present review is complementary to several
reviews recently appeared in the literature on different aspects just
touched upon by us and that will be cited in the appropriate sections. 

%%%%%%%%%%%%%%%%%%%% SESSION 2 %%%%%%%%%%%%%%%%%%%%%%%%%%%%%
\section{Solar-type star formation and chemical
  complexity}\label{sec:star-form-chem}

\begin{figure*}
  \includegraphics[width=\textwidth]{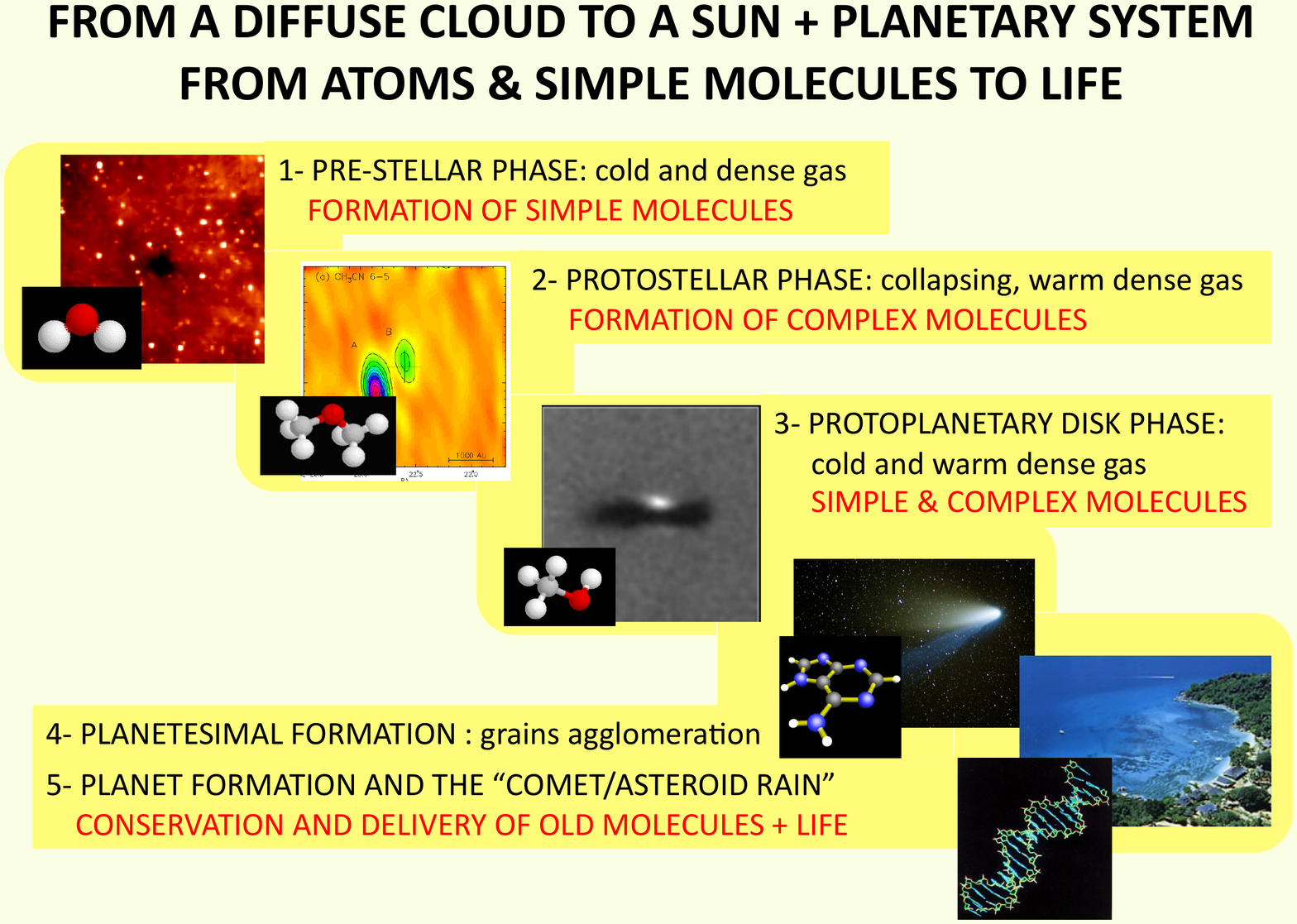}
  \caption{Star formation and chemical complexity. The formation of a
    star and a planetary system, like the Solar System, passes through
    five fundamental phases, marked in the sketch. }
\label{fig:starform-chem}
\end{figure*}

The formation of a Sun-like star and molecular complexity proceed hand
in hand. As the primordial cloud evolves into a protostellar envelope,
protoplanetary disk and planetary system, the chemical composition of
the gas becomes increasingly more complex. The five major phases of
the process that we think have formed the Earth are sketched in Figure
\ref{fig:starform-chem} and here listed.
%CC: should we insert in some way the giant planet interaction?
% Winn et al. 2010 show that the orbit of close-orbiting hot Jupiters
% are misaligned with the equatorial plane of the star, indicating
% that these planets have been scattered (rather than "simply"
% migrated) from outer orbits because of the interaction of multiple
% planets Weidenschilling \& Marzari 1996)
% 
\begin{itemize}
\item [Phase 1: ] {\bf Pre-stellar cores.}  These are the "small cold clouds" 
mentioned above. During this phase, matter
  slowly accumulates toward the center of the nebula. As a result, the
  density at the center increases while the temperature
  decreases. Atoms and molecules in the gas-phase freeze-out onto the
  cold surfaces of the sub-micron dust grains, forming the so-called
  icy grain mantles. Thanks to the mobility of the H atoms on the
  grain surfaces, hydrogenation of atoms and CO (the most abundant
  molecule, after H$_2$, in cold molecular gas) takes place, forming
  molecules such as water (H$_2$O), formaldehyde (H$_2$CO),
  methanol (CH$_3$OH) and other hydrogenated species.
\item [Phase 2: ]{\bf Protostellar envelopes.} The collapse proceeds,
  gravitational energy is converted into radiation and the envelope
  around the central object, the future star, warms up. The molecules
  frozen in grain mantles during the previous phase acquire
  mobility and likely form new, more complex species. When the
  temperature reaches the mantle sublimation temperature, in the 
  so-called hot corinos, the molecules in the mantles sublimate
  back in the gas-phase, where they react and form new, more complex,
  molecules. Simultaneously to the collapse, a fraction of matter is
  violently ejected outward in the form of highly supersonic
  collimated jets and molecular outflows. When the outflowing material
  encounters the quiescent gas of the envelope and the molecular cloud,
  it creates shocks, where the grain mantles and refractory grains are (partially)
  sputtered and vaporized. Once in the gas phase, molecules can be observed 
  via their rotational lines.
\item [Phase 3: ] {\bf Protoplanetary disks.} The envelope dissipates
  with time and eventually only a circumstellar disk remains, also
  called protoplanetary disk. In the hot regions, close to the central
  object, new complex molecules are synthesized by reactions between the
  species formed in the protostellar phase. In the cold regions of the
  disk, where the vast majority of matter resides, the molecules
  formed in the protostellar phase freeze-out again onto the grain mantles, 
  where part of the ice from the pre-stellar phase may still be present. 
  The process of "conservation and heritage" begins. 
\item [Phase 4: ] {\bf Planetesimal formation. } The sub-micron dust
  grains coagulate into larger rocks, called planetesimals, the seeds
  of the future planets, comets and asteroids. Some of the icy grain
  mantles are likely preserved while the grains glue together. At
  least part of the previous chemical history may be conserved in the
  building blocks of the Solar System rocky bodies.
\item [Phase 5: ] {\bf Planet formation.}  This is the last phase of
 rocky planet formation, with the embryos giant impact period and
  the formation of the Moon and Earth. The leftovers of the process,
  comets and asteroids, copiously "rain" on the primitive Earth,
  forming the oceans and the Earth second atmosphere. The heritage
  conserved in the ices trapped in the planetesimals and rocks is
  released onto the Earth. Life emerges sometime around 2 billion
  years after the Earth and Moon formation\footnote{The famous
    fossils of cyanobacteries of Australia and for long considered as
    the first traces of life dated 3.5 Myr
    (\cite{2002Natur.416...73S}), are interpreted as inorganic
    condensations (\cite{2003LPI....34.1677S};
    \cite{2002AsBio...2..353G}) and still source of intense debate
    (\cite{2011NatGe...4..240M}). Conversely, there is consensus on
    the rise of life about 2\,Gyr after the Earth formation, as testified by the rise in the
    O$_2$ abundance in the atmosphere (\cite{2010NatGe...3..522C}). }.
\end{itemize}  

The sections \ref{sec:calm-before-storm} to
\ref{sec:turb-assembl-plan} will review and discuss in detail the
chemistry in the first three phases of the process, those where the
heritage is likely accumulated. Box 1 briefly explains the data and
tools needed to interpret the observations and Table
\ref{tab:summary-phases} summarizes some key proprieties of the phase
1 to phase 3 objects.

% For tables use
\begin{table}
% table caption is above the table
  \caption{Summary of the proprieties of the objects in the first three phases of the solar-type star formation process, before planet formation.}
\label{tab:summary-phases}
\begin{tabular}{lcrrrl}
\hline\noalign{\smallskip}
Phase \& Object & Age & Radius  & Temp. & Density & Chemical processes\\
            & (yr)   & (AU)              &  (K)                  & (cm$^{-3}$) & \\
\noalign{\smallskip}\hline\noalign{\smallskip}
1- Pre-stellar core & $\sim 10^5$ & $\sim 10^4$ & 7--15 & $10^4$--$10^6$ & Ice formation \& \\
                                  &                        &                        &             &                                & molecular deuteration \\
2- Protostellar envelope:             & $10^4$--$10^5$ & $\sim 10^4$ &      &     &   \\
~~~~ Cold envelope &       & 100--$10^4$ & $\leq 100$& $10^5$--$10^7$ & Ice formation \& \\
                                  &         &                        &                    &                                & molecular deuteration \\
~~~~ Hot corinos    &        &  $\leq 100$   & $\geq 100$&  $\geq 10^7$ & Complex molecules \\
                                  &                       &                        &             &                                & formation\\ 
3- Protoplanetary disk: & $\sim 10^6$ &  $\sim 200$ &                     &                                & \\
~~~~ Outer midplane   &                       &  20--200 & 100--10     & $10^8$--$10^6$ &Ice formation \& \\
                                        &                       &                        &                     &                                & molecular deuteration \\
~~~~ Inner midplane   &                        &  $\leq 20$    & $\geq 100$&  $\geq 10^8$       & Complex molecules \\
                                  &                       &                        &             &                                & formation\\ 
\noalign{\smallskip}\hline
\end{tabular}
\end{table}

\begin{figure*}
  \includegraphics[width=\textwidth]{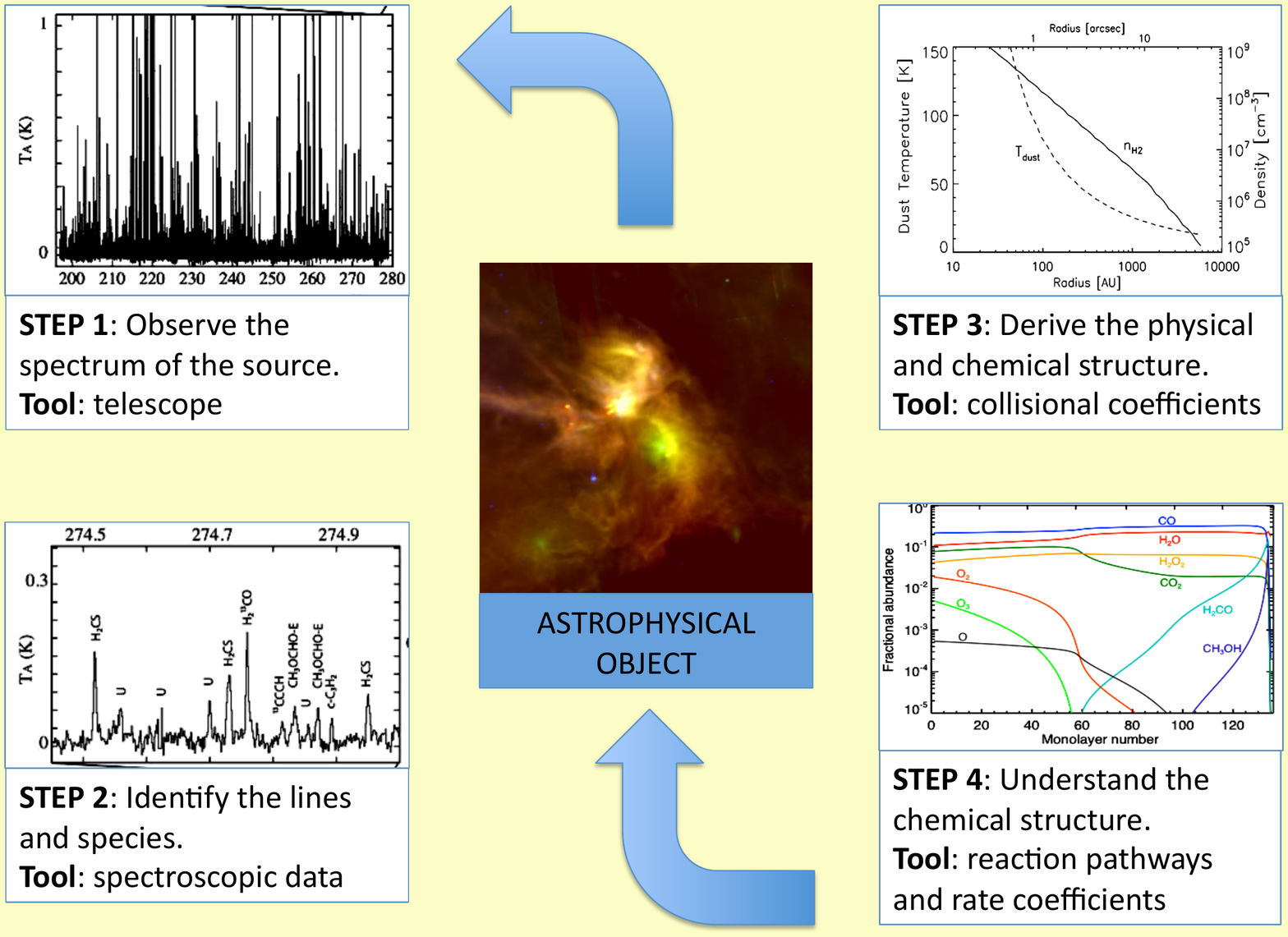}
  \caption{The four steps required to measure the chemical structure
    of an astrophysical object, as described in Box 1, including the
    tools needed to complete each step: (1) observations at the
    telescope, (2) identification of the lines and species, (3)
    derivation of the physical and chemical structure using radiative
    transfer codes, which require accurate collisional coefficients,
    and (4) chemical models.}
\label{fig:schema-obstheo}
\end{figure*}

\medskip
\noindent
\fbox{
\begin{minipage}{0.95\linewidth}
\bigskip
{\bf Box 1: Data needed to interpret the astronomical observations\\}
In order to derive the chemical composition of a celestial body from
line observations one needs to identify correctly the lines as due to
a specific molecule and to convert the observed line intensity into the 
species abundance. Then, to understand what these abundances mean one needs to
compare the observed with model predicted abundances.  The process,
sketched in Fig.\,\ref{fig:schema-obstheo} requires, therefore, data from 
different communities: i) spectroscopic data, to identify
the lines; ii) collisional coefficients, to convert them into abundances;
iii) chemical reactions to build up astrochemical models. The three
sets of data necessitate specific skills and enormous laboratory and
computational efforts. The available information is centralised in the
following databases:
\begin{itemize}
\item Spectroscopic databases:\\
  + JPL Molecular spectroscopy database (\cite{1998JQSRT..60..883P}):\\ 
     {\it http://spec.jpl.nasa.gov/home.html}\\
  + Cologne Database for Molecular Spectroscopy database
  (CDMS, \cite{2005JMoSt.742..215M}):\\
     {\it http://www.astro.uni-koeln.de/cdms/}\\
   + Splatalogue database for astronomical spectroscopy
   (SPLATALOGUE):\\
      {\it  http://splatalogue.net/}
    \item Collisional excitation databases:\\
      + Ro-vibrational collisional excitation database (BASECOL, \cite{2004sf2a.conf..525D}):\\
      {\it
        http://basecol.obspm.fr/index.php?page=pages/generalPages/home}\\
      + Leiden Atomic and Molecular Database (LAMDA, \cite{2005A&A...432..369S}):\\
      {\it http://home.strw.leidenuniv.nl/~moldata/}
\item Chemical reaction databases:\\
   + The KInetic Database for Astrochemistry (KIDA, \cite{2012ApJS..199...21W}):\\
   {\it http://kida.obs.u-bordeaux1.fr/ } \\
   + The Ohio State University (OSU) gas-phase and gas-grain chemical models (e.g. \cite{2008ApJ...682..283G}):\\
   {\it http://www.physics.ohio-state.edu/~eric/research.html } \\
   + The UMIST database for astrochemistry (UDFA, \cite{2007A&A...466.1197W}):\\
   {\it http://www.udfa.net/ }
\end{itemize}
We emphasize that the databases just collect the data which are
provided by several colleagues from all over the world. We would like
here to pay our tribute to Pierre Valiron and Fredrik Schoier, who
enormously contributed to the collisional excitation coefficients and
to the set up of the LAMDA database respectively, and who prematurely
passed away.
\bigskip
\end{minipage}
}

%%%%%%%%%%%%%%%%%%%% SESSION 3 %%%%%%%%%%%%%%%%%%%%%%%%%%%%%
\section{Pieces of the puzzle from the Solar System}\label{sec:pieces-puzzle}

A variety of information on the formation process of the Solar System
is provided to us by the small bodies believed to be the most pristine
objects of the Solar System: Kuiper Belt Objects (KBOs), comets,
meteorites and particularly carbonaceous chondrites, and
interplanetary dust particles (IDPs). Here we will review some
properties of these objects that can shed light on the formation
process when compared with what we know about other solar-type forming
stars in the Galaxy.  We emphasize that this summary is far from being
exhaustive and the reader is invited to look at the reviews cited in
the following subsections.

\subsection{Where does the terrestrial water come from?}\label{sec:where-terr-water}

We all know how fundamental water is for the terrestrial life. It is
the best solvent, allowing chemical
reactions to form large biotic molecules and to break down ATP
(Adenosine TriPhosphate), a process at the very base of the energy metabolism of
living cells.  Water had a fundamental role also on planet Earth, its history, evolution and equilibrium, for example allowing the magma to be viscous enough for tectonics to take place.

Sometimes the most obvious questions, like the one on the origin of
water or why the night sky is dark, do not have obvious answers.  The
explanation of the dark night sky had to wait for the discovery of the
expansion of the Universe, while the explanation of why Earth is so abundant
in water is still hotly debated.  But what are the facts? Two main
facts are fundamental pieces of this puzzle. The first one is
the quantity of terrestrial water, the second is its isotopic
composition.

Regarding the amount of water on Earth, we can easily measure it in
the Earth's crust where it is $\sim 3\times 10^{-4}$ the Earth mass
(\cite{2000E&PSL.181...33L}). It is much less obvious to measure it in
the mantle and core, where the vast majority of the Earth's mass resides
and where it is impossible to directly measure the volatile
components. Measurements of Earth's mantle water content are in fact
based on indirect evidence, mostly using noble gases as proxies for
the volatile hydrogen (\cite{1982PEPI...29..242F};
\cite{1983Natur.303..762A}), which implies assuming that the solar
abundance ratios are maintained in the Earth mantle. The most recent
estimates give a total amount of $\sim 2\times 10^{-3}$ Earth masses
(\cite{2012E&PSL.313...56M}), namely almost ten times more than
in the crust. It has to be noted, though, that Earth in the Archaean
was most likely more volatile-rich than in our days
(e.g. \cite{1996E&PSL.144..577K}).

The second fundamental piece of the puzzle is the HDO/H$_2$O ratio,
$1.5\times 10^{-4}$ in the terrestrial oceans, namely about ten times
larger than the elemental D/H ratio in the Solar Nebula
(\cite{1998SSRv...84..239G}). Direct measurements of the HDO/H$_2$O
ratio in the Earth mantle are impossible, but indirect ones seem to
suggest a slightly lower value than that of the oceans
(\cite{2012E&PSL.313...56M}).

The problem on the origin of the terrestrial water comes from the fact
that the planetesimals that built up the Earth, if they were located
at the same place where Earth is today, must have been dry. Therefore,
either water came later, when Earth was mostly formed, or the
planetesimals that formed the Earth were from a zone more distant than
1 AU. The first theory, also called ``late veneer'', was first
proposed by \cite{1992AdSpR..12....5D} and \cite{1995Icar..116..215O}
and postulates that water is mostly delivered to Earth from comets,
especially during the Late Heavy Bombardment
(\cite{2000Icar..148..508D}; \cite{2005Natur.435..466G}). For almost a
decade, the theory had the problem, though, that the HDO/H$_2$O
abundance ratio in the six comets where it had been measured is about
a factor of two too high (\cite{2009EM&P..105..167J}; see \S
\ref{sec:isotopic-anomalies} and Fig. \ref{fig:D-15N}). However, new
Herschel measurements are changing the situation. The measure on the
103P/Hartley2 comet gives exactly the terrestrial value
(\cite{2011Natur.478..218H}) whereas measurements toward C/2009 P1
give again a larger HDO/H$_2$O value, $2\times10^{-4}$
(\cite{2012A&A...544L..15B}). The other possibility is that Earth was
partly built from water-rich planetesimals from the outer zone
(\cite{2000M&PS...35.1309M}). Two arguments are in favor of this
theory. First, the HDO/H$_2$O ratio of carbonaceous chondrites is very
similar to the terrestrial one (1.3--1.8$\times 10^{-4}$,
\cite{2003SSRv..106...87R}; see \S \ref{sec:isotopic-anomalies} and
Fig. \ref{fig:D-15N}). Second, numerical simulations of the young
Solar System from several authors predict that up to 10\% of the Earth
may have been formed by planetesimals from the outer asteroid belt,
providing enough water to Earth (e.g. \cite{2000M&PS...35.1309M};
\cite{2009Icar..203..644R}). The same simulations tend to exclude the
cometary delivery as a major contribution. However, as any model, the predictions
are subject to a number of uncertainties, a major one being how much
water is in the outer asteroid belt planetesimals
(\cite{2012A&A...537A..73L}).

Finally, the question on the origin of Earth's water is somewhat
linked to the question on the origin of the Earth's atmosphere. Even
though the methods are different, also for the
Earth's atmosphere it is discussed a cometary delivery versus a meteoritic
origin. Likely, in this case, both sources are necessary
(e.g. \cite{2003Icar..165..326D}).

We emphasize the key role played, in both theories, by the HDO/H$_2$O
ratio in the terrestrial water, comets and asteroids. In the following
sections of this review, we will see why, when and how water becomes
enriched of deuterium.

\subsection{Molecular species in comets and KBOs}\label{sec:comets-kbo}

Several molecular species have been detected in comets since decades
and in KBOs since the last decade. Here we briefly summarise which
species have been detected and recommend to the interested reader the
reviews by \cite{2011ARA&A..49..471M} and \cite{2011IAUS..280..261B},
and \cite{2012AREPS..40..467B} on the comets and KBOs, respectively.
\\

\begin{figure*}
  \includegraphics[width=\textwidth]{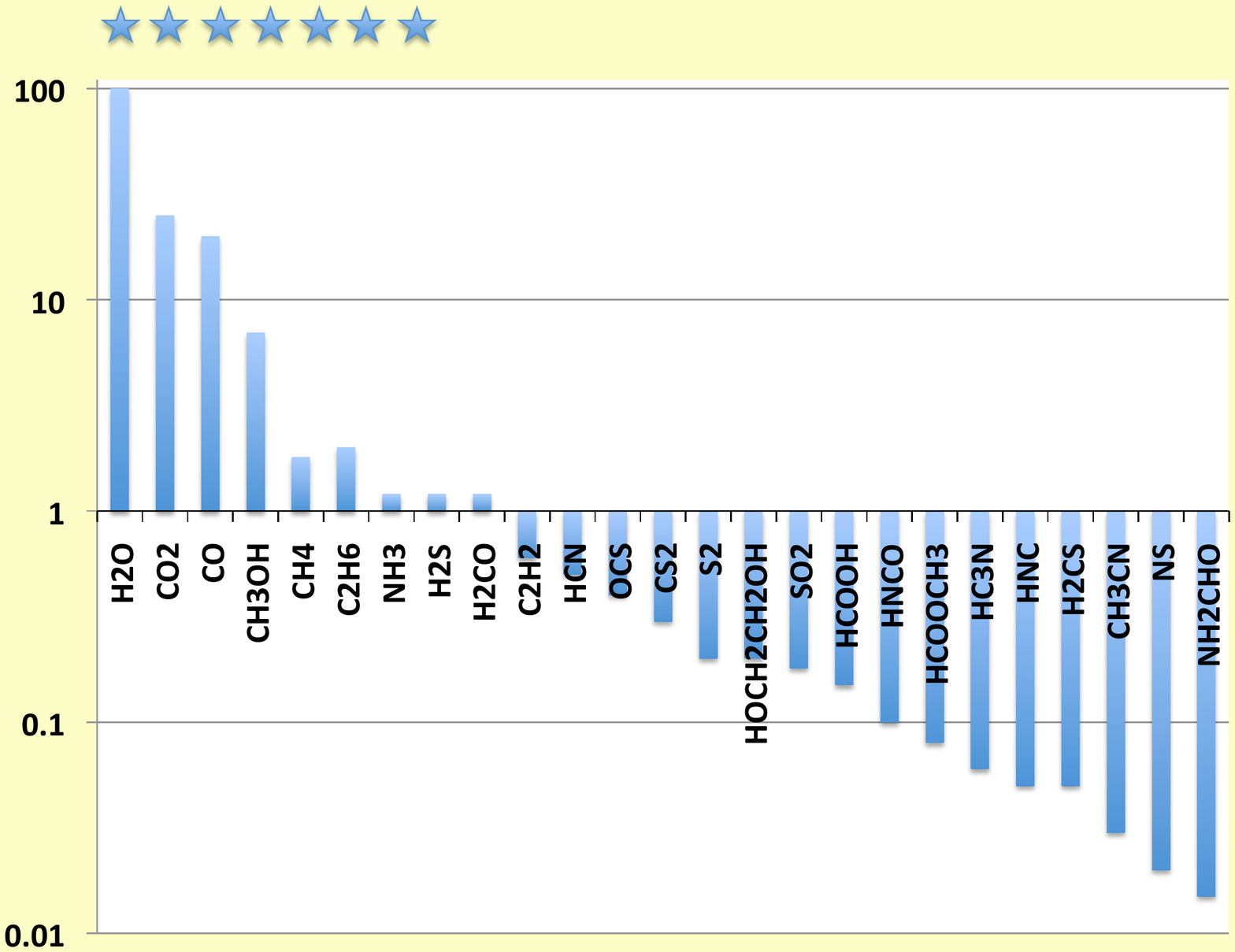}
  \caption{Abundances of molecular species in comets, with respect to H$_2$O. The species with
    an asterisk have also been detected in KBOs.}
\label{fig:species-comets}
\end{figure*}

\noindent {\it Comets:} Two dozens of molecular species have been
identified in various comets by several authors
(e.g. \cite{2002EM&P...90..323B}; \cite{2009EM&P..105..267C}). More
specifically:
\\
i) H$_2$O, CO, CO$_2$, CH$_4$, C$_2$H$_2$, C$_2$H$_6$, CH$_3$OH,
H$_2$CO, NH$_3$, HCN, HNC, CH$_3$CN and H$_2$S have been
detected in more than 10 comets;\\
ii) HCOOH, HNCO, HC$_3$N, OCS and S$_2$ have been detected in more
than 1 comet;\\
iii) HOCH$_2$CH$_2$OH, HCOOCH$_3$, CH$_3$CHO, NH$_2$CHO, SO$_2$,
H$_2$CS have been observed in one comet, Hale-Bopp.\\
Not all species are considered primary species, namely species present in the sublimated ices. Some, like HNC, are product species, namely they are the products of chemical reactions involving the primary species once ejected in the gas. Other species, like H$_2$CO and CO, have contributions from both primary and product species. The measured abundances
are summarised in Fig. \ref{fig:species-comets}. To this list one has
to add the recent detection of glycine, the simplest of amino acids,
in the 81P/Wild2 comet by the mission STARDUST
(\cite{2009M&PS...44.1323E}).
\\

\noindent {\it KBOs:} KBOs are the objects beyond Neptun's orbit, at
an heliocentric distance between 30 and 50 AU, and are thought to hold
precious information on the pristine chemical composition of the Solar
Nebula at those distances. Being relatively small objects, they are
difficult to study. However, in the last decade, important progress has
been made. Briefly, the six large KBOs where spectroscopic
observations could be obtained showed the presence in their atmosphere
of H$_2$O, CH$_4$, N$_2$ and CO, even though with different
proportions from object to object (e.g. \cite{2005AJ....130.1291B};
\cite{2007DPS....39.4910S}; \cite{2012AJ....143..146B}). In addition,
ethane (C$_2$H$_6$), believed to be the result of CH$_4$ photolysis
processes caused by the solar wind and cosmic rays, has been detected
in Makemake (\cite{2006ApJ...653..792B}). In other smaller KBOs,
spectroscopic observations showed the presence of water, ammonia
and likely methanol ices (\cite{2011A&ARv..19...48B};
\cite{2012AJ....143..146B}).

From Fig. \ref{fig:species-comets}, it is clear that the most abundant
species in comets (H$_2$O, CO, CO$_2$, CH$_4$, NH$_3$, CH$_3$OH and
C$_2$H$_6$) are observed also in KBOs and, in turn, the species
observed in KBOs are the most abundant species in comets.
Formaldehyde and hydrogen sulfide have abundances in comets comparable
to CH$_4$ and NH$_3$. Their non detection in KBOs may, however, be due
to observational effects only.

\subsection{Organics in meteorites and IDPs}\label{sec:organics}

Carbonaceous chondrites are rich in carbon, which constitutes about
1--4\% of this kind of meteorites. Organic carbon is present in two
forms, following the methods to extract the organic material:
insoluble organic matter (IOM) and soluble organic matter (SOM).

IOM is mainly ($\geq$70\%) constituted of organic compounds with a
relatively complex structure (nanoglobules, venatures...). The
compounds are made of small aromatic units (with up to six rings)
linked by branched aliphatic linkages shorter than seven carbon atoms
(e.g.\cite{2005LPI....36.1350R};
\cite{2012M&PS...47..345L}). Similarly, IDPs contain about 10-12 \%
carbon, mostly in organic material, including aromatic and aliphatic
compounds (\cite{1993GeCoA..57.1551T}; \cite{2004GeCoA..68.2577K}).

SOM is principally made of carboxylic acids, aliphatic and aromatic
hydrocarbons, and amino acids (e.g. \cite{2001Sci...293.2236P}). In
the Murchison meteorite, they represent $\sim 50$\%, $\sim 25$\% and
$\sim 10$\%, respectively, of the organic soluble matter. Of particular
interest, amino acids with no known terrestrial distribution have been
found in meteorites. In addition, a sub-group of amino acids shows a
small but significant L-enantiomeric excesses
(e.g.\cite{2003GeCoA..67.1589P}), namely one of the two chiral forms
is more abundant than the other, a characteristic of chiral
biomolecules in terrestrial life.

\subsection{The hydrogen and nitrogen isotopic anomalies}\label{sec:isotopic-anomalies}
One direct evidence of the link between pristine small Solar System bodies
like carbonaceous chondrite, IDPs and comets, and the
first phases of the Sun formation (phases 1 to 3 in Fig.
\ref{fig:starform-chem}) comes from the presence of the so-called
isotopic anomalies. Among the five most abundant elements in the
Universe (H, He, O, C and N), three present large anomalies, namely
they have isotopic values more than twice different than in the Solar
Nebula: hydrogen, oxygen and nitrogen (while carbon also shows
different values but to a lesser extent). Each of them brings
different information. Here, we briefly review the
information provided by the hydrogen and nitrogen isotopes. Oxygen
isotopic anomalies are discussed in \S \ref{sec:violent-start}.
\\

\begin{figure*}
  \includegraphics[width=\textwidth]{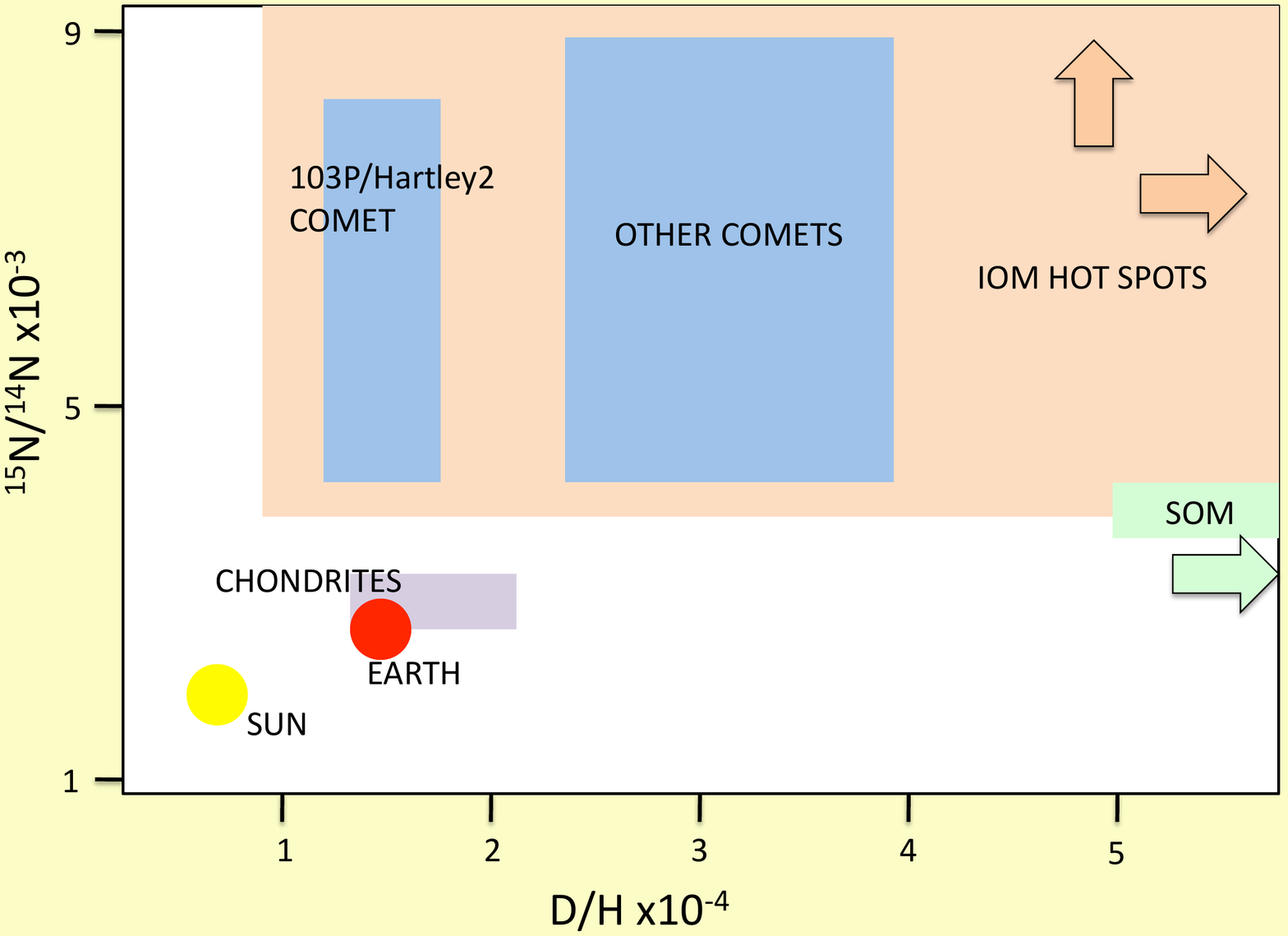}
  \caption{$^{15}$N/$^{14}$N versus D/H in comets, chondrites, hot
    spots in the IOM of meteorites and IDPs, SOM in meteorites, Earth
    and Sun.}
\label{fig:D-15N}
\end{figure*}

\noindent
{\it Deuterium in comets}: 
The first and most important isotopic anomaly, the deuterium
enrichment of terrestrial water, has been already discussed in 
\S \ref{sec:where-terr-water}. We also briefly
mentioned that comets show a D-enrichment one to two times the one of
the terrestrial oceans (Fig. \ref{fig:D-15N}). Regardless whether
comets substantially contributed or not to the terrestrial water, the
relatively high abundance of deuterated water can help us to understand when and
where comets formed and, consequently, how the Solar System formed. So far, the HDO/H$_2$O
ratio has been observed in seven comets from the Oort Cloud, the most
recent being the C/2009 P1 comet, and in one, 103P/Hartley2, from the
Jupiter family comets. In the first six comets, HDO/H$_2$O has been
measured to be $\sim 3\times 10^{-4}$ (\cite{2002Natur.416..403M};
\cite{1998Icar..133..147B}), in C/2009 P1 it is $2\times10^{-4}$
(\cite{2012A&A...544L..15B}), and in 103P/Hartley2 it is
$1.5\times10^{-4}$ (\cite{2011Natur.478..218H}). If, on the one hand,
this last measurement has brought back to life the debated late veneer
theory (\S \ref{sec:where-terr-water}), it has also challenged the
present view of where these comets are formed. In fact, according to
the widely-accepted theory, comets from the Oort Cloud and the Jupiter
family were likely formed in the Uranus-Neptune zone
(\cite{2004come.book..153D}), even though the Oort Cloud comets may
also originate from the Jupiter-Saturn region
(\cite{2008A&A...492..251B}). The HDO/H$_2$O ratio is an almost direct
measure of the temperature where the comet is formed and larger
heliocentric distances are expected to correspond to colder regions.
Therefore, one would expect that comets in the Oort Cloud present a
similar or lower HDO/H$_2$O ratio than the Jupiter family comets,
contrary to what is measured. Dedicated models support this simple
intuitive argument (e.g. \cite{2007EM&P..100...43H};
\cite{2011ApJ...734L..30K}; \cite{2012LPI....43.1937P}).  Therefore,
either comet formation theory is not correct in this aspect (for
example a new theory postulates that Oort Cloud comets are captured from
nearby stars; \cite{2010Sci...329..187L}), or the temperature in the
Solar Nebula was not monotonically decreasing with increasing
heliocentric distance.  This is in principle possible during the
accreting disk phase where viscosity may have created warm regions
(e.g. \cite{2012M&PS...47...99Y}).  This will be further discussed in
\S \ref{sec:assembl-some-piec}. So far for water, but deuterium
enrichment is also observed in HCN, in one comet
(\cite{1998Sci...279.1707M}), and it is about 10 times larger than the water
D-enrichment. This difference is not necessarily a problem as it may
just outline the different chemical formation pathway of these two
species, as explained in \S\ref{sec:origin_water}.
\\

\noindent
{\it Deuterium in carbonaceous chondrites and IDPs}: 
The bulk of carbonaceous chondrites contains hydrated silicates and
hydrous carbon with a D/H ratio = 1.2--2.2 $\times 10^{-4}$
(e.g. \cite{2003SSRv..106...87R}), very similar to that of the
terrestrial oceans.  However, D-enrichment, similar to that 
measured in comets and even higher, has also been found in the so
called ``hot spots'', namely micrometer-scale regions with positive
isotope anomalies, in the IOM of chondrites and IDPs. These hot spots
are in fact so named because of the enrichment of D and $^{15}$N, and
are systematically found in small regions of organic material. The D-enrichment 
in carbonaceous chondrites and IDPs is very variable, with
regions having D/H $\sim 8\times10^{-5}$ close to the Solar Nebula
value, and others having D/H up to $\sim 10^{-2}$
(\cite{2007GeCoA..71.4380A}; \cite{2009ApJ...698.2087R}). High
spatial resolution measurements suggest that the largest D-enrichment
is associated with organic radicals (\cite{2009ApJ...698.2087R}). Similarly,
molecules in the soluble organic matter component show enhanced
abundances of D-species with respect to H-species, at a level of D/H
up to almost $10^{-2}$ (\cite{2005GeCoA..69..599P}).
\\

\noindent
{\it $^{15}$N in comets}: 
Several measurements of the bulk of the $^{14}$N/$^{15}$N in the Solar
Nebula, from observations of NH$_3$ in Jupiter
(\cite{2001ApJ...553L..77O}; \cite{2004Icar..172...50F}) and the solar
wind (\cite{2010GeCoA..74..340M}) give a value of $\sim440$,
consistent with standard stellar nucleosynthesis models. In comets,
though, the $^{14}$N/$^{15}$N ratio measured in CN and HCN species is
more than a factor 2 lower, around 150 (\cite{2003Sci...301.1522A};
\cite{2009A&A...503..613M}). The origin of this $^{15}$N
enrichment in comets has puzzled astrochemists for years. One
possibility is that this is a direct heritage of the pre-stellar core
phase (\S \ref{sec:calm-before-storm}) or proto-planetary disk phase
(\S \ref{sec:turb-assembl-plan}). Finally, it is also possible that
$^{15}$N has been injected in the material forming the Solar System by
the explosion of a nearby supernova (\S \ref{sec:violent-start}).
\\

\noindent {\it $^{15}$N in chondrites and IDPs}: The $^{14}$N/$^{15}$N
as measured in TiN in a pristine condensate Ca-Al-rich inclusion
of a carbonaceous chondrites is very similar to the Solar Nebula
value ($\sim440$, \cite{2007ApJ...656L..33M}).  However,
the $^{14}$N/$^{15}$N in the IOM material of carbonaceous chondrites
and IDPs is low, up to $\sim 50$ (\cite{2009M&PSA..72.5178B};
\cite{2010GeCoA..74.6590B}; \cite{2012M&PS...47..525M}), as low as in
comets and significantly lower than in the Solar Nebula and the interstellar
medium.  Similar $^{15}$N enrichment has been reported in two amino
acids (\cite{2009GeCoA..73.2150P}). Therefore, the same question on
the origin of the $^{15}$N enrichment in comets applies to the organic
material in carbonaceous chondrites and IDPs.
\\

\noindent {\it A common origin for the D- and $^{15}$N-enrichment in
  comets and the organic material in carbonaceous chondrites and
  IDPs?}  Since comets and the organic material in chondrites and IDPs
are enriched in both D and $^{15}$N, the question whether the
enrichment has a common origin is a natural one
(e.g. \cite{2010ApJ...722.1342A}). Against this hypothesis is that
D-enriched spots in chondrites and IDPs do not coincide spatially with
$^{15}$N-enriched ones (\cite{2010ApJ...722.1342A};
\cite{2006M&PSA..41.5259R}).  Similarly, while the
D-enrichment differs by a factor two in 103P/Hartley2 and the other
six comets, the $^{15}$N-enrichment is practically the same in all
comets (Fig. \ref{fig:D-15N}). Therefore, very likely D- and $^{15}$N-enrichments do not have a common origin (see also \cite{2012ApJ...757L..11W}).

\subsection{A violent start in a crowded violent
  environment}\label{sec:violent-start}

Short-lived nuclides\footnote{Short-lived nuclides are the
  radionuclides with half-lives shorter than about 10 Myr.} present at
the formation of the Solar System and now disappeared, and isotopic
oxygen anomalies in meteorites tell us
that the Solar System had a violent start in a violent
environment. First, the young Sun irradiated the forming planetary
system with a strong wind of energetic particles.  Second, the Sun was
likely born in a large cluster of stars where one or more massive
stars exploded. All this is based on anomalies with respect to
the ``normal'' values of the abundances of these elements, which can
only be firmly known by assessing what is the normality in other
forming stars and, therefore, it is an important piece of the puzzle
to mention here. 
%In addition, we will discuss the evidence that we
%have that other Sun-like protostars have a similar violent start.
\\

\noindent
{\it A violent start:}
It is now well-known that young solar-type stars are bright X-rays
emitters, about $10^3$ times brighter than the present day Sun
(\cite{1999ARA&A..37..363F}; \cite{2005ApJS..160..390P}). It is very
likely that, together with X-rays, H and He nuclei with energies
larger than 10 MeV are also emitted in large quantities in the early
stages of star formation (e.g. \cite{1998ApJ...506..898L}). The Sun
likely passed through a similar violent phase and irradiated the
forming planetary system with energetic particles (sometimes also
referred as ``early solar cosmic rays'').  Extinct short-lived
nuclides bring traces of this violent past. Specifically, the enhanced
abundances of $^{10}$Be, $^7$Li and $^{21}$Ne
(\cite{2000Sci...289.1334M}; \cite{2012M&PSA..75.5192C}) can only be
explained by spallation reactions of solar energetic particles with O
and C atoms of the Solar Nebula. Similarly, other short-lived
nuclides, $^{36}$Cl, $^{53}$Mg and $^{41}$Ca, are now explained in
terms of irradiation from the early Sun
(e.g. \cite{2002Sci...298.2182M}; \cite{2008ApJ...680..781G}).
\\

\noindent
{\it A crowded violent environment:} 
Several lines of evidence converge toward a picture where the Sun was
born in a cluster of at least 1000 stars (see the review by
\cite{2010ARA&A..48...47A}). Likely, within this cluster, some were
massive stars and some exploded a little before or during the formation
of the Solar System. Since its discovery in meteorites, $^{26}$Al
(\cite{2000GeCoA..64.3913K}; \cite{2009Sci...325..985V}) became one of
the proofs, indeed highly debated for decades, that the Solar System
was polluted with material ejected from a nearby type II supernova,
whose progenitor mass is $\sim$25 M$_\odot$
(\cite{1977Icar...30..447C}; \cite{2008ApJ...680..781G}). Support to
this hypothesis was added by the discovery of $^{60}$Fe
(\cite{1998AMR....11..103K}), but the value of the $^{60}$Fe excess
with respect to the Galactic one has been revised since and nowadays
it is believed to be close to zero (\cite{2011ApJ...741...71M}). As a
consequence, theories based on $^{60}$Fe have to be taken with a grain
of salt (see the review by \cite{2011AREPS..39..351D}). Recently, the
anomalous $^{18}$O/$^{17}$O in meteorites, 5.2$\pm$0.2 (see the
compilation in \cite{2011ApJ...729...43Y}), with respect to the
Galactic one, 4.1$\pm$0.1 (\cite{2008A&A...487..237W}) has also been
taken as a proof of the injection of material from a type II supernova
exploded just before the birth of the Solar System.

%%%%%%%%%%%%%%%%%%% Paola %%%%%%%%%%%%%%%%%%%%%%%%%%%%%%%%%%%%%%%%%%%%%
\section{The calm before the storm: pre-stellar cores}   % Paola, about 10 pages, 3 (max 4) subsections
\label{sec:calm-before-storm}

Stars like our Sun form in slowly rotating and collapsing magnetized dense cloud cores (e.g.  \cite{1993ApJ...406..528G}; \cite{2008ApJ...680..457T}). Dense cores not associated with stars are called ''starless cores" and they represent the initial conditions in the process of star formation (\cite{1987ARA&A..25...23S}). They are the starting point of our journey. These objects have average volume densities at least one order of magnitude larger than the surrounding medium, have typical kinetic and dust temperatures of 10\,K and their internal energy is dominated by thermal motions (see review by \cite{2007ARA&A..45..339B}).  Not all starless cores give birth to stars, though. Some of them reach configurations close to hydrostatic equilibrium and display kinematic features consistent with oscillations (\cite{2003ApJ...586..286L}). Others show expanding motions (\cite{2004A&A...416..191T}). This class of starless cores typically displays a relatively flat density distribution, with central densities below 10$^5$ H$_2$ molecules cm$^{-3}$. This is the critical density for gas cooling by gas-dust collisions (\cite{2001ApJ...557..736G}) and it represents the "dividing line" for dynamical stability. Starless cores with central densities below this critical density are thermally subcritical (\cite{2008ApJ...683..238K}) and they may disperse back into the interstellar medium. When the central densities of H$_2$ molecules overcome $\simeq$10$^5$\,cm$^{-3}$, starless cores become thermally supercritical and gravitational forces take over. These are the so-called {\em pre-stellar cores}, first identified by \cite{1994MNRAS.268..276W} in the sub-millimeter continuum and then chemically and kinematically labelled by \cite{2005ApJ...619..379C} using millimeter spectroscopy. It is within pre-stellar cores that future star and planetary systems will form. 

\subsection{Freeze-out,  deuterium fractionation and the ionization fraction}
\label{freeze}

\begin{figure}[ht]
% Use the relevant command to insert your figure file.
% For example, with the graphicx package use
  \includegraphics[angle=270,width=1\textwidth]{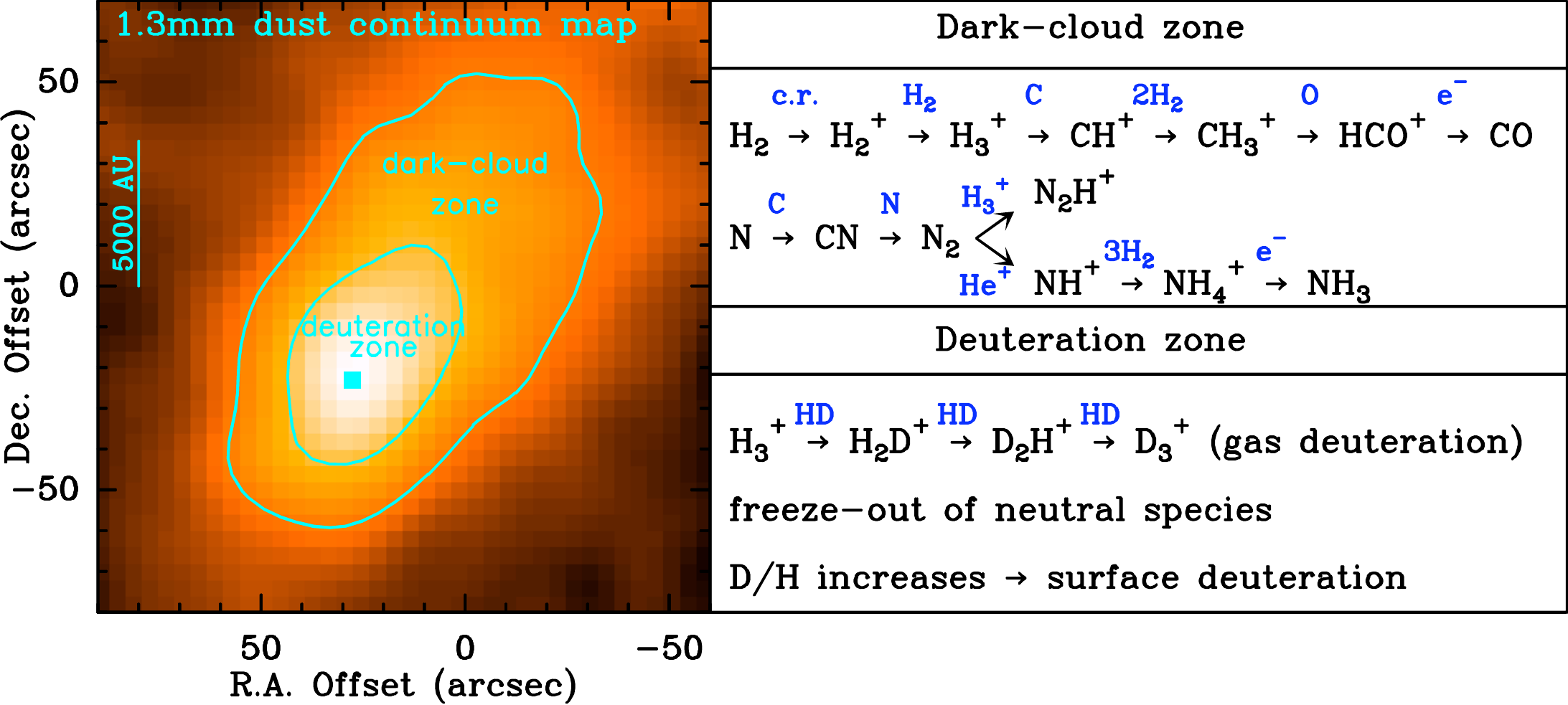}
% figure caption is below the figure
\caption{The chemical zones of the prototypical pre-stellar core L1544, embedded in the Taurus Molecular Cloud Complex, at a distance of 140\,pc. The background color image is the 1.3mm dust continuum emission map obtained with the IRAM-30m antenna (\cite{1999MNRAS.305..143W}). The cyan contours show the different chemical zones, with the corresponding main chemical processes listed in the right panel.  Blue labels indicate reaction partners.}
\label{fig:1}       % Give a unique label
\end{figure}

Pre-stellar cores span a range of number densities which goes from a few times 10$^3$\,cm$^{-3}$ toward the outer edges, where they merge with the surrounding molecular cloud, to about 10$^7$\,cm$^{-3}$ within the central 1,000\,AU (e.g. \cite{2001ApJ...557..193E}; \cite{2010MNRAS.402.1625K}), where the gas and dust temperature drops to 6-7\,K (\cite{2007A&A...470..221C}; \cite{2007A&A...467..179P}). These gradients in physical properties affect the chemical structure. Fig. \ref{fig:1} schematically shows the main chemical processes in the two-zones of the prototypical pre-stellar core L1544, embedded in the Taurus molecular cloud.  In the outer part of the core (between about 7,000 and 15,000\,AU), the gas density is $\simeq$10$^4$\,cm$^{-3}$ and the temperature $\simeq$10\,K.  "Classical" dark-cloud chemistry is at work, with ion-molecule reactions (\cite{1973ApJ...185..505H}) dominating the carbon chemistry, and neutral-neutral reactions which start the transformation of nitrogen atoms into N$_2$ (e.g. \cite{2010A&A...513A..41H}). These reactions form the "popular" species CO, N$_2$H$^+$ and NH$_3$, which are widely used to study cloud structures and kinematics. \\

{\it Freeze-out.} Within the central 7,000\,AU, the density increases above 10$^5$\,cm$^{-3}$, the temperature drops below 10\,K and species heavier than He tend to disappear from the gas-phase due to the process of freeze-out (the adsorption of species onto dust grain surfaces). CO freeze-out has been measured in starless and pre-stellar cores at a 80-90\% level (\cite{1998ApJ...507L.171W}; \cite{1999ApJ...523L.165C}; \cite{2002A&A...389L...6B}; \cite{2002MNRAS.337L..17R}). Nitrogen-bearing species have also been found to deplete from the gas-phase, although not as much as CO (e.g. \cite{2002ApJ...570L.101B}; \cite{2006A&A...455..577T}; \cite{2010ApJ...708.1002F}). The reason for this differential freeze-out has to be found in the fact that N-bearing species, such as N$_2$H$^+$ and NH$_3$, experience larger production rates when neutral species (in particular CO) start to disappear from the gas-phase. The freeze-out is a natural consequence of the quiescent nature of pre-stellar cores: once species land on a grain surfaces, they cannot thermally evaporate (because dust temperatures are $T_{\rm dust} \leq 10$\,K, typical binding energies are $E_{\rm B} \ge$  1000\,K and the thermal evaporation rate is $\propto exp[-E_{\rm B}/(k\,T_{\rm dust})]$) and they cannot photodesorb as interstellar photons cannot penetrate within pre-stellar cores (whose central regions have visual extinctions larger than 50\,mag).  Only a small fraction of the adsorbed species can return in the gas-phase via non-thermal desorption mechanisms mainly driven by cosmic-rays, such as dust impulsive heating due to cosmic-ray bombardment (e.g. \cite{1985A&A...144..147L}) and photodesorption due to the Far-UV (FUV) field produced by cosmic-ray impacts with H$_2$ molecules (\cite{1983ApJ...267..603P}; \cite{1989ApJ...347..289G}; \cite{2004A&A...415..203S}), although molecular hydrogen formation (\cite{1998MNRAS.298..562W}; \cite{2007MNRAS.382..733R}) and surface reactions involving radicals (\cite{1982A&A...109L..12D}) may also play a role. Desorption of mantle species by FUV photons has been included in the chemical-dynamical models of L1544, to explain the recent Herschel detection of water vapor in the center of this prototypical pre-stellar core (\cite{2012arXiv1208.5998C}). Freeze-out time scales ($t_{\rm freeze-out}$ $\propto10^9/n_{\rm H}$\,yr, where $n_{\rm H}$ is the total number density of hydrogen nuclei, \cite{1985MNRAS.217..413J}) are significantly shorter than the dynamical (free-fall) time scale ($t_{\rm free-fall} \propto 4\times 10^7 / \sqrt{n_{\rm H}}$, \cite{1978ppim.book.....S}), so dust grains are expected to build thick icy mantles during the pre-stellar phase of the star formation process (\S\ref{sec:ice-formation}).  \\

{\it Deuterium fractionation.} In the cold environments of pre-stellar
cores, another important process takes place: deuterium
fractionation. The starting point is the exothermic reaction between
H$_3^+$ and HD, which produces H$_2$D$^+$ and H$_2$ (H$_3^+$ + HD
$\rightarrow$ H$_2$D$^+$ + H$_2$ + 230\,K,
\cite{1974ApJ...188...35W}). This reaction cannot proceed from right
to left when the kinetic temperature is below $\simeq$20\,K and if a
large fraction of H$_2$ molecules is in para form, as expected in cold
and dense cores (\cite{2006A&A...449..621F};
\cite{2009A&A...494..623P}; \cite{2009A&A...506.1243T}). Therefore,
the H$_2$D$^+$/H$_3^+$ abundance ratio becomes significantly larger
than the D elemental abundance with respect to H.  When the freeze-out
of neutral species (especially CO and O, which are the main
destruction partners of H$_2$D$^+$) becomes important, deuterium
fractionation is further enhanced (\cite{1984ApJ...287L..47D}). In
fact, the deuteration zone of Fig. \ref{fig:1} is the region where the
brightest line of ortho-H$_2$D$^+$ has ever been detected
(\cite{2003A&A...403L..37C}).  This deuteration "jump" allows multiply
deuterated forms of H$_3^+$ to thrive (\cite{2004ApJ...606L.127V};
\cite{2011A&A...526A..31P}) and their dissociative recombinations with
electrons liberate D atoms, locally increasing the D/H ratio to values
larger than 0.1 (\cite{2003ApJ...591L..41R}).  The large D/H ratio in
the gas-phase implies efficient deuteration of surface species (in
particular CO), with the consequent production of deuterated and
doubly deuterated formaldehyde as well as singly, triply and doubly
deuterated methanol (e.g. \cite{1983A&A...119..177T};
\cite{1997ApJ...482L.203C}; \cite{2002P&SS...50.1257C};
\cite{2012ApJ...748L...3T}; \cite{2012A&A...submitted}). HDCO, D$_2$CO
and CH$_2$DOH have been detected in pre-stellar cores
(\cite{2003ApJ...585L..55B}; \cite{2011A&A...527A..39B}), while doubly
and triply deuterated methanol have been detected in the envelope of
young stellar objects (\cite{2002A&A...393L..49P};
\cite{2004A&A...416..159P}, see \S\ref{sec:violent-origin-sun}).
\\

{\it The ionization fraction.} Deuterated species are the main probes of the central regions of pre-stellar cores, the future stellar cradles. Their observation allows us to trace the kinematics (e.g. \cite{2005A&A...439..195V}; \cite{2007A&A...470..221C}) and, together with the non-deuterated isotopologue, to measure the elusive electron number density $n({\rm e}^-)$, which plays a crucial role in the dynamical evolution of the cloud.  In fact, electrons and ions gyrate around magnetic field lines which permeate the clouds, and decouple from the bulk motions. During the gravitational collapse, neutral species slip through magnetic field lines and collide with molecular ions in a process called ambipolar diffusion (\cite{1979ApJ...228..475M}; \cite{1987ARA&A..25...23S}). Depending on the fraction of ions present in the gas-phase, neutral-ion collisions can significantly slow down the collapse compared to free-fall. How do we measure the ionization degree? Using simple steady-state chemistry of (easy-to-observe) molecular ions,  such as HCO$^+$ and DCO$^+$, which form from the reaction of CO with H$_3^+$ and H$_2$D$^+$ and are destroyed by electrons, it is easy to arrive at analytic expressions relating the observed DCO$^+$/HCO$^+$ abundance ratio to  $n({\rm e}^-)$ (\cite{1977ApJ...217L.165G}; \cite{1979ApJ...234..876W}). Using time dependent chemical codes, \cite{1998ApJ...499..234C} and \cite{1999ApJ...512..724B} obtained values of $x({\rm e}^-)$ ($\equiv n({\rm e}^-)/n({\rm H_2})$) between 10$^{-8}$ and 10$^{-6}$. Given that the time scale for ambipolar diffusion is $t_{\rm AD}$ $\simeq$ $2.5 \times 10^{13} x({\rm e}^-)$\,yr (\cite{1978ppim.book.....S}), the above measurements imply values of $t_{\rm AD}$ $\simeq$ 2.5$\times10^5$ and 2.5$\times 10^{7}$\,yr, factors of 2-200 larger than $t_{\rm free-fall}$ for pre-stellar cores with an average $n_{\rm H}$ = 10$^5$\,cm$^{-3}$.  \\

{\it $^{15}$N-fractionation.}  On the one hand, no significant $^{15}$N  fractionation (compared to the Solar Nebula value of $\sim$440, see \S\ref{sec:isotopic-anomalies}) has been found in NH$_3$ ($^{14}$N/$^{15}$N $\simeq$350-850,  \cite{2009A&A...498L...9G}; 334$\pm$50, \cite{2010ApJ...710L..49L}) toward pre-stellar cores and protostellar envelopes, and in N$_2$H$^+$ ($^{14}$N/$^{15}$N = 446$\pm$71, \cite{2010A&A...510L...5B}) toward the prototypical pre-stellar core L1544. On the other hand, \cite{2012LPI....43.2618M} and Hily-Blant et al. (submitted) found significant $^{15}$N-enrichment  in HCN toward pre-stellar cores (between 70 and 380). Similar values have been found by \cite{2012ApJ...744..194A} in HNC observations of star-forming regions across the Galaxy. It is interesting to point out here that the $^{15}$N-fractionation observed in comets (\S\ref{sec:isotopic-anomalies}) has been measured for CN and HCN ($^{14}$N/$^{15}$N $\sim$130-170, \cite{2008ApJ...679L..49B}). This differential $^{15}$N fractionation for amines and nitriles has been recently reproduced in chemical models of dense clouds by \cite{2012ApJ...757L..11W}, who suggest that the processes able to reproduce the observed differentiation could be at the origin of the poor correlation between D- and $^{15}$N-fractionation observed in some primitive material in our Solar System (\S\ref{sec:isotopic-anomalies}). Thus,  a further link between pre-stellar core chemistry and the Solar System composition has been found (see \S\ref{sec:assembl-some-piec}).

\subsection{Ice formation and evolution}
\label{sec:ice-formation}

\begin{figure}[ht]
% Use the relevant command to insert your figure file.
% For example, with the graphicx package use
  \includegraphics[angle=0,width=1\textwidth]{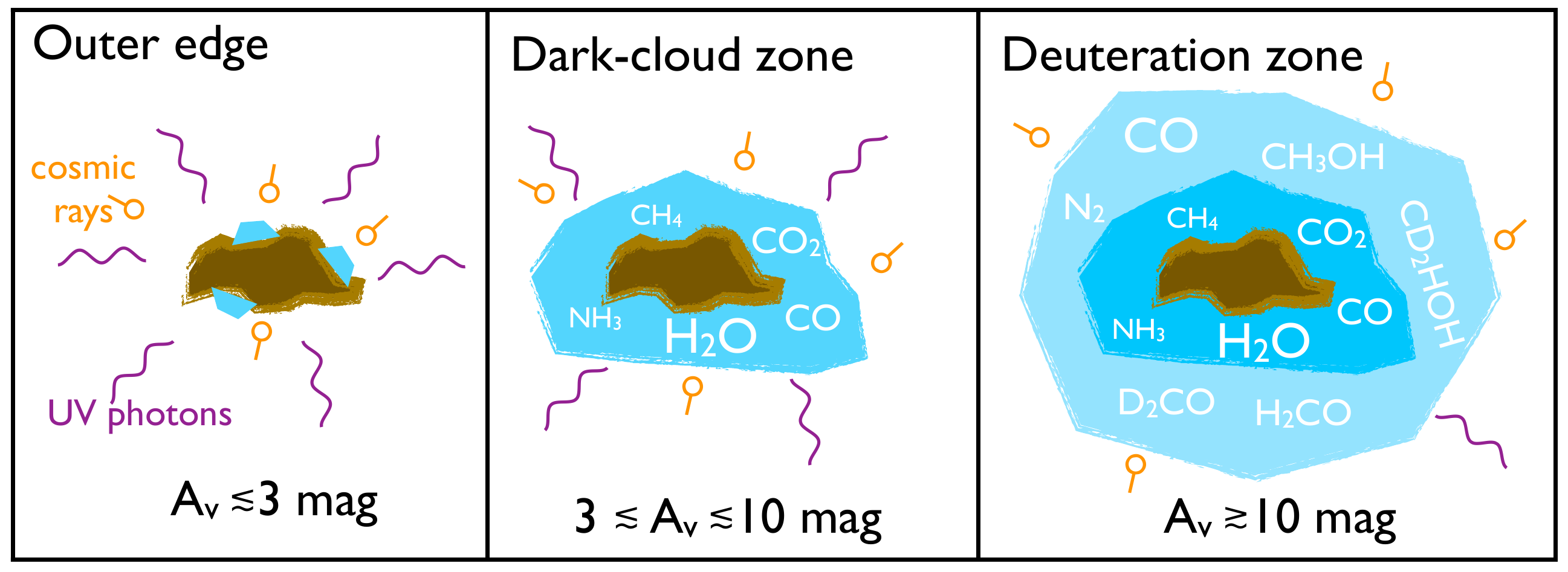}
% figure caption is below the figure
\caption{Ice mantle evolution within a pre-stellar core, from the outer-edge, where the core merges with the surrounding molecular cloud, to the dark-cloud zone and deuteration zone as depicted in Fig.\,\ref{fig:1}.  Ice mantles become thicker and richer in complex organic molecules moving toward the center of a pre-stellar core, where star- and planet-formation takes place.}
\label{fig:2}       % Give a unique label
\end{figure}

Interstellar dust grains are crucial for the chemical and physical evolution of interstellar clouds and for our astrochemical origins. First of all, hydrogen atoms can quickly scan their surfaces, meet and form volatile  H$_2$ molecules at rates large enough to defeat H$_2$ photodissociation due to the interstellar radiation field (\cite{1971ApJ...163..155H}; \cite{1999A&A...344..681P}; \cite{2002ApJ...575L..29C}; \cite{2005MNRAS.361..565C}). Thus, dust grains are responsible for the transition of interstellar gas in our Galaxy (as well as in external galaxies) from atomic to molecular - the first step toward chemical complexity. Secondly, they are efficient absorbers of the FUV photons, so that they act as "UV-filters", protecting molecules within clouds from the UV destructive action. Thirdly, they catalyze the formation of important species, in particular H$_2$O, with such high efficiency that more than 30\% of oxygen atoms are locked into water ice as soon as the visual extinction reaches values $\geq$3\,mag (e.g. \cite{2000ApJS..128..603M}; \cite{2009ApJ...690.1497H}; \cite{2010ApJ...710.1009W}; \cite{2011ApJ...731....9C}). Finally, they become the main gas coolants in the central regions of pre-stellar cores, where the densities are above $\simeq$10$^5$\,cm$^{-3}$, the temperatures fall below 10\,K and species heavier than He (including important coolants such as CO) are mostly frozen onto their surfaces. In such conditions, the freeze-out rate will become even more extreme and dust grains should develop thick ice mantles. How thick? A simple estimate can be made considering that levels of CO freeze-out of about 90\% are seen within the central pre-stellar core regions (see \S\ref{freeze}). Assuming that all species heavier than Helium are affected by a similar amount of freeze-out (including nitrogen; \cite{2010A&A...513A..41H}), then in clouds with total hydrogen density of 2$\times$10$^{6}$\,cm$^{-3}$, the total number density of heavy species frozen onto dust grains is about 1.3$\times$10$^3$\,cm$^{-3}$. Further assuming that they are combined in molecules with two heavy elements on average (e.g. CO, CH$_3$OH, CO$_2$, H$_2$O), the total number of solid species will be about 660\,cm$^{-3}$. Now, we just need to divide this number by the total number of sites on an average grain with radius 0.1\,$\mu$m ($\simeq$10$^6$; \cite{1992ApJS...82..167H}) to have the number of monolayers ($\simeq$250). Considering a monolayer thickness of about 1\,\AA, the total mantle thickness is then 2.5$\times$10$^{-6}$\,cm, or about a quarter of the grain radius. Such thick mantles boost dust coagulation (\cite{1994A&A...291..943O}). 

What are the main chemical processes on the surface of dust grains?
Our understanding is based on (i) observations of absorption features
along the line of sight of stars located behind molecular clouds or
protostars embedded in dense cores (e.g. \cite{2011ApJ...742...28W}
and references therein), and on (ii) laboratory work
(e.g. \cite{2002ApJ...571L.173W}; \cite{2002ApJ...577..265H};
\cite{2008CPL...456...27M}; \cite{2008ApJ...686.1474I};
\cite{2009A&A...505..629F}; \cite{2010A&A...512A..30D}).  From these
studies, we now know that surface reactions are mainly association
reaction: oxygen is transformed into water via successive association
reactions with hydrogen (e.g. O+H $\rightarrow$ OH; OH+H $\rightarrow$
H$_2$O, but see \S\ref{sec:origin_water} for more pathways to water
ice); similarly, CO is transformed first into formaldehyde, H$_2$CO,
and then into methanol, CH$_3$OH, via two and four association
reactions, respectively; atomic nitrogen saturates into ammonia,
NH$_3$. Other important processes are photoprocesses and cosmic-ray
bombardments. Photoprocesses are experimentally found to promote the
formation of organic species more complex than CH$_3$OH
(\cite{1996A&A...312..289G}; \cite{2007ApJ...661..899B};
\cite{2009A&A...504..891O}; \cite{2010ApJ...718..832O}) up to amino
acids (e.g. \cite{2002Natur.416..401B}; \cite{2002Natur.416..403M};
\cite{2004A&A...413..209M}) and allow solid species to return into the
gas-phase (\cite{2009A&A...496..281O};
\cite{2009ApJ...693.1209O}). Cosmic-rays, unlike UV photons, traverse
dense cores relatively unhampered, although their flux may be reduced by a factor
of a few by the mirroring effect of magnetic fields
(\cite{2011A&A...530A.109P}). When colliding with dust grains, they
can alter mantle compositions (e.g. \cite{2000ApJ...542..890P};
\cite{2009A&A...493.1017I}; \cite{2010A&A...519A..22M};
\cite{2012A&A...543A.155S}; \cite{2012A&A...544A..30B};
\cite{2012MNRAS.423.2209P}). Cosmic rays also play a crucial role in
molecular desorption, as mentioned in the previous section.  Surface
chemistry is one of the most challenging disciplines in
astrochemistry, but in the recent years several models have been
successful in reproducing the observed abundance of some simple and complex species
(e.g. \cite{2008ApJ...674..984A}; \cite{2009ApJ...700L..43G};
\cite{2009ApJ...690.1497H}; \cite{2009A&A...508..275C};
\cite{2011ApJ...741L..34C}; \cite{2012A&A...538A..42T};
\cite{2012ApJ...748L...3T}).

The picture that has emerged from the combination of observations, laboratory work and modeling is sketched in Fig.\ref{fig:2}, which shows the evolution of a dust grain mantle from the outer edge to the central regions of a pre-stellar core embedded in a molecular cloud bathed by the interstellar radiation field (with reference to Fig.\ref{fig:1} to locate the various zones). At the outer edge of the pre-stellar core,  photoprocesses are important and the ice mantles are just beginning to form. Here, oxygen atoms are transformed into water, carbon (still not locked in CO) into methane (CH$_4$) and nitrogen into ammonia. Water dominates the mantle composition (probably reflecting the larger cosmic abundance of oxygen relative to C and N).  Moving toward the dark-cloud zone (where the pre-stellar core merges with the molecular cloud within which it is embedded), UV photons are absorbed by dust grains, CO becomes the second most abundance molecule (after H$_2$) and the mantle starts to accumulate CO. CO$_2$ also starts to form, either via cosmic-ray bombardment (\cite{2009A&A...493.1017I}) and/or via the CO+OH reaction (\cite{2010ApJ...712L.174O}; \cite{2011MNRAS.413.2281I}; \cite{2011ApJ...735..121N}; \cite{2011ApJ...735...15G}). Here, the limited amount of CO freeze-out limits the degree of deuteration to levels of $\leq$ a few \% (as measured from the observed DCO$^+$/HCO$^+$ abundance ratio; e.g. \cite{2002ApJ...565..344C}). Deeper into the pre-stellar core, CO molecules are mostly in solid form, deuteration processes are dominant and the D/H ratio reaches values above 0.1 (see \S\ref{freeze}). When freeze-out is dominant, the main reactive species landing on dust grain surfaces are atomic H and D. Thus, CO is not only hydrogenated into formaldehyde and methanol, but also deuterated. Large amounts of deuterated and multiply deuterated H$_2$CO and CH$_3$OH are produced (see \S\ref{freeze}).  

\subsubsection{The origin of water}
\label{sec:origin_water}

Extra attention is given here to the production of water, because of its dominant presence in interstellar ices and its crucial role in our astrochemical origins. Recent measurements of water vapor toward a pre-stellar core with the Herschel Space Observatory and the use of chemical/dynamical/radiative transfer models, allowed \cite{2012arXiv1208.5998C} to measure a total mass of water vapor of 0.5\,Earth masses within the central 10,000\,AU and predicted about 2.6\,Jupiter masses of water ice (thus, plenty of ice to boost dust coagulation and the formation of giant planets via core accretion models, e.g. \cite{1996Icar..124...62P}). From observations of water ices in molecular clouds (e.g. \cite{2011ApJ...742...28W} and references therein), it is now well established that water ice forms on the surface of dust grains in regions of molecular clouds where the visual extinction is at least 3\,mag (when the impinging radiation field is close to the average Galactic value, called the Habing field; larger extinctions are needed for stronger fields). For lower extinction values, the interstellar UV field does not allow dust grain surfaces to accumulate a significant amount of water molecules, as they are efficiently photodesorbed (\cite{2009ApJ...693.1209O}). Laboratory work shows that H$_2$O can form via hydrogenation of atomic oxygen (\cite{1998ApJ...498..710H}; \cite{2010A&A...512A..30D}; \cite{2011ApJ...741L...9J}), molecular oxygen (\cite{2008ApJ...686.1474I}; \cite{2008CPL...456...27M}), ozone (\cite{2009ApJ...705L.195M}; \cite{2011JChPh.134h4504R}) and via OH + H$_2$ at 10\,K (\cite{2012ApJ...749...67O}).  As the abundance of water ice in molecular clouds, within which pre-stellar cores form, is already large (close to 10$^{-4}$ w.r.t. H$_2$ molecules; e.g. \cite{1991A&ARv...2..167W}), we now generally believe that the main production of water happens {\em before} the formation of a pre-stellar core, as also found by chemical models (e.g. \cite{2008ApJ...674..984A}; \cite{2009ApJ...690.1497H}; \cite{2010A&A...522A..74C}; \cite{2012A&A...538A..42T}).  This suggests that also the production of heavy water must be regulated by the molecular cloud characteristics. This is an important point, as the HDO/H$_2$O ratio is well measured on Earth, comets and asteroids (\S\ref{sec:where-terr-water}), as well as in star-forming regions (\S\ref{sec:water-and-deut}).
%($\simeq$0.015\%, the so-called standard mean ocean water, SMOW), in Oort cloud comets and asteroids ($\simeq$0.02-0.05\%, Altwegg \& Bockel{\'e}e-Morvan 2003; Mumma \& Charnley 2011 and references therein), in the Jupiter-family comet 103P/Hartley 2 (0.016$\pm$0.002\%; Hartogh et al. 2011), in the hot cores surrounding newly born high-mass stars ($\simeq$0.01\%; Gensheimer et al. 1996) and in the hot corinos around low-mass young stellar objects (between $\simeq$0.06 and 6\%,  Jorgensen \& van Dishoeck 2010; Liu et al. 2011; Coutens et al. 2012). 
Therefore, one could use our current understanding of surface chemistry and the observed HDO/H$_2$O abundance ratios in star-forming regions to find the link between interstellar chemistry and the Solar System. 

Cazaux et al. (\cite{2011ApJ...741L..34C}) predict that significant
variations in the HDO/H$_2$O ratio can be attributed to small
variations of the dust temperature at the time of ice formation.  In
particular, if the dust temperature is lower than $\simeq$15\,K,
the HDO/H$_2$O ratio is predicted to be $\leq$0.01\%, because, in
these conditions, a large fraction of the dust surface is covered by
H$_2$ molecules, allowing the reaction of H$_2$ + O to proceed despite
the large barrier of 3000\,K 
%%% CC: Oba et al did observe H2O (and HDO, D2O) formation from OH+H2,
%%%and measured relative rates of their deuteration reactions... I
%%%therefore commented this following sentence
 (\cite{2012ApJ...749...67O} did not find
evidence in the laboratory that this reaction is indeed proceeding,
but more laboratory work is ongoing to assess this). 
The HDO/H$_2$O ratio in these conditions simply reflects the HD/H$_2$
ratio, always close to the interstellar D/H value
($\simeq$1.5$\times$10$^{-5}$, \cite{2003ApJ...587..235O}).  For dust
temperatures above $\simeq$15\,K, H$_2$ molecules do not stay on the
dust surface for long (as their evaporation rate becomes an
increasingly large fraction of their accretion rate) and water
formation will mostly happen via the reaction of oxygen with atomic
hydrogen.  As the gas-phase D/H ratio sharply increases above the
cosmic deuterium abundance when ice formation takes place (see Figure
1 of \cite{2011ApJ...741L..34C}), then the HDO/H$_2$O ratio can be as
large as a few \%. In this scenario, our Solar System formed in a
pre-stellar core embedded in a molecular cloud with dust temperature
slightly above 15\,K. Taquet et
al. (\cite{2012A&A...538A..42T}), using a multilayered formation
mechanism of ice mantles (\cite{2012A&A...538A..42T}), find that water
is formed first on dust surfaces and that the HDO/H$_2$O ratio depends
on the (poorly constrained) ortho:para ratio of H$_2$, on the cloud
volume density and, to a lesser extent, on the dust temperature and
visual extinction.  However, water deuteration can also occur in the
gas-phase: Thi et al. (\cite{2010MNRAS.407..232T}) found that
significant deuteration levels ([HDO]/[H$_2$O] $\simeq$
10$^{-3}$-10$^{-2}$) can be produced without surface reactions and at
high temperature ($T > 100$\,K), in the inner regions of
protoplanetary disks (\S\ref{sec:naked_disks}). The fractionation
occurs because of the difference in activation energy between
deuteration enrichment and the back reactions.

\subsection{Complex organic molecules}

%\begin{figure}[ht]
% Use the relevant command to insert your figure file.
% For example, with the graphicx package use
%  \includegraphics[angle=0,width=1\textwidth]{aar_f3_new.pdf}
% figure caption is below the figure
%\caption{Pathways to the formation of complex molecules on dust grain surfaces: (1) formation (or storage) of radicals in the dark-cloud zone, and (2) formation of O-bearing complex molecules in the deuteration zone of a pre-stellar core (see Fig.\,\ref{fig:1}).  [DA MODIFICARE / TOGLIERE? ] } 
%\label{fig:3}       % Give a unique label
%\end{figure}

In the freezing cold of dark clouds and pre-stellar cores, active
gas-phase and surface chemistry produce complex organic molecules (COMs).
Since the '80s, organic molecules have been discovered in the TMC-1
dark cloud, part of the Taurus Molecular Cloud complex: methyl cyanide
(CH$_3$CN; \cite{1983ApJ...272..149M}), methylcyanoacetylene
(CH$_3$C$_3$N, \cite{1984ApJ...276L..25B}), acetaldehyde (CH$_3$CHO;
\cite{1985ApJ...290..609M}), ketene (CH$_2$CO;
\cite{1989ApJ...342..871I}), methanol (CH$_3$OH;
\cite{1988A&A...195..281F}), methylcyanodiacetylene (CH$_3$C$_5$N,
\cite{2006ApJ...647..412S}), methyltriacetylene (CH$_3$C$_6$H,
\cite{2006ApJ...643L..37R}), propylene (CH$_2$CHCH$_3$,
\cite{2007ApJ...665L.127M}), methyldiacetylene (CH$_3$C$_4$H),
cyanopolyynes (HC$_{\rm 2n+1}$N, $n$=0,1,...,5) and C$_{\rm 2n+1}$N
radicals (\cite{1984A&A...134L..11W}; \cite{1992ApJ...394..539H};
\cite{1998FaDi..109..205O}; \cite{2004PASJ...56...69K}) and the
negative ions C$_6$H$^-$, C$_8$H$^-$ (\cite{2006ApJ...652L.141M};
\cite{2007ApJ...664L..43B}).  Complex organics have also been found in
two pre-stellar cores: L183 (CH$_3$CHO; \cite{1985ApJ...290..609M};
HCOOH, \cite{2007ApJ...655L..37R}) and L1689B (CH$_3$CHO, HCOOCH$_3$,
CH$_3$OCHO, CH$_2$CO, \cite{2012A&A...541L..12B}). The chemistry of
C-bearing species such as cyanopolyynes and CH$_3$C$_5$N can be
understood if the gas-phase is carbon-rich (C/O $\simeq$ 1.2;
\cite{2006A&A...451..551W}) or if polycyclic aromatic hydrocarbons
(PAHs) are included in the chemistry (with a standard C/O abundance
ratio of $\simeq$0.4, \cite{2008ApJ...680..371W}). More problematic is
the explanation of complex O-bearing species, such as methanol, which
require surface chemistry. Garrod et al. (\cite{2007A&A...467.1103G})
assumed that the energy released during the formation process could be
at least partially used for the surface species to desorb upon
formation, reconciling observations with theory for CH$_3$OH and
propylene (if the desorption of this species is
efficient). Oxygen-bearing species more complex than methanol can also
be formed on the surface of low temperature dust grains if a source of
UV photons is present (\S\ref{sec:ice-formation}). For example, in the
laboratory experiments of \cite{2009A&A...504..891O}, it has been
shown that the photodissociation of CH$_3$OH produces radicals such as
CH$_3$ and CH$_3$O (recently discovered in a dark cloud by Cernicharo et al., in press),
which can then recombine to form CH$_3$OCH$_3$ or
react with CHO (probably produced by the photodissociation of solid
CH$_4$ and H$_2$O, see below) to form CH$_3$CHO and HCOOCH$_3$,
respectively. Interstellar UV photons are expected to be important up
to values of visual extinction of $\simeq$3\,mag
(e.g. \cite{2009ApJ...690.1497H}), where CO is not yet significantly
frozen onto dust grains (see Fig. \ref{fig:1}).  Deeper into pre-stellar cores, a
significantly more tenuous field of UV photons can be produced by the
collisions of cosmic-rays with H$_2$ molecules
(\cite{1983ApJ...267..603P}; \cite{1989ApJ...347..289G}).
%%% CC: I added/modified the following sentence
It is not yet clear if this cosmic-ray induced field is able (i) to
produce enough radicals, (ii) to furnish them enough energy to 
move on the surface, recombine and form
complex molecules, and (iii) to release them into the gas-phase where they
are observed (see also the discussion in
\cite{2012A&A...538A..42T}). 
%%% CC: end inclusion
Consequently, it is not yet clear whether models are able to reproduce
the abundances of complex molecules observed by
\cite{2012A&A...541L..12B}. 
%%% CC: I personally would prefer to drop this following sentence. You
%%% decide, of course
%It is however possible that some radicals
%(CHO, CH$_3$ and OH) could have been "stored" in the icy mantle soon
%after the formation of water and methane ice (Fig. \ref{fig:2}) and
%then used deeper into the pre-stellar core after a significant
%fraction of frozen CO is converted into methanol.

In summary, possible first steps toward the formation of COMs in the ice (before the switch-on of the protostar) are: \\
(1) {\it Production and storage of radicals.} In the molecular cloud
within which the pre-stellar core forms, at $A_{\rm V}$
$\simeq$3\,mag, interstellar UV photons can still partially dissociate
important ice components (H$_2$O and CH$_4$) and some of the products
can be trapped within the ice, which already contains significant
fractions of water (e.g. \cite{2011ApJ...731....9C}). Alternatively,
because of the multilayered nature of icy mantles, radicals can be
stored in the inner layers during mantle formation
(\cite{2012A&A...538A..42T}). \\
 (2) {\it Radical-radical reactions.} As
the density increases and CO starts to freeze-out onto the first
water-dominated ice layers, the CO is transformed into CH$_3$OH more
and more with increasing freeze-out (given that with the freeze-out of
CO and O, the H/O and H/CO abundance ratios in the gas-phase increase,
as the number density of H atoms is kept about constant to
1\,cm$^{-3}$ by the cosmic-ray dissociation and surface re-formation of
H$_2$ molecules). The energy released during the formation of methanol
is partially used by methanol itself to evaporate and partially
released as heat on the icy surface, allowing some of the previously
trapped radicals to move. The new radicals produced in the
dissociation of CH$_3$OH by cosmic-ray-induced UV photons (and
probably some of the intermediate compounds produced during the
CO$\rightarrow$CH$_3$OH conversion) will then participate in the
formation of the observed complex organic molecules
(e.g. \cite{2009A&A...504..891O}). As for the case of methanol, the
energy released in the process of formation of these COMs can be partially used
to return in the gas-phase. The impulsive heating of dust grains due to the impact of heavy 
cosmic rays (\cite{1985A&A...144..147L}) may also temporarily enhance the mobility of the stored radicals, allowing complex molecule formation.\\
%%% CC: I added this sentence
We emphasize that the above steps remain highly speculative and more
experimental and theoretical work is necessary to better understand
the grain surface chemistry processes.
%%% end
 Given that the observed COMs
are building blocks of biologically important species, this once again
underlines the importance of pre-stellar cores for the first steps
toward our astrochemical origins.

%%%%%%%%%%%%%%%%%%% Cecilia %%%%%%%%%%%%%%%%%%%%%%%%%%%%%%%%%%%%%%%%%%%%%

\section{The cocoon phase: protostars}   % Cecilia, about 10 pages, 3 (max 4) subsections
\label{sec:violent-origin-sun}

\begin{figure*}
  \includegraphics[width=\textwidth]{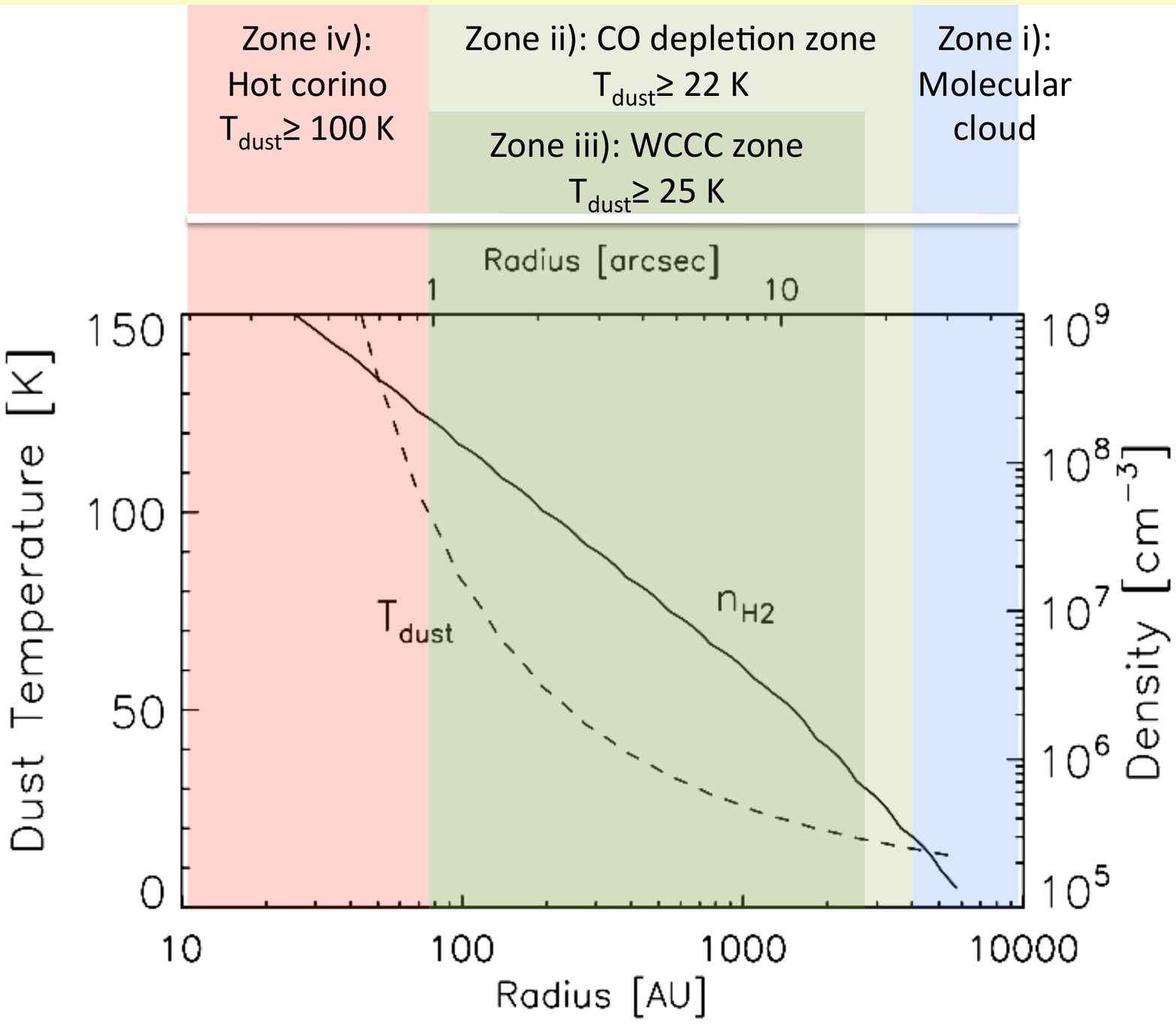}
  \caption{Temperature and density profile of IRAS16293-2422, the
    prototype for chemical studies in Class 0 sources (from
    \cite{2010A&A...519A..65C}). The colored boxes represent the four
    chemical zones described in the text: i) molecular cloud, ii) CO
    depletion, iii) Warm-Carbon-Chain-Chemistry (WCCC); iv) hot
    corino.}
\label{fig:i16293-structure}
\end{figure*}

Once the collapse starts, the gravitational energy released at the
center of the infalling envelope is converted into radiation. During
the first phases of star formation, this is the main source of the
protostar luminosity $L_*$ and it is given by $ L_* = {G M_*
  \dot{M}}/{R_*}$, where $M_*$ and $R_*$ are the mass and radius of
the central object, and $\dot{M}$ is the mass accretion rate. The
approximate structure of the envelope, as derived by observations of
the continuum and line emission (e.g. \cite{2000A&A...355.1129C};
 \cite{2002A&A...389..908J}; \cite{2006ApJS..167..256R}) is reported in
Fig. \ref{fig:i16293-structure}. Both the density and temperature
increase toward the center. Similarly, the velocity of the infalling
gas increases with decreasing distance from the center with an
$r^{-1/2}$ power law, although part of the envelope may not be 
collapsing yet. The infall motion has proved difficult to
disentangle from the outflow motions, but high spatial and spectral
resolution observations recently obtained with ALMA\footnotemark
\footnotetext{The Atacama Large Millimeter/submillimeter Array.} have succeeded to probe
it unmistakably toward IRAS16293-2422
(\cite{2012A&A...544L...7P}). Finally, new Herschel observations
provide a much more complicated picture where, at least in some
sources, the cavity created by the outflowing gas is illuminated and
heated by the UV photons of the central star, making the interpretation
of the observed lines not straightforward (\cite{2012A&A...537A..55V}).

\subsection{The chemical composition of protostellar envelopes: a powerful tool to
  understand the present and the past}\label{sec:prot-chem-comp}

Chemistry has been recognised to be a powerful diagnostic tool in
several fields of astrophysics to understand the present and the past
of the studied object. For example, at large scale, the chemical
enrichment in stars throughout the Milky Way tells us about different
star populations and ages, and, consequently, how the Milky Way formed
(e.g. \cite{2012A&ARv..20...50G}). Similarly, at much smaller scales,
the chemical composition in protostellar envelopse tell us about their present
status and past history.

Figure \ref{fig:i16293-structure} shows the approximate and very
simplified density and temperature profiles of a typical protostellar envelope. To a scale of $\geq 100$ AU, a roughly spherical envelope heated by the internal new
born star this is probably a correct description. However, at smaller scales, the envelope is not spherical, because of the presence of a circumstellar disk (\S
\ref{sec:turb-assembl-plan}) and the presence of multiple sources, as
in the case of IRAS16293-2422 and NGC1333-IRAS4
(e.g. \cite{1989ApJ...337..858W}), among the two most studied examples
of solar-type protostars. Nonetheless, from a chemical point of view,
four major zones can be identified (Fig. \ref{fig:i16293-structure}): i) an outer zone, with
the same chemical composition as that of the placental molecular cloud; ii) a
CO depleted zone, usually called cold envelope, with the chemistry is
very similar to that of pre-stellar cores (\S\ref{sec:calm-before-storm}); iii) a CH$_4$ ice sublimation region, where the chemistry is dominated by the warm carbon chain chemistry,
called WCCC, triggered by sublimation of the methane from the grain
mantles; iv) the hot corino zone, where the chemistry is dominated by
the water-matrix grain mantle sublimation and hot gas chemistry. The
transition between zones ii) to iv) is determined by the dust
temperature, which governs the sublimation of the icy mantles, whereas
the CO depleted region depends on the density and age of the
protostellar envelope. In the following we summarise the characteristics of the
four zones.

\begin{itemize}
\item [Zone i)] The chemical composition in this zone is similar to
  typical molecular clouds, with no particularly important freeze-out 
  of species. Whether this zone is present or not in a protostellar envelope
  depends on the envelope density and age, which determines the
  existence of zone ii).
\item [Zone ii)] As described in \S \ref{sec:calm-before-storm}, if
  the density and age of the envelope are high enough, molecules
  freeze-out onto dust surfaces. Important for the various reasons
  again described in \S \ref{sec:calm-before-storm} is the region
  where CO freezes out, defined by a dust temperature lower than
  about 22 K. J{\o}rgensen et al. (\cite{2005A&A...435..177J}) found that
  a large fraction of Class 0 and Class I protostars have CO-depleted
  regions in their envelopes, typically where the density is larger than $\sim 10^5$
  cm$^{-3}$. Models of the chemistry in young protostellar envelopes provide a
  theoretical interpretation to these observations (e.g. \cite{2005ApJ...631..351L}).
\item [Zone iii)] When the dust temperature exceeds the methane
  sublimation temperature, $\sim 25$ K, the chemistry is governed by
  the injection of methane in the gas-phase, if the CH$_4$ abundance is
  larger than $\sim 10^{-7}$.  In this case, CH$_4$ becomes a major
  destruction partner for C$^+$, starting the
  efficient formation of C-chain molecules in the relatively warm
  (30--60 K) gas (\cite{2008ApJ...674..984A};
  \cite{2008ApJ...681.1385H}; \cite{2011ApJ...743..182H}). So far,
  only a few protostellar envelopes with very abundant
  C-chain molecules have been discovered.  L1527 is the prototype of this class of
  sources, called Warm-Carbon-Chain-Chemistry (WCCC) sources
  (\cite{2008ApJ...672..371S}; \cite{2010ApJ...718L..49S};
  \cite{2010ApJ...718L..49S}). Note that the abundance of methane has
  been indirectly inferred in those sources by modelling the observed
  C-chain molecules, as gaseous CH$_4$ does not have observable
  rotational transitions.
\item [Zone iv)] When the dust temperature exceeds about 100 K, the
  grain mantles evaporate and all species trapped in them are
  released in the gas-phase, giving rise to a rich chemistry, first
  discovered in high-mass protostellar envelopes and called hot core chemistry
  (e.g. \cite{1987ApJ...315..621B}), and successively unveiled in low-mass
  protostellar envelopes (\cite{2003ApJ...593L..51C}). However, as it will be
  discussed in detail in \S\ref{sec:compl-chem-hot}, the chemical composition of 
  low- and high- mass cores is not identical.
\end{itemize}
The transition zones in Fig.\,ref{fig:i16293-structure} are, of course, approximate, as laboratory experiments show that ice sublimation is a complex process where molecules are released into
the gas through several steps at different dust temperatures
(e.g. \cite{2004MNRAS.354.1141V}). Also, the outflows emanating from
the central objects open up cavities which are
directly illuminated by the UV photons of the new born star
(e.g. \cite{2009A&A...507.1425V};
\cite{2012A&A...542A..86Y};\cite{2012A&A...537A..55V}). In these
cases, large Photon-Dominated-Regions (PDRs) may dominate and mask the
molecular emission from the various zones, depending on the extent of
the cavity.

As already mentioned, the presence of the WCCC zone (zone iii) depends on the
abundance of methane in the dust mantles. Methane is formed, as the
vast majority of the grain mantles, during the pre-stellar phase (\S
\ref{sec:calm-before-storm}). Specifically, it is believed to
form by hydrogenation of neutral carbon. However, in typical molecular clouds, neutral carbon is a rare species because of the efficient formation of CO. Therefore, to have a
large quantity of iced CH$_4$, one needs particular conditions, namely
a relatively high abundance of neutral carbon in the gas-phase. This
occurs when the transition from the diffuse cloud to molecular cloud
is very fast, and a substantial fraction of carbon atoms freeze-out
into the grain mantles before the CO formation is achieved
(e.g. \cite{2011ApJ...743..182H}). Therefore, the presence of a WCCC zone may be a signature of fast collapse (\cite{2008ApJ...672..371S}), for example triggered by a
shock from a nearby forming star or two encountering diffuse
clouds. Alternatively, if the pre-stellar core is
embedded in a relatively tenuous cloud, CO photodissociation could
still play a role and led to a large amount of methane ice.
 Unfortunately, the limited number of observations do not allow
us to go much further in the interpretation of this peculiar
chemistry, and more studies are needed to fully exploit it.
In the same vein, the chemical composition in the hot corino zone, as
well as the observed molecular deuteration, are all largely influenced
by the pre-stellar phase. These cases will be discussed in detail in
the following paragraphs.

Last, a potentially powerful diagnostic is provided by the relative
abundances of isomers of the same generic formula. Since the
interstellar chemistry is dominated by kinetics, different isomers
have in principle the imprint of the different chemical formation
routes. Therefore, the isomer relative abundances help understanding
the reactions at work and, consequently, how well we understand the
interstellar chemistry. A puzzling and interesting example is provided
by the isocyanic acid (HNCO) and its isomers fulminic acid (HCNO) and
cyanic acid (HOCN), which have zero energies respectively 71 and 25
kcal/mol above HNCO. In cold gas, the HNCO/HCNO and HNCO/HOCN
abundance ratios are about 50, whereas in warm gas HNCO/HCNO is about
50 and HNCO/HOCN more than 5 times larger
(\cite{2010A&A...516A.105M}). The difference of abundances between the
different isomers is thought to be due to the different chemical
routes of formation and destructions
(\cite{2010ApJ...725.2101Q}). However, the available gas-phase and
gas-grain+gas-phase models have some difficulties in reproducing the
observations and the results very much depend on the assumption made on the CH$_2$ + NO
reaction rate coefficient. Even more puzzling, these models do not explain the observed
difference in the HCNO/HOCN ratio between cold and warm
sources. Marcelino et al. (\cite{2010A&A...516A.105M}) speculate the
presence of a mechanism that converts HCNO into HOCN, despite the
large energy barrier necessary for the isomerisation. On the other
hand, Lattelais et al. (\cite{2009ApJ...696L.133L}) already noted that a
pseudo-isomerisation seems to occur to the majority of species where
different isomers have been detected. They studied 14 species and 32
isomers and found that the larger the energy difference, the larger
the abundance ratio between the most stable species and its isomer, with a few
exceptions. They called it the ``minimum energy principle'' and its
origin is still unclear, as the isomerisation barriers are generally
very large and different isomers are formed from different ``mother''
species.

Similar arguments on the diagnostic value applies for the
isotopologues of a species. Nice examples are provided by the CCS and
CCH studies by Sakai and collaborators
(e.g. \cite{2007ApJ...663.1174S};
\cite{2010A&A...512A..31S}). Studying the abundance ratio of
$^{13}$CCS/C$^{13}$CS and $^{13}$CCH/C$^{13}$CH, they constrained
the formation routes of CCS and CCH and demonstrated that the
$^{12}$C/$^{13}$C depends on the position of the carbon in the chain.

\subsection{The chemical complexity in hot
  corinos}\label{sec:compl-chem-hot}

In the 90s, several abundant complex organic molecules (COMs) were
discovered in an unbiased spectral survey of the prototype massive star forming 
region, the Orion Molecular Cloud (\cite{1987ApJ...315..621B}). Soon after, a
similar rich chemistry was observed in several other massive
protostellar envelopes. The proprieties of the line emission indicate that these
COMs reside in compact ($\leq 0.01$ pc), dense ($\geq 10^7$ cm$^{-3}$)
and hot ($\geq 100$ K) regions, soon called ``hot cores''. A simple
and obvious interpretation is that the observed rich chemistry is due
to the sublimation of some species from the grain mantles, called
``mother'' or ``primary'' species, and the synthesis of others, called
``daughter'' or ``secondary'' species, thanks to the high gas
temperature (e.g. \cite{1992ApJ...399L..71C}). Almost two decades
later, similar results were obtained toward the envelope of the prototype low-mass 
protostar IRAS16293-2422 (\cite{2000A&A...357L...9C};
\cite{2003ApJ...593L..51C}), where several COMs were detected. Since
then, more low-mass hot cores have been discovered and, to distinguish
them from the high-mass hot cores, they were called hot corinos
(\cite{2004ApJ...615..354B}; \cite{2004ApJ...617L..69B};
\cite{2007A&A...463..601B}; \cite{2006ApJ...636L.145L};
\cite{2012ApJ...757L...4J}; see also the review by
\cite{2009ARA&A..47..427H}). Hot corinos differ from hot cores not
only for the smaller sizes, lower temperatures and densities, but also
chemically. In fact, when normalized to methanol or formaldehyde, hot
corinos have typically one order of magnitude more abundant COMs (such
as HCOOCH$_3$ or CH$_3$OCH$_3$) than hot cores
(\cite{2007prpl.conf...47C}; \cite{2007A&A...463..601B};
\cite{2009ARA&A..47..427H}; \cite{2011ApJ...740...14O};
\cite{2012ApJ...744..131C}). The difference in the richness and COMs
abundances between hot cores and hot corinos is likely due to various
factors. Among them, two certainly play a major role: i) the gas
temperature, which governs the neutral-neutral reactions that often
possess large activation energy barriers; ii) the composition of the sublimated
ices, governed by the past pre-stellar history (\S\ref{sec:calm-before-storm}).

\begin{figure*}
  \includegraphics[width=0.9\textwidth]{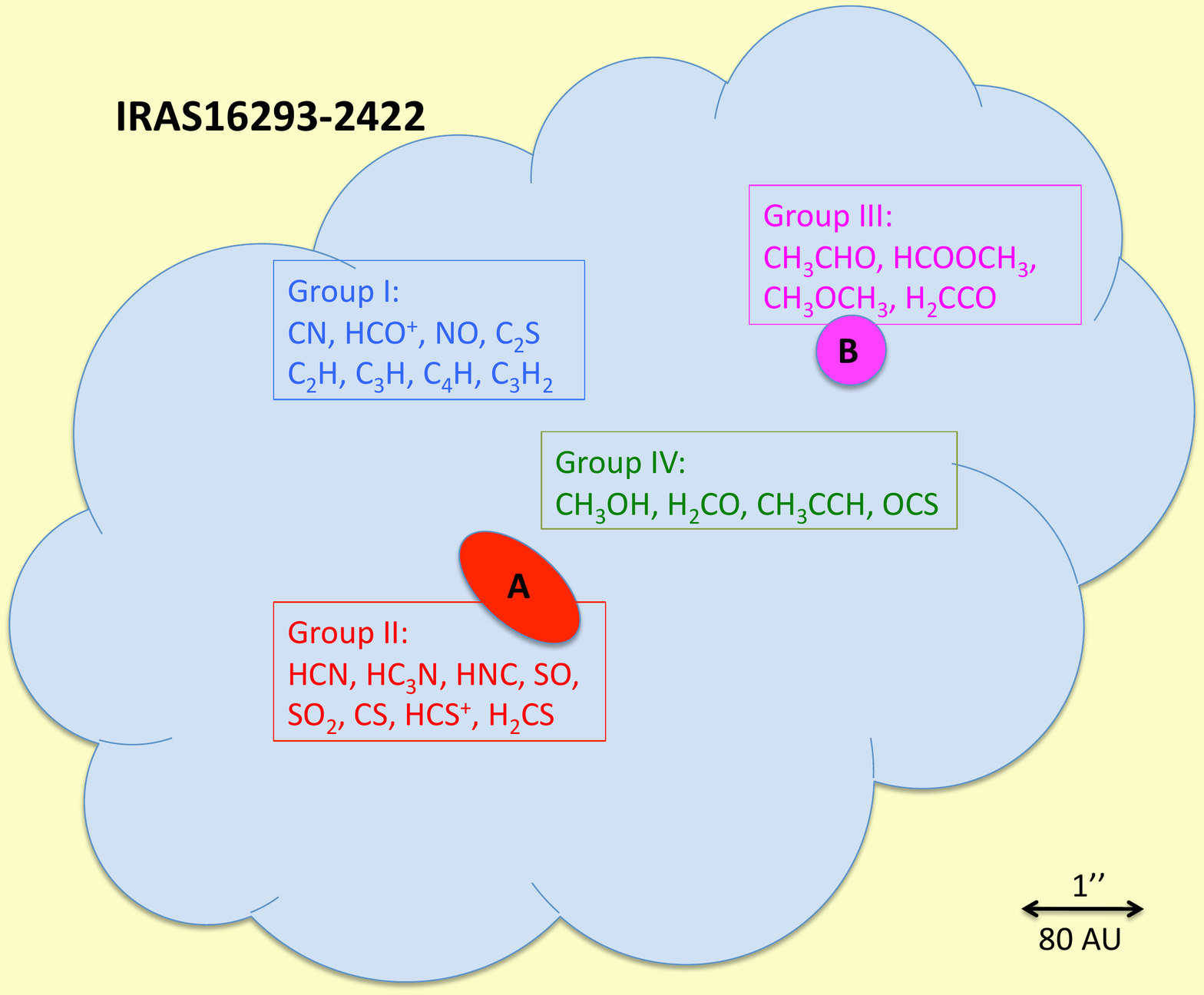}
  \caption{Sketch of the chemical composition of the protostellar envelope of
    IRAS16293-2422, a protobinary system composed of two sources, A
    and B, as marked. The four boxes list the species in the different
    components of the system: species in Group I are associated with
    the cold envelope surrounding A and B; species in Group II are
    associated with source A and in Group III with source B; species
    in Group IV are present in the cold envelope and the two sources.}
\label{fig:i16293-chemistry}
\end{figure*}

In addition to being weak line emitters and small objects, the study
of hot corinos is also complicated by the fact that low-mass
protostars are often binary or multiple systems (as in the case of high-mass 
protostars). The hot corino
prototype IRAS16293-2422 is in fact a binary system and the two
objects composing it, called A and B in the literature, show
definitively a different chemistry (see for example the recent
articles by \cite{2011A&A...532A..23C} and \cite{2011A&A...534A.100J},
and reference therein). To illustrate this aspect,
Fig. \ref{fig:i16293-chemistry} shows a sketch of the chemical
composition of IRAS16293-2422, based on the analysis of the
single-dish unbiased spectral millimeter and sub-millimeter survey carried out by
\cite{2011A&A...532A..23C} and confirmed by the submillimeter
interferometric unbiased survey of \cite{2011A&A...534A.100J}.
Four groups of species are identified:\\
{\it Group I:} Millimeter lines from simple molecules, like CN and
HCO$^+$, are dominated by the cold envelope. Also, emission from
simple carbon-chains are associated with the cold envelope (see the
discussion on their chemistry in \S
\ref{sec:prot-chem-comp}).\\
{\it Group II:} Source A is rich in N- and S- bearing molecules.\\
{\it Group III:} Source B is rich in O-bearing COMs.\\
{\it Group IV:} Molecules like CH$_3$OH, H$_2$CO, CH$_3$CCH and OCS
emit low-lying lines in the cold envelope and high-lying lines in the
two sources A and B.

The obvious question is: why are source A and B so chemically
different? They must have had a similar composition of the sublimated
ices, as they belong to the same core, so that the difference is probably
originating from the different evolutionary status caused by the
difference in mass of the two objects (\cite{2004ApJ...617L..69B};
\cite{2011A&A...532A..23C}; \cite{2012A&A...544L...7P};
\cite{2012ApJ...757L...4J}). However, so far no attempt has appeared 
in the literature to theoretically model the two sources to understand
what exactly causes the observed chemical differences.

Finally, as mentioned in \S \ref{sec:calm-before-storm}, COMs are
predicted to be formed on grain surfaces. Four fundamental steps
are involved: i) freeze-out of atoms and simple molecules (such as O and
CO) on the grain surface; ii) successive additions of H atoms
to form hydrogenated species (such as CH$_3$OH); iii) formation and
trapping of radicals, such as CH$_3$, on the grain surfaces; iv)
combination of radicals to form COMs in the warm-up period. While
laboratory experiments and quantum chemistry calculations have tested
and quantified the second step, the third step is still a matter of
debate. Garrod \& Herbst \cite{2006A&A...457..927G} and subsequent
works from the same authors assume that the radicals are formed from
the secondary UV photons emitted by the interaction of cosmic rays
with H$_2$ molecules. Specifically, it is assumed that UV photons
break iced species like CH$_3$OH into radicals like CH$_3$ and that
the broken pieces remain frozen on the grains, which may not be
necessarily the case. On the other hand, Taquet et
al. \cite{2012A&A...538A..42T} showed that radicals can indeed be
trapped in the grain mantles without the intervention of UV photons,
just because of the intrinsic layered structure of the forming mantle.

\begin{figure*}
  \includegraphics[width=0.9\textwidth]{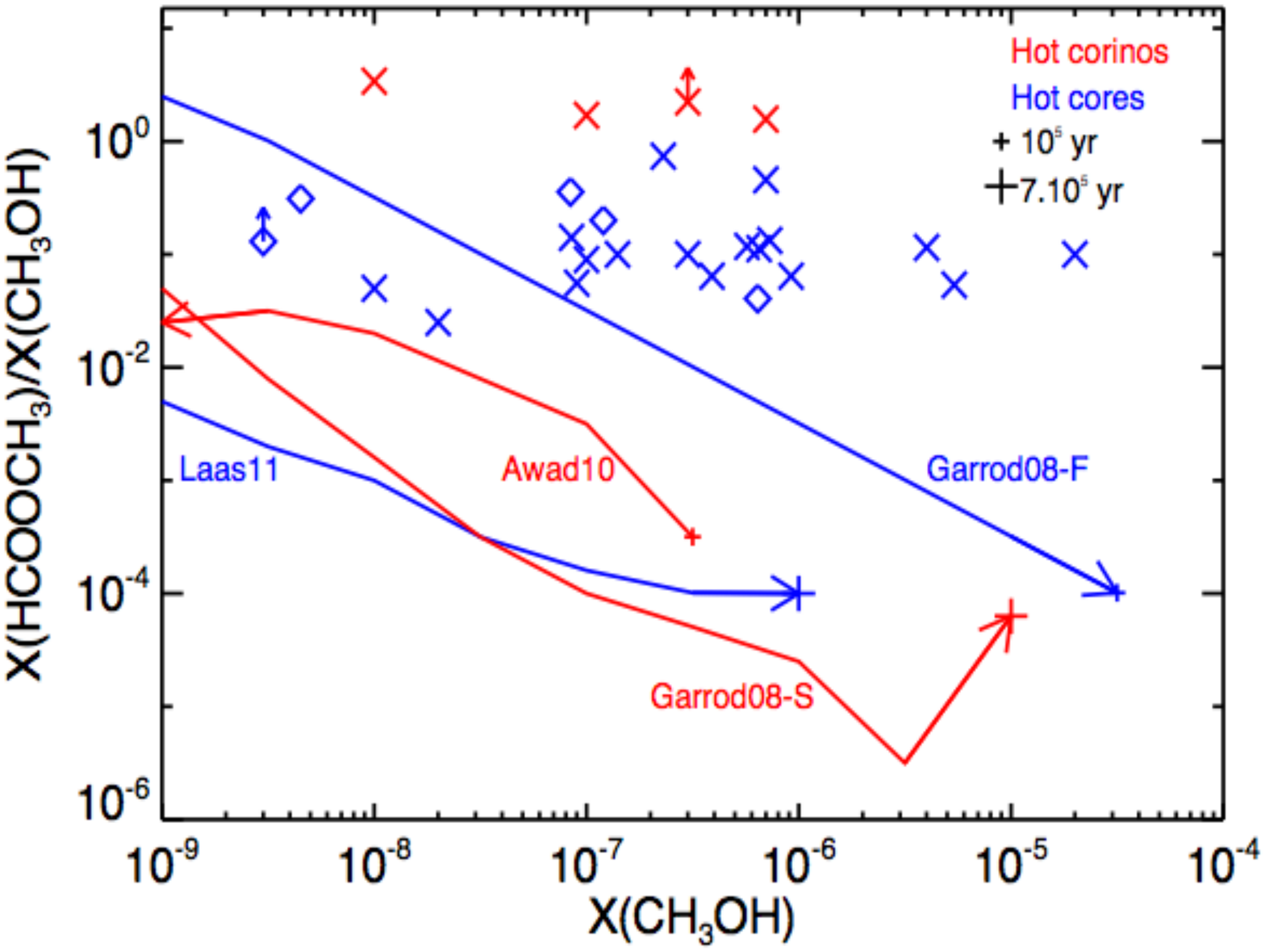}
  \caption{Observed (red and blue crosses: hot corinos and cores) and
    predicted (continuum lines) methyl formate normalized to methanol
    abundance as function of methanol abundance. The blue lines refer
    to models of hot cores (\cite{2008ApJ...682..283G} and
    \cite{2011ApJ...728...71L}) and red lines to models of hot corinos
    (\cite{2008ApJ...682..283G} and
    \cite{2010MNRAS.407.2511A}). Figure from
    \cite{2012A&A...538A..42T}, with permission. It is probably safe
    to assume that the plotted values are correct within one order of
    magnitude.}
\label{fig:coms}
\end{figure*}

However, it is important to emphasize that, whatever is the possible
origin of the radicals, models still fail to reproduce the observed
amount of COMs. For example, Fig. \ref{fig:coms} shows the
comparison between the observed and predicted methyl formate abundance
normalized to the methanol one. Published models are off by at least
one order of magnitude. Considering that COMs are also observed
in pre-stellar cores (see \S \ref{sec:calm-before-storm}) and outflows
(\S \ref{sec:molec-compl-outfl}), something basic on how COMs are
formed in the ISM must still escape our understanding.

\subsection{The chemical complexity in molecular
  outflows}\label{sec:molec-compl-outfl}

The birth of a star is accompanied by a violent and substantial
ejection of material simultaneous to the accretion toward the central
object. The process has an enormous importance in the star formation
process because i) it allows the infalling matter to lose angular
momentum and accrete onto the central object, and ii) the ejected material
interacts with the surroundings, deeply modifying it and completely
destroying, in some cases, the parental cloud (e.g. \cite{1998A&A...334..269L};
\cite{2008ApJ...683..255S}; \cite{2011ApJ...742..105A}; Lopez-Sepulcre et
al., submitted). The ejected material creates shocks at the interface
between the outflowing jet and the quiescent material. Those shocks
are chemically rich sites, showing a chemical composition very similar
to hot cores/corinos. In fact, in the shocks, dust grains are
sputtered and vaporized releasing the mantle components and part of their refractory material
into the gas phase. Moreover, shocked regions become hot enough to allow neutral-neutral reactions to take over and produce complex molecules.  In the following, we will only
review the studies on the chemical composition of the outflow shocks,
leaving out the many and important questions on the physical structure
of the shock and the acceleration mechanisms of the jet. 

Although several molecular outflows have been observed and mapped in
the past three decades, the study of their molecular complexity
started much later. Bachiller et al. \cite{1997ApJ...487L..93B} were
the first to show the chemical structure of L1157-B1, considered
nowadays a prototype for the studies of molecular complexity in
molecular outflows.  Toward this source, not only relatively simple
complex molecules, like methanol, have been detected
(\cite{2010A&A...518L.112C}), but also molecules considered hot
cores/corinos tracers, like methyl formate (HCOOCH$_3$), ethanol
(C$_2$H$_5$OH), formic acid (HCOOH) and methyl cyanide (CH$_3$CN)
(\cite{2008ApJ...681L..21A}). High spatial resolution observations
show that emission of these species is concentrated in a small region
associated with the violent shocks at the head of the outflowing
material (\cite{2009A&A...507L..25C}). The presence of COMs in
molecular outflows strongly suggests that these species were part of
the sputtered icy mantles (as the time elapsed since the shock is too
short for any gas-phase route to build up COMs) and provides us with
another piece of the puzzle regarding their formation. The abundances
normalized to methanol are at least one order of magnitude lower in
molecular outflows than in hot corinos.

It is worth noticing the presence of species not even detected in
other sources, like the phosphorus nitride (PN), whose abundance is
only a few times $10^{-10}$ with respect to H$_2$
(\cite{2011PASJ...63L..37Y}). In fact, molecular outflows can be
considered, for some aspects, unique laboratories to understand 
interstellar medium chemistry. For example,  hydrogen
chloride (HCl) has been recently detected with the
Herschel Space Observatory in L1157-B1
(\cite{2012ApJ...744..164C}). The measured abundance is 3--6
$\times10^{-9}$, practically the same value as in high- and low-mass
protostellar envelopes (e.g. \cite{2010ApJ...723..218P}) and about 200 times lower than the
Cl elemental abundance. This is a puzzling result, as chemical models
predict that HCl would be the major reservoir of chlorine and
observational evidence suggests that L1157-B1 is a shock site where
grains are sputtered/vaporized and mantles almost entirely destroyed, as 
also suggested by the large fraction of silicon found in the gas-phase as SiO.
Therefore, the low measured HCl abundance raises the question "where is
chlorine?". It is not in the mantle, but not even in the vaporized refractory material of dust
grains where silicates reside. Is then chlorine in a significantly more refractory component than silicates? Which one? All questions that will need more observations to be answered.

\subsection{Water and deuterated water}
\label{sec:water-and-deut}

Water and deuterated water are special species, because of the hints
on the Earth and Solar System formation that they bring
(\S\ref{sec:pieces-puzzle}) and because water plays a leading role in
the thermal and chemical evolution of protostellar envelopes
(\cite{1996ApJ...471..400C}; \cite{1997ApJ...489..122D};
\cite{2011PASP..123..138V}). However, since water lines can only be
observed from out-of-the atmosphere telescopes, the water content in
the envelope of solar-type protostars has been estimated only recently.

{\it Water abundance in hot corinos.} The first estimates based on the
Infrared Space Observatory (ISO) suggested that the water abundance in
the hot corino region is only a few times $10^{-6}$
(e.g. \cite{2000A&A...355.1129C}). The more recent observations
obtained with Herschel, with a much better spatial and spectral
resolution, have confirmed that first claim with an increased
reliability and in a larger number of sources
(\cite{2010A&A...521L..30K}; \cite{2012A&A...542A...8K};
\cite{2012A&A...537A..55V}; \cite{2012A&A...539A.132C}). If, on the
one hand, these observations confirm the old theoretical predictions
that water should be abundant in the innermost and warmest regions of
the envelopes surrounding Class 0 protostars
(\cite{1996ApJ...471..400C}; \cite{2009A&A...506.1229C}), they also
raise the question why the measured water abundance is much lower than
that expected, $\sim 10^{-4}$, based on the ice measurements
(\S\ref{sec:calm-before-storm}). Finally, interferometric
observations have shown that a compact H$_2^{18}$O emitting region is
associated with the hot corinos/disk of a few Class 0 sources
(\cite{2010ApJ...725L.172J} ; \cite{2012A&A...541A..39P}).\\

{\it Water abundance in molecular outflows.} Again, the first
estimates of the water abundance in molecular outflows were obtained
with ISO and gave abundances varying from $\sim 10^{-5}$ to $\sim
10^{-4}$ (\cite{1996A&A...315L.181L}; \cite{2000A&A...360..297N};
\cite{2000A&A...359..148B}). Water in outflows was also the target of
the Submillimeter Wave Astronomy Satellite (SWAS) and Odin satellite, 
which were tuned on the H$_2$O ground-state
transition at 557 GHz (\cite{2008ApJ...674.1015F};
\cite{2009A&A...507.1455B}; \cite{2002A&A...395..657B}). More
recently, the new Herschel observations are providing a mine of new
information, allowing us to map the water emission along the outflow
and to distinguish the water content in low to high velocity
shocks. The Herschel maps show bright water emission at the shock
sites of the molecular outflows (\cite{2000A&A...360..297N};
\cite{2012A&A...539L...3B}; \cite{2010A&A...521L..30K};
\cite{2011A&A...533A..80B}; \cite{2012A&A...546A..29B}). The study of
the water abundance as a function of the velocity of the shock then
shows that high velocity shocks are associated with larger water
abundances (\cite{2010A&A...518L.113L}; \cite{2011A&A...531L...1K};
\cite{2011A&A...533A..80B}; \cite{2012A&A...538A..45S};
\cite{2012A&A...537A..98V}; \cite{2012A&A...539L...3B}), as predicted
by the C-shock models (\cite{1996ApJ...456..611K}). These models
predict that H$_2$O is formed in the gas-phase via reactions with
large activation barriers (e.g. O + H$_2$ and OH + H$_2$; see also
\cite{1989ApJ...342..306H}). Finally, interferometric observations show
that the dense shock very close to the central source produces a large
quantity of water (\cite{2011A&A...527L...3L}).
\\

\begin{figure*}
  \includegraphics[width=0.9\textwidth]{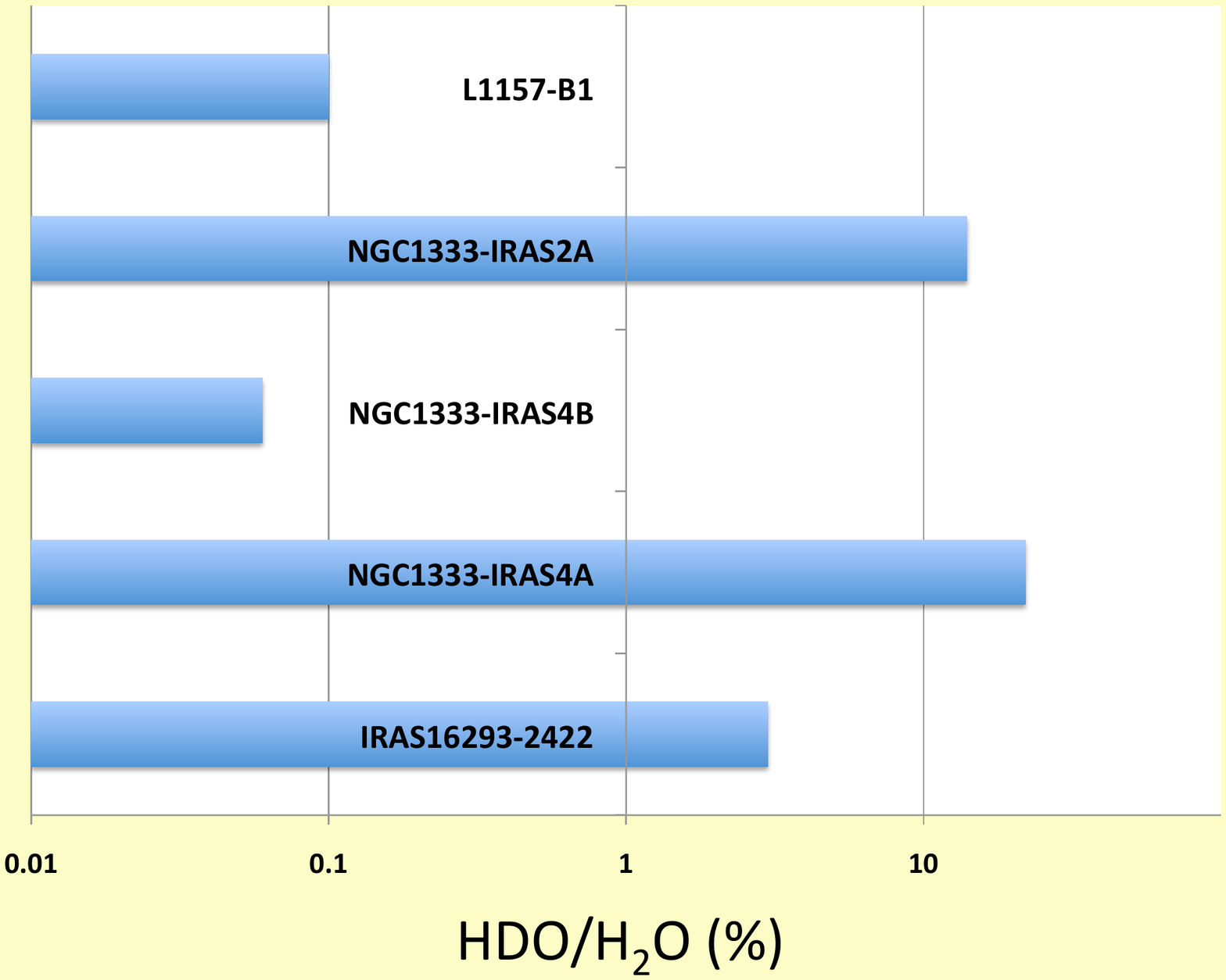}
  \caption{HDO/H$_2$O abundance ration in the envelope of Class 0 protostars and the
    L1157-B1 outflow shock. References: IRAS16293-2422
    (\cite{2012A&A...539A.132C}), NGC1333- IRAS2A
    (\cite{2012dA&A...submitted}), NGC1333- IRAS4B
    (\cite{2010ApJ...710L..72J}), NGC1333- IRAS4A
    (\cite{2012dA&A...submitted}), L1157-B1
    (\cite{2012ApJ...757L...9C}). The differences in the
    HDO/H$_2$O abundance ratio probably reflect the different conditions,
    density and temperature, at the time when ice mantles formed (see
    \S\ref{sec:calm-before-storm} and \S \ref{sec:orig-deut-molec}).}
\label{fig:hdo-sources}
\end{figure*}
{\it Deuterated water.} The HDO abundance and HDO/H$_2$O abundance
ratio have been measured toward a handful of hot corinos, with
different techniques. From single dish telescopes (IRAM 30m and ISO
first, then Herschel) the HDO/H$_2$O has been estimated to be 3\% in
IRAS16293-2422 (\cite{2005A&A...431..547P};
\cite{2012A&A...539A.132C}) and $\geq$1\% in NGC1333-IRAS2A
(\cite{2011A&A...527A..19L}). Estimates obtained with interferometric
observations of HDO and H$_2^{18}$O lines give $\leq$0.06\% in
NGC1333- IRAS4B (\cite{2010ApJ...725L.172J}), and 14\% and 22\%
toward NGC1333- IRAS2A and NGC1333- IRAS4A, respectively
(\cite{2012dA&A...submitted}). Note that the interferometric
observations provide a direct, almost model-independent, estimate of
the HDO/H$_2$O abundance ratio as they do measure the extent of the
emission and use the rare H$_2^{18}$O isotopologue reducing the
problem of line opacity. In summary, the HDO/H$_2$O ratio has been
measured toward four hot corinos: in three of them it is larger than
a few percent, whereas in NGC1333- IRAS4B it is at least one order of
magnitude lower. Herschel observations have also allowed, for the
first time, to estimate the HDO/H$_2$O in a molecular outflow shock,
L1157-B1 (0.4--2$\times 10^{-3}$, \cite{2012ApJ...757L...9C}), a 
likely direct measure of the deuteration in the ice. The situation is
summarised in Fig. \ref{fig:hdo-sources}. The differences
in the HDO/H$_2$O abundance ratio probably reflect the different conditions,
density and temperature, when the ice was formed (see \S
\ref{sec:calm-before-storm} and \S \ref{sec:orig-deut-molec}).\\

{\it Doubly deuterated water.} Although it has a very low abundance,
D$_2$O has an important diagnostic power as it sets very tight
constraints to models of water formation. So far, thanks to Herschel,
D$_2$O/H$_2$O has been measured only toward the cold envelope of
IRAS16293-2422, with the observations of both the para and ortho forms
of D$_2$O (\cite{2007ApJ...659L.137B};
\cite{2010A&A...521L..31V}). The D$_2$O/H$_2$O abundance ratio
results to be 1--4$\times10^{-3}$ (\cite{2012A&A...539A.132C}). Similarly,
the para-D$_2$O/H$_2$O toward the hot corino is $\sim5\times10^{-5}$
(\cite{2007ApJ...659L.137B}). Assuming an ortho-to-para ratio equal to 2 gives
D$_2$O/H$_2$O$\sim 10^{-4}$.  For example, comparison with the model
by \cite{2012A&A...submitted} indicates that the bulk of water was formed
on grains when the cloud/envelope temperature was 10 K and the density
between $10^4$ and $10^5$ cm$^{-3}$. In other words, when the
density at the center of the IRAS16293-2422 pre-stellar cloud reached
$10^6$ cm$^{-3}$, the oxygen not locked into CO was almost entirely
already converted into water.

\subsection{Deuteration of other species}\label{sec:deut-other-spec}

\begin{figure*}
  \includegraphics[width=0.9\textwidth]{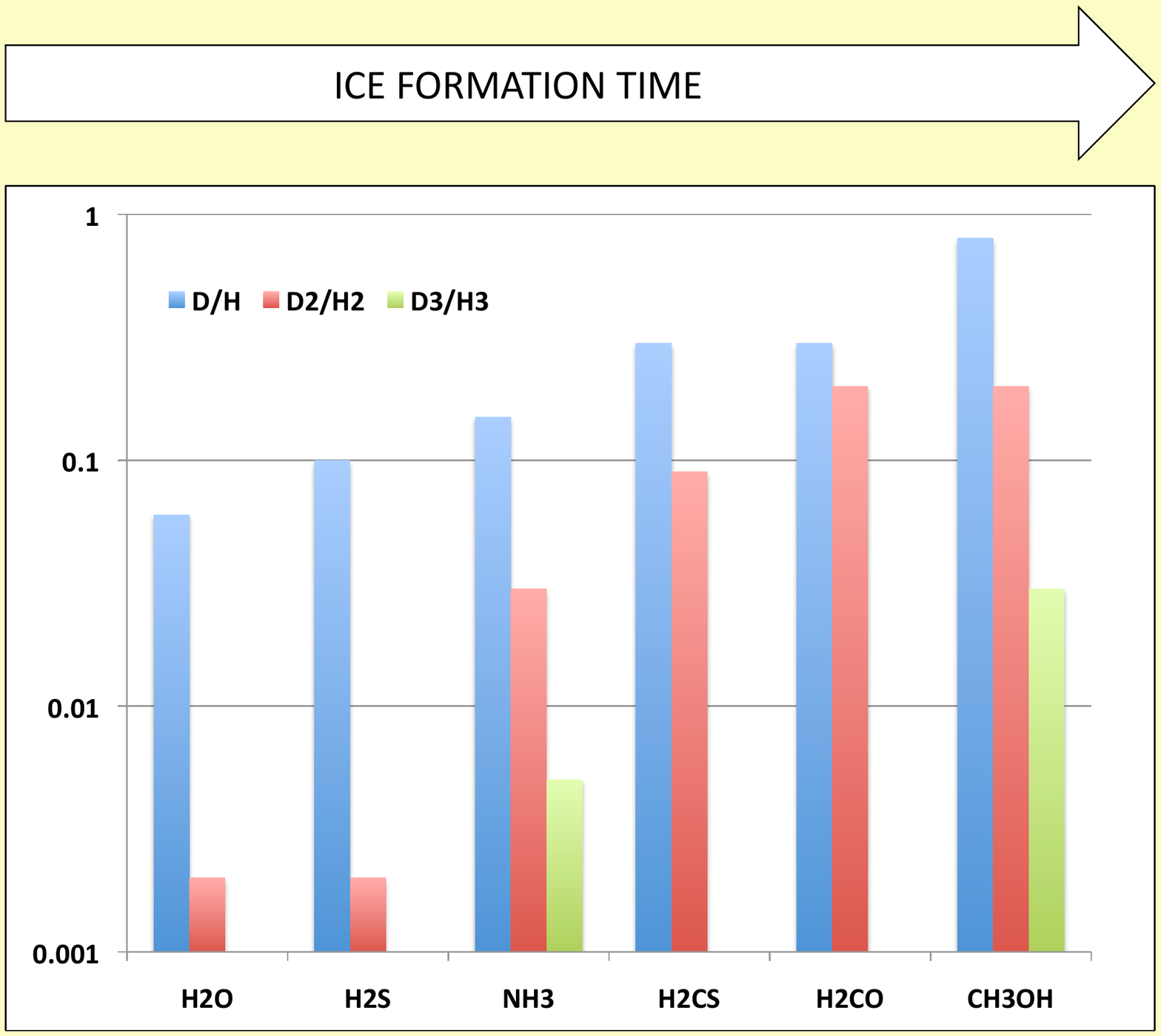}
  \caption{Measured deuteration ratios of singly, doubly and triply
    deuterated isotopologues. Based on the modelling of the formation
    of H$_2$O, H$_2$CO and CH$_3$OH (\cite{2011ApJ...741L..34C};
    \cite{2012A&A...submitted}) we speculate that the increasing
    deuteration reflects the formation time of the species on the
    ices. References: H$_2$O: \cite{2011A&A...527A..19L},
    \cite{2012A&A...539A.132C}, \cite{2012dA&A...submitted},
    \cite{2007ApJ...659L.137B}, \cite{2010A&A...521L..31V}; H$_2$S:
    \cite{2003ApJ...593L..97V}; NH$_3$:
    \cite{2001ApJ...552L.163L},\cite{2002A&A...388L..53V}; H$_2$CS:
    \cite{2005ApJ...620..308M}; H$_2$CO: \cite{1998A&A...338L..43C};
    \cite{2006A&A...453..949P}; CH$_3$OH:
    \cite{2002A&A...393L..49P},\cite{2004A&A...416..159P},
    \cite{2006A&A...453..949P}. }
\label{fig:deuteration}
\end{figure*}

As water, several molecules present large deuteration factors in low
mass protostellar envelopes and molecular outflows (e.g. \cite{2006A&A...453..949P} and
\cite{2012ApJ...757L...9C} respectively). Figure \ref{fig:deuteration}
presents a graphic summary of the observations of species with
detected doubly or triply deuterated isotopologues. The deuterated
ratios are extremely high, with enhancements of the D/H of up to 13
orders of magnitude with respect to the elemental D/H abundance
ratio. Given the conditions in the envelopes of the protostars (\S
\ref{sec:prot-chem-comp} and Fig. \ref{fig:i16293-structure}), the
observed deuteration is mostly an inherited product of the pre-stellar
phase (\S \ref{sec:calm-before-storm}). Furthermore, for the typical
physical condition where the deuterated molecules have been detected, the
measured deuteration ratios likely reflect the deuteration on the
grain mantles (e.g. \cite{1997ApJ...482L.203C}).

We emphasize here that the deuteration ratio is not the same for all
species. As mentioned in \S \ref{sec:calm-before-storm}, the lower
deuteration ratio of water with respect to formaldehyde and methanol
probably reflects the different epoch in which the bulk of the iced
species has been formed (during the pre-stellar phase). Specifically,
water is (mostly) formed before formaldehyde, and methanol is the last
in the sequence (\cite{2011ApJ...741L..34C}; \cite{2012ApJ...748L...3T};
\cite{2012A&A...submitted}).  Even though not specific modelling has
been published for all observed deuterated species, we 
speculate that the sequence in the figure represents a
temporal sequence of the species formation.
\begin{figure*}
  \includegraphics[width=0.9\textwidth]{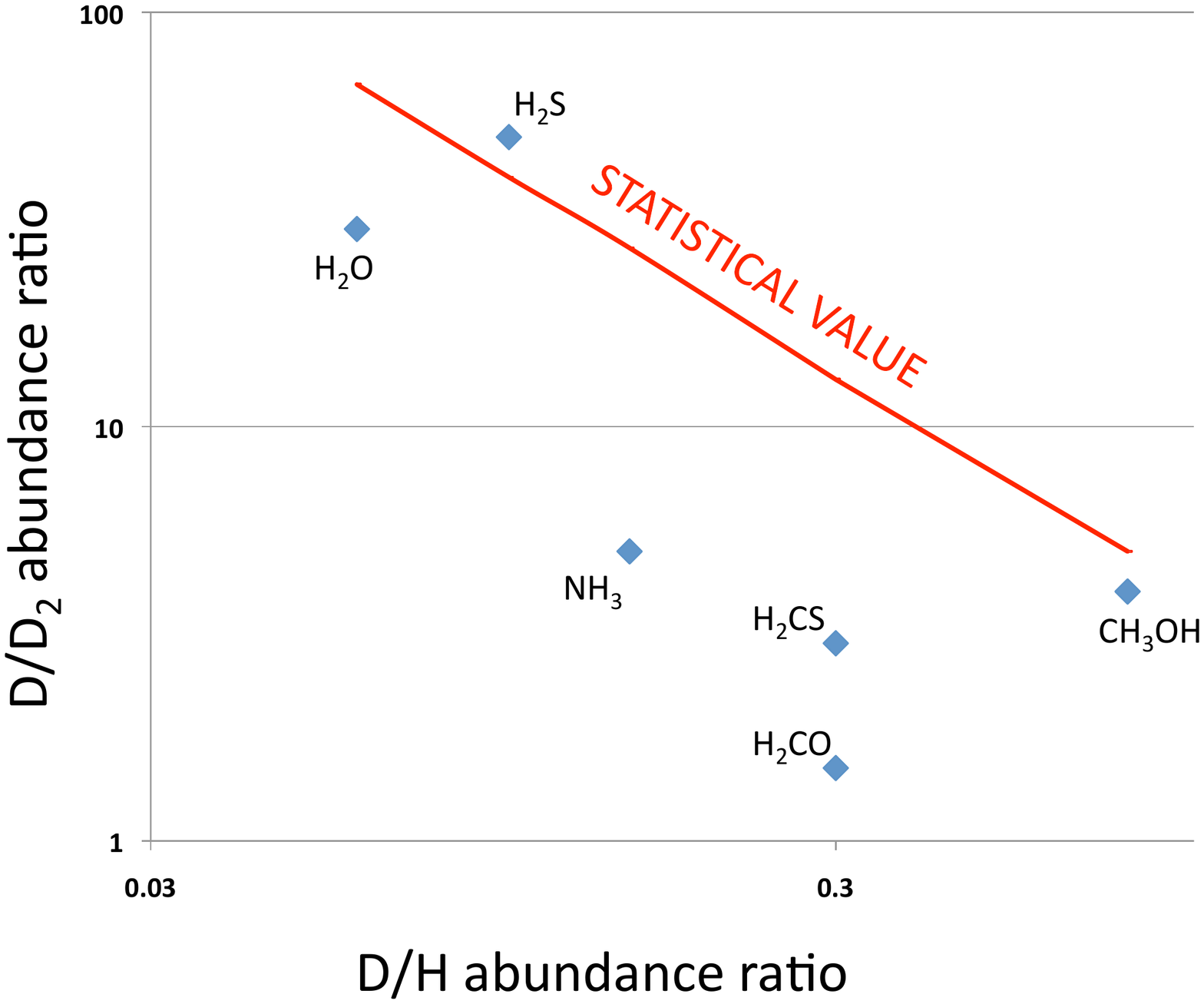}
  \caption{Measured ratios of singly to doubly deuterated
    isotopologues of the species marked in the plot. References as in
    Fig. \ref{fig:deuteration}.}.
\label{fig:d2-vs-d}
\end{figure*}

Finally, the comparison between the singly and doubly deuterated
isotopologues provides some interesting additional information. First,
if the deuterium atoms were purely statistically distributed, namely
just proportional to the D/H ratio, then it would hold:
D-species/D$_2$-species = 4 (D-species/H-species)$^{-1}$.  As shown in
Fig. \ref{fig:d2-vs-d}, this is not the case for the measured
deuteration of H$_2$O, NH$_3$, H$_2$CS and H$_2$CO. As also noted by
\cite{2007ApJ...659L.137B}, this points to a change of the atomic D/H
ratio during the formation of those species or to an origin from gas
phase reactions.  On the contrary, the statistical relation is roughly
valid for H$_2$S and CH$_3$OH. This suggests that these two species
have been formed on the grain surfaces in a very short time, when the
atomic D/H ratio can be considered roughly constant.

In summary, the observed deuteration ratios tell us that H$_2$O,
NH$_3$, H$_2$CS and H$_2$CO were formed at various stages during the
star formation process, with different values of the atomic gas D/H
ratio. On the other hand, H$_2$S and CH$_3$OH were formed in a shorter
time range. This behavior roughly agrees with the models of the
formation of H$_2$O, H$_2$CO and CH$_3$OH on grains, that predict that
methanol is only formed at very late time, whereas water and
formaldehyde are formed over a larger period of time
(\cite{2011ApJ...741L..34C}; \cite{2012ApJ...748L...3T};
\cite{2012A&A...538A..42T}). It is however possible that the species
not close to the statistical value are, at least in part, gas-phase
products.

As a final remark, it is important to emphasize that low- and
high-mass protostellar envelopes present important differences in the molecular
deuteration. A clear example is provided by the CH$_2$DOH/CH$_3$OD
abundance ratio, which is at least one order of magnitude larger in 
low-mass than in high-mass protostellar envelopes (\cite{2011A&A...528L..13R};
\cite{2012A&A...543A.152P}).

%%%%%%%%%%%%%%%%%%% Paola %%%%%%%%%%%%%%%%%%%%%%%%%%%%%%%%%%%%%%%%%%%%%
\section{Toward planet formation:  protoplanetary disks}   % Paola, about 10 pages, 3 (max 4) subsections
\label{sec:turb-assembl-plan}

Starless and pre-stellar cores present evidence of overall (slow) rotation (\cite{1986ApJ...303..356A}; \cite{1993ApJ...406..528G}; \cite{2002ApJ...572..238C}), thus they possess an initial angular momentum. As a natural consequence of angular momentum conservation, the collapse of pre-stellar cores produces flattened structures which harbor the future protoplanetary disks. Even non-rotating collapsing cores are expected to produce flattened structures in the presence of magnetic fields, as explained in the following. As ionized particles within the core are linked to the magnetic field lines, while neutrals only feel the gravitational field, a drag between ions and neutrals is established during the collapse phase (see \S\ref{sec:calm-before-storm}).  Galli \& Shu (\cite{1993ApJ...417..220G}; \cite{1993ApJ...417..243G}) found that during the collapse of a singular isothermal sphere (i.e. an unstable spherical cloud with a density profile proportional to $r^{-2}$, thus with a singularity in the center, \cite{1977ApJ...214..488S}), the magnetic field, dragged by the flow, deflects  the infalling gas toward the midplane, forming a large ($\simeq$2000\,AU) "pseudodisk". 
The magnetic field lines, initially parallel, are shaped as an hourglass, consistent with observations of polarization maps of the dust continuum emission toward young stellar objects (e.g. \cite{2006Sci...313..812G}; \cite{2011A&A...535A..44F}). The twisting of magnetic field lines in the pseudodisk acts as a "magnetic break", in the sense that it slows down the rotation by transferring angular momentum from the inner regions (which tend to rotate faster for angular momentum conservation) of the pseudodisk toward its outer parts (\cite{1994ApJ...432..720B}). Indeed, magnetic breaking is so efficient, that disks cannot form at all in ideal magneto-hydrodynamic (IMHD\footnotemark \footnotetext{IMHD assumes that the mass to magnetic-flux ratio is constant, which implies that magnetic field lines follow the gas motions, i.e. the magnetic field is "frozen" into the neutral medium.}) simulations of collapsing cores (e.g. \cite{2003ApJ...599..363A}; \cite{2008ApJ...681.1356M}; \cite{2008A&A...477....9H}).  More recently, the inclusion of non-ideal MHD effects, in particular the Hall effect\footnotemark \footnotetext{The Hall effect mainly operates at volume densities between 10$^8$ and 10$^{11}$\,cm$^{-3}$ (\cite{2004Ap&SS.292..317W}), where the more massive charged particles (ions and charged dust grains) decouple from the magnetic field  and collisionally-couple with the neutral gas.} (\cite{2012MNRAS.422..261B}; \cite{2011ApJ...733...54K}), has helped to avoid this so-called magnetic breaking catastrophe, allowing disks of about 100\,AU to form (even without initial rotation of the collapsing cloud, \cite{2012MNRAS.422..261B}). This has also been shown in simulations by \cite{2011PASJ...63..555M}, who found rapid growth to $\geq$100\,AU of the circumstellar disk when depletion of the infalling envelope is taken into account, and by \cite{2012A&A...543A.128J}, who explored the case of magnetic fields non-aligned with the rotation axis and found less efficient angular momentum transport, allowing the formation of $\simeq$100-200\,AU disks, with masses as large as 10\% the original core mass.   These characteristics are similar to the young self-gravitating protoplanetary disks (we refer to them as "embedded disks") which can become gravitationally unstable (e.g. \cite{1994ApJ...436..335L}; \cite{1997Sci...276.1836B}; \cite{2007prpl.conf..607D} and references therein; \cite{2008ApJ...685.1193B}; \cite{2011ApJ...729..146V}) and which represent the starting point of our final journey toward the formation of a planetary system. Here we will focus on the chemical evolution (see \cite{2011ARA&A..49..195A} and \cite{2011ARA&A..49...67W} for comprehensive reviews on the physical characteristics and evolution of protoplanetary disks). 

\subsection{Embedded disks: chemistry at the dawn of planet formation}

\begin{figure}[ht]
% Use the relevant command to insert your figure file.
% For example, with the graphicx package use
  \includegraphics[angle=0,width=1\textwidth]{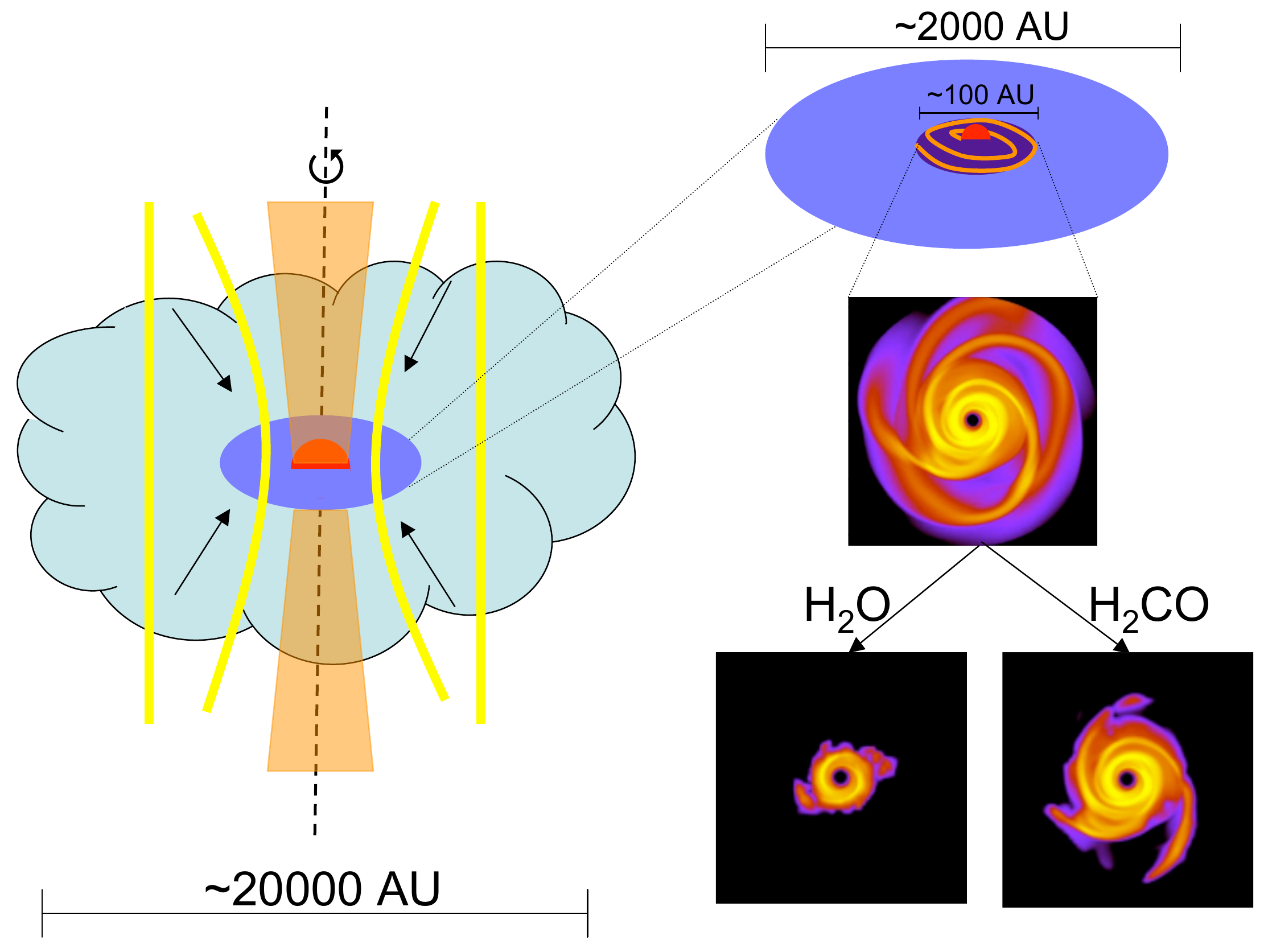}
% figure caption is below the figure
\caption{The earliest stages of a protoplanetary disk. Magnetic fields (yellow curves) and the initial rotation of the pre-stellar core lead to the formation of a flattened structure (the "pseudodisk", size ~2000\,AU) surrounding the accreting protostar. In the central few hundred AU, the embedded disk can be self-gravitating, develop spiral structure and experience fragmentation. Molecules such as H$_2$O and H$_2$CO are good tracers of these central regions of young embedded disk (\cite{2011MNRAS.417.2950I}).}
\label{fig:4-1}       % Give a unique label
\end{figure}

Young disks are embedded within the thick and massive envelopes of Class 0 sources (see \S\ref{sec:violent-origin-sun}).  Therefore, they are not  easy to study and it is hard to put constraints on theoretical predictions. Indirect evidence of young disks in Class 0 sources is given by the presence of collimated outflows, observed with millimeter and sub-millimeter telescopes (\S\ref{sec:violent-origin-sun}).  ALMA will of course revolutionize this field. After the pioneer work by, e.g., \cite{1995ApJ...449L.139C}, \cite{2000MNRAS.319..154B} and \cite{2000ApJ...529..477L}, further steps toward the characterization of these embedded disks have been made by \cite{2007ApJ...659..479J}, \cite{2009A&A...507..861J} and \cite{2011ApJS..195...21E}. With the help of interferometric observations (able to filter out the surrounding envelopes), these authors found evidence of compact embedded disks in Class 0 sources, with masses ranging from 0.04 to 1.7\,M$_{\odot}$. Choi et al. (\cite{2007ApJ...667L.183C}) observed NH$_3$ with the Very Large Array (VLA) and found a 130\,AU circumstellar disk around NGC1333 IRAS4A2. With the IRAM Plateau de Bure Interferometer (PdBI), \cite{2010ApJ...710L..72J} measured water vapor (H$_2^{18}$O) in the inner 25\,AU of the NGC1333 IRAS4B disk, suggesting the presence of a thin warm layer containing about 25 Earth masses of material. Toward the same object, \cite{2010ApJ...725L.172J} also set a stringent upper limit on the HDO/H$_2$O abundance ratio to 6$\times$10$^{-4}$ (\S\ref{sec:violent-origin-sun}).  Pineda et al. (\cite{2012A&A...544L...7P}) observed methyl formate with ALMA toward IRAS16293-2422, a binary Class 0 source in Ophiuchus (\S\ref{sec:violent-origin-sun}), and found the first evidence of infall toward source B and evidence of rotation toward source A, consistent with an almost edge-on disk (see also \cite{2005ApJ...621L.133R}).  If confirmed, this could be the first chemically and kinematically characterized embedded disk (discovered with a complex organic molecule!).  

What are the chemical model predictions of these embedded disks?
Visser et al. (\cite{2009A&A...495..881V}, \cite{2011A&A...534A.132V}) have been the first to self-consistently follow the chemistry in a two-dimensional axisymmetric model of a collapsing (initially) spherical and slowly rotating cloud, on its way toward the formation of a protoplanetary disk. The material infalling in the equatorial plane, within the centrifugal radius\footnotemark \footnotetext{The radius at which the gravitational force is balanced by the centrifugal force.}, forms the disk, whose evolution is also considered assuming no mixing. The disk-envelope boundary and the outflow cavities are well defined. Detailed predictions are given about the ice and gas-phase composition of the cloud-disk system at different evolutionary phases. At the end of the collapse phase, they find that disks can be divided in zones with different chemical history, which will ultimately affect the composition of comets formed in different zones.  Different results are found by \cite{2011MNRAS.417.2950I}, who used the hydrodynamic simulations of a young and relatively massive (0.39\,M$_{\odot}$) disk by \cite{2009ApJ...695L..53B} as input in their gas-phase and simple surface chemistry network. Boley's disk resembles in mass and size the embedded disk mentioned above, it is non-axysimmetric and present complex spiral and physical structure, with shocks moving with the spirals arms (see Figure\,\ref{fig:4-1}, middle panel in the left, which reports the gas column density map). No accretion of material from the envelope and no outflow is considered. Despite these assumptions, the  disk structure is complex and its physical characteristics are continuously stirred by the rotating spiral arms. Because of this continuous mixing, \cite{2011MNRAS.417.2950I} found no separated chemical zones as in the case of \cite{2011A&A...534A.132V}, but they identified species able to trace the inner regions of the disk (such as H$_2$O, HNO and NH$_3$) and those tracing the spiral arms (e.g. H$_2$CO and HCO$^+$). Examples of these column density maps are given in Fig. \ref{fig:4-1} (bottom right panels), which also summarizes the various physical mechanisms to be considered for a comprehensive study of the earliest stages of star formation: the collapsing envelope of a Class 0 source (red semicircle) under the influence of magnetic fields (yellow lines and curves in the figure), the pseudodisk (blue), the central embedded disk (violet) and the outflow (orange) driven by the central protostar (red semicircle). Furuya et al.(\cite{2012arXiv1207.6693F}) studied the chemical evolution of a molecular core toward the formation of the first hydrostatic core (protostellar precursor) using three-dimensional radiation hydrodynamic simulations.  They show that after a first destruction of molecules, simple species such as CO, H$_2$O and N$_2$ reform and more complex molecules (CH$_3$OH and HCOOCH$_3$) can trace the first hydrostatic  core, on its way to becoming a protostar. ALMA observations are needed to disentangle the various phenomena at work during the earliest stages of star formation, to test model predictions of collapsing magnetized pre-stellar cores and to unveil the physical and chemical structure of the embedded disks, precursors to the protoplanetary disks which will be reviewed in the next sections.

\subsection{"Naked" protoplanetary disks}
\label{sec:naked_disks}

\begin{figure}[ht]
% Use the relevant command to insert your figure file.
% For example, with the graphicx package use
  \includegraphics[angle=0,width=1\textwidth]{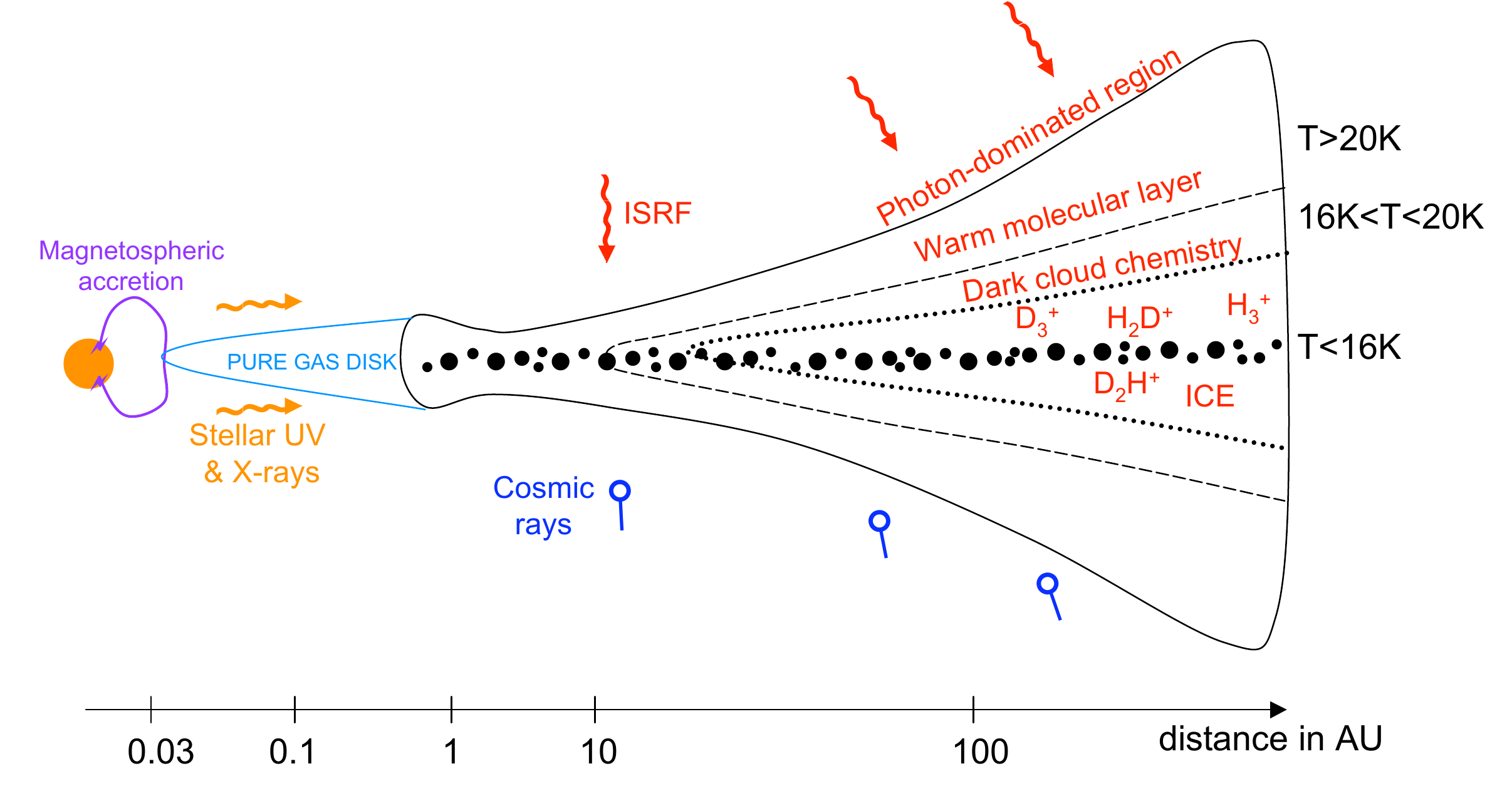}
% figure caption is below the figure
\caption{Schematic structure of a "naked" protoplanetary disk, adapted from \cite{2011ApJ...743..152O},  \cite{2010ARA&A..48..205D}, \cite{2011IAUS..280..114S},  \cite{2007prpl.conf..555D} and \cite{2007prpl.conf..751B}. The various regions are labeled. The black dots with various sizes represent the coagulated dust in the disk midplane. See text for details.}
\label{fig:4-2}       % Give a unique label
\end{figure}

The embedded phase of disks does not last long. After about 0.5\,Myr since the birth of the protostar/disk/outflow system, the parent core envelope quickly disperses and the disk enters a new phase which lasts several Myr (\cite{2011ARA&A..49...67W}). This is the T\,Tauri (or Class II) phase.  The disk mass is now only a few \% the stellar mass (\cite{2011ARA&A..49...67W}) and the motions are expected to be Keplerian.  Despite being "naked" disks, thus easier to observe than during the earlier embedded phase, the physical and chemical processes at work are complex and more (interferometric) data are sorely needed to fully understand them. Figure \ref{fig:4-2} shows a schematic picture of a T\,Tauri disk, compiled from a combination of figures found in \cite{2011ApJ...743..152O},  \cite{2010ARA&A..48..205D}, \cite{2011IAUS..280..114S},  \cite{2007prpl.conf..555D} and \cite{2007prpl.conf..751B}: (1) within the central ~1 AU from the star, a pure gas disk and the dust inner rim are present. This zone is mainly probed by Br-$\gamma$ lines (e.g. \cite{2003ApJ...597L.149M}; \cite{2007A&A...464...43M}; \cite{2007A&A...464...55T}; \cite{2012ApJ...748....6G}), H$_2$ (\cite{2004ApJ...614L.133B}; \cite{2012ApJ...756..171F}), as well as near-infrared lines of CO, H$_2$O, OH (\cite{2008ApJ...676L..49S}; \cite{2008Sci...319.1504C}; \cite{2010ApJ...722L.173P}) and simple organic molecules (\cite{2012ApJ...747...92M}). (2) Moving away from the central star, one finds the "puffed-up" inner dust wall (\cite{2001A&A...371..186N}), clearly seen in the near-infared continuum, where the higher temperature affects its vertical scale height, which is set by hydrostatic equilibrium. Within the central few AU, mid-infrared emission of H$_2$O, CO and the organic molecules HCN and C$_2$H$_2$ have been measured (\cite{2006ApJ...636L.145L}; \cite{2008Sci...319.1504C}).  Carr \& Najita (\cite{2008Sci...319.1504C}) note that the HCN/H$_2$O abundance ratio is largest in the most massive disks and speculate that this may be indication of the sequestration of H$_2$O in the outer disk during the process of planetesimal formation. It is interesting to note that toward the disks surrounding the intermediate mass ($\simeq$ 2 $<$ M/M$_{\odot}$ $<$8) Herbig Ae/Be stars, no organic molecules have been detected (\cite{2010ApJ...720..887P}; \cite{2011ApJ...731..130S}) and water is only seen in the far-infared at larger radii ($\simeq$15-20\,AU; \cite{2012A&A...544L...9F}), probably due to the larger UV fluxes compared to T\,Tauri stars. Beyond the "wall",  the disk is thought to have a layered structure.  (3) A photon-dominated region (PDR) is present all around the disk, which is exposed to the stellar and interstellar UV field, as well as the stellar X-rays. Here, forbidden line emission from the well-known PDR coolants, [CII]158\,$\mu$m, [OI]63\,$\mu$m and 145\,$\mu$m,  are observed (\cite{2010A&A...518L.129S}; \cite{2012A&A...545A..44P}), although the [CII]158\,$\mu$m and the [OI]145\,$\mu$m are not always detected (\cite{2010A&A...518L.127M}; \cite{2010A&A...518L.125T}). (4) A warm molecular layer. Just below the PDR zone, molecules survive, although photochemistry is still playing an important role (\cite{2010ApJ...714.1511H};\cite{2012arXiv1209.0591A}). The gas and dust are warm and radical and ions dominate the gas composition (\cite{2011IAUS..280..114S}). (5) A dark-cloud chemistry zone, where the temperature drops below 20\,K, molecular freeze-out becomes important and simple species typically found in dark clouds are detected: CO isotopologues with evidence of depletion (\cite{1996A&A...309..493D}; \cite{2007prpl.conf..495D}; \cite{2004ApJ...616L..11Q}),  CN, HCN, HNC, CS, HCO$^+$, C$_2$H and H$_2$CO (\cite{1997A&A...317L..55D}; \cite{2001A&A...377..566V}; \cite{2004A&A...425..955T}; \cite{2012A&A...537A..60C}), N$_2$H$^+$ (\cite{2007A&A...464..615D}), SO (\cite{2010A&A...524A..19F}), CS (\cite{2011A&A...535A.104D}), DCO$^+$ (\cite{2003A&A...400L...1V}), H$_2$D$^+$ (\cite{2004ApJ...607L..51C}), HDO (\cite{2005ApJ...631L..81C}, but see \cite{2006A&A...448L...5G}), HC$_3$N (\cite{2012ApJ...756...58C}).   Qi et al. (\cite{2008ApJ...681.1396Q}) spatially resolved the emission of DCO$^+$ and measured the deuterium fraction across the disk of TW Hydrae, finding a range between 0.01 and 0.1, with a peak around 70\,AU. They also measured the DCN/HCN abundance ratio, $\simeq$0.02, similar to that measured in the jets of material coming from the nucleus of comet Hale-Bopp (\cite{1998Sci...279.1707M}).   \"Oberg et al. (\cite{2010ApJ...720..480O}) used the Sub-Millimeter Array (SMA) to image disks of six Taurus sources  with spectral type from M1 to A4, finding similar intensities of CN and HCN lines in T\,Tauri and Herbig Ae stars, but a significantly different chemical richness:  deuterated molecules, N$_2$H$^+$ and H$_2$CO were only detected toward T\,Tauri star disks, implying a lack of long-lived cold regions in the disks of the more massive Herbig Ae stars (see also \cite{2011ApJ...734...98O}). Water vapor in the cold outer disk has been detected toward TW Hydrae by \cite{2011Sci...334..338H} with Herschel, revealing a hidden large reservoir of water ice at large radii (between 100 and 200\,AU).  Indeed, ice features have been detected in the direction of edge-on protoplanetary disks  by \cite{2007ApJ...667..303T} and \cite{2009ApJ...690L.110H}. More recently, \cite{2012A&A...538A..57A} measured with the AKARI satellite several ice features in edge-on Class II disks, including a faint HDO feature, which allowed them to measure a solid HDO/H$_2$O abundance ratio between 2\% and 22\% (significantly larger than the HDO/H$_2$O ratio measured in comets and in star-forming regions; see \S\ref{sec:pieces-puzzle} and \S\ref{sec:violent-origin-sun}). (6) The midplane, characterized by cold and dense regions, with large amounts of molecular freeze-out, where only light species can survive (\cite{2011ApJ...743..152O}), in analogy with the central $\simeq$1000\,AU of pre-stellar cores (\S\ref{sec:calm-before-storm}).

Several chemical models of this protoplanetary disk phase, with various degrees of complexity, have been developed:  X-ray chemistry (\cite{1997ApJ...480..344G}; \cite{2008ApJ...676..518M}; \cite{2005A&A...440..949S}), surface chemistry (e.g. \cite{2000ApJ...544..903W}), accretion flows (\cite{1999ApJ...519..705A}; \cite{2004A&A...415..643I}), thermal balance (\cite{2004ApJ...613..424G}), grain growth (\cite{2006ApJ...642.1152A}; \cite{2011ApJ...727...76V}), UV continuum and Ly$\alpha$ radiation (\cite{2003ApJ...591L.159B}; \cite{2011ApJ...726...29F}), turbulence-driven diffusion (\cite{1995ApJ...440..674X}; \cite{2006ApJ...644.1202W}),  viscous accretion, turbulence mixing and disk winds (\cite{2009A&A...493L..49H}; \cite{2011ApJ...731..115H}), photochemistry and wavelength-dependent reaction cross sections, (\cite{2012ApJ...747..114W}), comprehensive physical, chemical and radiative transfer modeling (\cite{2008ApJ...683..287G}; \cite{2010MNRAS.405L..26W}; \cite{2011A&A...532A..85K}). Despite the advances in chemical complexity, large uncertainties are still present on several reaction rates (\cite{2008ApJ...672..629V}) and collisional coefficients, so that laboratory studies and theoretical investigations are still sorely needed to improve the reliability of modern astrochemical models.  Moreover, the large uncertainties in the process of dust evolution and coagulation in disks are also shaking our understanding of the disk chemical structure. Laboratory experiments (e.g. \cite{2010A&A...513A..56G}; \cite{2012arXiv1208.3095S}), numerical simulations (e.g. \cite{2011A&A...534A..73Z}) and theoretical work (e.g. \cite{2007prpl.conf..783D}; \cite{2012A&A...544L..16W}) are fundamental to progress in this field and an effort has to be made to link dust coagulation models with astrochemistry. 

As schematically shown in Fig.\,\ref{fig:4-2}, in the midplane the dust settles and coagulates with its thick icy mantles and larger grains tend to settle first (e.g. \cite{2007A&A...473..457D}). The differential dust settling and the presence of some degree of turbulence mixing, maintains a population of small dust grains in the upper layers of the disk (see also \cite{1999ApJ...527..893D}). This includes polycyclic aromatic hydrocarbons (PAHs), ubiquitous in active star-forming regions (\cite{2005pcim.book.....T}) and also present in protoplanetary disks, especially around the intermediate-mass Herbig Ae/Be stars (e.g. \cite{2004A&A...427..179H}; \cite{2004A&A...426..151A}; \cite{2008ApJ...684..411K}; see also \cite{2011EAS....46..271K} for a recent review of PAH in disks). PAH features have been detected in only 8\% of the less massive T\,Tauri stars (\cite{2006A&A...459..545G}).  PAHs are not only important from an organic and pre-biotic chemistry point of view, but also for the physical structure of disks, as they can be photoionized, releasing energetic photons which heat the gas, thus maintaining flared disk structures (\cite{2011EAS....46..271K}). Moreover, PAHs boost the formation of H$_2$ molecules (\cite{2004A&A...414..531H}), thus the atomic-to-molecular transition in the upper disk atmospheres. Habart et al. (\cite{2006A&A...449.1067H}) spatially resolved the 3.3\,$\mu$m PAH feature toward Herbig Ae/Be stars, finding that the emission originates from within 30\,AU of the star. In T\,Tauri stars, the less intense stellar UV field makes the detection of PAH features more difficult (as PAH features are excited by photons).  Visser et al. (\cite{2007A&A...466..229V}) predict that PAHs in T\,Tauri disks can survive much closer to the star (down to about 0.01\,AU for a 50-carbon atoms PAH)  compared to the Herbig disks (down to 5\,AU for PAHs with 96 carbon atoms). However, \cite{2010A&A...511A...6S} include the effects of extreme UV and X-ray components in their models and find very efficient PAH destruction also in T\,Tauri stars; by taking into account typical X-ray luminosities,  \cite{2012A&A...543A..25S} are able to reproduce the different PAH detection probabilities observed in T Tauri and Herbig Ae disks. Fedele et al. (\cite{2008A&A...491..809F}) found PAH emission co-spatial with the [OI]63\,$\mu$m line, i.e. in the photon-dominated zone of the disk of a Herbig star. As UV photons can break the weaker C-H bonds in PAHs and their carbon skeleton can also brake above a certain threshold of energy intake (\cite{1989ApJ...345..230G}),  the presence of PAHs in the upper atmosphere of disks hints at some replenishing mechanism, possibly vertical mixing (\cite{2012A&A...543A..25S}), which maintains a population of small grains mixed with the gas (\cite{2007A&A...473..457D}). Habart et al. (\cite{2004A&A...427..179H}) suggest that the observed PAHs are evaporated from the icy grain mantles within the disk, while others consider them as the result of fragmentation of larger grains (\cite{2006ApJ...646..288R}). The mixing of PAHs within the icy mantles of dust grains, could provide an interesting starting point for the formation of more complex molecules, once dust grains start to coagulate and form larger bodies (\cite{2011A&A...525A..93B}; \cite{2011A&A...529A..46B}).

\subsection{From debris to icy worlds} 

\begin{figure}[ht]
% Use the relevant command to insert your figure file.
% For example, with the graphicx package use
  \includegraphics[angle=0,width=1\textwidth]{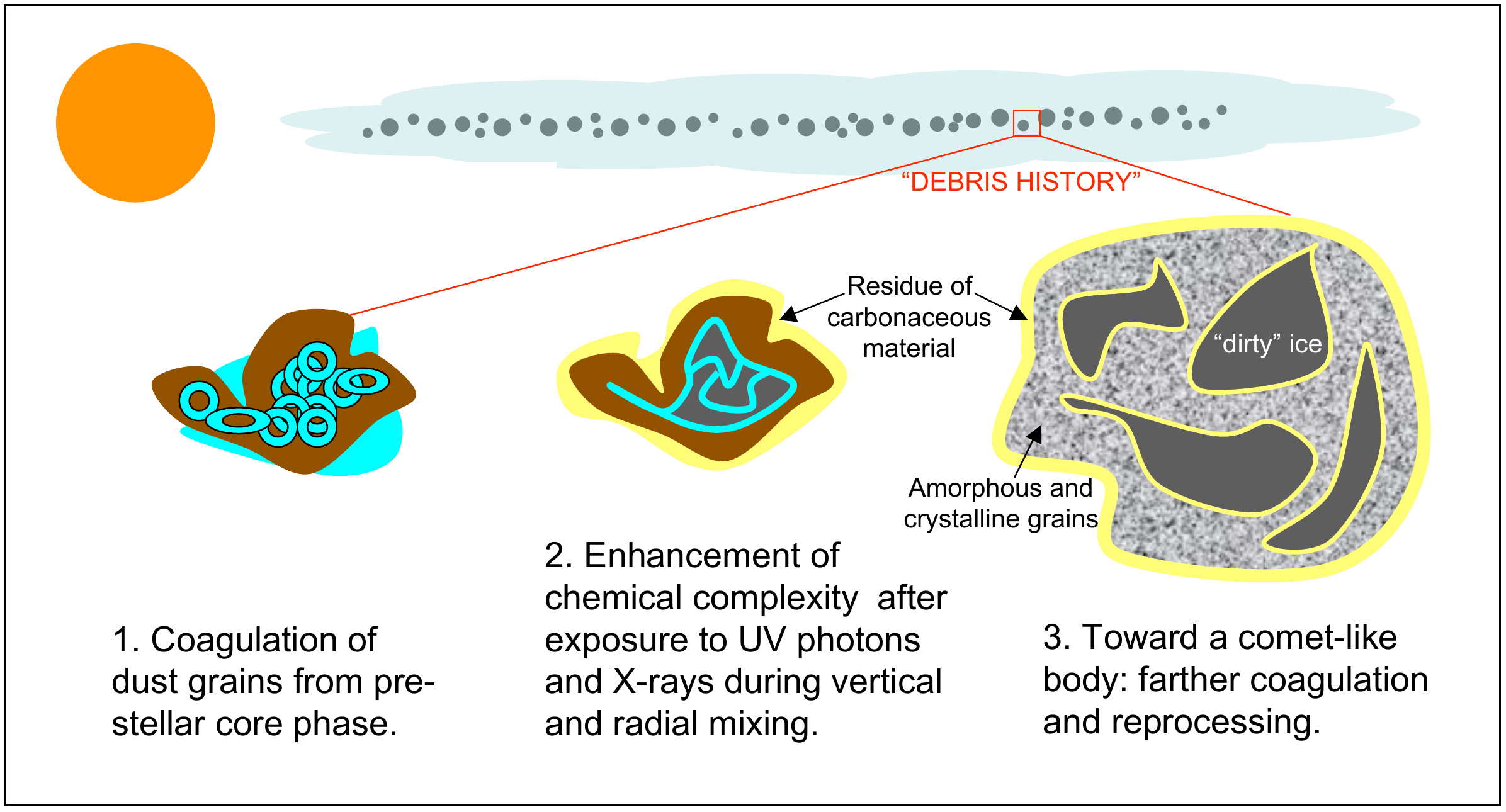}
% figure caption is below the figure
\caption{Sketch of a debris disk and a speculative history of a debris: 1. Dust grains with their icy mantles (blue rings) coagulate and form a fluffy structure, on top of which more ice can adsorb/form; 2. Exposure to UV photons and X-rays changes the inner structure of the grain and allows the surface icy mantle to be reprocesses and form a refractory carbonaceous material; 3. Further coagulation will lead to small rocks, composed by a mixture of the fluffy grains in 1 \& 2 with their refractory organic material and "dirty" ice, glued together by a mixture of amorphous and crystalline dust material.}
\label{fig:4-3}       % Give a unique label
\end{figure}

The transition between protoplanetary disks and planetary system is far from being understood. As \cite{2011ARA&A..49...67W} pointed out, "exactly how and when protoplanetary disks evolve into planetary debris disks remains an open question". In protoplanetary disks, there is plenty of evidence of dust grain growth (\cite{1990AJ.....99..924B}; \cite{2001ApJ...553..321D}; \cite{2002ApJ...567.1183W}; \cite{2003A&A...403..323T}; \cite{2005ApJ...626L.109W}; \cite{2007ApJ...659..705A};  \cite{2006A&A...451..951I}; \cite{2009ApJ...697.1305C}; \cite{2010A&A...515A..77L}; \cite{2011ApJ...736..135L}), dust settling (\cite{2003A&A...400..559D}; \cite{2005AJ....129..935C}; \cite{2006ApJ...638..314D}; \cite{2006ApJS..165..568F}), dust processing (e.g. presence of crystalline silicates; \cite{2005ApJ...622..404K}; \cite{2007prpl.conf..767N}; \cite{2009ApJS..182..477S}; \cite{2007ApJ...661..361M}; \cite{2012A&A...542A..90O}; \cite{2012MNRAS.420.2603R}),  inner holes (probably carved by a planet or by photoevaporation, \cite{2005ApJ...630L.185C}; \cite{2009ApJ...698..131H}; \cite{2009ApJ...700.1502A}; \cite{2011ApJ...732...42A}; \cite{2012ApJ...752...75C}).  Debris disks are also observed, with their poor gas content and with evidence of large grains and/or planets (\cite{2008ARA&A..46..339W}; \cite{2011ApJ...740...38H}; \cite{2012A&A...539L...6R}). Despite all these measurements, the story behind grain growth and planetesimal formation remains obscure (see also previous sub-section). For example, one of the biggest challenges for planet formation theories is the so-called "meter-size barrier", where models show destructive collisions and rapid inward migration of meter-sized solids (\cite{1977MNRAS.180...57W}; \cite{2011ARA&A..49...67W}). Nevertheless, the presence of large grains in protoplanetary disks and the structure of our Solar System tell us that dust grains coagulate and evolve toward rocks, comets, asteroids, planetesimals, planets and moons. There are connections between the petrology observed in protoplanetary disks and that in our Solar System bodies. In fact, crystalline grains detected in comets (\cite{1999ApJ...517.1034W}; \cite{2004ApJ...612L..77W}; \cite{2008M&PS...43..261Z}, who also suggest aqueous alterations in the comet P81/Wild2) are mostly made out of Mg-rich olivine grains, consistent with observations of gas-rich T\,Tauri disks. Fe-rich grains have been observed in several interplanetary dust particles (IDPs, e.g. \cite{2011Icar..212..896B}) and recently in warm debris disks (\cite{2012A&A...542A..90O}). Such Fe-rich grains may be due to a secondary alteration of the disk mineralogy (see also \cite{2007ApJ...656.1223N}), probably originated within large differentiated bodies (as in the case of the S-type asteroid recently studied with the Hayabusa re-entry module; \cite{2011Sci...333.1113N}).  In this scenario, planetesimals form with internal temperatures large enough (from the decay of short-lived radionuclides) to allow the melting and gravitational segregation of silica and metals.  Destructive collisions among these planetesimals would then contribute to the production of the Fe-rich particles found in IDPs and in warm debris disks and to the replenishment of small dust grains in our Solar System as well as in exo-zodiacal belts. 

Let us now retrace the history of a dust grain during the process of star and planet formation. The starting point has to be found within dense cores, where dust grains have thick icy mantles (see \S\ref{sec:calm-before-storm} and Fig.\,\ref{fig:2}) and show some evidence of coagulation (e.g. \cite{2010MNRAS.402.1625K}; \cite{2010Sci...329.1622P}), also found soon after protostellar birth, in Class 0 objects (\cite{2007ApJ...659..479J}; \cite{2009ApJ...696..841K}; \cite{2012ApJ...756..168C}).  As we have seen in previous sections, these dust mantles are rich in water and simple organic material and the chemical complexity in ices appears to increase with dynamical evolution.  Figure\,\ref{fig:4-3} show a schematic possible scenario of the formation of a debris in the late stages of evolution of protoplanetary disks: (1) soon after the formation of the protoplanetary disk, dust grains coagulate and become fluffy aggregates of the original icy dust grains. They may go through shocks during the early "stirring" of the embedded self-gravitating disks (Fig.\,\ref{fig:4-1}). (2) During the "naked" T\,Tauri phase, some vertical and radial mixing may expose the fluffy aggregate to stellar and interstellar UV photons and stellar X-rays, so that icy material on the surface can partially be photodesorbed and partially reprocessed, with the production of radicals and formation of an organic residue on the surface (the yellow layer in the figure) and formation of complex organic molecules  in the ice trapped within the aggregate.  (3) Further processing and coagulation (including some crystalline dust reformed in the inner parts of the disk) could then lead to a cometary-like body, where "dirty ice" (i.e. ice mixed with complex organic molecules) is a major component.

We are now ready to attempt assembling some pieces of the puzzle. 
 
 %%%%%%%%%%%%%%%%%%%% SESSION x %%%%%%%%%%%%%%%%%%%%%%%%%%%%%
\section{Putting together some pieces of the puzzle}\label{sec:assembl-some-piec}

\subsection{Molecules in comets and solar-type protostars}\label{sec:molec-comets-solar}

\begin{figure*}
  \includegraphics[width=\textwidth]{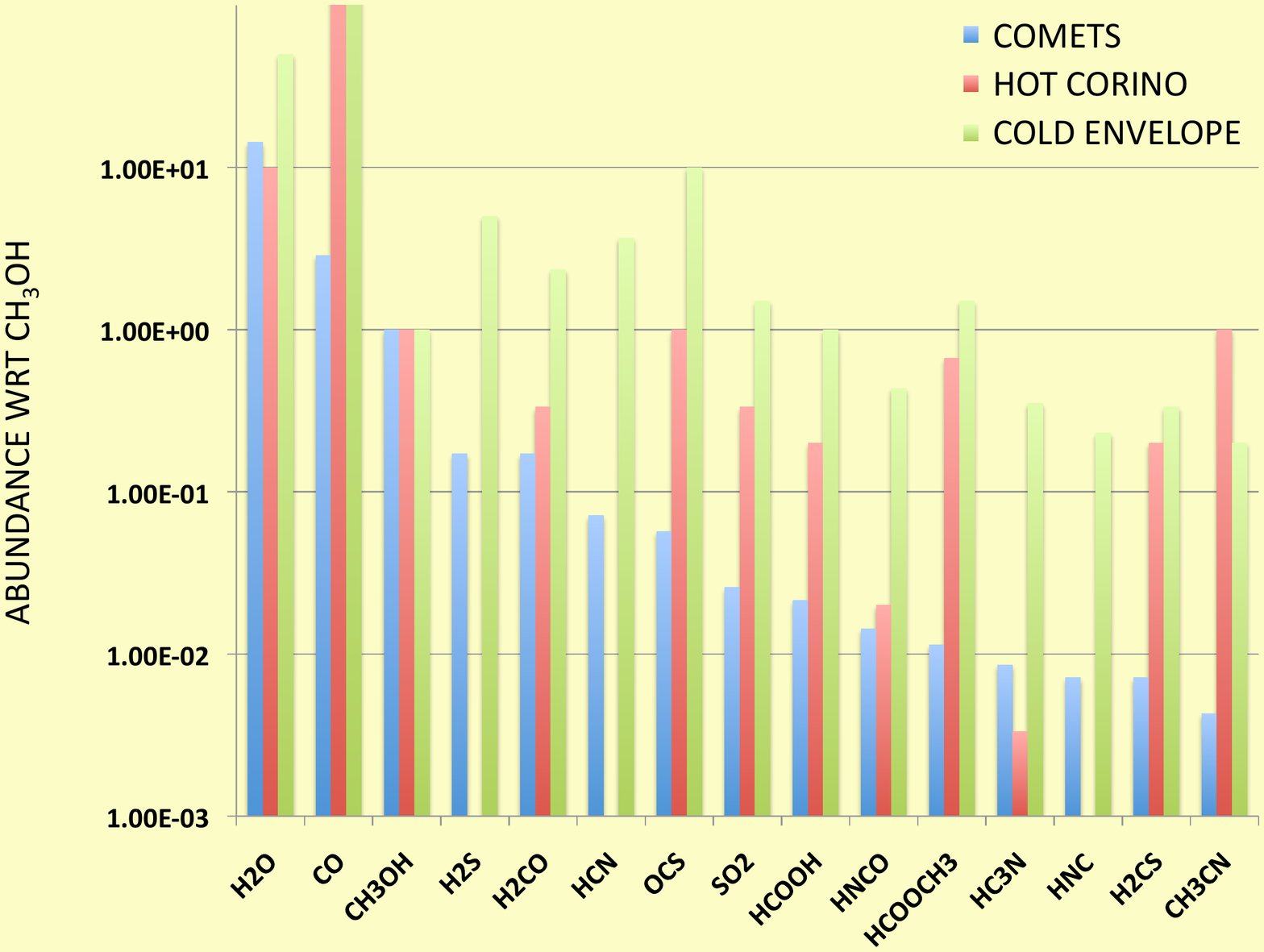}
  \caption{Abundances (with respect to CH$_3$OH) of molecules detected
    in comets (blue) and in the hot corino (red) and cold envelope
    (green) of IRAS16293-2422. References: for comets
    \cite{2011ARA&A..49..471M} and\cite{2011IAUS..280..261B}; for
    IRAS16293-2422 \cite{2000A&A...357L...9C},
    \cite{2002A&A...390.1001S} and \cite{2003ApJ...593L..51C}.}
\label{fig:comets-hotcorino}
\end{figure*}

All molecules detected in comets are also observed in star-forming
regions. However, the measured abundances in comets and Sun-like star
formation regions are not the same. This is clearly shown in
Fig. \ref{fig:comets-hotcorino}, which reports the abundances,
normalized to the methanol abundance, of species detected in various
comets (see the reviews \cite{2011ARA&A..49..471M}
and \cite{2011IAUS..280..261B}, and \S \ref{sec:comets-kbo}) and those
in the hot corino and cold envelope of IRAS16293-2422 (see \S
\ref{sec:prot-chem-comp}). A similar plot is obtained also if the
normalisation is done with respect to water rather than methanol. In
general, species are more abundant with respect to methanol (and
water) in IRAS16293-2422, both in the cold envelope and the hot
corino, than in comets by more than a factor of ten. In other words,
the chemistry in comets seems to be less rich than in both the cold
envelope and the hot corino of IRAS16293.
It is, therefore, probably fortuitous the rough correlation found in
the abundance of a fewer molecules in comets and hot cores
(e.g. \cite{2011IAUS..280..261B}).

Where does this difference come from? The molecules in the cold envelope
of IRAS16293-2422 are likely the product of gas-phase chemistry (but
see the comments in \S \ref{sec:calm-before-storm}) in cold gas, where
CO is largely frozen into the grain mantles. Therefore, the systematic
difference between the molecular abundances in comets and the cold
envelope may point to different physical conditions, likely warmer, at
the time of the comet formation. Similarly, the molecules in the hot
corino are thought to mostly reflect the composition of the grain
mantles during the pre-collapse phase, so that the difference in this
case also suggests warmer conditions of the material when the cometary
ices were formed. There are, however, also other possibilities. It is
possible that the cometary ices have undergone a massive reprocessing
of the molecular composition due to the long irradiation from cosmic
rays and solar wind particles and UV irradiation. Or it is possible
that our Sun's progenitor, in fact, did not resemble the IRAS16293-2422
protostar, which is rather isolated, whereas the proto-Sun likely was
born in a crowded and much harsher environment (\S
\ref{sec:violent-start}).  Our census of the molecular composition in
comets and in protostellar objects thought to be similar to the
proto-Sun is still too poor to have a definitive answer.
%CC: Paola, I would reemphasise this point in the conclusive remarks

\subsection{Origin of deuterated molecules in comets and chondrites}\label{sec:orig-deut-molec}

\begin{figure*}
  \includegraphics[width=\textwidth]{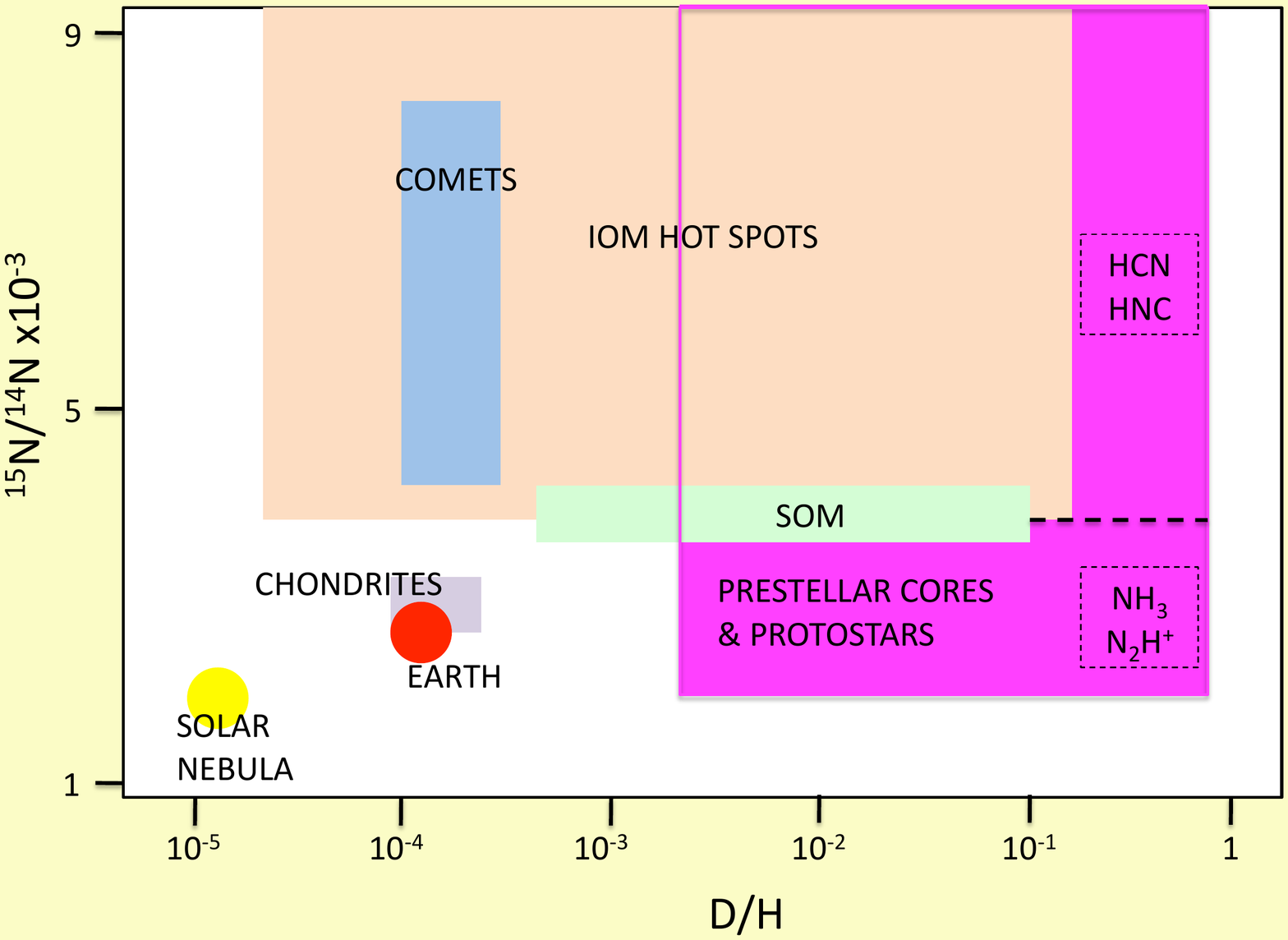}
  \caption{$^{15}$N/$^{14}$N versus D/H in comets, chondrites, hot
    spots in the IOM of meteorites and IDPs, SOM in meteorites, Earth,
    Solar Nebula and pre-stellar cores and protostellar envelopes.}
\label{fig:ism-solsystem}
\end{figure*}

For a long period it has been though that there is a link between the
chemistry in comets, chondrites and interstellar medium, especially
because of the enhanced abundance of deuterated molecules
(Fig. \ref{fig:ism-solsystem}). It is possible that the
link is not direct, meaning that it may not be due to the passage of the
molecular deuteration from one phase to the next, during the formation of 
the Solar System. However, the link certainly exists
because the chemistry regulating the molecular deuteration is
common to all phases and it has to do with the low temperatures
occurring during the star and planet formation.

Two key parameters play a major role in the molecular deuteration,
regardless of the details which depend on the specific molecule: the
ratios of H$_2$D$^+$/H$^+_3$ and of the atomic D/H in the gas (\S
\ref{freeze} and \S \ref{sec:deut-other-spec}). 
\begin{figure*}
  \includegraphics[width=\textwidth]{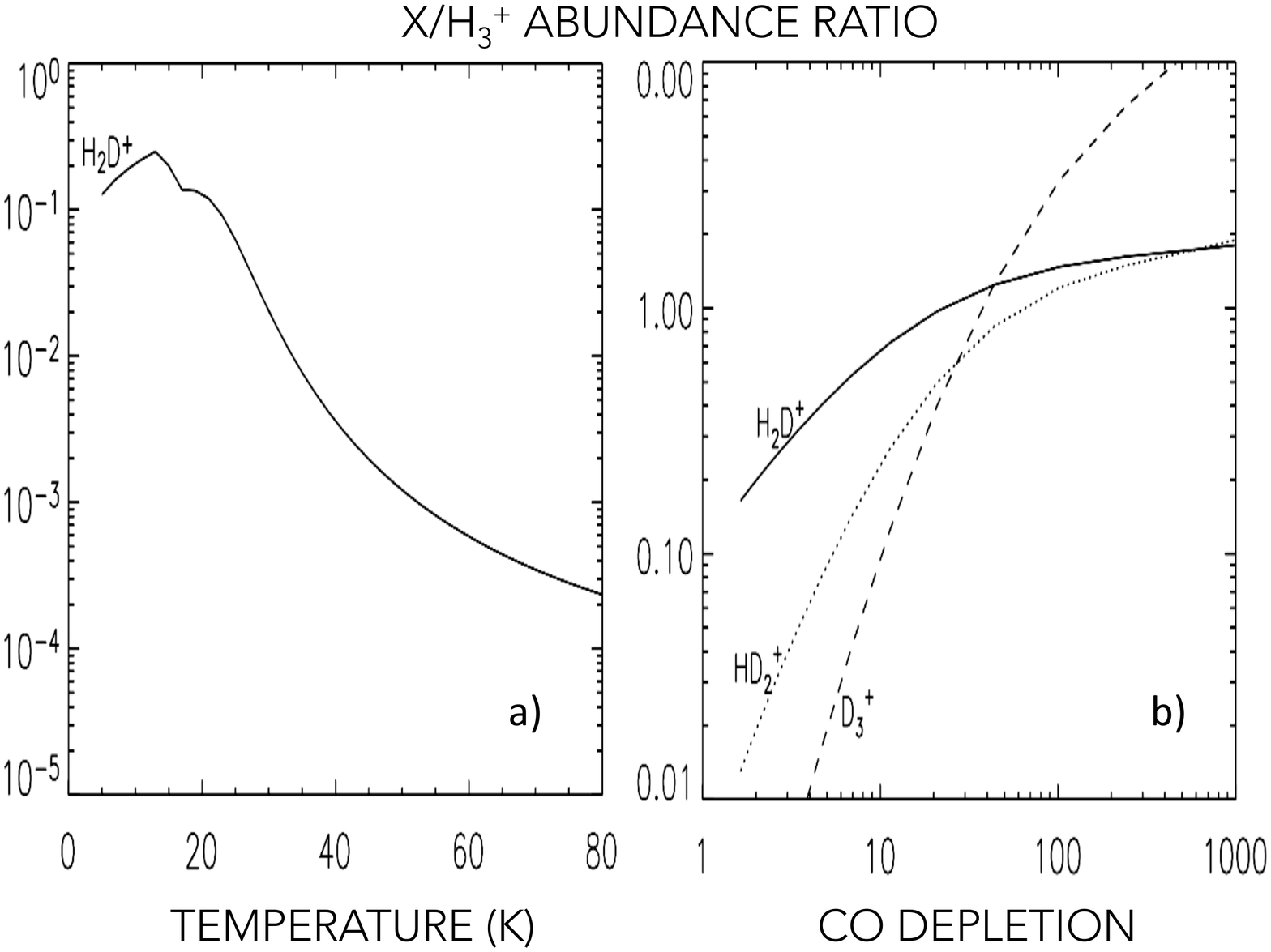}
  \caption{Theoretical abundances of the H$^+_3$ isotopologues in the
    gas. The left panel (a) reports the plot of the ratio
    H$_2$D$^+$/H$^+_3$ as function of the gas temperature, for gas
    with no CO depletion. The right panel (b) shows the abundance ratio of
    all the H$^+_3$ isotopologues as a function of CO depletion, for
    a gas with temperature 10 K. In both cases, the gas density is
    assumed $10^5$ cm$^{-3}$ (adapted from Ceccarelli \& Dominik
    2005).}
\label{fig:h2dp}
\end{figure*}
Figure \ref{fig:h2dp} shows how the H$_2$D$^+$/H$^+_3$ ratio depends
on the gas temperature. Another important parameter for this ratio is
the abundance of gaseous CO, as CO is a major destroyer of molecular
ions, being the most abundant heavy-atom-bearing neutral molecule. In
cold and dense regions, CO may freeze-out onto the grain mantles and
disappear, therefore, from the gas-phase (\S \ref{freeze}).  Figure
\ref{fig:h2dp} also shows the dependence of the
H$_2$D$^+$/H$^+_3$ and the other isotopologues of H$^+_3$ as a
function of the CO depletion, namely how much the CO abundance is
reduced with respect to the standard molecular cloud value. In general,
molecular deuteration exceeding 10\% requires not only cold gas but
also a substantial drop of the CO abundance in the gas-phase.

\begin{figure*}
  \includegraphics[width=\textwidth]{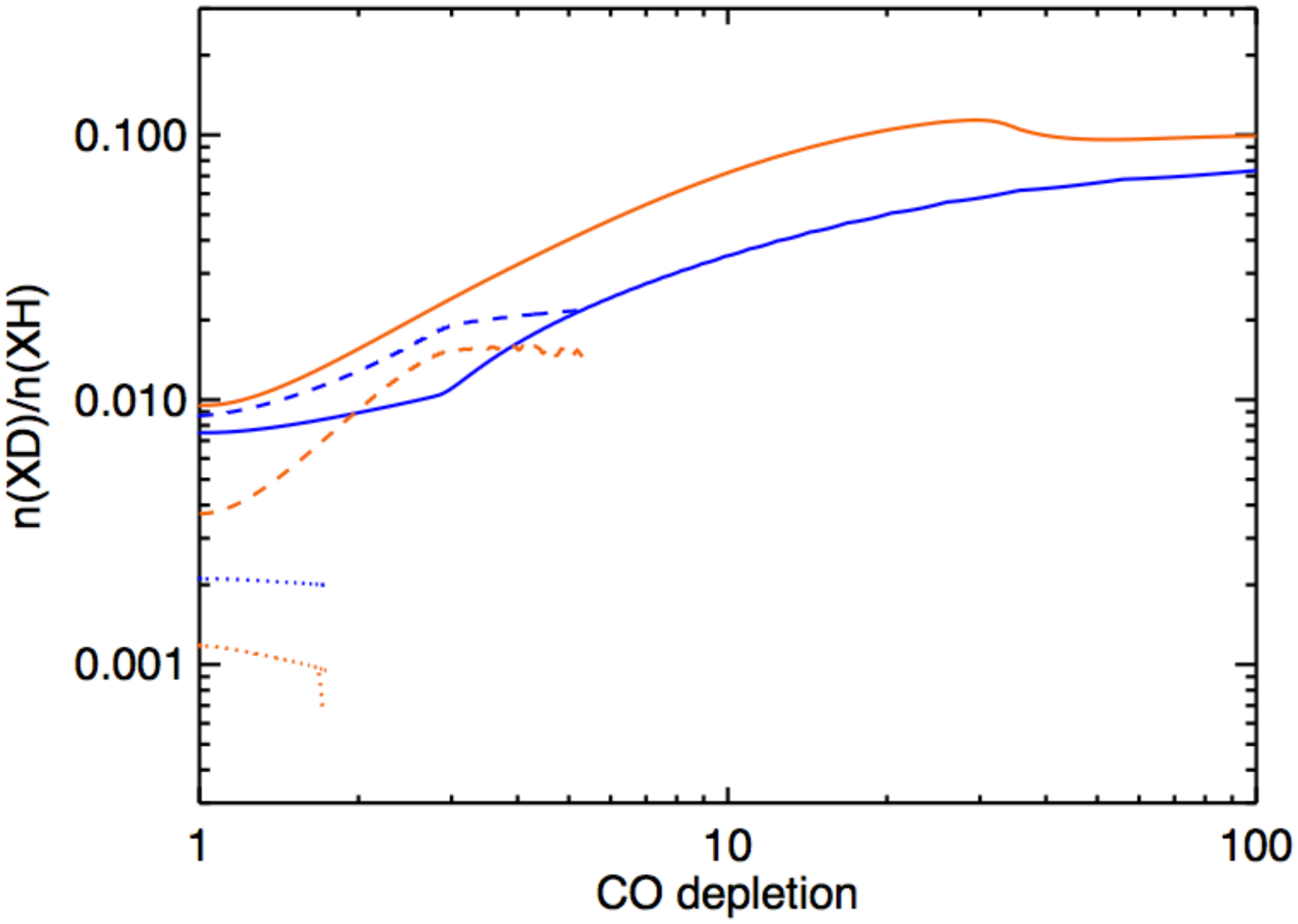}
  \caption{Theoretical abundances of the HDO/H$_2$O (blue curves) and
    D/H (orange curves) as a function of the CO depletion, for three
    different densities and incident visual extinction. Dotted lines:
    $10^4$ cm$^{-3}$ and 2 mag; dashed lines: $10^4$ cm$^{-3}$ and 4
    mag; solid lines: $10^5$ cm$^{-3}$ and 10 mag. Courtesy of
    (\cite{2012A&A...538A..42T}).}
\label{fig:DoverH}
\end{figure*}

Similarly, Fig. \ref{fig:DoverH} shows how the gaseous D/H ratio,
which governs the molecular deuteration of grain-surface product
molecules (\S \ref{sec:ice-formation}), varies with the CO depletion in
different situations. Also in this case, large ($\geq 10\%$) molecular
deuteration can only be achieved in cold gas deprived of CO.

Therefore, the relatively low water deuteration measured on
Earth, in comets and in chondrites with respect to the values measured
in the pre-stellar and protostellar phases suggest 
substantial remixing of water ices since their first
formation. How much of the first ices remains in the terrestrial,
cometary and chondritic water is difficult to say but cannot be
substantial. On the contrary, the extremely large deuteration found in
the chondritic organic material, both soluble and insoluble,
testifies either the preservation of molecular species since the
very first stages (where not only the temperature was very low but also
the gas was deprived of CO, \S \ref{freeze}), or the presence of
similar conditions in some zones of the Solar System up to a late
stage of the protoplanetary disk phase.

\subsection{The $^{15}$N-enrichment in comets and chondrites}\label{sec:15n-enrichm-comets}

In our Solar System, the $^{15}$N enhancement spans a large range of
values. As seen in \S\ref{sec:isotopic-anomalies}, $^{14}$N/$^{15}$N
is around 150 in comets, $<$300 in "primitive" material (such as IDPs
and carbonaceous chondrites), $<$100 in interplanetary dust particles
(IDPs) and carbonaceous chondrites "hotspots", where the largest D-
and $^{15}$N fractions have been measured
(e.g. \cite{2000Natur.404..968M}), $\sim$450 in Jupiter's atmosphere
(representative of the protosolar value), 272 on Earth. In pre-stellar
cores (\S\ref{sec:calm-before-storm}), current data show a
differential $^{15}$N-fractionation between amine- and nitrile-bearing
species, with the largest $^{15}$N enhancement found in the latter
(between 70 and 380 in HCN and HNC; \cite{2012M&PSA..75.5226B}), while
no significant enhancement is found in NH$_3$ and N$_2$H$^+$, both
highly D-fractionated. The situation is summarised in
Fig. \ref{fig:ism-solsystem}. Thus, D- and $^{15}$N-fractionations do
not go hand in hand in all species within pre-stellar cores, resembling
the mixed level of correlation between D- and $^{15}$N-enrichments in
primitive material of our Solar System.  Moreover, the cometary
$^{15}$N-enhancement has been measured in HCN and CN, again suggesting
a pre-stellar core origin. It will be interesting to find out if amines
will ever experience significant $^{15}$N at all during the pre-
and/or proto- stellar phase.  Wirstr{\"o}m et
al. (\cite{2012ApJ...757L..11W}) predict large $^{15}$N-fractionation
in NH$_3$ at late times ($>$a few million years), so maybe only
relatively long-lived dense cores will have the chance to have both
amines and nitriles highly $^{15}$N-fractionated.  However, this has
all to be tested with observations, which should be extended also to
the starless cores found in massive star-forming regions
(e.g. \cite{2012A&A...544A.146W}) to check if environmental conditions
affect the fractionation process. As $^{15}$N-enhancement has been
measured in IOMs and amino acids trapped in carbonaceous chondrites
(\S\ref{sec:isotopic-anomalies}), amines in pre-stellar cores should
also be able to experience significant enhancement, if a link between
these two extreme stages has to be found (\cite{2012M&PSA..75.5226B}).
However, we do not know if further processes within icy mantles or
within the "rocks" made out of coagulated icy dust grains, during the
proto-Sun stage, would affect the $^{15}$N-fractionation in IOM and
amino acids found in carbonaceous chondrites. Experiments are needed
here.

%%%%%%%%%%%%%%%%%%% both %%%%%%%%%%%%%%%%%%%%%%%%%%%%%%%%%%%%%%%%%%%%%
\section{Concluding remarks}  
\label{sec:conclusive-remarks}
Our chemical heritage is hidden in the large amount of information
obtained by observations of Solar System bodies and star and planet-forming regions, and it needs to be deciphered.  In this review,
glimpses of links between the present star and planet formation in our
Galaxy and the remote past of our Solar System have been given.  A
summary of these glimpses is given here, together with comments/open
questions and suggestions for future developments:
\begin{itemize}
\item [$\bullet$] {\it Water on Earth.} The total amount of water on
  Earth is  $\geq 2\times$10$^{-3}$ Earth masses, a small fraction of
  the total amount of water vapor measured in a pre-stellar core ($\simeq$800
  Earth masses) and in a protoplanetary disk ($\simeq$1.5 Earth
  masses), and a negligible factor of the deduced water ice mass in the pre-stellar core
  and protoplanetary disk (at least three orders of magnitude larger than the water vapor mass).
  Thus, a large water reservoir was originally available to
  seed a large number of Solar System bodies, as in fact observed in
  moons, comets, KBOs and asteroids. Tracing the formation, storage
  and delivery of water will require more observations as well as a
  better understanding of the icy-dust coagulation process during the
  pre-stellar and protoplanetary-disk phases.
\item [$\bullet$] {\it Complex organic molecules, COMs.} Organic
  material in meteorites and IDPs is organized in aromatic and
  aliphatic compounds, carboxylic acids and amino acids, including
  those found in all living beings on Earth. PAHs, hydrocarbons and
  complex organic molecules observed in star-forming regions are a
  simpler version (building blocks) of the organic material found in
  Solar System bodies. Thus, interstellar COMs may have contributed to
  the formation of organic matter during the processing of coagulated
  icy dust particles once the proto-Sun was born. More experiments are
  needed to confirm this statement.  Moreover, COMs observed in comets
  have abundances relative to methanol more than a factor of ten lower
  than those measured in hot corinos, possibly suggesting that the hot
  corino conditions and chemical history may not be representative of
  our Solar System. It is also possible that the difference arises
  because of a substantial reprocessing of the protostellar material
  during the protoplanetary disk phase. Finally, the chemical
  composition of the envelopes of solar-type protostars in crowded environments
  populated by massive stars, as suggested in the case of the
  proto-Sun, has not so far been studied, due to the sensitivity of
  the available instrumentation. The advent of ALMA should
   clarify this aspect in the near future.
\item [$\bullet$] {\it D-fractionation in water.} The HDO/H$_2$O
  abundance ratio measured in comets is between 1 and 2 times that
  measured in our oceans (1.5$\times$10$^{-4}$), whereas in cold
  envelopes and hot corinos a larger spread of this ratio has been
  found: from $\leq$ 6$\times$10$^{-4}$ to 0.2. Is the D-fraction in
  water set at the beginning of the star formation process or is it
  modified during the various phases of star and planet formation? 
  Or does the water D-enrichment observed in hot
  corinos probe only the outer layers of the ices, those that sublimate
  first, while the bulk of the ice, less deuterated as
  probably inherited in the previous phases, remains frozen and hidden?
  Is this the ice that we observe in comets? And which process forms
  water in comets if it is not inherited? Is it surface chemistry,
  like in pre-stellar objects, or gas-phase chemistry? In order to answer all
  these questions, we will need to measure the HDO/H$_2$O in different
  stages of the protostellar evolution and use observations to constrain detailed chemical 
  models, where gas-phase and surface processes are linked, spanning a broad range 
  of physical conditions.  
\item [$\bullet$] {\it D-fractionation in other molecules.} The
  D-fractionation is an active process in the cold gas of molecular
  clouds and becomes one of the dominant chemical processes within
  pre-stellar cores, in regions where abundant neutral species (mainly
  CO and O) freeze-out onto dust grains. Here, the increase of the D/H
  elemental ratio in the gas-phase is thought to be responsible for
  the efficient deuteration of methanol, which happens on the surface
  of dust grains (as no gas-phase routes are available). Organic
  molecules such as methanol and formaldehyde observed in star-forming
  regions (in particular toward low-mass protostellar envelopes) display a
  D-fraction orders of magnitude higher than that measured in
  water. This differential D-fractionation of water and organics is
  also measured in comets, where DCN/HCN $\sim$ 10 times
  HDO/H$_2$O. Similarly, in the hot spots of carbonaceous chondrites
  and IDPs, the D/H associated with organic radicals reaches values as
  high as 1\%, suggesting, in this case, a direct link with the pre-
  and proto-stellar phases of the Solar System's formation.
\item [$\bullet$] {\it $^{15}$N-fractionation.} $^{15}$N-enrichments
  in HCN and HNC are measured in comets, with values similar to those
  observed in pre-stellar cores and Galactic star-forming regions. No
  significant $^{15}$N-fractionation is measured in NH$_3$ and
  N$_2$H$^+$, which are, on the other hand, highly deuterated during
  the pre-stellar core phase. Thus, no significant correlation is expected
  between D- and $^{15}$N-fractionated material in Solar System
  bodies, as in fact measured. More observations of $^{15}$N
  isotopologues of NH$_3$ and N$_2$H$^+$ in a larger sample of
  pre-stellar cores (also including massive pre-stellar cores) are
  needed to look for possible fractionations of these species and to
  understand if different environmental conditions affect the
  fractionation process.
\end{itemize}

To further advance in this field, different communities need to join
the effort and work together. In particular, the star formation and
Solar System communities should continuously exchange new results and
information on new measurements, experiments and theoretical
developments. At the same time, laboratory work on molecular
spectroscopy to identify the observed lines, on rate coefficients to
understand chemical pathways, on surface chemistry and dust
coagulation to understand ice formation and ice/dust evolution, as
well as calculations of collisional coefficients required for
radiative transfer studies are all necessary for a correct
interpretation of observations. Finally, theoretical and observational
astrophysicists and astrochemists should work together to make sure
that, on the one hand, the best physical and dynamical model is used
as input for astrochemical modeling and, on the other hand, the best
physical parameters derived from the combination of observations,
astrochemistry and radiative transfer, are used as input in the
physical/dynamical models of star and planet-forming regions. To
understand our origins, we cannot work alone!

%%%%%%%%%%%%%%%%%%% References %%%%%%%%%%%%%%%%%%%%%%%%%%%%%%%%%%%%%%%%%%%%%
\begin{acknowledgements}
  We wish to thank our many colleagues for enlightening discussions over the years: L. Bizzocchi,
  A. Boley, S. Bottinelli, E. Caux, S. Cazaux, C. Codella,  A. Crapsi, L. Dore, T. Douglas, F. Fontani, E. Herbst, A. Faure, J. Henshaw,  P. Hily-Blant, J. Ilee, I. Jimenez-Serra, T. Hartquist, C. Kahane, E. Keto, B. Lelfoch, P. Myers, J. Pineda,  M. Spaans, M. Tafalla, V. Taquet, J. Tan, C. Vastel, M. Walmsley and L. Wiesenfeld.
  In addition, C.Ceccarelli acknowledges the stimulating discussions
  with M. Chaussidon, B. Marty and F. Robert on the meteorites.
 P. Caselli acknowledges the financial support of successive rolling grants awarded by the UK Science and Technology Funding Council.  C. Ceccarelli acknowledges the financial support from the
  Agence Nationale pour la Recherche (ANR), France (project FORCOMS,
  contracts ANR-08-BLAN-022, and the CNES (Centre National d'Etudes
  Spatiales) .

\end{acknowledgements}

% BibTeX users please use one of
%\bibliographystyle{spbasic}      % basic style, author-year citations
\bibliographystyle{spmpsci}      % mathematics and physical sciences
\bibliography{references}   % name your BibTeX data base

\begin{thebibliography}{100}
\providecommand{\url}[1]{{#1}}
\providecommand{\urlprefix}{URL }
\expandafter\ifx\csname urlstyle\endcsname\relax
  \providecommand{\doi}[1]{DOI~\discretionary{}{}{}#1}\else
  \providecommand{\doi}{DOI~\discretionary{}{}{}\begingroup
  \urlstyle{rm}\Url}\fi

\bibitem{2004A&A...426..151A}
{Acke}, B., {van den Ancker}, M.E.: {ISO spectroscopy of disks around Herbig
  Ae/Be stars}.
\newblock \aap \textbf{426}, 151--170 (2004).
\newblock \doi{10.1051/0004-6361:20040400}

\bibitem{2010ARA&A..48...47A}
{Adams}, F.C.: {The Birth Environment of the Solar System}.
\newblock \araa \textbf{48}, 47--85 (2010).
\newblock \doi{10.1146/annurev-astro-081309-130830}

\bibitem{2012ApJ...744..194A}
{Adande}, G.R., {Ziurys}, L.M.: {Millimeter-wave Observations of CN and HNC and
  Their $^{15}$N Isotopologues: A New Evaluation of the $^{14}$N/$^{15}$N Ratio
  across the Galaxy}.
\newblock \apj \textbf{744}, 194 (2012).
\newblock \doi{10.1088/0004-637X/744/2/194}

\bibitem{2012A&A...538A..57A}
{Aikawa}, Y., {Kamuro}, D., {Sakon}, I., {Itoh}, Y., {Terada}, H., {Noble},
  J.A., {Pontoppidan}, K.M., {Fraser}, H.J., {Tamura}, M., {Kandori}, R.,
  {Kawamura}, A., {Ueno}, M.: {AKARI observations of ice absorption bands
  towards edge-on young stellar objects}.
\newblock \aap \textbf{538}, A57 (2012).
\newblock \doi{10.1051/0004-6361/201015999}

\bibitem{2006ApJ...642.1152A}
{Aikawa}, Y., {Nomura}, H.: {Physical and Chemical Structure of Protoplanetary
  Disks with Grain Growth}.
\newblock \apj \textbf{642}, 1152--1162 (2006).
\newblock \doi{10.1086/501114}

\bibitem{1999ApJ...519..705A}
{Aikawa}, Y., {Umebayashi}, T., {Nakano}, T., {Miyama}, S.M.: {Evolution of
  Molecular Abundances in Protoplanetary Disks with Accretion Flow}.
\newblock \apj \textbf{519}, 705--725 (1999).
\newblock \doi{10.1086/307400}

\bibitem{2008ApJ...674..984A}
{Aikawa}, Y., {Wakelam}, V., {Garrod}, R.T., {Herbst}, E.: {Molecular Evolution
  and Star Formation: From Prestellar Cores to Protostellar Cores}.
\newblock \apj \textbf{674}, 984--996 (2008).
\newblock \doi{10.1086/524096}

\bibitem{2010ApJ...722.1342A}
{Ale@ARTICLE{2006MPSA..41.5327B, author = {{Busemann}, H. and {Alexander},
  C.~M.~O. and {Nittler}, L.~R. and {Zega}, T.~J. and {Stroud}, R.~M. and
  {Bajt}, S. and {Cody}, G.~D. and {Yabuta}, H.}, title = "{Correlated Analyses
  of D- and 15N-rich Carbon Grains from CR2 Chondrite EET 92042}", journal =
  {Meteoritics and Planetary Science Supplement}, year = 2006, month = sep,
  volume = 41, pages = {5327}, adsurl =
  {http://adsabs.harvard.edu/abs/2006M%26PSA..41.5327B}, adsnote = {Provided by
  the SAO/NASA Astrophysics Data System} } on}, J.: {Multiple Origins of
  Nitrogen Isotopic Anomalies in Meteorites and Comets}.
\newblock \apj \textbf{722}, 1342--1351 (2010).
\newblock \doi{10.1088/0004-637X/722/2/1342}

\bibitem{2007GeCoA..71.4380A}
{Alexander}, C.M.O.., {Fogel}, M., {Yabuta}, H., {Cody}, G.D.: {The origin and
  evolution of chondrites recorded in the elemental and isotopic compositions
  of their macromolecular organic matter}.
\newblock \gca \textbf{71}, 4380--4403 (2007).
\newblock \doi{10.1016/j.gca.2007.06.052}

\bibitem{1983Natur.303..762A}
{Allegre}, C.J., {Staudacher}, T., {Sarda}, P., {Kurz}, M.: {Constraints on
  evolution of earth's mantle from rare gas systematics}.
\newblock \nat \textbf{303}, 762--766 (1983).
\newblock \doi{10.1038/303762a0}

\bibitem{2003ApJ...599..363A}
{Allen}, A., {Li}, Z.Y., {Shu}, F.H.: {Collapse of Magnetized Singular
  Isothermal Toroids. II. Rotation and Magnetic Braking}.
\newblock \apj \textbf{599}, 363--379 (2003).
\newblock \doi{10.1086/379243}

\bibitem{2007ApJ...659..705A}
{Andrews}, S.M., {Williams}, J.P.: {High-Resolution Submillimeter Constraints
  on Circumstellar Disk Structure}.
\newblock \apj \textbf{659}, 705--728 (2007).
\newblock \doi{10.1086/511741}

\bibitem{2011ApJ...732...42A}
{Andrews}, S.M., {Wilner}, D.J., {Espaillat}, C., {Hughes}, A.M., {Dullemond},
  C.P., {McClure}, M.K., {Qi}, C., {Brown}, J.M.: {Resolved Images of Large
  Cavities in Protoplanetary Transition Disks}.
\newblock \apj \textbf{732}, 42 (2011).
\newblock \doi{10.1088/0004-637X/732/1/42}

\bibitem{2009ApJ...700.1502A}
{Andrews}, S.M., {Wilner}, D.J., {Hughes}, A.M., {Qi}, C., {Dullemond}, C.P.:
  {Protoplanetary Disk Structures in Ophiuchus}.
\newblock \apj \textbf{700}, 1502--1523 (2009).
\newblock \doi{10.1088/0004-637X/700/2/1502}

\bibitem{2011ApJ...742..105A}
{Arce}, H.G., {Borkin}, M.A., {Goodman}, A.A., {Pineda}, J.E., {Beaumont},
  C.N.: {A Bubbling Nearby Molecular Cloud: COMPLETE Shells in Perseus}.
\newblock \apj \textbf{742}, 105 (2011).
\newblock \doi{10.1088/0004-637X/742/2/105}

\bibitem{2008ApJ...681L..21A}
{Arce}, H.G., {Santiago-Garc{\'{\i}}a}, J., {J{\o}rgensen}, J.K., {Tafalla},
  M., {Bachiller}, R.: {Complex Molecules in the L1157 Molecular Outflow}.
\newblock \apjl \textbf{681}, L21--L24 (2008).
\newblock \doi{10.1086/590110}

\bibitem{2012arXiv1209.0591A}
{Aresu}, G., {Meijerink}, R., {Kamp}, I., {Spaans}, M., {Thi}, W.F., {Woitke},
  P.: {FUV and X-ray irradiated protoplanetary disks: a grid of models II - Gas
  diagnostic line emission}.
\newblock ArXiv e-prints  (2012)

\bibitem{2011ARA&A..49..195A}
{Armitage}, P.J.: {Dynamics of Protoplanetary Disks}.
\newblock \araa \textbf{49}, 195--236 (2011).
\newblock \doi{10.1146/annurev-astro-081710-102521}

\bibitem{2003Sci...301.1522A}
{Arpigny}, C., {Jehin}, E., {Manfroid}, J., {Hutsem{\'e}kers}, D., {Schulz},
  R., {St{\"u}we}, J.A., {Zucconi}, J.M., {Ilyin}, I.: {Anomalous Nitrogen
  Isotope Ratio in Comets}.
\newblock Science \textbf{301}, 1522--1525 (2003).
\newblock \doi{10.1126/science.1086711}

\bibitem{1986ApJ...303..356A}
{Arquilla}, R., {Goldsmith}, P.F.: {A detailed examination of the kinematics of
  rotating dark clouds}.
\newblock \apj \textbf{303}, 356--374 (1986).
\newblock \doi{10.1086/164082}

\bibitem{2010MNRAS.407.2511A}
{Awad}, Z., {Viti}, S., {Collings}, M.P., {Williams}, D.A.: {Warm cores around
  regions of low-mass star formation}.
\newblock \mnras \textbf{407}, 2511--2518 (2010).
\newblock \doi{10.1111/j.1365-2966.2010.17077.x}

\bibitem{1997ApJ...487L..93B}
{Bachiller}, R., {Perez Gutierrez}, M.: {Shock Chemistry in the Young Bipolar
  Outflow L1157}.
\newblock \apjl \textbf{487}, L93 (1997).
\newblock \doi{10.1086/310877}

\bibitem{2002A&A...389L...6B}
{Bacmann}, A., {Lefloch}, B., {Ceccarelli}, C., {Castets}, A., {Steinacker},
  J., {Loinard}, L.: {The degree of CO depletion in pre-stellar cores}.
\newblock \aap \textbf{389}, L6--L10 (2002).
\newblock \doi{10.1051/0004-6361:20020652}

\bibitem{2003ApJ...585L..55B}
{Bacmann}, A., {Lefloch}, B., {Ceccarelli}, C., {Steinacker}, J., {Castets},
  A., {Loinard}, L.: {CO Depletion and Deuterium Fractionation in Prestellar
  Cores}.
\newblock \apjl \textbf{585}, L55--L58 (2003).
\newblock \doi{10.1086/374263}

\bibitem{2012A&A...541L..12B}
{Bacmann}, A., {Taquet}, V., {Faure}, A., {Kahane}, C., {Ceccarelli}, C.:
  {Detection of complex organic molecules in a prestellar core: a new challenge
  for astrochemical models}.
\newblock \aap \textbf{541}, L12 (2012).
\newblock \doi{10.1051/0004-6361/201219207}

\bibitem{2005AJ....130.1291B}
{Barucci}, M.A., {Belskaya}, I.N., {Fulchignoni}, M., {Birlan}, M.: {Taxonomy
  of Centaurs and Trans-Neptunian Objects}.
\newblock \aj \textbf{130}, 1291--1298 (2005).
\newblock \doi{10.1086/431957}

\bibitem{2011A&ARv..19...48B}
{Barucci}, M.A., {Dotto}, E., {Levasseur-Regourd}, A.C.: {Space missions to
  small bodies: asteroids and cometary nuclei}.
\newblock \aapr \textbf{19}, 48 (2011).
\newblock \doi{10.1007/s00159-011-0048-2}

\bibitem{1994ApJ...432..720B}
{Basu}, S., {Mouschovias}, T.C.: {Magnetic braking, ambipolar diffusion, and
  the formation of cloud cores and protostars. 1: Axisymmetric solutions}.
\newblock \apj \textbf{432}, 720--741 (1994).
\newblock \doi{10.1086/174611}

\bibitem{1990AJ.....99..924B}
{Beckwith}, S.V.W., {Sargent}, A.I., {Chini}, R.S., {Guesten}, R.: {A survey
  for circumstellar disks around young stellar objects}.
\newblock \aj \textbf{99}, 924--945 (1990).
\newblock \doi{10.1086/115385}

\bibitem{2012A&A...539L...3B}
{Benedettini}, M., {Busquet}, G., {Lefloch}, B., {Codella}, C., {Cabrit}, S.,
  {Ceccarelli}, C., {Giannini}, T., {Nisini}, B., {Vasta}, M., {Cernicharo},
  J., {Lorenzani}, A., {di Giorgio}, A.M.: {The CHESS survey of the L1157-B1
  shock: the dissociative jet shock as revealed by Herschel-PACS}.
\newblock \aap \textbf{539}, L3 (2012).
\newblock \doi{10.1051/0004-6361/201118732}

\bibitem{2000A&A...359..148B}
{Benedettini}, M., {Giannini}, T., {Nisini}, B., {Tommasi}, E., {Lorenzetti},
  D., {Di Giorgio}, A.M., {Saraceno}, P., {Smith}, H.A., {White}, G.J.: {The
  ISO spectroscopic view of the HH 24-26 region}.
\newblock \aap \textbf{359}, 148--158 (2000)

\bibitem{2002A&A...395..657B}
{Benedettini}, M., {Viti}, S., {Giannini}, T., {Nisini}, B., {Goldsmith}, P.F.,
  {Saraceno}, P.: {Comparing SWAS and ISO observations of water in outflows}.
\newblock \aap \textbf{395}, 657--662 (2002).
\newblock \doi{10.1051/0004-6361:20021303}

\bibitem{2006ApJ...653..792B}
{Bennett}, C.J., {Jamieson}, C.S., {Osamura}, Y., {Kaiser}, R.I.: {Laboratory
  Studies on the Irradiation of Methane in Interstellar, Cometary, and Solar
  System Ices}.
\newblock \apj \textbf{653}, 792--811 (2006).
\newblock \doi{10.1086/508561}

\bibitem{2007ApJ...661..899B}
{Bennett}, C.J., {Kaiser}, R.I.: {On the Formation of Glycolaldehyde
  (HCOCH$_{2}$OH) and Methyl Formate (HCOOCH$_{3}$) in Interstellar Ice
  Analogs}.
\newblock \apj \textbf{661}, 899--909 (2007).
\newblock \doi{10.1086/516745}

\bibitem{2003ApJ...591L.159B}
{Bergin}, E., {Calvet}, N., {D'Alessio}, P., {Herczeg}, G.J.: {The Effects of
  UV Continuum and Ly{$\alpha$} Radiation on the Chemical Equilibrium of T
  Tauri Disks}.
\newblock \apjl \textbf{591}, L159--L162 (2003).
\newblock \doi{10.1086/377148}

\bibitem{2004ApJ...614L.133B}
{Bergin}, E., {Calvet}, N., {Sitko}, M.L., {Abgrall}, H., {D'Alessio}, P.,
  {Herczeg}, G.J., {Roueff}, E., {Qi}, C., {Lynch}, D.K., {Russell}, R.W.,
  {Brafford}, S.M., {Perry}, R.B.: {A New Probe of the Planet-forming Region in
  T Tauri Disks}.
\newblock \apjl \textbf{614}, L133--L136 (2004).
\newblock \doi{10.1086/425865}

\bibitem{2007prpl.conf..751B}
{Bergin}, E.A., {Aikawa}, Y., {Blake}, G.A., {van Dishoeck}, E.F.: {The
  Chemical Evolution of Protoplanetary Disks}.
\newblock Protostars and Planets V pp. 751--766 (2007)

\bibitem{2002ApJ...570L.101B}
{Bergin}, E.A., {Alves}, J., {Huard}, T., {Lada}, C.J.: {N$_{2}$H$^{+}$ and
  C$^{18}$O Depletion in a Cold Dark Cloud}.
\newblock \apjl \textbf{570}, L101--L104 (2002).
\newblock \doi{10.1086/340950}

\bibitem{1999ApJ...512..724B}
{Bergin}, E.A., {Plume}, R., {Williams}, J.P., {Myers}, P.C.: {The Ionization
  Fraction in Dense Molecular Gas. II. Massive Cores}.
\newblock \apj \textbf{512}, 724--739 (1999).
\newblock \doi{10.1086/306791}

\bibitem{2007ARA&A..45..339B}
{Bergin}, E.A., {Tafalla}, M.: {Cold Dark Clouds: The Initial Conditions for
  Star Formation}.
\newblock \araa \textbf{45}, 339--396 (2007).
\newblock \doi{10.1146/annurev.astro.45.071206.100404}

\bibitem{2011A&A...527A..39B}
{Bergman}, P., {Parise}, B., {Liseau}, R., {Larsson}, B.: {Deuterated
  formaldehyde in {$\rho$} Ophiuchi A}.
\newblock \aap \textbf{527}, A39 (2011).
\newblock \doi{10.1051/0004-6361/201015012}

\bibitem{2002Natur.416..401B}
{Bernstein}, M.P., {Dworkin}, J.P., {Sandford}, S.A., {Cooper}, G.W.,
  {Allamandola}, L.J.: {Racemic amino acids from the ultraviolet photolysis of
  interstellar ice analogues}.
\newblock \nat \textbf{416}, 401--403 (2002)

\bibitem{2002EM&P...90..323B}
{Biver}, N., {Bockel{\'e}e-Morvan}, D., {Crovisier}, J., {Colom}, P., {Henry},
  F., {Moreno}, R., {Paubert}, G., {Despois}, D., {Lis}, D.C.: {Chemical
  Composition Diversity Among 24 Comets Observed At Radio Wavelengths}.
\newblock Earth Moon and Planets \textbf{90}, 323--333 (2002)

\bibitem{2010A&A...510L...5B}
{Bizzocchi}, L., {Caselli}, P., {Dore}, L.: {Detection of N$^{15}$NH$^{+}$ in
  L1544}.
\newblock \aap \textbf{510}, L5 (2010).
\newblock \doi{10.1051/0004-6361/200913835}

\bibitem{2012A&A...546A..29B}
{Bjerkeli}, P., {Liseau}, R., {Larsson}, B., {Rydbeck}, G., {Nisini}, B.,
  {Tafalla}, M., {Antoniucci}, S., {Benedettini}, M., {Bergman}, P., {Cabrit},
  S., {Giannini}, T., {Melnick}, G., {Neufeld}, D., {Santangelo}, G., {van
  Dishoeck}, E.F.: {H$_{2}$O line mapping at high spatial and spectral
  resolution. Herschel observations of the VLA 1623 outflow}.
\newblock \aap \textbf{546}, A29 (2012).
\newblock \doi{10.1051/0004-6361/201219776}

\bibitem{2011A&A...533A..80B}
{Bjerkeli}, P., {Liseau}, R., {Nisini}, B., {Tafalla}, M., {Benedettini}, M.,
  {Bergman}, P., {Dionatos}, O., {Giannini}, T., {Herczeg}, G., {Justtanont},
  K., {Larsson}, B., {McOey}, C., {Olberg}, M., {Olofsson}, A.O.H.: {Herschel
  observations of the Herbig-Haro objects HH 52-54}.
\newblock \aap \textbf{533}, A80 (2011).
\newblock \doi{10.1051/0004-6361/201116846}

\bibitem{2009A&A...507.1455B}
{Bjerkeli}, P., {Liseau}, R., {Olberg}, M., {Falgarone}, E., {Frisk}, U.,
  {Hjalmarson}, {\AA}., {Klotz}, A., {Larsson}, B., {Olofsson}, A.O.H.,
  {Olofsson}, G., {Ristorcelli}, I., {Sandqvist}, A.: {Odin observations of
  water in molecular outflows and shocks}.
\newblock \aap \textbf{507}, 1455--1466 (2009).
\newblock \doi{10.1051/0004-6361/200912064}

\bibitem{1987ApJ...315..621B}
{Blake}, G.A., {Sutton}, E.C., {Masson}, C.R., {Phillips}, T.G.: {Molecular
  abundances in OMC-1 - The chemical composition of interstellar molecular
  clouds and the influence of massive star formation}.
\newblock \apj \textbf{315}, 621--645 (1987).
\newblock \doi{10.1086/165165}

\bibitem{2011IAUS..280..261B}
{Bockel{\'e}e-Morvan}, D.: {An Overview of Comet Composition}.
\newblock In: IAU Symposium, \emph{IAU Symposium}, vol. 280, pp. 261--274
  (2011).
\newblock \doi{10.1017/S1743921311025038}

\bibitem{2008ApJ...679L..49B}
{Bockel{\'e}e-Morvan}, D., {Biver}, N., {Jehin}, E., {Cochran}, A.L.,
  {Wiesemeyer}, H., {Manfroid}, J., {Hutsem{\'e}kers}, D., {Arpigny}, C.,
  {Boissier}, J., {Cochran}, W., {Colom}, P., {Crovisier}, J., {Milutinovic},
  N., {Moreno}, R., {Prochaska}, J.X., {Ramirez}, I., {Schulz}, R., {Zucconi},
  J.M.: {Large Excess of Heavy Nitrogen in Both Hydrogen Cyanide and Cyanogen
  from Comet 17P/Holmes}.
\newblock \apjl \textbf{679}, L49--L52 (2008).
\newblock \doi{10.1086/588781}

\bibitem{2012A&A...544L..15B}
{Bockel{\'e}e-Morvan}, D., {Biver}, N., {Swinyard}, B., {de Val-Borro}, M.,
  {Crovisier}, J., {Hartogh}, P., {Lis}, D.C., {Moreno}, R., {Szutowicz}, S.,
  {Lellouch}, E., {Emprechtinger}, M., {Blake}, G.A., {Courtin}, R., {Jarchow},
  C., {Kidger}, M., {K{\"u}ppers}, M., {Rengel}, M., {Davis}, G.R., {Fulton},
  T., {Naylor}, D., {Sidher}, S., {Walker}, H.: {Herschel measurements of the
  D/H and $^{16}$O/$^{18}$O ratios in water in the Oort-cloud comet C/2009 P1
  (Garradd)}.
\newblock \aap \textbf{544}, L15 (2012).
\newblock \doi{10.1051/0004-6361/201219744}

\bibitem{1998Icar..133..147B}
{Bockelee-Morvan}, D., {Gautier}, D., {Lis}, D.C., {Young}, K., {Keene}, J.,
  {Phillips}, T., {Owen}, T., {Crovisier}, J., {Goldsmith}, P.F., {Bergin},
  E.A., {Despois}, D., {Wootten}, A.: {Deuterated Water in Comet C/1996 B2
  (Hyakutake) and Its Implications for the Origin of Comets}.
\newblock \icarus \textbf{133}, 147--162 (1998).
\newblock \doi{10.1006/icar.1998.5916}

\bibitem{2012A&A...544A..30B}
{Boduch}, P., {Domaracka}, A., {Fulvio}, D., {Langlinay}, T., {Lv}, X.Y.,
  {Palumbo}, M.E., {Rothard}, H., {Strazzulla}, G.: {Chemistry induced by
  energetic ions in water ice mixed with molecular nitrogen and oxygen}.
\newblock \aap \textbf{544}, A30 (2012).
\newblock \doi{10.1051/0004-6361/201219365}

\bibitem{2009ApJ...695L..53B}
{Boley}, A.C.: {The Two Modes of Gas Giant Planet Formation}.
\newblock \apjl \textbf{695}, L53--L57 (2009).
\newblock \doi{10.1088/0004-637X/695/1/L53}

\bibitem{2008ApJ...685.1193B}
{Boley}, A.C., {Durisen}, R.H.: {Gravitational Instabilities, Chondrule
  Formation, and the FU Orionis Phenomenon}.
\newblock \apj \textbf{685}, 1193--1209 (2008).
\newblock \doi{10.1086/591013}

\bibitem{2012M&PSA..75.5226B}
{Bonal}, L., {Hily-Blant}, P., {Faure}, A., {Quirico}, E.: {Highly Variable
  15N-Enrichments in Solar System Reflect Different Routes of Interstellar N
  Isotopic Fractionation}.
\newblock Meteoritics and Planetary Science Supplement \textbf{75}, 5226 (2012)

\bibitem{2010GeCoA..74.6590B}
{Bonal}, L., {Huss}, G.R., {Krot}, A.N., {Nagashima}, K., {Ishii}, H.A.,
  {Bradley}, J.P.: {Highly $^{15}$N-enriched chondritic clasts in the
  CB/CH-like meteorite Isheyevo}.
\newblock \gca \textbf{74}, 6590--6609 (2010).
\newblock \doi{10.1016/j.gca.2010.08.017}

\bibitem{2009M&PSA..72.5178B}
{Bonal}, L., {Huss}, G.R., {Nagashima}, K., {Krot}, A.N.: {Hydrogen Isotopic
  Composition of 15N-rich Clasts in the CB/CH-like Chondrite Isheyevo}.
\newblock Meteoritics and Planetary Science Supplement \textbf{72}, 5178 (2009)

\bibitem{1997Sci...276.1836B}
{Boss}, A.P.: {Giant planet formation by gravitational instability.}
\newblock Science \textbf{276}, 1836--1839 (1997).
\newblock \doi{10.1126/science.276.5320.1836}

\bibitem{2004ApJ...615..354B}
{Bottinelli}, S., {Ceccarelli}, C., {Lefloch}, B., {Williams}, J.P., {Castets},
  A., {Caux}, E., {Cazaux}, S., {Maret}, S., {Parise}, B., {Tielens}, A.G.G.M.:
  {Complex Molecules in the Hot Core of the Low-Mass Protostar NGC 1333 IRAS
  4A}.
\newblock \apj \textbf{615}, 354--358 (2004).
\newblock \doi{10.1086/423952}

\bibitem{2004ApJ...617L..69B}
{Bottinelli}, S., {Ceccarelli}, C., {Neri}, R., {Williams}, J.P., {Caux}, E.,
  {Cazaux}, S., {Lefloch}, B., {Maret}, S., {Tielens}, A.G.G.M.:
  {Near-Arcsecond Resolution Observations of the Hot Corino of the Solar-Type
  Protostar IRAS 16293-2422}.
\newblock \apjl \textbf{617}, L69--L72 (2004).
\newblock \doi{10.1086/426964}

\bibitem{2007A&A...463..601B}
{Bottinelli}, S., {Ceccarelli}, C., {Williams}, J.P., {Lefloch}, B.: {Hot
  corinos in NGC 1333-IRAS4B and IRAS2A}.
\newblock \aap \textbf{463}, 601--610 (2007).
\newblock \doi{10.1051/0004-6361:20066242}

\bibitem{2011A&A...529A..46B}
{Bouwman}, J., {Cuppen}, H.M., {Steglich}, M., {Allamandola}, L.J., {Linnartz},
  H.: {Photochemistry of polycyclic aromatic hydrocarbons in cosmic water ice.
  II. Near UV/VIS spectroscopy and ionization rates}.
\newblock \aap \textbf{529}, A46 (2011).
\newblock \doi{10.1051/0004-6361/201015762}

\bibitem{2011A&A...525A..93B}
{Bouwman}, J., {Mattioda}, A.L., {Linnartz}, H., {Allamandola}, L.J.:
  {Photochemistry of polycyclic aromatic hydrocarbons in cosmic water ice. I.
  Mid-IR spectroscopy and photoproducts}.
\newblock \aap \textbf{525}, A93 (2011).
\newblock \doi{10.1051/0004-6361/201015059}

\bibitem{2012MNRAS.422..261B}
{Braiding}, C.R., {Wardle}, M.: {The Hall effect in star formation}.
\newblock \mnras \textbf{422}, 261--281 (2012).
\newblock \doi{10.1111/j.1365-2966.2012.20601.x}

\bibitem{2008A&A...492..251B}
{Brasser}, R.: {A two-stage formation process for the Oort comet cloud and its
  implications}.
\newblock \aap \textbf{492}, 251--255 (2008).
\newblock \doi{10.1051/0004-6361:200810452}

\bibitem{1984ApJ...276L..25B}
{Broten}, N.W., {MacLeod}, J.M., {Avery}, L.W., {Irvine}, W.M., {Hoglund}, B.,
  {Friberg}, P., {Hjalmarson}, A.: {The detection of interstellar
  methylcyanoacetylene}.
\newblock \apjl \textbf{276}, L25--L29 (1984).
\newblock \doi{10.1086/184181}

\bibitem{2000MNRAS.319..154B}
{Brown}, D.W., {Chandler}, C.J., {Carlstrom}, J.E., {Hills}, R.E., {Lay}, O.P.,
  {Matthews}, B.C., {Richer}, J.S., {Wilson}, C.D.: {A submillimetre survey for
  protostellar accretion discs using the JCMT-CSO interferometer}.
\newblock \mnras \textbf{319}, 154--162 (2000).
\newblock \doi{10.1046/j.1365-8711.2000.03805.x}

\bibitem{2012AREPS..40..467B}
{Brown}, M.E.: {The Compositions of Kuiper Belt Objects}.
\newblock Annual Review of Earth and Planetary Sciences \textbf{40}, 467--494
  (2012).
\newblock \doi{10.1146/annurev-earth-042711-105352}

\bibitem{2012AJ....143..146B}
{Brown}, M.E., {Schaller}, E.L., {Fraser}, W.C.: {Water Ice in the Kuiper
  Belt}.
\newblock \aj \textbf{143}, 146 (2012).
\newblock \doi{10.1088/0004-6256/143/6/146}

\bibitem{2011Icar..212..896B}
{Brunetto}, R., {Borg}, J., {Dartois}, E., {Rietmeijer}, F.J.M., {Grossemy},
  F., {Sandt}, C., {Le Sergeant D'Hendecourt}, L., {Rotundi}, A., {Dumas}, P.,
  {Djouadi}, Z., {Jamme}, F.: {Mid-IR, Far-IR, Raman micro-spectroscopy, and
  FESEM-EDX study of IDP L2021C5: Clues to its origin}.
\newblock \icarus \textbf{212}, 896--910 (2011).
\newblock \doi{10.1016/j.icarus.2011.01.038}

\bibitem{2007ApJ...664L..43B}
{Br{\"u}nken}, S., {Gupta}, H., {Gottlieb}, C.A., {McCarthy}, M.C., {Thaddeus},
  P.: {Detection of the Carbon Chain Negative Ion C$_{8}$H$^{-}$ in TMC-1}.
\newblock \apjl \textbf{664}, L43--L46 (2007).
\newblock \doi{10.1086/520703}

\bibitem{2007ApJ...659L.137B}
{Butner}, H.M., {Charnley}, S.B., {Ceccarelli}, C., {Rodgers}, S.D., {Pardo},
  J.R., {Parise}, B., {Cernicharo}, J., {Davis}, G.R.: {Discovery of
  Interstellar Heavy Water}.
\newblock \apjl \textbf{659}, L137--L140 (2007).
\newblock \doi{10.1086/517883}

\bibitem{2005AJ....129..935C}
{Calvet}, N., {Brice{\~n}o}, C., {Hern{\'a}ndez}, J., {Hoyer}, S., {Hartmann},
  L., {Sicilia-Aguilar}, A., {Megeath}, S.T., {D'Alessio}, P.: {Disk Evolution
  in the Orion OB1 Association}.
\newblock \aj \textbf{129}, 935--946 (2005).
\newblock \doi{10.1086/426910}

\bibitem{2005ApJ...630L.185C}
{Calvet}, N., {D'Alessio}, P., {Watson}, D.M., {Franco-Hern{\'a}ndez}, R.,
  {Furlan}, E., {Green}, J., {Sutter}, P.M., {Forrest}, W.J., {Hartmann}, L.,
  {Uchida}, K.I., {Keller}, L.D., {Sargent}, B., {Najita}, J., {Herter}, T.L.,
  {Barry}, D.J., {Hall}, P.: {Disks in Transition in the Taurus Population:
  Spitzer IRS Spectra of GM Aurigae and DM Tauri}.
\newblock \apjl \textbf{630}, L185--L188 (2005).
\newblock \doi{10.1086/491652}

\bibitem{1977Icar...30..447C}
{Cameron}, A.G.W., {Truran}, J.W.: {The supernova trigger for formation of the
  solar system}.
\newblock \icarus \textbf{30}, 447--461 (1977).
\newblock \doi{10.1016/0019-1035(77)90101-4}

\bibitem{2008Sci...319.1504C}
{Carr}, J.S., {Najita}, J.R.: {Organic Molecules and Water in the Planet
  Formation Region of Young Circumstellar Disks}.
\newblock Science \textbf{319}, 1504-- (2008).
\newblock \doi{10.1126/science.1153807}

\bibitem{2002ApJ...572..238C}
{Caselli}, P., {Benson}, P.J., {Myers}, P.C., {Tafalla}, M.: {Dense Cores in
  Dark Clouds. XIV. N$_{2}$H$^{+}$ (1-0) Maps of Dense Cloud Cores}.
\newblock \apj \textbf{572}, 238--263 (2002).
\newblock \doi{10.1086/340195}

\bibitem{2012arXiv1208.5998C}
{Caselli}, P., {Keto}, E., {Bergin}, E.A., {Tafalla}, M., {Aikawa}, Y.,
  {Douglas}, T., {Pagani}, L., {Yildiz}, U.A., {van der Tak}, F.F.S.,
  {Walmsley}, C.M., {Codella}, C., {Nisini}, B., {Kristensen}, L.E., {van
  Dishoeck}, E.F.: {First detection of water vapor in a pre-stellar core}.
\newblock ArXiv e-prints  (2012)

\bibitem{2002P&SS...50.1257C}
{Caselli}, P., {Stantcheva}, T., {Shalabiea}, O., {Shematovich}, V.I.,
  {Herbst}, E.: {Deuterium fractionation on interstellar grains studied with
  modified rate equations and a Monte Carlo approach}.
\newblock \planss \textbf{50}, 1257--1266 (2002).
\newblock \doi{10.1016/S0032-0633(02)00092-2}

\bibitem{2003A&A...403L..37C}
{Caselli}, P., {van der Tak}, F.F.S., {Ceccarelli}, C., {Bacmann}, A.:
  {Abundant H$_{2}$D$^{+}$ in the pre-stellar core L1544}.
\newblock \aap \textbf{403}, L37--L41 (2003).
\newblock \doi{10.1051/0004-6361:20030526}

\bibitem{1999ApJ...523L.165C}
{Caselli}, P., {Walmsley}, C.M., {Tafalla}, M., {Dore}, L., {Myers}, P.C.: {CO
  Depletion in the Starless Cloud Core L1544}.
\newblock \apjl \textbf{523}, L165--L169 (1999).
\newblock \doi{10.1086/312280}

\bibitem{1998ApJ...499..234C}
{Caselli}, P., {Walmsley}, C.M., {Terzieva}, R., {Herbst}, E.: {The Ionization
  Fraction in Dense Cloud Cores}.
\newblock \apj \textbf{499}, 234 (1998).
\newblock \doi{10.1086/305624}

\bibitem{2002ApJ...565..344C}
{Caselli}, P., {Walmsley}, C.M., {Zucconi}, A., {Tafalla}, M., {Dore}, L.,
  {Myers}, P.C.: {Molecular Ions in L1544. II. The Ionization Degree}.
\newblock \apj \textbf{565}, 344--358 (2002).
\newblock \doi{10.1086/324302}

\bibitem{2011A&A...532A..23C}
{Caux}, E., {Kahane}, C., {Castets}, A., {Coutens}, A., {Ceccarelli}, C.,
  {Bacmann}, A., {Bisschop}, S., {Bottinelli}, S., {Comito}, C., {Helmich},
  F.P., {Lefloch}, B., {Parise}, B., {Schilke}, P., {Tielens}, A.G.G.M., {van
  Dishoeck}, E., {Vastel}, C., {Wakelam}, V., {Walters}, A.: {TIMASSS: the IRAS
  16293-2422 millimeter and submillimeter spectral survey. I. Observations,
  calibration, and analysis of the line kinematics}.
\newblock \aap \textbf{532}, A23 (2011).
\newblock \doi{10.1051/0004-6361/201015399}

\bibitem{2011ApJ...741L..34C}
{Cazaux}, S., {Caselli}, P., {Spaans}, M.: {Interstellar Ices as Witnesses of
  Star Formation: Selective Deuteration of Water and Organic Molecules
  Unveiled}.
\newblock \apjl \textbf{741}, L34 (2011).
\newblock \doi{10.1088/2041-8205/741/2/L34}

\bibitem{2010A&A...522A..74C}
{Cazaux}, S., {Cobut}, V., {Marseille}, M., {Spaans}, M., {Caselli}, P.: {Water
  formation on bare grains: When the chemistry on dust impacts interstellar
  gas}.
\newblock \aap \textbf{522}, A74 (2010).
\newblock \doi{10.1051/0004-6361/201014026}

\bibitem{2002ApJ...575L..29C}
{Cazaux}, S., {Tielens}, A.G.G.M.: {Molecular Hydrogen Formation in the
  Interstellar Medium}.
\newblock \apjl \textbf{575}, L29--L32 (2002).
\newblock \doi{10.1086/342607}

\bibitem{2003ApJ...593L..51C}
{Cazaux}, S., {Tielens}, A.G.G.M., {Ceccarelli}, C., {Castets}, A., {Wakelam},
  V., {Caux}, E., {Parise}, B., {Teyssier}, D.: {The Hot Core around the
  Low-mass Protostar IRAS 16293-2422: Scoundrels Rule!}
\newblock \apjl \textbf{593}, L51--L55 (2003).
\newblock \doi{10.1086/378038}

\bibitem{2007prpl.conf...47C}
{Ceccarelli}, C., {Caselli}, P., {Herbst}, E., {Tielens}, A.G.G.M., {Caux}, E.:
  {Extreme Deuteration and Hot Corinos: The Earliest Chemical Signatures of
  Low-Mass Star Formation}.
\newblock Protostars and Planets V pp. 47--62 (2007)

\bibitem{2000A&A...355.1129C}
{Ceccarelli}, C., {Castets}, A., {Caux}, E., {Hollenbach}, D., {Loinard}, L.,
  {Molinari}, S., {Tielens}, A.G.G.M.: {The structure of the collapsing
  envelope around the low-mass protostar IRAS 16293-2422}.
\newblock \aap \textbf{355}, 1129--1137 (2000)

\bibitem{1998A&A...338L..43C}
{Ceccarelli}, C., {Castets}, A., {Loinard}, L., {Caux}, E., {Tielens},
  A.G.G.M.: {Detection of doubly deuterated formaldehyde towards the
  low-luminosity protostar IRAS 16293-2422}.
\newblock \aap \textbf{338}, L43--L46 (1998)

\bibitem{2005ApJ...631L..81C}
{Ceccarelli}, C., {Dominik}, C., {Caux}, E., {Lefloch}, B., {Caselli}, P.:
  {Discovery of Deuterated Water in a Young Protoplanetary Disk}.
\newblock \apjl \textbf{631}, L81--L84 (2005).
\newblock \doi{10.1086/497028}

\bibitem{2004ApJ...607L..51C}
{Ceccarelli}, C., {Dominik}, C., {Lefloch}, B., {Caselli}, P., {Caux}, E.:
  {Detection of H$_{2}$D$^{+}$: Measuring the Midplane Degree of Ionization in
  the Disks of DM Tauri and TW Hydrae}.
\newblock \apjl \textbf{607}, L51--L54 (2004).
\newblock \doi{10.1086/421461}

\bibitem{1996ApJ...471..400C}
{Ceccarelli}, C., {Hollenbach}, D.J., {Tielens}, A.G.G.M.: {Far-Infrared Line
  Emission from Collapsing Protostellar Envelopes}.
\newblock \apj \textbf{471}, 400 (1996).
\newblock \doi{10.1086/177978}

\bibitem{2000A&A...357L...9C}
{Ceccarelli}, C., {Loinard}, L., {Castets}, A., {Tielens}, A.G.G.M., {Caux},
  E.: {The hot core of the solar-type protostar IRAS 16293-2422: H\_2CO
  emission}.
\newblock \aap \textbf{357}, L9--L12 (2000)

\bibitem{1995ApJ...449L.139C}
{Chandler}, C.J., {Koerner}, D.W., {Sargent}, A.I., {Wood}, D.O.S.: {Dust
  Emission from Protostars: The Disk and Envelope of HH 24 MMS}.
\newblock \apjl \textbf{449}, L139 (1995).
\newblock \doi{10.1086/309644}

\bibitem{2012ApJ...756...58C}
{Chapillon}, E., {Dutrey}, A., {Guilloteau}, S., {Pi{\'e}tu}, V., {Wakelam},
  V., {Hersant}, F., {Gueth}, F., {Henning}, T., {Launhardt}, R., {Schreyer},
  K., {Semenov}, D.: {Chemistry in Disks. VII. First Detection of HC$_{3}$N in
  Protoplanetary Disks}.
\newblock \apj \textbf{756}, 58 (2012).
\newblock \doi{10.1088/0004-637X/756/1/58}

\bibitem{2012A&A...537A..60C}
{Chapillon}, E., {Guilloteau}, S., {Dutrey}, A., {Pi{\'e}tu}, V., {Gu{\'e}lin},
  M.: {Chemistry in disks. VI. CN and HCN in protoplanetary disks}.
\newblock \aap \textbf{537}, A60 (2012).
\newblock \doi{10.1051/0004-6361/201116762}

\bibitem{1992ApJ...399L..71C}
{Charnley}, S.B., {Tielens}, A.G.G.M., {Millar}, T.J.: {On the molecular
  complexity of the hot cores in Orion A - Grain surface chemistry as 'The last
  refuge of the scoundrel'}.
\newblock \apjl \textbf{399}, L71--L74 (1992).
\newblock \doi{10.1086/186609}

\bibitem{1997ApJ...482L.203C}
{Charnley}, S.B., {Tielens}, A.G.G.M., {Rodgers}, S.D.: {Deuterated Methanol in
  the Orion Compact Ridge}.
\newblock \apjl \textbf{482}, L203 (1997).
\newblock \doi{10.1086/310697}

\bibitem{2012M&PSA..75.5192C}
{Chaussidon}, M., {Srinivasan}, G.: {New Constraints on the Origin of
  Short-Lived 10Be in the Early Solar System}.
\newblock Meteoritics and Planetary Science Supplement \textbf{75}, 5192 (2012)

\bibitem{2012ApJ...756..168C}
{Chiang}, H.F., {Looney}, L.W., {Tobin}, J.J.: {The Envelope and Embedded Disk
  around the Class 0 Protostar L1157-mm: Dual-wavelength Interferometric
  Observations and Modeling}.
\newblock \apj \textbf{756}, 168 (2012).
\newblock \doi{10.1088/0004-637X/756/2/168}

\bibitem{2011ApJ...731....9C}
{Chiar}, J.E., {Pendleton}, Y.J., {Allamandola}, L.J., {Boogert}, A.C.A.,
  {Ennico}, K., {Greene}, T.P., {Geballe}, T.R., {Keane}, J.V., {Lada}, C.J.,
  {Mason}, R.E., {Roellig}, T.L., {Sandford}, S.A., {Tielens}, A.G.G.M.,
  {Werner}, M.W., {Whittet}, D.C.B., {Decin}, L., {Eriksson}, K.: {Ices in the
  Quiescent IC 5146 Dense Cloud}.
\newblock \apj \textbf{731}, 9 (2011).
\newblock \doi{10.1088/0004-637X/731/1/9}

\bibitem{2007ApJ...667L.183C}
{Choi}, M., {Tatematsu}, K., {Park}, G., {Kang}, M.: {Ammonia Imaging of the
  Disks in the NGC 1333 IRAS 4A Protobinary System}.
\newblock \apjl \textbf{667}, L183--L186 (2007).
\newblock \doi{10.1086/522116}

\bibitem{2012ApJ...752...75C}
{Cieza}, L.A., {Mathews}, G.S., {Williams}, J.P., {M{\'e}nard}, F.C., {Kraus},
  A.L., {Schreiber}, M.R., {Romero}, G.A., {Orellana}, M., {Ireland}, M.J.:
  {Submillimeter Array Observations of the RX J1633.9-2442 Transition Disk:
  Evidence for Multiple Planets in the Making}.
\newblock \apj \textbf{752}, 75 (2012).
\newblock \doi{10.1088/0004-637X/752/1/75}

\bibitem{2009A&A...507L..25C}
{Codella}, C., {Benedettini}, M., {Beltr{\'a}n}, M.T., {Gueth}, F., {Viti}, S.,
  {Bachiller}, R., {Tafalla}, M., {Cabrit}, S., {Fuente}, A., {Lefloch}, B.:
  {Methyl cyanide as tracer of bow shocks in L1157-B1}.
\newblock \aap \textbf{507}, L25--L28 (2009).
\newblock \doi{10.1051/0004-6361/200913340}

\bibitem{2012ApJ...744..164C}
{Codella}, C., {Ceccarelli}, C., {Bottinelli}, S., {Salez}, M., {Viti}, S.,
  {Lefloch}, B., {Cabrit}, S., {Caux}, E., {Faure}, A., {Vasta}, M.,
  {Wiesenfeld}, L.: {First Detection of Hydrogen Chloride Toward Protostellar
  Shocks}.
\newblock \apj \textbf{744}, 164 (2012).
\newblock \doi{10.1088/0004-637X/744/2/164}

\bibitem{2012ApJ...757L...9C}
{Codella}, C., {Ceccarelli}, C., {Lefloch}, B., {Fontani}, F., {Busquet}, G.,
  {Caselli}, P., {Kahane}, C., {Lis}, D., {Taquet}, V., {Vasta}, M., {Viti},
  S., {Wiesenfeld}, L.: {The Herschel and IRAM CHESS Spectral Surveys of the
  Protostellar Shock L1157-B1: Fossil Deuteration}.
\newblock \apjl \textbf{757}, L9 (2012).
\newblock \doi{10.1088/2041-8205/757/1/L9}

\bibitem{2010A&A...518L.112C}
{Codella}, C., {Lefloch}, B., {Ceccarelli}, C., {Cernicharo}, J., {Caux}, E.,
  {Lorenzani}, A., {Viti}, S., {Hily-Blant}, P., {Parise}, B., {Maret}, S.,
  {Nisini}, B., {Caselli}, P., {Cabrit}, S., {Pagani}, L., {Benedettini}, M.,
  {Boogert}, A., {Gueth}, F., {Melnick}, G., {Neufeld}, D., {Pacheco}, S.,
  {Salez}, M., {Schuster}, K., {Bacmann}, A., {Baudry}, A., {Bell}, T.,
  {Bergin}, E.A., {Blake}, G., {Bottinelli}, S., {Castets}, A., {Comito}, C.,
  {Coutens}, A., {Crimier}, N., {Dominik}, C., {Demyk}, K., {Encrenaz}, P.,
  {Falgarone}, E., {Fuente}, A., {Gerin}, M., {Goldsmith}, P., {Helmich}, F.,
  {Hennebelle}, P., {Henning}, T., {Herbst}, E., {Jacq}, T., {Kahane}, C.,
  {Kama}, M., {Klotz}, A., {Langer}, W., {Lis}, D., {Lord}, S., {Pearson}, J.,
  {Phillips}, T., {Saraceno}, P., {Schilke}, P., {Tielens}, X., {van der Tak},
  F., {van der Wiel}, M., {Vastel}, C., {Wakelam}, V., {Walters}, A.,
  {Wyrowski}, F., {Yorke}, H., {Borys}, C., {Delorme}, Y., {Kramer}, C.,
  {Larsson}, B., {Mehdi}, I., {Ossenkopf}, V., {Stutzki}, J.: {The CHESS
  spectral survey of star forming regions: Peering into the protostellar shock
  L1157-B1. I. Shock chemical complexity}.
\newblock \aap \textbf{518}, L112 (2010).
\newblock \doi{10.1051/0004-6361/201014582}

\bibitem{2012ApJ...744..131C}
{Cordiner}, M.A., {Charnley}, S.B., {Wirstr{\"o}m}, E.S., {Smith}, R.G.:
  {Organic Chemistry of Low-mass Star-forming Cores. I. 7 mm Spectroscopy of
  Chamaeleon MMS1}.
\newblock \apj \textbf{744}, 131 (2012).
\newblock \doi{10.1088/0004-637X/744/2/131}

\bibitem{2009ApJ...697.1305C}
{Cortes}, S.R., {Meyer}, M.R., {Carpenter}, J.M., {Pascucci}, I., {Schneider},
  G., {Wong}, T., {Hines}, D.C.: {Grain Growth and Global Structure of the
  Protoplanetary Disk Associated with the Mature Classical T Tauri Star, PDS
  66}.
\newblock \apj \textbf{697}, 1305--1315 (2009).
\newblock \doi{10.1088/0004-637X/697/2/1305}

\bibitem{2012A&A...539A.132C}
{Coutens}, A., {Vastel}, C., {Caux}, E., {Ceccarelli}, C., {Bottinelli}, S.,
  {Wiesenfeld}, L., {Faure}, A., {Scribano}, Y., {Kahane}, C.: {A study of
  deuterated water in the low-mass protostar IRAS 16293-2422}.
\newblock \aap \textbf{539}, A132 (2012).
\newblock \doi{10.1051/0004-6361/201117627}

\bibitem{2005ApJ...619..379C}
{Crapsi}, A., {Caselli}, P., {Walmsley}, C.M., {Myers}, P.C., {Tafalla}, M.,
  {Lee}, C.W., {Bourke}, T.L.: {Probing the Evolutionary Status of Starless
  Cores through N$_{2}$H$^{+}$ and N$_{2}$D$^{+}$ Observations}.
\newblock \apj \textbf{619}, 379--406 (2005).
\newblock \doi{10.1086/426472}

\bibitem{2007A&A...470..221C}
{Crapsi}, A., {Caselli}, P., {Walmsley}, M.C., {Tafalla}, M.: {Observing the
  gas temperature drop in the high-density nucleus of L 1544}.
\newblock \aap \textbf{470}, 221--230 (2007).
\newblock \doi{10.1051/0004-6361:20077613}

\bibitem{2009A&A...506.1229C}
{Crimier}, N., {Ceccarelli}, C., {Lefloch}, B., {Faure}, A.: {Physical
  structure and water line spectrum predictions of the intermediate mass
  protostar OMC2-FIR4}.
\newblock \aap \textbf{506}, 1229--1241 (2009).
\newblock \doi{10.1051/0004-6361/200911651}

\bibitem{2010A&A...519A..65C}
{Crimier}, N., {Ceccarelli}, C., {Maret}, S., {Bottinelli}, S., {Caux}, E.,
  {Kahane}, C., {Lis}, D.C., {Olofsson}, J.: {The solar type protostar
  IRAS16293-2422: new constraints on the physical structure}.
\newblock \aap \textbf{519}, A65 (2010).
\newblock \doi{10.1051/0004-6361/200913112}

\bibitem{2009EM&P..105..267C}
{Crovisier}, J., {Biver}, N., {Bockel{\'e}e-Morvan}, D., {Boissier}, J.,
  {Colom}, P., {Lis}, D.C.: {The Chemical Diversity of Comets: Synergies
  Between Space Exploration and Ground-based Radio Observations}.
\newblock Earth Moon and Planets \textbf{105}, 267--272 (2009).
\newblock \doi{10.1007/s11038-009-9293-z}

\bibitem{2005MNRAS.361..565C}
{Cuppen}, H.M., {Herbst}, E.: {Monte Carlo simulations of H$_{2}$ formation on
  grains of varying surface roughness}.
\newblock \mnras \textbf{361}, 565--576 (2005).
\newblock \doi{10.1111/j.1365-2966.2005.09189.x}

\bibitem{2009A&A...508..275C}
{Cuppen}, H.M., {van Dishoeck}, E.F., {Herbst}, E., {Tielens}, A.G.G.M.:
  {Microscopic simulation of methanol and formaldehyde ice formation in cold
  dense cores}.
\newblock \aap \textbf{508}, 275--287 (2009).
\newblock \doi{10.1051/0004-6361/200913119}

\bibitem{2010NatGe...3..522C}
{Czaja}, A.D.: {Early Earth: Microbes and the rise of oxygen}.
\newblock Nature Geoscience \textbf{3}, 522--523 (2010).
\newblock \doi{10.1038/ngeo929}

\bibitem{2001ApJ...553..321D}
{D'Alessio}, P., {Calvet}, N., {Hartmann}, L.: {Accretion Disks around Young
  Objects. III. Grain Growth}.
\newblock \apj \textbf{553}, 321--334 (2001).
\newblock \doi{10.1086/320655}

\bibitem{2006ApJ...638..314D}
{D'Alessio}, P., {Calvet}, N., {Hartmann}, L., {Franco-Hern{\'a}ndez}, R.,
  {Serv{\'{\i}}n}, H.: {Effects of Dust Growth and Settling in T Tauri Disks}.
\newblock \apj \textbf{638}, 314--335 (2006).
\newblock \doi{10.1086/498861}

\bibitem{1999ApJ...527..893D}
{D'Alessio}, P., {Calvet}, N., {Hartmann}, L., {Lizano}, S., {Cant{\'o}}, J.:
  {Accretion Disks around Young Objects. II. Tests of Well-mixed Models with
  ISM Dust}.
\newblock \apj \textbf{527}, 893--909 (1999).
\newblock \doi{10.1086/308103}

\bibitem{1984ApJ...287L..47D}
{Dalgarno}, A., {Lepp}, S.: {Deuterium fractionation mechanisms in interstellar
  clouds}.
\newblock \apjl \textbf{287}, L47--L50 (1984).
\newblock \doi{10.1086/184395}

\bibitem{2003Icar..165..326D}
{Dauphas}, N.: {The dual origin of the terrestrial atmosphere}.
\newblock \icarus \textbf{165}, 326--339 (2003).
\newblock \doi{10.1016/S0019-1035(03)00198-2}

\bibitem{2011AREPS..39..351D}
{Dauphas}, N., {Chaussidon}, M.: {A Perspective from Extinct Radionuclides on a
  Young Stellar Object: The Sun and Its Accretion Disk}.
\newblock Annual Review of Earth and Planetary Sciences \textbf{39}, 351--386
  (2011).
\newblock \doi{10.1146/annurev-earth-040610-133428}

\bibitem{2000Icar..148..508D}
{Dauphas}, N., {Robert}, F., {Marty}, B.: {The Late Asteroidal and Cometary
  Bombardment of Earth as Recorded in Water Deuterium to Protium Ratio}.
\newblock \icarus \textbf{148}, 508--512 (2000).
\newblock \doi{10.1006/icar.2000.6489}

\bibitem{1992AdSpR..12....5D}
{Delsemme}, A.H.: {Cometary origin of carbon and water on the terrestrial
  planets}.
\newblock Advances in Space Research \textbf{12}, 5--12 (1992).
\newblock \doi{10.1016/0273-1177(92)90147-P}

\bibitem{1982A&A...109L..12D}
{D'Hendecourt}, L.B., {Allamandola}, L.J., {Baas}, F., {Greenberg}, J.M.:
  {Interstellar grain explosions - Molecule cycling between gas and dust}.
\newblock \aap \textbf{109}, L12--L14 (1982)

\bibitem{2007prpl.conf..783D}
{Dominik}, C., {Blum}, J., {Cuzzi}, J.N., {Wurm}, G.: {Growth of Dust as the
  Initial Step Toward Planet Formation}.
\newblock Protostars and Planets V pp. 783--800 (2007)

\bibitem{2004come.book..153D}
{Dones}, L., {Weissman}, P.R., {Levison}, H.F., {Duncan}, M.J.: {Oort cloud
  formation and dynamics}, pp. 153--174 (2004)

\bibitem{1997ApJ...489..122D}
{Doty}, S.D., {Neufeld}, D.A.: {Models for Dense Molecular Cloud Cores}.
\newblock \apj \textbf{489}, 122 (1997).
\newblock \doi{10.1086/304764}

\bibitem{2004sf2a.conf..525D}
{Dubernet}, M.L., {Cernicharo}, J., {Daniel}, F., {Debray}, B., {Faure}, A.,
  {Feautrier}, N., {Flower}, D., {Grosjean}, A., {Roueff}, E., {Spielfiedel},
  A., {Stoecklin}, T., {Valiron}, P.: {Ro-vibrational Collisional Excitation
  Database: BASECOL - http://www.obspm.fr/basecol}.
\newblock In: F.~{Combes}, D.~{Barret}, T.~{Contini}, F.~{Meynadier},
  L.~{Pagani} (eds.) SF2A-2004: Semaine de l'Astrophysique Francaise, p. 525
  (2004)

\bibitem{2003A&A...400..559D}
{Duch{\^e}ne}, G., {M{\'e}nard}, F., {Stapelfeldt}, K., {Duvert}, G.: {A
  layered edge-on circumstellar disk around HK Tau B}.
\newblock \aap \textbf{400}, 559--565 (2003).
\newblock \doi{10.1051/0004-6361:20021906}

\bibitem{2010A&A...512A..30D}
{Dulieu}, F., {Amiaud}, L., {Congiu}, E., {Fillion}, J.H., {Matar}, E.,
  {Momeni}, A., {Pirronello}, V., {Lemaire}, J.L.: {Experimental evidence for
  water formation on interstellar dust grains by hydrogen and oxygen atoms}.
\newblock \aap \textbf{512}, A30 (2010).
\newblock \doi{10.1051/0004-6361/200912079}

\bibitem{2007A&A...473..457D}
{Dullemond}, C.P., {Henning}, T., {Visser}, R., {Geers}, V.C., {van Dishoeck},
  E.F., {Pontoppidan}, K.M.: {Dust sedimentation in protoplanetary disks with
  polycyclic aromatic hydrocarbons}.
\newblock \aap \textbf{473}, 457--466 (2007).
\newblock \doi{10.1051/0004-6361:20077581}

\bibitem{2007prpl.conf..555D}
{Dullemond}, C.P., {Hollenbach}, D., {Kamp}, I., {D'Alessio}, P.: {Models of
  the Structure and Evolution of Protoplanetary Disks}.
\newblock Protostars and Planets V pp. 555--572 (2007)

\bibitem{2010ARA&A..48..205D}
{Dullemond}, C.P., {Monnier}, J.D.: {The Inner Regions of Protoplanetary
  Disks}.
\newblock \araa \textbf{48}, 205--239 (2010).
\newblock \doi{10.1146/annurev-astro-081309-130932}

\bibitem{2007prpl.conf..607D}
{Durisen}, R.H., {Boss}, A.P., {Mayer}, L., {Nelson}, A.F., {Quinn}, T.,
  {Rice}, W.K.M.: {Gravitational Instabilities in Gaseous Protoplanetary Disks
  and Implications for Giant Planet Formation}.
\newblock Protostars and Planets V pp. 607--622 (2007)

\bibitem{1996A&A...309..493D}
{Dutrey}, A., {Guilloteau}, S., {Duvert}, G., {Prato}, L., {Simon}, M.,
  {Schuster}, K., {Menard}, F.: {Dust and gas distribution around T Tauri stars
  in Taurus-Auriga. I. Interferometric 2.7mm continuum and \^{}13\^{}CO J=1-0
  observations}.
\newblock \aap \textbf{309}, 493--504 (1996)

\bibitem{1997A&A...317L..55D}
{Dutrey}, A., {Guilloteau}, S., {Guelin}, M.: {Chemistry of protosolar-like
  nebulae: The molecular content of the DM Tau and GG Tau disks.}
\newblock \aap \textbf{317}, L55--L58 (1997)

\bibitem{2007prpl.conf..495D}
{Dutrey}, A., {Guilloteau}, S., {Ho}, P.: {Interferometric Spectroimaging of
  Molecular Gas in Protoplanetary Disks}.
\newblock Protostars and Planets V pp. 495--506 (2007)

\bibitem{2007A&A...464..615D}
{Dutrey}, A., {Henning}, T., {Guilloteau}, S., {Semenov}, D., {Pi{\'e}tu}, V.,
  {Schreyer}, K., {Bacmann}, A., {Launhardt}, R., {Pety}, J., {Gueth}, F.:
  {Chemistry in disks. I. Deep search for N2H$^{+}$ in the protoplanetary disks
  around LkCa 15, MWC 480, and DM Tauri}.
\newblock \aap \textbf{464}, 615--623 (2007).
\newblock \doi{10.1051/0004-6361:20065385}

\bibitem{2011A&A...535A.104D}
{Dutrey}, A., {Wakelam}, V., {Boehler}, Y., {Guilloteau}, S., {Hersant}, F.,
  {Semenov}, D., {Chapillon}, E., {Henning}, T., {Pi{\'e}tu}, V., {Launhardt},
  R., {Gueth}, F., {Schreyer}, K.: {Chemistry in disks. V. Sulfur-bearing
  molecules in the protoplanetary disks surrounding LkCa15, MWC480, DM Tauri,
  and GO Tauri}.
\newblock \aap \textbf{535}, A104 (2011).
\newblock \doi{10.1051/0004-6361/201116931}

\bibitem{2009M&PS...44.1323E}
{Elsila}, J.E., {Glavin}, D.P., {Dworkin}, J.P.: {Cometary glycine detected in
  samples returned by Stardust}.
\newblock Meteoritics and Planetary Science \textbf{44}, 1323--1330 (2009).
\newblock \doi{10.1111/j.1945-5100.2009.tb01224.x}

\bibitem{2011ApJS..195...21E}
{Enoch}, M.L., {Corder}, S., {Duch{\^e}ne}, G., {Bock}, D.C., {Bolatto}, A.D.,
  {Culverhouse}, T.L., {Kwon}, W., {Lamb}, J.W., {Leitch}, E.M., {Marrone},
  D.P., {Muchovej}, S.J., {P{\'e}rez}, L.M., {Scott}, S.L., {Teuben}, P.J.,
  {Wright}, M.C.H., {Zauderer}, B.A.: {Disk and Envelope Structure in Class 0
  Protostars. II. High-resolution Millimeter Mapping of the Serpens Sample}.
\newblock \apjs \textbf{195}, 21 (2011).
\newblock \doi{10.1088/0067-0049/195/2/21}

\bibitem{2001ApJ...557..193E}
{Evans} II, N.J., {Rawlings}, J.M.C., {Shirley}, Y.L., {Mundy}, L.G.: {Tracing
  the Mass during Low-Mass Star Formation. II. Modeling the Submillimeter
  Emission from Preprotostellar Cores}.
\newblock \apj \textbf{557}, 193--208 (2001).
\newblock \doi{10.1086/321639}

\bibitem{2012A&A...544L...9F}
{Fedele}, D., {Bruderer}, S., {van Dishoeck}, E.F., {Herczeg}, G.J., {Evans},
  N.J., {Bouwman}, J., {Henning}, T., {Green}, J.: {Warm H$_{2}$O and OH in the
  disk around the Herbig star HD 163296}.
\newblock \aap \textbf{544}, L9 (2012).
\newblock \doi{10.1051/0004-6361/201219615}

\bibitem{2008A&A...491..809F}
{Fedele}, D., {van den Ancker}, M.E., {Acke}, B., {van der Plas}, G., {van
  Boekel}, R., {Wittkowski}, M., {Henning}, T., {Bouwman}, J., {Meeus}, G.,
  {Rafanelli}, P.: {The structure of the protoplanetary disk surrounding three
  young intermediate mass stars. II. Spatially resolved dust and gas
  distribution}.
\newblock \aap \textbf{491}, 809--820 (2008).
\newblock \doi{10.1051/0004-6361:200810126}

\bibitem{1999ARA&A..37..363F}
{Feigelson}, E.D., {Montmerle}, T.: {High-Energy Processes in Young Stellar
  Objects}.
\newblock \araa \textbf{37}, 363--408 (1999).
\newblock \doi{10.1146/annurev.astro.37.1.363}

\bibitem{1982PEPI...29..242F}
{Fisher}, D.E.: {Implications of terrestrial Ar-40/Ar-36 for atmospheric and
  mantle evolutionary models}.
\newblock Physics of the Earth and Planetary Interiors \textbf{29}, 242--251
  (1982).
\newblock \doi{10.1016/0031-9201(82)90015-2}

\bibitem{2006A&A...449..621F}
{Flower}, D.R., {Pineau Des For{\^e}ts}, G., {Walmsley}, C.M.: {The importance
  of the ortho:para H$_{2}$ ratio for the deuteration of molecules during
  pre-protostellar collapse}.
\newblock \aap \textbf{449}, 621--629 (2006).
\newblock \doi{10.1051/0004-6361:20054246}

\bibitem{2011ApJ...726...29F}
{Fogel}, J.K.J., {Bethell}, T.J., {Bergin}, E.A., {Calvet}, N., {Semenov}, D.:
  {Chemistry of a Protoplanetary Disk with Grain Settling and Ly{$\alpha$}
  Radiation}.
\newblock \apj \textbf{726}, 29 (2011).
\newblock \doi{10.1088/0004-637X/726/1/29}

\bibitem{2004Icar..172...50F}
{Fouchet}, T., {Irwin}, P.G.J., {Parrish}, P., {Calcutt}, S.B., {Taylor}, F.W.,
  {Nixon}, C.A., {Owen}, T.: {Search for spatial variation in the jovian
  $^{15}$N/$^{14}$N ratio from Cassini/CIRS observations}.
\newblock \icarus \textbf{172}, 50--58 (2004).
\newblock \doi{10.1016/j.icarus.2003.11.011}

\bibitem{2012ApJ...756..171F}
{France}, K., {Schindhelm}, E., {Herczeg}, G.J., {Brown}, A., {Abgrall}, H.,
  {Alexander}, R.D., {Bergin}, E.A., {Brown}, J.M., {Linsky}, J.L., {Roueff},
  E., {Yang}, H.: {A Hubble Space Telescope Survey of H$_{2}$ Emission in the
  Circumstellar Environments of Young Stars}.
\newblock \apj \textbf{756}, 171 (2012).
\newblock \doi{10.1088/0004-637X/756/2/171}

\bibitem{2008ApJ...674.1015F}
{Franklin}, J., {Snell}, R.L., {Kaufman}, M.J., {Melnick}, G.J., {Neufeld},
  D.A., {Hollenbach}, D.J., {Bergin}, E.A.: {SWAS Observations of Water in
  Molecular Outflows}.
\newblock \apj \textbf{674}, 1015--1031 (2008).
\newblock \doi{10.1086/524924}

\bibitem{2011A&A...535A..44F}
{Frau}, P., {Galli}, D., {Girart}, J.M.: {Comparing star formation models with
  interferometric observations of the protostar NGC 1333 IRAS 4A. I.
  Magnetohydrodynamic collapse models}.
\newblock \aap \textbf{535}, A44 (2011).
\newblock \doi{10.1051/0004-6361/201117813}

\bibitem{1988A&A...195..281F}
{Friberg}, P., {Hjalmarson}, A., {Madden}, S.C., {Irvine}, W.M.: {Methanol in
  dark clouds}.
\newblock \aap \textbf{195}, 281--289 (1988)

\bibitem{2010ApJ...708.1002F}
{Friesen}, R.K., {Di Francesco}, J., {Shimajiri}, Y., {Takakuwa}, S.: {The
  Initial Conditions of Clustered Star Formation. II. N$_{2}$H$^{+}$
  Observations of the Ophiuchus B Core}.
\newblock \apj \textbf{708}, 1002--1024 (2010).
\newblock \doi{10.1088/0004-637X/708/2/1002}

\bibitem{2009A&A...505..629F}
{Fuchs}, G.W., {Cuppen}, H.M., {Ioppolo}, S., {Romanzin}, C., {Bisschop}, S.E.,
  {Andersson}, S., {van Dishoeck}, E.F., {Linnartz}, H.: {Hydrogenation
  reactions in interstellar CO ice analogues. A combined
  experimental/theoretical approach}.
\newblock \aap \textbf{505}, 629--639 (2009).
\newblock \doi{10.1051/0004-6361/200810784}

\bibitem{2010A&A...524A..19F}
{Fuente}, A., {Cernicharo}, J., {Ag{\'u}ndez}, M., {Bern{\'e}}, O.,
  {Goicoechea}, J.R., {Alonso-Albi}, T., {Marcelino}, N.: {Molecular content of
  the circumstellar disk in AB Aurigae. First detection of SO in a
  circumstellar disk}.
\newblock \aap \textbf{524}, A19 (2010).
\newblock \doi{10.1051/0004-6361/201014905}

\bibitem{2006ApJS..165..568F}
{Furlan}, E., {Hartmann}, L., {Calvet}, N., {D'Alessio}, P.,
  {Franco-Hern{\'a}ndez}, R., {Forrest}, W.J., {Watson}, D.M., {Uchida}, K.I.,
  {Sargent}, B., {Green}, J.D., {Keller}, L.D., {Herter}, T.L.: {A Survey and
  Analysis of Spitzer Infrared Spectrograph Spectra of T Tauri Stars in
  Taurus}.
\newblock \apjs \textbf{165}, 568--605 (2006).
\newblock \doi{10.1086/505468}

\bibitem{2012arXiv1207.6693F}
{Furuya}, K., {Aikawa}, Y., {Tomida}, K., {Matsumoto}, T., {Saigo}, K.,
  {Tomisaka}, K., {Hersant}, F., {Wakelam}, V.: {Chemistry in the First
  Hydrostatic Core Stage Adopting Three-Dimensional Radiation Hydrodynamic
  Simulations}.
\newblock ArXiv e-prints  (2012)

\bibitem{1993ApJ...417..220G}
{Galli}, D., {Shu}, F.H.: {Collapse of Magnetized Molecular Cloud Cores. I.
  Semianalytical Solution}.
\newblock \apj \textbf{417}, 220 (1993).
\newblock \doi{10.1086/173305}

\bibitem{1993ApJ...417..243G}
{Galli}, D., {Shu}, F.H.: {Collapse of Magnetized Molecular Cloud Cores. II.
  Numerical Results}.
\newblock \apj \textbf{417}, 243 (1993).
\newblock \doi{10.1086/173306}

\bibitem{2002AsBio...2..353G}
{Garc{\'{\i}}a Ruiz}, J.M., {Carnerup}, A., {Christy}, A.G., {Welham}, N.J.,
  {Hyde}, S.T.: {Morphology: An Ambiguous Indicator of Biogenicity}.
\newblock Astrobiology \textbf{2}, 353--369 (2002).
\newblock \doi{10.1089/153110702762027925}

\bibitem{2006A&A...457..927G}
{Garrod}, R.T., {Herbst}, E.: {Formation of methyl formate and other organic
  species in the warm-up phase of hot molecular cores}.
\newblock \aap \textbf{457}, 927--936 (2006).
\newblock \doi{10.1051/0004-6361:20065560}

\bibitem{2011ApJ...735...15G}
{Garrod}, R.T., {Pauly}, T.: {On the Formation of CO$_{2}$ and Other
  Interstellar Ices}.
\newblock \apj \textbf{735}, 15 (2011).
\newblock \doi{10.1088/0004-637X/735/1/15}

\bibitem{2009ApJ...700L..43G}
{Garrod}, R.T., {Vasyunin}, A.I., {Semenov}, D.A., {Wiebe}, D.S., {Henning},
  T.: {A New Modified-Rate Approach For Gas-Grain Chemistry: Comparison with a
  Unified Large-Scale Monte Carlo Simulation}.
\newblock \apjl \textbf{700}, L43--L46 (2009).
\newblock \doi{10.1088/0004-637X/700/1/L43}

\bibitem{2007A&A...467.1103G}
{Garrod}, R.T., {Wakelam}, V., {Herbst}, E.: {Non-thermal desorption from
  interstellar dust grains via exothermic surface reactions}.
\newblock \aap \textbf{467}, 1103--1115 (2007).
\newblock \doi{10.1051/0004-6361:20066704}

\bibitem{2008ApJ...682..283G}
{Garrod}, R.T., {Weaver}, S.L.W., {Herbst}, E.: {Complex Chemistry in
  Star-forming Regions: An Expanded Gas-Grain Warm-up Chemical Model}.
\newblock \apj \textbf{682}, 283--302 (2008).
\newblock \doi{10.1086/588035}

\bibitem{2006A&A...459..545G}
{Geers}, V.C., {Augereau}, J.C., {Pontoppidan}, K.M., {Dullemond}, C.P.,
  {Visser}, R., {Kessler-Silacci}, J.E., {Evans} II, N.J., {van Dishoeck},
  E.F., {Blake}, G.A., {Boogert}, A.C.A., {Brown}, J.M., {Lahuis}, F.,
  {Mer{\'{\i}}n}, B.: {C2D Spitzer-IRS spectra of disks around T Tauri stars.
  II. PAH emission features}.
\newblock \aap \textbf{459}, 545--556 (2006).
\newblock \doi{10.1051/0004-6361:20064830}

\bibitem{1998SSRv...84..239G}
{Geiss}, J., {Gloeckler}, G.: {Abundances of Deuterium and Helium-3 in the
  Protosolar Cloud}.
\newblock \ssr \textbf{84}, 239--250 (1998)

\bibitem{1996A&A...312..289G}
{Gerakines}, P.A., {Schutte}, W.A., {Ehrenfreund}, P.: {Ultraviolet processing
  of interstellar ice analogs. I. Pure ices.}
\newblock \aap \textbf{312}, 289--305 (1996)

\bibitem{2009A&A...498L...9G}
{Gerin}, M., {Marcelino}, N., {Biver}, N., {Roueff}, E., {Coudert}, L.H.,
  {Elkeurti}, M., {Lis}, D.C., {Bockel{\'e}e-Morvan}, D.: {Detection of
  $^{15}$NH$\{$2$\}$D in dense cores: a new tool for measuring the
  $^{14}$N/$^{15}$N ratio in the cold ISM}.
\newblock \aap \textbf{498}, L9--L12 (2009).
\newblock \doi{10.1051/0004-6361/200911759}

\bibitem{2006Sci...313..812G}
{Girart}, J.M., {Rao}, R., {Marrone}, D.P.: {Magnetic Fields in the Formation
  of Sun-Like Stars}.
\newblock Science \textbf{313}, 812--814 (2006).
\newblock \doi{10.1126/science.1129093}

\bibitem{1997ApJ...480..344G}
{Glassgold}, A.E., {Najita}, J., {Igea}, J.: {X-Ray Ionization of
  Protoplanetary Disks}.
\newblock \apj \textbf{480}, 344 (1997).
\newblock \doi{10.1086/303952}

\bibitem{2001ApJ...557..736G}
{Goldsmith}, P.F.: {Molecular Depletion and Thermal Balance in Dark Cloud
  Cores}.
\newblock \apj \textbf{557}, 736--746 (2001).
\newblock \doi{10.1086/322255}

\bibitem{2005Natur.435..466G}
{Gomes}, R., {Levison}, H.F., {Tsiganis}, K., {Morbidelli}, A.: {Origin of the
  cataclysmic Late Heavy Bombardment period of the terrestrial planets}.
\newblock \nat \textbf{435}, 466--469 (2005).
\newblock \doi{10.1038/nature03676}

\bibitem{1993ApJ...406..528G}
{Goodman}, A.A., {Benson}, P.J., {Fuller}, G.A., {Myers}, P.C.: {Dense cores in
  dark clouds. VIII - Velocity gradients}.
\newblock \apj \textbf{406}, 528--547 (1993).
\newblock \doi{10.1086/172465}

\bibitem{2004ApJ...613..424G}
{Gorti}, U., {Hollenbach}, D.: {Models of Chemistry, Thermal Balance, and
  Infrared Spectra from Intermediate-Aged Disks around G and K Stars}.
\newblock \apj \textbf{613}, 424--447 (2004).
\newblock \doi{10.1086/422406}

\bibitem{2008ApJ...683..287G}
{Gorti}, U., {Hollenbach}, D.: {Line Emission from Gas in Optically Thick Dust
  Disks around Young Stars}.
\newblock \apj \textbf{683}, 287--303 (2008).
\newblock \doi{10.1086/589616}

\bibitem{2012ApJ...748....6G}
{Goto}, M., {Carmona}, A., {Linz}, H., {Stecklum}, B., {Henning}, T., {Meeus},
  G., {Usuda}, T.: {Kinematics of Ionized Gas at 0.01 AU of TW Hya}.
\newblock \apj \textbf{748}, 6 (2012).
\newblock \doi{10.1088/0004-637X/748/1/6}

\bibitem{2008ApJ...680..781G}
{Gounelle}, M., {Meibom}, A.: {The Origin of Short-lived Radionuclides and the
  Astrophysical Environment of Solar System Formation}.
\newblock \apj \textbf{680}, 781--792 (2008).
\newblock \doi{10.1086/587613}

\bibitem{2012A&ARv..20...50G}
{Gratton}, R.G., {Carretta}, E., {Bragaglia}, A.: {Multiple populations in
  globular clusters. Lessons learned from the Milky Way globular clusters}.
\newblock \aapr \textbf{20}, 50 (2012).
\newblock \doi{10.1007/s00159-012-0050-3}

\bibitem{1989ApJ...347..289G}
{Gredel}, R., {Lepp}, S., {Dalgarno}, A., {Herbst}, E.: {Cosmic-ray-induced
  photodissociation and photoionization rates of interstellar molecules}.
\newblock \apj \textbf{347}, 289--293 (1989).
\newblock \doi{10.1086/168117}

\bibitem{1977ApJ...217L.165G}
{Guelin}, M., {Langer}, W.D., {Snell}, R.L., {Wootten}, H.A.: {Observations of
  DCO/plus/ - The electron abundance in dark clouds}.
\newblock \apjl \textbf{217}, L165--L168 (1977).
\newblock \doi{10.1086/182562}

\bibitem{1989ApJ...345..230G}
{Guhathakurta}, P., {Draine}, B.T.: {Temperature fluctuations in interstellar
  grains. I - Computational method and sublimation of small grains}.
\newblock \apj \textbf{345}, 230--244 (1989).
\newblock \doi{10.1086/167899}

\bibitem{2006A&A...448L...5G}
{Guilloteau}, S., {Pi{\'e}tu}, V., {Dutrey}, A., {Gu{\'e}lin}, M.: {Deuterated
  molecules in DM Tauri: DCO$^{+}$, but no HDO}.
\newblock \aap \textbf{448}, L5--L8 (2006).
\newblock \doi{10.1051/0004-6361:200600005}

\bibitem{2010A&A...513A..56G}
{G{\"u}ttler}, C., {Blum}, J., {Zsom}, A., {Ormel}, C.W., {Dullemond}, C.P.:
  {The outcome of protoplanetary dust growth: pebbles, boulders, or
  planetesimals?. I. Mapping the zoo of laboratory collision experiments}.
\newblock \aap \textbf{513}, A56 (2010).
\newblock \doi{10.1051/0004-6361/200912852}

\bibitem{2004A&A...414..531H}
{Habart}, E., {Boulanger}, F., {Verstraete}, L., {Walmsley}, C.M., {Pineau des
  For{\^e}ts}, G.: {Some empirical estimates of the H$_{2}$ formation rate in
  photon-dominated regions}.
\newblock \aap \textbf{414}, 531--544 (2004).
\newblock \doi{10.1051/0004-6361:20031659}

\bibitem{2004A&A...427..179H}
{Habart}, E., {Natta}, A., {Kr{\"u}gel}, E.: {PAHs in circumstellar disks
  around Herbig Ae/Be stars}.
\newblock \aap \textbf{427}, 179--192 (2004).
\newblock \doi{10.1051/0004-6361:20035916}

\bibitem{2006A&A...449.1067H}
{Habart}, E., {Natta}, A., {Testi}, L., {Carbillet}, M.: {Spatially resolved
  PAH emission in the inner disks of Herbig Ae/Be stars}.
\newblock \aap \textbf{449}, 1067--1075 (2006).
\newblock \doi{10.1051/0004-6361:20052994}

\bibitem{2011Natur.478..218H}
{Hartogh}, P., {Lis}, D.C., {Bockel{\'e}e-Morvan}, D., {de Val-Borro}, M.,
  {Biver}, N., {K{\"u}ppers}, M., {Emprechtinger}, M., {Bergin}, E.A.,
  {Crovisier}, J., {Rengel}, M., {Moreno}, R., {Szutowicz}, S., {Blake}, G.A.:
  {Ocean-like water in the Jupiter-family comet 103P/Hartley 2}.
\newblock \nat \textbf{478}, 218--220 (2011).
\newblock \doi{10.1038/nature10519}

\bibitem{1992ApJS...82..167H}
{Hasegawa}, T.I., {Herbst}, E., {Leung}, C.M.: {Models of gas-grain chemistry
  in dense interstellar clouds with complex organic molecules}.
\newblock \apjs \textbf{82}, 167--195 (1992).
\newblock \doi{10.1086/191713}

\bibitem{2011ApJ...743..182H}
{Hassel}, G.E., {Harada}, N., {Herbst}, E.: {Carbon-chain Species in Warm-up
  Models}.
\newblock \apj \textbf{743}, 182 (2011).
\newblock \doi{10.1088/0004-637X/743/2/182}

\bibitem{2008ApJ...681.1385H}
{Hassel}, G.E., {Herbst}, E., {Garrod}, R.T.: {Modeling the Lukewarm Corino
  Phase: Is L1527 unique?}
\newblock \apj \textbf{681}, 1385--1395 (2008).
\newblock \doi{10.1086/588185}

\bibitem{2011ApJ...731..115H}
{Heinzeller}, D., {Nomura}, H., {Walsh}, C., {Millar}, T.J.: {Chemical
  Evolution of Protoplanetary Disks - The Effects of Viscous Accretion,
  Turbulent Mixing, and Disk Winds}.
\newblock \apj \textbf{731}, 115 (2011).
\newblock \doi{10.1088/0004-637X/731/2/115}

\bibitem{2008A&A...477....9H}
{Hennebelle}, P., {Fromang}, S.: {Magnetic processes in a collapsing dense
  core. I. Accretion and ejection}.
\newblock \aap \textbf{477}, 9--24 (2008).
\newblock \doi{10.1051/0004-6361:20078309}

\bibitem{2010ApJ...714.1511H}
{Henning}, T., {Semenov}, D., {Guilloteau}, S., {Dutrey}, A., {Hersant}, F.,
  {Wakelam}, V., {Chapillon}, E., {Launhardt}, R., {Pi{\'e}tu}, V., {Schreyer},
  K.: {Chemistry in Disks. III. Photochemistry and X-ray Driven Chemistry
  Probed by the Ethynyl Radical (CCH) in DM Tau, LkCa 15, and MWC 480}.
\newblock \apj \textbf{714}, 1511--1520 (2010).
\newblock \doi{10.1088/0004-637X/714/2/1511}

\bibitem{1973ApJ...185..505H}
{Herbst}, E., {Klemperer}, W.: {The Formation and Depletion of Molecules in
  Dense Interstellar Clouds}.
\newblock \apj \textbf{185}, 505--534 (1973).
\newblock \doi{10.1086/152436}

\bibitem{2009ARA&A..47..427H}
{Herbst}, E., {van Dishoeck}, E.F.: {Complex Organic Interstellar Molecules}.
\newblock \araa \textbf{47}, 427--480 (2009).
\newblock \doi{10.1146/annurev-astro-082708-101654}

\bibitem{2009A&A...493L..49H}
{Hersant}, F., {Wakelam}, V., {Dutrey}, A., {Guilloteau}, S., {Herbst}, E.:
  {Cold CO in circumstellar disks. On the effects of photodesorption and
  vertical mixing}.
\newblock \aap \textbf{493}, L49--L52 (2009).
\newblock \doi{10.1051/0004-6361:200811082}

\bibitem{2010A&A...513A..41H}
{Hily-Blant}, P., {Walmsley}, M., {Pineau Des For{\^e}ts}, G., {Flower}, D.:
  {Nitrogen chemistry and depletion in starless cores}.
\newblock \aap \textbf{513}, A41 (2010).
\newblock \doi{10.1051/0004-6361/200913200}

\bibitem{1992ApJ...394..539H}
{Hirahara}, Y., {Suzuki}, H., {Yamamoto}, S., {Kawaguchi}, K., {Kaifu}, N.,
  {Ohishi}, M., {Takano}, S., {Ishikawa}, S.I., {Masuda}, A.: {Mapping
  observations of sulfur-containing carbon-chain molecules in Taurus Molecular
  Cloud 1 (TMC-1)}.
\newblock \apj \textbf{394}, 539--551 (1992).
\newblock \doi{10.1086/171605}

\bibitem{1998ApJ...498..710H}
{Hiraoka}, K., {Miyagoshi}, T., {Takayama}, T., {Yamamoto}, K., {Kihara}, Y.:
  {Gas-Grain Processes for the Formation of CH 4 and H 2O: Reactions of H Atoms
  with C, O, and CO in the Solid Phase at 12 K}.
\newblock \apj \textbf{498}, 710 (1998).
\newblock \doi{10.1086/305572}

\bibitem{2002ApJ...577..265H}
{Hiraoka}, K., {Sato}, T., {Sato}, S., {Sogoshi}, N., {Yokoyama}, T.,
  {Takashima}, H., {Kitagawa}, S.: {Formation of Formaldehyde by the Tunneling
  Reaction of H with Solid CO at 10 K Revisited}.
\newblock \apj \textbf{577}, 265--270 (2002).
\newblock \doi{10.1086/342132}

\bibitem{2011Sci...334..338H}
{Hogerheijde}, M.R., {Bergin}, E.A., {Brinch}, C., {Cleeves}, L.I., {Fogel},
  J.K.J., {Blake}, G.A., {Dominik}, C., {Lis}, D.C., {Melnick}, G., {Neufeld},
  D., {Pani{\'c}}, O., {Pearson}, J.C., {Kristensen}, L., {Y{\i}ld{\i}z}, U.A.,
  {van Dishoeck}, E.F.: {Detection of the Water Reservoir in a Forming
  Planetary System}.
\newblock Science \textbf{334}, 338-- (2011).
\newblock \doi{10.1126/science.1208931}

\bibitem{2009ApJ...690.1497H}
{Hollenbach}, D., {Kaufman}, M.J., {Bergin}, E.A., {Melnick}, G.J.: {Water,
  O$_{2}$, and Ice in Molecular Clouds}.
\newblock \apj \textbf{690}, 1497--1521 (2009).
\newblock \doi{10.1088/0004-637X/690/2/1497}

\bibitem{1989ApJ...342..306H}
{Hollenbach}, D., {McKee}, C.F.: {Molecule formation and infrared emission in
  fast interstellar shocks. III - Results for J shocks in molecular clouds}.
\newblock \apj \textbf{342}, 306--336 (1989).
\newblock \doi{10.1086/167595}

\bibitem{1971ApJ...163..155H}
{Hollenbach}, D., {Salpeter}, E.E.: {Surface Recombination of Hydrogen
  Molecules}.
\newblock \apj \textbf{163}, 155 (1971).
\newblock \doi{10.1086/150754}

\bibitem{2009ApJ...690L.110H}
{Honda}, M., {Inoue}, A.K., {Fukagawa}, M., {Oka}, A., {Nakamoto}, T., {Ishii},
  M., {Terada}, H., {Takato}, N., {Kawakita}, H., {Okamoto}, Y.K., {Shibai},
  H., {Tamura}, M., {Kudo}, T., {Itoh}, Y.: {Detection of Water Ice Grains on
  the Surface of the Circumstellar Disk Around HD 142527}.
\newblock \apjl \textbf{690}, L110--L113 (2009).
\newblock \doi{10.1088/0004-637X/690/2/L110}

\bibitem{2007EM&P..100...43H}
{Horner}, J., {Mousis}, O., {Hersant}, F.: {Constraints on the Formation
  Regions of Comets from their D:H Ratios}.
\newblock Earth Moon and Planets \textbf{100}, 43--56 (2007).
\newblock \doi{10.1007/s11038-006-9096-4}

\bibitem{2009ApJ...698..131H}
{Hughes}, A.M., {Andrews}, S.M., {Espaillat}, C., {Wilner}, D.J., {Calvet}, N.,
  {D'Alessio}, P., {Qi}, C., {Williams}, J.P., {Hogerheijde}, M.R.: {A
  Spatially Resolved Inner Hole in the Disk Around GM Aurigae}.
\newblock \apj \textbf{698}, 131--142 (2009).
\newblock \doi{10.1088/0004-637X/698/1/131}

\bibitem{2011ApJ...740...38H}
{Hughes}, A.M., {Wilner}, D.J., {Andrews}, S.M., {Williams}, J.P., {Su},
  K.Y.L., {Murray-Clay}, R.A., {Qi}, C.: {Resolved Submillimeter Observations
  of the HR 8799 and HD 107146 Debris Disks}.
\newblock \apj \textbf{740}, 38 (2011).
\newblock \doi{10.1088/0004-637X/740/1/38}

\bibitem{2011MNRAS.417.2950I}
{Ilee}, J.D., {Boley}, A.C., {Caselli}, P., {Durisen}, R.H., {Hartquist}, T.W.,
  {Rawlings}, J.M.C.: {Chemistry in a gravitationally unstable protoplanetary
  disc}.
\newblock \mnras \textbf{417}, 2950--2961 (2011).
\newblock \doi{10.1111/j.1365-2966.2011.19455.x}

\bibitem{2004A&A...415..643I}
{Ilgner}, M., {Henning}, T., {Markwick}, A.J., {Millar}, T.J.: {Transport
  processes and chemical evolution in steady accretion disk flows}.
\newblock \aap \textbf{415}, 643--659 (2004).
\newblock \doi{10.1051/0004-6361:20034061}

\bibitem{2008ApJ...686.1474I}
{Ioppolo}, S., {Cuppen}, H.M., {Romanzin}, C., {van Dishoeck}, E.F.,
  {Linnartz}, H.: {Laboratory Evidence for Efficient Water Formation in
  Interstellar Ices}.
\newblock \apj \textbf{686}, 1474--1479 (2008).
\newblock \doi{10.1086/591506}

\bibitem{2009A&A...493.1017I}
{Ioppolo}, S., {Palumbo}, M.E., {Baratta}, G.A., {Mennella}, V.: {Formation of
  interstellar solid CO$\{$\_2$\}$ after energetic processing of icy grain
  mantles}.
\newblock \aap \textbf{493}, 1017--1028 (2009).
\newblock \doi{10.1051/0004-6361:200809769}

\bibitem{2011MNRAS.413.2281I}
{Ioppolo}, S., {van Boheemen}, Y., {Cuppen}, H.M., {van Dishoeck}, E.F.,
  {Linnartz}, H.: {Surface formation of CO$_{2}$ ice at low temperatures}.
\newblock \mnras \textbf{413}, 2281--2287 (2011).
\newblock \doi{10.1111/j.1365-2966.2011.18306.x}

\bibitem{1989ApJ...342..871I}
{Irvine}, W.M., {Friberg}, P., {Kaifu}, N., {Kawaguchi}, K., {Kitamura}, Y.,
  {Matthews}, H.E., {Minh}, Y., {Saito}, S., {Ukita}, N., {Yamamoto}, S.:
  {Observations of some oxygen-containing and sulfur-containing organic
  molecules in cold dark clouds}.
\newblock \apj \textbf{342}, 871--875 (1989).
\newblock \doi{10.1086/167643}

\bibitem{2006A&A...451..951I}
{Isella}, A., {Testi}, L., {Natta}, A.: {Large dust grains in the inner region
  of circumstellar disks}.
\newblock \aap \textbf{451}, 951--959 (2006).
\newblock \doi{10.1051/0004-6361:20054647}

\bibitem{2009EM&P..105..167J}
{Jehin}, E., {Manfroid}, J., {Hutsem{\'e}kers}, D., {Arpigny}, C., {Zucconi},
  J.M.: {Isotopic Ratios in Comets: Status and Perspectives}.
\newblock Earth Moon and Planets \textbf{105}, 167--180 (2009).
\newblock \doi{10.1007/s11038-009-9322-y}

\bibitem{2011ApJ...741L...9J}
{Jing}, D., {He}, J., {Brucato}, J., {De Sio}, A., {Tozzetti}, L., {Vidali},
  G.: {On Water Formation in the Interstellar Medium: Laboratory Study of the
  O+D Reaction on Surfaces}.
\newblock \apjl \textbf{741}, L9 (2011).
\newblock \doi{10.1088/2041-8205/741/1/L9}

\bibitem{1985MNRAS.217..413J}
{Jones}, A.P., {Williams}, D.A.: {Time-dependent sticking coefficients and
  mantle growth on interstellar grains}.
\newblock \mnras \textbf{217}, 413--421 (1985)

\bibitem{2012A&A...543A.128J}
{Joos}, M., {Hennebelle}, P., {Ciardi}, A.: {Protostellar disk formation and
  transport of angular momentum during magnetized core collapse}.
\newblock \aap \textbf{543}, A128 (2012).
\newblock \doi{10.1051/0004-6361/201118730}

\bibitem{2007ApJ...659..479J}
{J{\o}rgensen}, J.K., {Bourke}, T.L., {Myers}, P.C., {Di Francesco}, J., {van
  Dishoeck}, E.F., {Lee}, C.F., {Ohashi}, N., {Sch{\"o}ier}, F.L., {Takakuwa},
  S., {Wilner}, D.J., {Zhang}, Q.: {PROSAC: A Submillimeter Array Survey of
  Low-Mass Protostars. I. Overview of Program: Envelopes, Disks, Outflows, and
  Hot Cores}.
\newblock \apj \textbf{659}, 479--498 (2007).
\newblock \doi{10.1086/512230}

\bibitem{2011A&A...534A.100J}
{J{\o}rgensen}, J.K., {Bourke}, T.L., {Nguyen Luong}, Q., {Takakuwa}, S.:
  {Arcsecond resolution images of the chemical structure of the low-mass
  protostar IRAS 16293-2422. An overview of a large molecular line survey from
  the Submillimeter Array}.
\newblock \aap \textbf{534}, A100 (2011).
\newblock \doi{10.1051/0004-6361/201117139}

\bibitem{2012ApJ...757L...4J}
{J{\o}rgensen}, J.K., {Favre}, C., {Bisschop}, S.E., {Bourke}, T.L., {van
  Dishoeck}, E.F., {Schmalzl}, M.: {Detection of the Simplest Sugar,
  Glycolaldehyde, in a Solar-type Protostar with ALMA}.
\newblock \apjl \textbf{757}, L4 (2012).
\newblock \doi{10.1088/2041-8205/757/1/L4}

\bibitem{2002A&A...389..908J}
{J{\o}rgensen}, J.K., {Sch{\"o}ier}, F.L., {van Dishoeck}, E.F.: {Physical
  structure and CO abundance of low-mass protostellar envelopes}.
\newblock \aap \textbf{389}, 908--930 (2002).
\newblock \doi{10.1051/0004-6361:20020681}

\bibitem{2005A&A...435..177J}
{J{\o}rgensen}, J.K., {Sch{\"o}ier}, F.L., {van Dishoeck}, E.F.: {Molecular
  freeze-out as a tracer of the thermal and dynamical evolution of pre- and
  protostellar cores}.
\newblock \aap \textbf{435}, 177--182 (2005).
\newblock \doi{10.1051/0004-6361:20042092}

\bibitem{2010ApJ...725L.172J}
{J{\o}rgensen}, J.K., {van Dishoeck}, E.F.: {The HDO/H$_{2}$O Ratio in Gas in
  the Inner Regions of a Low-mass Protostar}.
\newblock \apjl \textbf{725}, L172--L175 (2010).
\newblock \doi{10.1088/2041-8205/725/2/L172}

\bibitem{2010ApJ...710L..72J}
{J{\o}rgensen}, J.K., {van Dishoeck}, E.F.: {Water Vapor in the Inner 25 AU of
  a Young Disk Around a Low-Mass Protostar}.
\newblock \apjl \textbf{710}, L72--L76 (2010).
\newblock \doi{10.1088/2041-8205/710/1/L72}

\bibitem{2009A&A...507..861J}
{J{\o}rgensen}, J.K., {van Dishoeck}, E.F., {Visser}, R., {Bourke}, T.L.,
  {Wilner}, D.J., {Lommen}, D., {Hogerheijde}, M.R., {Myers}, P.C.: {PROSAC: a
  submillimeter array survey of low-mass protostars. II. The mass evolution of
  envelopes, disks, and stars from the Class 0 through I stages}.
\newblock \aap \textbf{507}, 861--879 (2009).
\newblock \doi{10.1051/0004-6361/200912325}

\bibitem{2004PASJ...56...69K}
{Kaifu}, N., {Ohishi}, M., {Kawaguchi}, K., {Saito}, S., {Yamamoto}, S.,
  {Miyaji}, T., {Miyazawa}, K., {Ishikawa}, S.I., {Noumaru}, C., {Harasawa},
  S., {Okuda}, M., {Suzuki}, H.: {A 8.8--50GHz Complete Spectral Line Survey
  toward TMC-1 I. Survey Data}.
\newblock \pasj \textbf{56}, 69--173 (2004)

\bibitem{2011EAS....46..271K}
{Kamp}, I.: {Evolution of PAHs in Protoplanetary Disks}.
\newblock In: C.~{Joblin}, A.G.G.M. {Tielens} (eds.) EAS Publications Series,
  \emph{EAS Publications Series}, vol.~46, pp. 271--283 (2011).
\newblock \doi{10.1051/eas/1146029}

\bibitem{2011A&A...532A..85K}
{Kamp}, I., {Woitke}, P., {Pinte}, C., {Tilling}, I., {Thi}, W.F., {Menard},
  F., {Duchene}, G., {Augereau}, J.C.: {Continuum and line modelling of discs
  around young stars. II. Line diagnostics for GASPS from the DENT grid}.
\newblock \aap \textbf{532}, A85 (2011).
\newblock \doi{10.1051/0004-6361/201016399}

\bibitem{1996ApJ...456..611K}
{Kaufman}, M.J., {Neufeld}, D.A.: {Far-Infrared Water Emission from
  Magnetohydrodynamic Shock Waves}.
\newblock \apj \textbf{456}, 611 (1996).
\newblock \doi{10.1086/176683}

\bibitem{2011ApJ...734L..30K}
{Kavelaars}, J.J., {Mousis}, O., {Petit}, J.M., {Weaver}, H.A.: {On the
  Formation Location of Uranus and Neptune as Constrained by Dynamical and
  Chemical Models of Comets}.
\newblock \apjl \textbf{734}, L30 (2011).
\newblock \doi{10.1088/2041-8205/734/2/L30}

\bibitem{1996E&PSL.144..577K}
{Kawamoto}, T.: {Experimental constraints on differentiation and H$_{2}$O
  abundance of calc-alkaline magmas}.
\newblock Earth and Planetary Science Letters \textbf{144}, 577--589 (1996).
\newblock \doi{10.1016/S0012-821X(96)00182-3}

\bibitem{2008ApJ...684..411K}
{Keller}, L.D., {Sloan}, G.C., {Forrest}, W.J., {Ayala}, S., {D'Alessio}, P.,
  {Shah}, S., {Calvet}, N., {Najita}, J., {Li}, A., {Hartmann}, L., {Sargent},
  B., {Watson}, D.M., {Chen}, C.H.: {PAH Emission from Herbig Ae/Be Stars}.
\newblock \apj \textbf{684}, 411--429 (2008).
\newblock \doi{10.1086/589818}

\bibitem{2004GeCoA..68.2577K}
{Keller}, L.P., {Messenger}, S., {Flynn}, G.J., {Clemett}, S., {Wirick}, S.,
  {Jacobsen}, C.: {The nature of molecular cloud material in interplanetary
  dust}.
\newblock \gca \textbf{68}, 2577--2589 (2004).
\newblock \doi{10.1016/j.gca.2003.10.044}

\bibitem{2005ApJ...622..404K}
{Kessler-Silacci}, J.E., {Hillenbrand}, L.A., {Blake}, G.A., {Meyer}, M.R.:
  {8-13 {$\mu$}m Spectroscopy of Young Stellar Objects: Evolution of the
  Silicate Feature}.
\newblock \apj \textbf{622}, 404--429 (2005).
\newblock \doi{10.1086/427793}

\bibitem{2008ApJ...683..238K}
{Keto}, E., {Caselli}, P.: {The Different Structures of the Two Classes of
  Starless Cores}.
\newblock \apj \textbf{683}, 238--247 (2008).
\newblock \doi{10.1086/589147}

\bibitem{2010MNRAS.402.1625K}
{Keto}, E., {Caselli}, P.: {Dynamics and depletion in thermally supercritical
  starless cores}.
\newblock \mnras \textbf{402}, 1625--1634 (2010).
\newblock \doi{10.1111/j.1365-2966.2009.16033.x}

\bibitem{2000GeCoA..64.3913K}
{Kita}, N.T., {Nagahara}, H., {Togashi}, S., {Morishita}, Y.: {A short duration
  of chondrule formation in the solar nebula: evidence from \^{}2\^{}6Al in
  Semarkona ferromagnesian chondrules}.
\newblock \gca \textbf{64}, 3913--3922 (2000).
\newblock \doi{10.1016/S0016-7037(00)00488-9}

\bibitem{1998AMR....11..103K}
{Kita}, N.T., {Togashi}, S., {Morishita}, Y., {Terashima}, S., {Yurimoto}, H.:
  {Search for $^{60}$Ni excesses in MET-78008 ureilite: An ion microprobe
  study}.
\newblock Antarctic Meteorite Research \textbf{11}, 103 (1998)

\bibitem{2011ApJ...733...54K}
{Krasnopolsky}, R., {Li}, Z.Y., {Shang}, H.: {Disk Formation in Magnetized
  Clouds Enabled by the Hall Effect}.
\newblock \apj \textbf{733}, 54 (2011).
\newblock \doi{10.1088/0004-637X/733/1/54}

\bibitem{2012A&A...542A...8K}
{Kristensen}, L.E., {van Dishoeck}, E.F., {Bergin}, E.A., {Visser}, R.,
  {Y{\i}ld{\i}z}, U.A., {San Jose-Garcia}, I., {J{\o}rgensen}, J.K., {Herczeg},
  G.J., {Johnstone}, D., {Wampfler}, S.F., {Benz}, A.O., {Bruderer}, S.,
  {Cabrit}, S., {Caselli}, P., {Doty}, S.D., {Harsono}, D., {Herpin}, F.,
  {Hogerheijde}, M.R., {Karska}, A., {van Kempen}, T.A., {Liseau}, R.,
  {Nisini}, B., {Tafalla}, M., {van der Tak}, F., {Wyrowski}, F.: {Water in
  star-forming regions with Herschel (WISH). II. Evolution of 557 GHz
  1$_{10}$-1$_{01}$ emission in low-mass protostars}.
\newblock \aap \textbf{542}, A8 (2012).
\newblock \doi{10.1051/0004-6361/201118146}

\bibitem{2011A&A...531L...1K}
{Kristensen}, L.E., {van Dishoeck}, E.F., {Tafalla}, M., {Bachiller}, R.,
  {Nisini}, B., {Liseau}, R., {Y{\i}ld{\i}z}, U.A.: {Water in low-mass
  star-forming regions with Herschel (WISH-LM). High-velocity H$_{2}$O bullets
  in L1448-MM observed with HIFI}.
\newblock \aap \textbf{531}, L1 (2011).
\newblock \doi{10.1051/0004-6361/201116975}

\bibitem{2010A&A...521L..30K}
{Kristensen}, L.E., {Visser}, R., {van Dishoeck}, E.F., {Y{\i}ld{\i}z}, U.A.,
  {Doty}, S.D., {Herczeg}, G.J., {Liu}, F.C., {Parise}, B., {J{\o}rgensen},
  J.K., {van Kempen}, T.A., {Brinch}, C., {Wampfler}, S.F., {Bruderer}, S.,
  {Benz}, A.O., {Hogerheijde}, M.R., {Deul}, E., {Bachiller}, R., {Baudry}, A.,
  {Benedettini}, M., {Bergin}, E.A., {Bjerkeli}, P., {Blake}, G.A., {Bontemps},
  S., {Braine}, J., {Caselli}, P., {Cernicharo}, J., {Codella}, C., {Daniel},
  F., {de Graauw}, T., {di Giorgio}, A.M., {Dominik}, C., {Encrenaz}, P.,
  {Fich}, M., {Fuente}, A., {Giannini}, T., {Goicoechea}, J.R., {Helmich}, F.,
  {Herpin}, F., {Jacq}, T., {Johnstone}, D., {Kaufman}, M.J., {Larsson}, B.,
  {Lis}, D., {Liseau}, R., {Marseille}, M., {McCoey}, C., {Melnick}, G.,
  {Neufeld}, D., {Nisini}, B., {Olberg}, M., {Pearson}, J.C., {Plume}, R.,
  {Risacher}, C., {Santiago-Garc{\'{\i}}a}, J., {Saraceno}, P., {Shipman}, R.,
  {Tafalla}, M., {Tielens}, A.G.G.M., {van der Tak}, F., {Wyrowski}, F.,
  {Beintema}, D., {de Jonge}, A., {Dieleman}, P., {Ossenkopf}, V., {Roelfsema},
  P., {Stutzki}, J., {Whyborn}, N.: {Water in low-mass star-forming regions
  with Herschel . HIFI spectroscopy of NGC 1333}.
\newblock \aap \textbf{521}, L30 (2010).
\newblock \doi{10.1051/0004-6361/201015100}

\bibitem{2009ApJ...696..841K}
{Kwon}, W., {Looney}, L.W., {Mundy}, L.G., {Chiang}, H.F., {Kemball}, A.J.:
  {Grain Growth and Density Distribution of the Youngest Protostellar Systems}.
\newblock \apj \textbf{696}, 841--852 (2009).
\newblock \doi{10.1088/0004-637X/696/1/841}

\bibitem{2011ApJ...728...71L}
{Laas}, J.C., {Garrod}, R.T., {Herbst}, E., {Widicus Weaver}, S.L.:
  {Contributions from Grain Surface and Gas Phase Chemistry to the Formation of
  Methyl Formate and Its Structural Isomers}.
\newblock \apj \textbf{728}, 71 (2011).
\newblock \doi{10.1088/0004-637X/728/1/71}

\bibitem{2003ApJ...586..286L}
{Lada}, C.J., {Bergin}, E.A., {Alves}, J.F., {Huard}, T.L.: {The Dynamical
  State of Barnard 68: A Thermally Supported, Pulsating Dark Cloud}.
\newblock \apj \textbf{586}, 286--295 (2003).
\newblock \doi{10.1086/367610}

\bibitem{2006ApJ...636L.145L}
{Lahuis}, F., {van Dishoeck}, E.F., {Boogert}, A.C.A., {Pontoppidan}, K.M.,
  {Blake}, G.A., {Dullemond}, C.P., {Evans} II, N.J., {Hogerheijde}, M.R.,
  {J{\o}rgensen}, J.K., {Kessler-Silacci}, J.E., {Knez}, C.: {Hot Organic
  Molecules toward a Young Low-Mass Star: A Look at Inner Disk Chemistry}.
\newblock \apjl \textbf{636}, L145--L148 (2006).
\newblock \doi{10.1086/500084}

\bibitem{2009ApJ...696L.133L}
{Lattelais}, M., {Pauzat}, F., {Ellinger}, Y., {Ceccarelli}, C.: {Interstellar
  Complex Organic Molecules and the Minimum Energy Principle}.
\newblock \apjl \textbf{696}, L133--L136 (2009).
\newblock \doi{10.1088/0004-637X/696/2/L133}

\bibitem{1994ApJ...436..335L}
{Laughlin}, G., {Bodenheimer}, P.: {Nonaxisymmetric evolution in protostellar
  disks}.
\newblock \apj \textbf{436}, 335--354 (1994).
\newblock \doi{10.1086/174909}

\bibitem{2012M&PS...47..345L}
{Le Guillou}, C., {Rouzaud}, J.N., {Bonal}, L., {Quirico}, E., {Derenne}, S.,
  {Remusat}, L.: {High resolution TEM of chondritic carbonaceous matter:
  Metamorphic evolution and heterogeneity}.
\newblock Meteoritics and Planetary Science \textbf{47}, 345--362 (2012).
\newblock \doi{10.1111/j.1945-5100.2012.01336.x}

\bibitem{2000E&PSL.181...33L}
{L{\'e}cuyer}, C., {Simon}, L., {Guy}, F.: {Comparison of carbon, nitrogen and
  water budgets on Venus and the Earth}.
\newblock Earth and Planetary Science Letters \textbf{181}, 33--40 (2000).
\newblock \doi{10.1016/S0012-821X(00)00195-3}

\bibitem{2005ApJ...631..351L}
{Lee}, J.E., {Evans} II, N.J., {Bergin}, E.A.: {Comparisons of an Evolutionary
  Chemical Model with Other Models}.
\newblock \apj \textbf{631}, 351--360 (2005).
\newblock \doi{10.1086/432531}

\bibitem{2011ApJ...736..135L}
{Lee}, N., {Williams}, J.P., {Cieza}, L.A.: {Protoplanetary Disk Masses in
  IC348: A Rapid Decline in the Population of Small Dust Grains After 1 Myr}.
\newblock \apj \textbf{736}, 135 (2011).
\newblock \doi{10.1088/0004-637X/736/2/135}

\bibitem{1998ApJ...506..898L}
{Lee}, T., {Shu}, F.H., {Shang}, H., {Glassgold}, A.E., {Rehm}, K.E.:
  {Protostellar Cosmic Rays and Extinct Radioactivities in Meteorites}.
\newblock \apj \textbf{506}, 898--912 (1998).
\newblock \doi{10.1086/306284}

\bibitem{2010A&A...518L.113L}
{Lefloch}, B., {Cabrit}, S., {Codella}, C., {Melnick}, G., {Cernicharo}, J.,
  {Caux}, E., {Benedettini}, M., {Boogert}, A., {Caselli}, P., {Ceccarelli},
  C., {Gueth}, F., {Hily-Blant}, P., {Lorenzani}, A., {Neufeld}, D., {Nisini},
  B., {Pacheco}, S., {Pagani}, L., {Pardo}, J.R., {Parise}, B., {Salez}, M.,
  {Schuster}, K., {Viti}, S., {Bacmann}, A., {Baudry}, A., {Bell}, T.,
  {Bergin}, E.A., {Blake}, G., {Bottinelli}, S., {Castets}, A., {Comito}, C.,
  {Coutens}, A., {Crimier}, N., {Dominik}, C., {Demyk}, K., {Encrenaz}, P.,
  {Falgarone}, E., {Fuente}, A., {Gerin}, M., {Goldsmith}, P., {Helmich}, F.,
  {Hennebelle}, P., {Henning}, T., {Herbst}, E., {Jacq}, T., {Kahane}, C.,
  {Kama}, M., {Klotz}, A., {Langer}, W., {Lis}, D., {Lord}, S., {Maret}, S.,
  {Pearson}, J., {Phillips}, T., {Saraceno}, P., {Schilke}, P., {Tielens}, X.,
  {van der Tak}, F., {van der Wiel}, M., {Vastel}, C., {Wakelam}, V.,
  {Walters}, A., {Wyrowski}, F., {Yorke}, H., {Bachiller}, R., {Borys}, C., {de
  Lange}, G., {Delorme}, Y., {Kramer}, C., {Larsson}, B., {Lai}, R., {Maiwald},
  F.W., {Martin-Pintado}, J., {Mehdi}, I., {Ossenkopf}, V., {Siegel}, P.,
  {Stutzki}, J., {Wunsch}, J.H.: {The CHESS spectral survey of star forming
  regions: Peering into the protostellar shock L1157-B1. II. Shock dynamics}.
\newblock \aap \textbf{518}, L113 (2010).
\newblock \doi{10.1051/0004-6361/201014630}

\bibitem{1998A&A...334..269L}
{Lefloch}, B., {Castets}, A., {Cernicharo}, J., {Langer}, W.D., {Zylka}, R.:
  {Cores and cavities in NGC 1333}.
\newblock \aap \textbf{334}, 269--279 (1998)

\bibitem{2011A&A...527L...3L}
{Lefloch}, B., {Cernicharo}, J., {Pacheco}, S., {Ceccarelli}, C.: {Shocked
  water in the Cepheus E protostellar outflow}.
\newblock \aap \textbf{527}, L3 (2011).
\newblock \doi{10.1051/0004-6361/201016247}

\bibitem{1985A&A...144..147L}
{Leger}, A., {Jura}, M., {Omont}, A.: {Desorption from interstellar grains}.
\newblock \aap \textbf{144}, 147--160 (1985)

\bibitem{2010Sci...329..187L}
{Levison}, H.F., {Duncan}, M.J., {Brasser}, R., {Kaufmann}, D.E.: {Capture of
  the Sun's Oort Cloud from Stars in Its Birth Cluster}.
\newblock Science \textbf{329}, 187-- (2010).
\newblock \doi{10.1126/science.1187535}

\bibitem{2012A&A...537A..73L}
{Licandro}, J., {Hargrove}, K., {Kelley}, M., {Campins}, H., {Ziffer}, J.,
  {Al{\'{\i}}-Lagoa}, V., {Fern{\'a}ndez}, Y., {Rivkin}, A.: {5-14 {$\mu$}m
  Spitzer spectra of Themis family asteroids}.
\newblock \aap \textbf{537}, A73 (2012).
\newblock \doi{10.1051/0004-6361/201118142}

\bibitem{2010ApJ...710L..49L}
{Lis}, D.C., {Wootten}, A., {Gerin}, M., {Roueff}, E.: {Nitrogen Isotopic
  Fractionation in Interstellar Ammonia}.
\newblock \apjl \textbf{710}, L49--L52 (2010).
\newblock \doi{10.1088/2041-8205/710/1/L49}

\bibitem{1996A&A...315L.181L}
{Liseau}, R., {Ceccarelli}, C., {Larsson}, B., {Nisini}, B., {White}, G.J.,
  {Ade}, P., {Armand}, C., {Burgdorf}, M., {Caux}, E., {Cerulli}, R., {Church},
  S., {Clegg}, P.E., {Digorgio}, A., {Furniss}, I., {Giannini}, T.,
  {Glencross}, W., {Gry}, C., {King}, K., {Lim}, T., {Lorenzetti}, D.,
  {Molinari}, S., {Naylor}, D., {Orfei}, R., {Saraceno}, P., {Sidher}, S.,
  {Smith}, H., {Spinoglio}, L., {Swinyard}, B., {Texier}, D., {Tommasi}, E.,
  {Trams}, N., {Unger}, S.: {Thermal H\_2\_O emission from the Herbig-Haro flow
  HH 54.}
\newblock \aap \textbf{315}, L181--L184 (1996)

\bibitem{2011A&A...527A..19L}
{Liu}, F.C., {Parise}, B., {Kristensen}, L., {Visser}, R., {van Dishoeck},
  E.F., {G{\"u}sten}, R.: {Water deuterium fractionation in the low-mass
  protostar NGC1333-IRAS2A}.
\newblock \aap \textbf{527}, A19 (2011).
\newblock \doi{10.1051/0004-6361/201015519}

\bibitem{2001ApJ...552L.163L}
{Loinard}, L., {Castets}, A., {Ceccarelli}, C., {Caux}, E., {Tielens},
  A.G.G.M.: {Doubly Deuterated Molecular Species in Protostellar Environments}.
\newblock \apjl \textbf{552}, L163--L166 (2001).
\newblock \doi{10.1086/320331}

\bibitem{2010A&A...515A..77L}
{Lommen}, D.J.P., {van Dishoeck}, E.F., {Wright}, C.M., {Maddison}, S.T.,
  {Min}, M., {Wilner}, D.J., {Salter}, D.M., {van Langevelde}, H.J., {Bourke},
  T.L., {van der Burg}, R.F.J., {Blake}, G.A.: {Grain growth across
  protoplanetary discs: 10 {$\mu$}m silicate feature versus millimetre slope}.
\newblock \aap \textbf{515}, A77 (2010).
\newblock \doi{10.1051/0004-6361/200913150}

\bibitem{2000ApJ...529..477L}
{Looney}, L.W., {Mundy}, L.G., {Welch}, W.J.: {Unveiling the Circumstellar
  Envelope and Disk: A Subarcsecond Survey of Circumstellar Structures}.
\newblock \apj \textbf{529}, 477--498 (2000).
\newblock \doi{10.1086/308239}

\bibitem{2011PASJ...63..555M}
{Machida}, M.N., {Inutsuka}, S.I., {Matsumoto}, T.: {Effect of Magnetic Braking
  on Circumstellar Disk Formation in a Strongly Magnetized Cloud}.
\newblock \pasj \textbf{63}, 555-- (2011)

\bibitem{2007A&A...464...43M}
{Malbet}, F., {Benisty}, M., {de Wit}, W.J., {Kraus}, S., {Meilland}, A.,
  {Millour}, F., {Tatulli}, E., {Berger}, J.P., {Chesneau}, O., {Hofmann},
  K.H., {Isella}, A., {Natta}, A., {Petrov}, R.G., {Preibisch}, T., {Stee}, P.,
  {Testi}, L., {Weigelt}, G., {Antonelli}, P., {Beckmann}, U., {Bresson}, Y.,
  {Chelli}, A., {Dugu{\'e}}, M., {Duvert}, G., {Gennari}, S., {Gl{\"u}ck}, L.,
  {Kern}, P., {Lagarde}, S., {Le Coarer}, E., {Lisi}, F., {Perraut}, K.,
  {Puget}, P., {Rantakyr{\"o}}, F., {Robbe-Dubois}, S., {Roussel}, A., {Zins},
  G., {Accardo}, M., {Acke}, B., {Agabi}, K., {Altariba}, E., {Arezki}, B.,
  {Aristidi}, E., {Baffa}, C., {Behrend}, J., {Bl{\"o}cker}, T., {Bonhomme},
  S., {Busoni}, S., {Cassaing}, F., {Clausse}, J.M., {Colin}, J., {Connot}, C.,
  {Delboulb{\'e}}, A., {Domiciano de Souza}, A., {Driebe}, T., {Feautrier}, P.,
  {Ferruzzi}, D., {Forveille}, T., {Fossat}, E., {Foy}, R., {Fraix-Burnet}, D.,
  {Gallardo}, A., {Giani}, E., {Gil}, C., {Glentzlin}, A., {Heiden}, M.,
  {Heininger}, M., {Hernandez Utrera}, O., {Kamm}, D., {Kiekebusch}, M., {Le
  Contel}, D., {Le Contel}, J.M., {Lesourd}, T., {Lopez}, B., {Lopez}, M.,
  {Magnard}, Y., {Marconi}, A., {Mars}, G., {Martinot-Lagarde}, G., {Mathias},
  P., {M{\`e}ge}, P., {Monin}, J.L., {Mouillet}, D., {Mourard}, D., {Nussbaum},
  E., {Ohnaka}, K., {Pacheco}, J., {Perrier}, C., {Rabbia}, Y., {Rebattu}, S.,
  {Reynaud}, F., {Richichi}, A., {Robini}, A., {Sacchettini}, M., {Schertl},
  D., {Sch{\"o}ller}, M., {Solscheid}, W., {Spang}, A., {Stefanini}, P.,
  {Tallon}, M., {Tallon-Bosc}, I., {Tasso}, D., {Vakili}, F., {von der
  L{\"u}he}, O., {Valtier}, J.C., {Vannier}, M., {Ventura}, N.: {Disk and wind
  interaction in the young stellar object <ASTROBJ>MWC 297</ASTROBJ> spatially
  resolved with AMBER/VLTI}.
\newblock \aap \textbf{464}, 43--53 (2007).
\newblock \doi{10.1051/0004-6361:20053924}

\bibitem{2012ApJ...747...92M}
{Mandell}, A.M., {Bast}, J., {van Dishoeck}, E.F., {Blake}, G.A., {Salyk}, C.,
  {Mumma}, M.J., {Villanueva}, G.: {First Detection of Near-infrared Line
  Emission from Organics in Young Circumstellar Disks}.
\newblock \apj \textbf{747}, 92 (2012).
\newblock \doi{10.1088/0004-637X/747/2/92}

\bibitem{2009A&A...503..613M}
{Manfroid}, J., {Jehin}, E., {Hutsem{\'e}kers}, D., {Cochran}, A., {Zucconi},
  J.M., {Arpigny}, C., {Schulz}, R., {St{\"u}we}, J.A., {Ilyin}, I.: {The CN
  isotopic ratios in comets}.
\newblock \aap \textbf{503}, 613--624 (2009).
\newblock \doi{10.1051/0004-6361/200911859}

\bibitem{2010A&A...516A.105M}
{Marcelino}, N., {Br{\"u}nken}, S., {Cernicharo}, J., {Quan}, D., {Roueff}, E.,
  {Herbst}, E., {Thaddeus}, P.: {The puzzling behavior of HNCO isomers in
  molecular clouds}.
\newblock \aap \textbf{516}, A105 (2010).
\newblock \doi{10.1051/0004-6361/200913806}

\bibitem{2007ApJ...665L.127M}
{Marcelino}, N., {Cernicharo}, J., {Ag{\'u}ndez}, M., {Roueff}, E., {Gerin},
  M., {Mart{\'{\i}}n-Pintado}, J., {Mauersberger}, R., {Thum}, C.: {Discovery
  of Interstellar Propylene (CH$_{2}$CHCH$_{3}$): Missing Links in Interstellar
  Gas-Phase Chemistry}.
\newblock \apjl \textbf{665}, L127--L130 (2007).
\newblock \doi{10.1086/521398}

\bibitem{2005ApJ...620..308M}
{Marcelino}, N., {Cernicharo}, J., {Roueff}, E., {Gerin}, M., {Mauersberger},
  R.: {Deuterated Thioformaldehyde in the Barnard 1 Cloud}.
\newblock \apj \textbf{620}, 308--320 (2005).
\newblock \doi{10.1086/426934}

\bibitem{2002Sci...298.2182M}
{Marhas}, K.K., {Goswami}, J.N., {Davis}, A.M.: {Short-Lived Nuclides in
  Hibonite Grains from Murchison: Evidence for Solar System Evolution}.
\newblock Science \textbf{298}, 2182--2185 (2002).
\newblock \doi{10.1126/science.1078322}

\bibitem{2011NatGe...4..240M}
{Marshall}, C.P., {Emry}, J.R., {Olcott Marshall}, A.: {Haematite
  pseudomicrofossils present in the 3.5-billion-year-old Apex Chert}.
\newblock Nature Geoscience \textbf{4}, 240--243 (2011).
\newblock \doi{10.1038/ngeo1084}

\bibitem{2012E&PSL.313...56M}
{Marty}, B.: {The origins and concentrations of water, carbon, nitrogen and
  noble gases on Earth}.
\newblock Earth and Planetary Science Letters \textbf{313}, 56--66 (2012).
\newblock \doi{10.1016/j.epsl.2011.10.040}

\bibitem{2010GeCoA..74..340M}
{Marty}, B., {Zimmermann}, L., {Burnard}, P.G., {Wieler}, R., {Heber}, V.S.,
  {Burnett}, D.L., {Wiens}, R.C., {Bochsler}, P.: {Nitrogen isotopes in the
  recent solar wind from the analysis of Genesis targets: Evidence for large
  scale isotope heterogeneity in the early solar system}.
\newblock \gca \textbf{74}, 340--355 (2010)

\bibitem{2010A&A...518L.127M}
{Mathews}, G.S., {Dent}, W.R.F., {Williams}, J.P., {Howard}, C.D., {Meeus}, G.,
  {Riaz}, B., {Roberge}, A., {Sandell}, G., {Vandenbussche}, B., {Duch{\^e}ne},
  G., {Kamp}, I., {M{\'e}nard}, F., {Montesinos}, B., {Pinte}, C., {Thi}, W.F.,
  {Woitke}, P., {Alacid}, J.M., {Andrews}, S.M., {Ardila}, D.R., {Aresu}, G.,
  {Augereau}, J.C., {Barrado}, D., {Brittain}, S., {Ciardi}, D.R., {Danchi},
  W., {Eiroa}, C., {Fedele}, D., {Grady}, C.A., {de Gregorio-Monsalvo}, I.,
  {Heras}, A., {Huelamo}, N., {Krivov}, A., {Lebreton}, J., {Liseau}, R.,
  {Martin-Zaidi}, C., {Mendigut{\'{\i}}a}, I., {Mora}, A., {Morales-Calderon},
  M., {Nomura}, H., {Pantin}, E., {Pascucci}, I., {Phillips}, N., {Podio}, L.,
  {Poelman}, D.R., {Ramsay}, S., {Rice}, K., {Riviere-Marichalar}, P.,
  {Solano}, E., {Tilling}, I., {Walker}, H., {White}, G.J., {Wright}, G.: {GAS
  in Protoplanetary Systems (GASPS). I. First results}.
\newblock \aap \textbf{518}, L127 (2010).
\newblock \doi{10.1051/0004-6361/201014595}

\bibitem{2012M&PS...47..525M}
{Matrajt}, G., {Messenger}, S., {Brownlee}, D., {Joswiak}, D.: {Diverse forms
  of primordial organic matter identified in interplanetary dust particles}.
\newblock Meteoritics and Planetary Science \textbf{47}, 525--549 (2012).
\newblock \doi{10.1111/j.1945-5100.2011.01310.x}

\bibitem{1985ApJ...290..609M}
{Matthews}, H.E., {Friberg}, P., {Irvine}, W.M.: {The detection of acetaldehyde
  in cold dust clouds}.
\newblock \apj \textbf{290}, 609--614 (1985).
\newblock \doi{10.1086/163018}

\bibitem{1983ApJ...272..149M}
{Matthews}, H.E., {Sears}, T.J.: {The detection of vinyl cyanide in TMC-1}.
\newblock \apj \textbf{272}, 149--153 (1983).
\newblock \doi{10.1086/161271}

\bibitem{2006ApJ...652L.141M}
{McCarthy}, M.C., {Gottlieb}, C.A., {Gupta}, H., {Thaddeus}, P.: {Laboratory
  and Astronomical Identification of the Negative Molecular Ion
  C$_{6}$H$^{-}$}.
\newblock \apjl \textbf{652}, L141--L144 (2006).
\newblock \doi{10.1086/510238}

\bibitem{2000Sci...289.1334M}
{McKeegan}, K.D., {Chaussidon}, M., {Robert}, F.: {Incorporation of Short-Lived
  $^{10}$Be in a Calcium-Aluminum-Rich Inclusion from the Allende Meteorite}.
\newblock Science \textbf{289}, 1334--1337 (2000).
\newblock \doi{10.1126/science.289.5483.1334}

\bibitem{2007ApJ...656L..33M}
{Meibom}, A., {Krot}, A.N., {Robert}, F., {Mostefaoui}, S., {Russell}, S.S.,
  {Petaev}, M.I., {Gounelle}, M.: {Nitrogen and Carbon Isotopic Composition of
  the Sun Inferred from a High-Temperature Solar Nebular Condensate}.
\newblock \apjl \textbf{656}, L33--L36 (2007).
\newblock \doi{10.1086/512052}

\bibitem{1998Sci...279.1707M}
{Meier}, R., {Owen}, T.C., {Jewitt}, D.C., {Matthews}, H.E., {Senay}, M.,
  {Biver}, N., {Bockelee-Morvan}, D., {Crovisier}, J., {Gautier}, D.:
  {Deuterium in Comet C/1995 O1 (Hale-Bopp): Detection of DCN}.
\newblock Science \textbf{279}, 1707 (1998).
\newblock \doi{10.1126/science.279.5357.1707}

\bibitem{2008ApJ...676..518M}
{Meijerink}, R., {Glassgold}, A.E., {Najita}, J.R.: {Atomic Diagnostics of
  X-Ray-Irradiated Protoplanetary Disks}.
\newblock \apj \textbf{676}, 518--531 (2008).
\newblock \doi{10.1086/527411}

\bibitem{2008ApJ...681.1356M}
{Mellon}, R.R., {Li}, Z.Y.: {Magnetic Braking and Protostellar Disk Formation:
  The Ideal MHD Limit}.
\newblock \apj \textbf{681}, 1356--1376 (2008).
\newblock \doi{10.1086/587542}

\bibitem{2007ApJ...661..361M}
{Mer{\'{\i}}n}, B., {Augereau}, J.C., {van Dishoeck}, E.F., {Kessler-Silacci},
  J., {Dullemond}, C.P., {Blake}, G.A., {Lahuis}, F., {Brown}, J.M., {Geers},
  V.C., {Pontoppidan}, K.M., {Comer{\'o}n}, F., {Frasca}, A., {Guieu}, S.,
  {Alcal{\'a}}, J.M., {Boogert}, A.C.A., {Evans} II, N.J., {D'Alessio}, P.,
  {Mundy}, L.G., {Chapman}, N.: {Abundant Crystalline Silicates in the Disk of
  a Very Low Mass Star}.
\newblock \apj \textbf{661}, 361--367 (2007).
\newblock \doi{10.1086/513092}

\bibitem{2000Natur.404..968M}
{Messenger}, S.: {Identification of molecular-cloud material in interplanetary
  dust particles}.
\newblock \nat \textbf{404}, 968--971 (2000)

\bibitem{2012LPI....43.2618M}
{Milam}, S.N., {Charnley}, S.B.: {Observations of Nitrogen Fractionation in
  Prestellar Cores: Nitriles Tracing Interstellar Chemistry}.
\newblock In: Lunar and Planetary Institute Science Conference Abstracts,
  \emph{Lunar and Planetary Inst. Technical Report}, vol.~43, p. 2618 (2012)

\bibitem{2008CPL...456...27M}
{Miyauchi}, N., {Hidaka}, H., {Chigai}, T., {Nagaoka}, A., {Watanabe}, N.,
  {Kouchi}, A.: {Formation of hydrogen peroxide and water from the reaction of
  cold hydrogen atoms with solid oxygen at 10 K}.
\newblock Chemical Physics Letters \textbf{456}, 27--30 (2008).
\newblock \doi{10.1016/j.cplett.2008.02.095}

\bibitem{2010A&A...519A..22M}
{Modica}, P., {Palumbo}, M.E.: {Formation of methyl formate after cosmic ion
  irradiation of icy grain mantles}.
\newblock \aap \textbf{519}, A22 (2010).
\newblock \doi{10.1051/0004-6361/201014101}

\bibitem{2009ApJ...705L.195M}
{Mokrane}, H., {Chaabouni}, H., {Accolla}, M., {Congiu}, E., {Dulieu}, F.,
  {Chehrouri}, M., {Lemaire}, J.L.: {Experimental Evidence for Water Formation
  Via Ozone Hydrogenation on Dust Grains at 10 K}.
\newblock \apjl \textbf{705}, L195--L198 (2009).
\newblock \doi{10.1088/0004-637X/705/2/L195}

\bibitem{2000M&PS...35.1309M}
{Morbidelli}, A., {Chambers}, J., {Lunine}, J.I., {Petit}, J.M., {Robert}, F.,
  {Valsecchi}, G.B., {Cyr}, K.E.: {Source regions and time scales for the
  delivery of water to Earth}.
\newblock Meteoritics and Planetary Science \textbf{35}, 1309--1320 (2000).
\newblock \doi{10.1111/j.1945-5100.2000.tb01518.x}

\bibitem{1979ApJ...228..475M}
{Mouschovias}, T.C.: {Ambipolar diffusion in interstellar clouds - A new
  solution}.
\newblock \apj \textbf{228}, 475--481 (1979).
\newblock \doi{10.1086/156868}

\bibitem{2011ApJ...741...71M}
{Moynier}, F., {Blichert-Toft}, J., {Wang}, K., {Herzog}, G.F., {Albarede}, F.:
  {The Elusive $^{60}$Fe in the Solar Nebula}.
\newblock \apj \textbf{741}, 71 (2011).
\newblock \doi{10.1088/0004-637X/741/2/71}

\bibitem{2004A&A...413..209M}
{Mu{\~n}oz Caro}, G.M., {Meierhenrich}, U., {Schutte}, W.A., {Thiemann},
  W.H.P., {Greenberg}, J.M.: {UV-photoprocessing of interstellar ice analogs:
  Detection of hexamethylenetetramine-based species}.
\newblock \aap \textbf{413}, 209--216 (2004).
\newblock \doi{10.1051/0004-6361:20031447}

\bibitem{2002Natur.416..403M}
{Mu{\~n}oz Caro}, G.M., {Meierhenrich}, U.J., {Schutte}, W.A., {Barbier}, B.,
  {Arcones Segovia}, A., {Rosenbauer}, H., {Thiemann}, W.H.P., {Brack}, A.,
  {Greenberg}, J.M.: {Amino acids from ultraviolet irradiation of interstellar
  ice analogues}.
\newblock \nat \textbf{416}, 403--406 (2002)

\bibitem{2005JMoSt.742..215M}
{M{\"u}ller}, H.S.P., {Schl{\"o}der}, F., {Stutzki}, J., {Winnewisser}, G.:
  {The Cologne Database for Molecular Spectroscopy, CDMS: a useful tool for
  astronomers and spectroscopists}.
\newblock \jmstr \textbf{742}, 215--227 (2005).
\newblock \doi{10.1016/j.molstruc.2005.01.027}

\bibitem{2011ARA&A..49..471M}
{Mumma}, M.J., {Charnley}, S.B.: {The Chemical Composition of Comets: Emerging
  Taxonomies and Natal Heritage}.
\newblock \araa \textbf{49}, 471--524 (2011).
\newblock \doi{10.1146/annurev-astro-081309-130811}

\bibitem{2000ApJS..128..603M}
{Murakawa}, K., {Tamura}, M., {Nagata}, T.: {1-4 Micron Spectrophotometry of
  Dust in the Taurus Dark Cloud: Water Ice Distribution in Heiles Cloud 2}.
\newblock \apjs \textbf{128}, 603--613 (2000).
\newblock \doi{10.1086/313387}

\bibitem{2003ApJ...597L.149M}
{Muzerolle}, J., {Calvet}, N., {Hartmann}, L., {D'Alessio}, P.: {Unveiling the
  Inner Disk Structure of T Tauri Stars}.
\newblock \apjl \textbf{597}, L149--L152 (2003).
\newblock \doi{10.1086/379921}

\bibitem{2011Sci...333.1113N}
{Nakamura}, T., {Noguchi}, T., {Tanaka}, M., {Zolensky}, M.E., {Kimura}, M.,
  {Tsuchiyama}, A., {Nakato}, A., {Ogami}, T., {Ishida}, H., {Uesugi}, M.,
  {Yada}, T., {Shirai}, K., {Fujimura}, A., {Okazaki}, R., {Sandford}, S.A.,
  {Ishibashi}, Y., {Abe}, M., {Okada}, T., {Ueno}, M., {Mukai}, T.,
  {Yoshikawa}, M., {Kawaguchi}, J.: {Itokawa Dust Particles: A Direct Link
  Between S-Type Asteroids and Ordinary Chondrites}.
\newblock Science \textbf{333}, 1113-- (2011).
\newblock \doi{10.1126/science.1207758}

\bibitem{2001A&A...371..186N}
{Natta}, A., {Prusti}, T., {Neri}, R., {Wooden}, D., {Grinin}, V.P.,
  {Mannings}, V.: {A reconsideration of disk properties in Herbig Ae stars}.
\newblock \aap \textbf{371}, 186--197 (2001).
\newblock \doi{10.1051/0004-6361:20010334}

\bibitem{2007prpl.conf..767N}
{Natta}, A., {Testi}, L., {Calvet}, N., {Henning}, T., {Waters}, R., {Wilner},
  D.: {Dust in Protoplanetary Disks: Properties and Evolution}.
\newblock Protostars and Planets V pp. 767--781 (2007)

\bibitem{2007ApJ...656.1223N}
{Nguyen}, A.N., {Stadermann}, F.J., {Zinner}, E., {Stroud}, R.M., {Alexander},
  C.M.O., {Nittler}, L.R.: {Characterization of Presolar Silicate and Oxide
  Grains in Primitive Carbonaceous Chondrites}.
\newblock \apj \textbf{656}, 1223--1240 (2007).
\newblock \doi{10.1086/510612}

\bibitem{2000A&A...360..297N}
{Nisini}, B., {Benedettini}, M., {Giannini}, T., {Codella}, C., {Lorenzetti},
  D., {di Giorgio}, A.M., {Richer}, J.S.: {Far infrared mapping of the gas
  cooling along the L1448 outflow}.
\newblock \aap \textbf{360}, 297--310 (2000)

\bibitem{2011ApJ...735..121N}
{Noble}, J.A., {Dulieu}, F., {Congiu}, E., {Fraser}, H.J.: {CO$_{2}$ Formation
  in Quiescent Clouds: An Experimental Study of the CO + OH Pathway}.
\newblock \apj \textbf{735}, 121 (2011).
\newblock \doi{10.1088/0004-637X/735/2/121}

\bibitem{2012ApJ...749...67O}
{Oba}, Y., {Watanabe}, N., {Hama}, T., {Kuwahata}, K., {Hidaka}, H., {Kouchi},
  A.: {Water Formation through a Quantum Tunneling Surface Reaction, OH +
  H$_{2}$, at 10 K}.
\newblock \apj \textbf{749}, 67 (2012).
\newblock \doi{10.1088/0004-637X/749/1/67}

\bibitem{2010ApJ...712L.174O}
{Oba}, Y., {Watanabe}, N., {Kouchi}, A., {Hama}, T., {Pirronello}, V.:
  {Experimental Study of CO$_{2}$ Formation by Surface Reactions of
  Non-energetic OH Radicals with CO Molecules}.
\newblock \apjl \textbf{712}, L174--L178 (2010).
\newblock \doi{10.1088/2041-8205/712/2/L174}

\bibitem{2009A&A...504..891O}
{{\"O}berg}, K.I., {Garrod}, R.T., {van Dishoeck}, E.F., {Linnartz}, H.:
  {Formation rates of complex organics in UV irradiated CH\_3OH-rich ices. I.
  Experiments}.
\newblock \aap \textbf{504}, 891--913 (2009).
\newblock \doi{10.1051/0004-6361/200912559}

\bibitem{2009ApJ...693.1209O}
{{\"O}berg}, K.I., {Linnartz}, H., {Visser}, R., {van Dishoeck}, E.F.:
  {Photodesorption of Ices. II. H$_{2}$O and D$_{2}$O}.
\newblock \apj \textbf{693}, 1209--1218 (2009).
\newblock \doi{10.1088/0004-637X/693/2/1209}

\bibitem{2010ApJ...720..480O}
{{\"O}berg}, K.I., {Qi}, C., {Fogel}, J.K.J., {Bergin}, E.A., {Andrews}, S.M.,
  {Espaillat}, C., {van Kempen}, T.A., {Wilner}, D.J., {Pascucci}, I.: {The
  Disk Imaging Survey of Chemistry with SMA. I. Taurus Protoplanetary Disk
  Data}.
\newblock \apj \textbf{720}, 480--493 (2010).
\newblock \doi{10.1088/0004-637X/720/1/480}

\bibitem{2011ApJ...734...98O}
{{\"O}berg}, K.I., {Qi}, C., {Fogel}, J.K.J., {Bergin}, E.A., {Andrews}, S.M.,
  {Espaillat}, C., {Wilner}, D.J., {Pascucci}, I., {Kastner}, J.H.: {Disk
  Imaging Survey of Chemistry with SMA. II. Southern Sky Protoplanetary Disk
  Data and Full Sample Statistics}.
\newblock \apj \textbf{734}, 98 (2011).
\newblock \doi{10.1088/0004-637X/734/2/98}

\bibitem{2011ApJ...743..152O}
{{\"O}berg}, K.I., {Qi}, C., {Wilner}, D.J., {Andrews}, S.M.: {The Ionization
  Fraction in the DM Tau Protoplanetary Disk}.
\newblock \apj \textbf{743}, 152 (2011).
\newblock \doi{10.1088/0004-637X/743/2/152}

\bibitem{2011ApJ...740...14O}
{{\"O}berg}, K.I., {van der Marel}, N., {Kristensen}, L.E., {van Dishoeck},
  E.F.: {Complex Molecules toward Low-mass Protostars: The Serpens Core}.
\newblock \apj \textbf{740}, 14 (2011).
\newblock \doi{10.1088/0004-637X/740/1/14}

\bibitem{2009A&A...496..281O}
{{\"O}berg}, K.I., {van Dishoeck}, E.F., {Linnartz}, H.: {Photodesorption of
  ices I: CO, N$_{2}$, and CO$_{2}$}.
\newblock \aap \textbf{496}, 281--293 (2009).
\newblock \doi{10.1051/0004-6361/200810207}

\bibitem{2010ApJ...718..832O}
{{\"O}berg}, K.I., {van Dishoeck}, E.F., {Linnartz}, H., {Andersson}, S.: {The
  Effect of H$_{2}$O on Ice Photochemistry}.
\newblock \apj \textbf{718}, 832--840 (2010).
\newblock \doi{10.1088/0004-637X/718/2/832}

\bibitem{1998FaDi..109..205O}
{Ohishi}, M., {Kaifu}, N.: {Chemical and physical evolution of dark clouds.
  Molecular spectral line survey toward TMC-1}.
\newblock Faraday Discussions \textbf{109}, 205 (1998).
\newblock \doi{10.1039/a801058g}

\bibitem{2003ApJ...587..235O}
{Oliveira}, C.M., {H{\'e}brard}, G., {Howk}, J.C., {Kruk}, J.W., {Chayer}, P.,
  {Moos}, H.W.: {Interstellar Deuterium, Nitrogen, and Oxygen Abundances toward
  GD 246, WD 2331-475, HZ 21, and Lanning 23: Results from the FUSE Mission}.
\newblock \apj \textbf{587}, 235--255 (2003).
\newblock \doi{10.1086/368019}

\bibitem{2012A&A...542A..90O}
{Olofsson}, J., {Juh{\'a}sz}, A., {Henning}, T., {Mutschke}, H., {Tamanai}, A.,
  {Mo{\'o}r}, A., {{\'A}brah{\'a}m}, P.: {Transient dust in warm debris disks.
  Detection of Fe-rich olivine grains}.
\newblock \aap \textbf{542}, A90 (2012).
\newblock \doi{10.1051/0004-6361/201118735}

\bibitem{1994A&A...291..943O}
{Ossenkopf}, V., {Henning}, T.: {Dust opacities for protostellar cores}.
\newblock \aap \textbf{291}, 943--959 (1994)

\bibitem{1995Icar..116..215O}
{Owen}, T., {Bar-Nun}, A.: {Comets, impacts and atmospheres.}
\newblock \icarus \textbf{116}, 215--226 (1995).
\newblock \doi{10.1006/icar.1995.1122}

\bibitem{2001ApJ...553L..77O}
{Owen}, T., {Mahaffy}, P.R., {Niemann}, H.B., {Atreya}, S., {Wong}, M.:
  {Protosolar Nitrogen}.
\newblock \apjl \textbf{553}, L77--L79 (2001).
\newblock \doi{10.1086/320501}

\bibitem{2011A&A...530A.109P}
{Padovani}, M., {Galli}, D.: {Effects of magnetic fields on the cosmic-ray
  ionization of molecular cloud cores}.
\newblock \aap \textbf{530}, A109 (2011).
\newblock \doi{10.1051/0004-6361/201116853}

\bibitem{2007A&A...467..179P}
{Pagani}, L., {Bacmann}, A., {Cabrit}, S., {Vastel}, C.: {Depletion and low gas
  temperature in the L183 (=L134N) prestellar core: the N2H\^{}+-N2D$^{+}$
  tool}.
\newblock \aap \textbf{467}, 179--186 (2007).
\newblock \doi{10.1051/0004-6361:20066670}

\bibitem{2010Sci...329.1622P}
{Pagani}, L., {Steinacker}, J., {Bacmann}, A., {Stutz}, A., {Henning}, T.: {The
  Ubiquity of Micrometer-Sized Dust Grains in the Dense Interstellar Medium}.
\newblock Science \textbf{329}, 1622-- (2010).
\newblock \doi{10.1126/science.1193211}

\bibitem{2009A&A...494..623P}
{Pagani}, L., {Vastel}, C., {Hugo}, E., {Kokoouline}, V., {Greene}, C.H.,
  {Bacmann}, A., {Bayet}, E., {Ceccarelli}, C., {Peng}, R., {Schlemmer}, S.:
  {Chemical modeling of <ASTROBJ>L183</ASTROBJ> (<ASTROBJ>L134N</ASTROBJ>): an
  estimate of the ortho/para H$\{$\_2$\}$ ratio}.
\newblock \aap \textbf{494}, 623--636 (2009).
\newblock \doi{10.1051/0004-6361:200810587}

\bibitem{2000ApJ...542..890P}
{Palumbo}, M.E., {Pendleton}, Y.J., {Strazzulla}, G.: {Hydrogen Isotopic
  Substitution Studies of the 2165 Wavenumber (4.62 Micron) ``XCN'' Feature
  Produced by Ion Bombardment}.
\newblock \apj \textbf{542}, 890--893 (2000).
\newblock \doi{10.1086/317061}

\bibitem{2011A&A...526A..31P}
{Parise}, B., {Belloche}, A., {Du}, F., {G{\"u}sten}, R., {Menten}, K.M.:
  {Extended emission of D$_{2}$H$^{+}$ in a prestellar core}.
\newblock \aap \textbf{526}, A31 (2011).
\newblock \doi{10.1051/0004-6361/201015475}

\bibitem{2004A&A...416..159P}
{Parise}, B., {Castets}, A., {Herbst}, E., {Caux}, E., {Ceccarelli}, C.,
  {Mukhopadhyay}, I., {Tielens}, A.G.G.M.: {First detection of
  triply-deuterated methanol}.
\newblock \aap \textbf{416}, 159--163 (2004).
\newblock \doi{10.1051/0004-6361:20034490}

\bibitem{2005A&A...431..547P}
{Parise}, B., {Caux}, E., {Castets}, A., {Ceccarelli}, C., {Loinard}, L.,
  {Tielens}, A.G.G.M., {Bacmann}, A., {Cazaux}, S., {Comito}, C., {Helmich},
  F., {Kahane}, C., {Schilke}, P., {van Dishoeck}, E., {Wakelam}, V.,
  {Walters}, A.: {HDO abundance in the envelope of the solar-type protostar
  IRAS 16293-2422}.
\newblock \aap \textbf{431}, 547--554 (2005).
\newblock \doi{10.1051/0004-6361:20041899}

\bibitem{2006A&A...453..949P}
{Parise}, B., {Ceccarelli}, C., {Tielens}, A.G.G.M., {Castets}, A., {Caux}, E.,
  {Lefloch}, B., {Maret}, S.: {Testing grain surface chemistry: a survey of
  deuterated formaldehyde and methanol in low-mass class 0 protostars}.
\newblock \aap \textbf{453}, 949--958 (2006).
\newblock \doi{10.1051/0004-6361:20054476}

\bibitem{2002A&A...393L..49P}
{Parise}, B., {Ceccarelli}, C., {Tielens}, A.G.G.M., {Herbst}, E., {Lefloch},
  B., {Caux}, E., {Castets}, A., {Mukhopadhyay}, I., {Pagani}, L., {Loinard},
  L.: {Detection of doubly-deuterated methanol in the solar-type protostar IRAS
  16293-2422}.
\newblock \aap \textbf{393}, L49--L53 (2002).
\newblock \doi{10.1051/0004-6361:20021131}

\bibitem{2010ApJ...723..218P}
{Peng}, R., {Yoshida}, H., {Chamberlin}, R.A., {Phillips}, T.G., {Lis}, D.C.,
  {Gerin}, M.: {A Comprehensive Survey of Hydrogen Chloride in the Galaxy}.
\newblock \apj \textbf{723}, 218--228 (2010).
\newblock \doi{10.1088/0004-637X/723/1/218}

\bibitem{2012A&A...543A.152P}
{Peng}, T.C., {Despois}, D., {Brouillet}, N., {Parise}, B., {Baudry}, A.:
  {Deuterated methanol in Orion BN/KL}.
\newblock \aap \textbf{543}, A152 (2012).
\newblock \doi{10.1051/0004-6361/201118310}

\bibitem{2012A&A...541A..39P}
{Persson}, M.V., {J{\o}rgensen}, J.K., {van Dishoeck}, E.F.: {Subarcsecond
  resolution observations of warm water toward three deeply embedded low-mass
  protostars}.
\newblock \aap \textbf{541}, A39 (2012).
\newblock \doi{10.1051/0004-6361/201117917}

\bibitem{2012LPI....43.1937P}
{Petit}, J.M., {Mousis}, O., {Kavelaars}, J.J.: {Formation Location of
  Enceladus and Comets from D/H Measurements}.
\newblock In: Lunar and Planetary Institute Science Conference Abstracts,
  \emph{Lunar and Planetary Institute Science Conference Abstracts}, vol.~43,
  p. 1937 (2012)

\bibitem{1998JQSRT..60..883P}
{Pickett}, H.M., {Poynter}, R.L., {Cohen}, E.A., {Delitsky}, M.L., {Pearson},
  J.C., {M{\"u}ller}, H.S.P.: {Submillimeter, millimeter and microwave spectral
  line catalog.}
\newblock \jqsrt \textbf{60}, 883--890 (1998).
\newblock \doi{10.1016/S0022-4073(98)00091-0}

\bibitem{2012MNRAS.423.2209P}
{Pilling}, S., {Andrade}, D.P.P., {da Silveira}, E.F., {Rothard}, H.,
  {Domaracka}, A., {Boduch}, P.: {Formation of unsaturated hydrocarbons in
  interstellar ice analogues by cosmic rays}.
\newblock \mnras \textbf{423}, 2209--2221 (2012).
\newblock \doi{10.1111/j.1365-2966.2012.21031.x}

\bibitem{2012A&A...544L...7P}
{Pineda}, J.E., {Maury}, A.J., {Fuller}, G.A., {Testi}, L.,
  {Garc{\'{\i}}a-Appadoo}, D., {Peck}, A.B., {Villard}, E., {Corder}, S.A.,
  {van Kempen}, T.A., {Turner}, J.L., {Tachihara}, K., {Dent}, W.: {The first
  ALMA view of IRAS 16293-2422. Direct detection of infall onto source B and
  high-resolution kinematics of source A}.
\newblock \aap \textbf{544}, L7 (2012).
\newblock \doi{10.1051/0004-6361/201219589}

\bibitem{1999A&A...344..681P}
{Pirronello}, V., {Liu}, C., {Roser}, J.E., {Vidali}, G.: {Measurements of
  molecular hydrogen formation on carbonaceous grains}.
\newblock \aap \textbf{344}, 681--686 (1999)

\bibitem{2009GeCoA..73.2150P}
{Pizzarello}, S., {Holmes}, W.: {Nitrogen-containing compounds in two CR2
  meteorites: $^{15}$N composition, molecular distribution and precursor
  molecules}.
\newblock \gca \textbf{73}, 2150--2162 (2009)

\bibitem{2005GeCoA..69..599P}
{Pizzarello}, S., {Huang}, Y.: {The deuterium enrichment of individual amino
  acids in carbonaceous meteorites: A case for the presolar distribution of
  biomolecule precursors}.
\newblock \gca \textbf{69}, 599--605 (2005).
\newblock \doi{10.1016/j.gca.2004.07.031}

\bibitem{2001Sci...293.2236P}
{Pizzarello}, S., {Huang}, Y., {Becker}, L., {Poreda}, R.J., {Nieman}, R.A.,
  {Cooper}, G., {Williams}, M.: {The Organic Content of the Tagish Lake
  Meteorite}.
\newblock Science \textbf{293}, 2236--2239 (2001).
\newblock \doi{10.1126/science.1062614}

\bibitem{2003GeCoA..67.1589P}
{Pizzarello}, S., {Zolensky}, M., {Turk}, K.A.: {Nonracemic isovaline in the
  Murchison meteorite: chiral distribution and mineral association}.
\newblock \gca \textbf{67}, 1589--1595 (2003).
\newblock \doi{10.1016/S0016-7037(02)01283-8}

\bibitem{2012A&A...545A..44P}
{Podio}, L., {Kamp}, I., {Flower}, D., {Howard}, C., {Sandell}, G., {Mora}, A.,
  {Aresu}, G., {Brittain}, S., {Dent}, W.R.F., {Pinte}, C., {White}, G.J.:
  {Herschel/PACS observations of young sources in Taurus: the far-infrared
  counterpart of optical jets}.
\newblock \aap \textbf{545}, A44 (2012).
\newblock \doi{10.1051/0004-6361/201219475}

\bibitem{1996Icar..124...62P}
{Pollack}, J.B., {Hubickyj}, O., {Bodenheimer}, P., {Lissauer}, J.J.,
  {Podolak}, M., {Greenzweig}, Y.: {Formation of the Giant Planets by
  Concurrent Accretion of Solids and Gas}.
\newblock \icarus \textbf{124}, 62--85 (1996).
\newblock \doi{10.1006/icar.1996.0190}

\bibitem{2010ApJ...722L.173P}
{Pontoppidan}, K.M., {Salyk}, C., {Blake}, G.A., {K{\"a}ufl}, H.U.: {Spectrally
  Resolved Pure Rotational Lines of Water in Protoplanetary Disks}.
\newblock \apjl \textbf{722}, L173--L177 (2010).
\newblock \doi{10.1088/2041-8205/722/2/L173}

\bibitem{2010ApJ...720..887P}
{Pontoppidan}, K.M., {Salyk}, C., {Blake}, G.A., {Meijerink}, R., {Carr}, J.S.,
  {Najita}, J.: {A Spitzer Survey of Mid-infrared Molecular Emission from
  Protoplanetary Disks. I. Detection Rates}.
\newblock \apj \textbf{720}, 887--903 (2010).
\newblock \doi{10.1088/0004-637X/720/1/887}

\bibitem{1983ApJ...267..603P}
{Prasad}, S.S., {Tarafdar}, S.P.: {UV radiation field inside dense clouds - Its
  possible existence and chemical implications}.
\newblock \apj \textbf{267}, 603--609 (1983).
\newblock \doi{10.1086/160896}

\bibitem{2005ApJS..160..390P}
{Preibisch}, T., {Feigelson}, E.D.: {The Evolution of X-Ray Emission in Young
  Stars}.
\newblock \apjs \textbf{160}, 390--400 (2005).
\newblock \doi{10.1086/432094}

\bibitem{2004ApJ...616L..11Q}
{Qi}, C., {Ho}, P.T.P., {Wilner}, D.J., {Takakuwa}, S., {Hirano}, N., {Ohashi},
  N., {Bourke}, T.L., {Zhang}, Q., {Blake}, G.A., {Hogerheijde}, M., {Saito},
  M., {Choi}, M., {Yang}, J.: {Imaging the Disk around TW Hydrae with the
  Submillimeter Array}.
\newblock \apjl \textbf{616}, L11--L14 (2004).
\newblock \doi{10.1086/421063}

\bibitem{2008ApJ...681.1396Q}
{Qi}, C., {Wilner}, D.J., {Aikawa}, Y., {Blake}, G.A., {Hogerheijde}, M.R.:
  {Resolving the Chemistry in the Disk of TW Hydrae. I. Deuterated Species}.
\newblock \apj \textbf{681}, 1396--1407 (2008).
\newblock \doi{10.1086/588516}

\bibitem{2010ApJ...725.2101Q}
{Quan}, D., {Herbst}, E., {Osamura}, Y., {Roueff}, E.: {Gas-grain Modeling of
  Isocyanic Acid (HNCO), Cyanic Acid (HOCN), Fulminic Acid (HCNO), and
  Isofulminic Acid (HONC) in Assorted Interstellar Environments}.
\newblock \apj \textbf{725}, 2101--2109 (2010).
\newblock \doi{10.1088/0004-637X/725/2/2101}

\bibitem{2006ApJ...646..288R}
{Rafikov}, R.R.: {Microwave Emission from Spinning Dust in Circumstellar
  Disks}.
\newblock \apj \textbf{646}, 288--296 (2006).
\newblock \doi{10.1086/504793}

\bibitem{2011A&A...528L..13R}
{Ratajczak}, A., {Taquet}, V., {Kahane}, C., {Ceccarelli}, C., {Faure}, A.,
  {Quirico}, E.: {The puzzling deuteration of methanol in low- to high-mass
  protostars}.
\newblock \aap \textbf{528}, L13 (2011).
\newblock \doi{10.1051/0004-6361/201016402}

\bibitem{2009Icar..203..644R}
{Raymond}, S.N., {O'Brien}, D.P., {Morbidelli}, A., {Kaib}, N.A.: {Building the
  terrestrial planets: Constrained accretion in the inner Solar System}.
\newblock \icarus \textbf{203}, 644--662 (2009).
\newblock \doi{10.1016/j.icarus.2009.05.016}

\bibitem{2002MNRAS.337L..17R}
{Redman}, M.P., {Rawlings}, J.M.C., {Nutter}, D.J., {Ward-Thompson}, D.,
  {Williams}, D.A.: {Molecular gas freeze-out in the pre-stellar core L1689B}.
\newblock \mnras \textbf{337}, L17--L21 (2002).
\newblock \doi{10.1046/j.1365-8711.2002.06106.x}

\bibitem{2006ApJ...643L..37R}
{Remijan}, A.J., {Hollis}, J.M., {Snyder}, L.E., {Jewell}, P.R., {Lovas}, F.J.:
  {Methyltriacetylene (CH$_{3}$C$_{6}$H) toward TMC-1: The Largest Detected
  Symmetric Top}.
\newblock \apjl \textbf{643}, L37--L40 (2006).
\newblock \doi{10.1086/504918}

\bibitem{2005LPI....36.1350R}
{Remusat}, L., {Palhol}, F., {Robert}, F., {Derenne}, S.: {Hydrogen Isotopic
  Composition of Aliphatic Linkages in Carbonaceous Chondrites Insoluble
  Organic Matter}.
\newblock In: S.~{Mackwell}, E.~{Stansbery} (eds.) 36th Annual Lunar and
  Planetary Science Conference, \emph{Lunar and Planetary Institute Science
  Conference Abstracts}, vol.~36, p. 1350 (2005)

\bibitem{2009ApJ...698.2087R}
{Remusat}, L., {Robert}, F., {Meibom}, A., {Mostefaoui}, S., {Delpoux}, O.,
  {Binet}, L., {Gourier}, D., {Derenne}, S.: {Proto-Planetary Disk Chemistry
  Recorded by D-Rich Organic Radicals in Carbonaceous Chondrites}.
\newblock \apj \textbf{698}, 2087--2092 (2009).
\newblock \doi{10.1088/0004-637X/698/2/2087}

\bibitem{2007ApJ...655L..37R}
{Requena-Torres}, M.A., {Marcelino}, N., {Jim{\'e}nez-Serra}, I.,
  {Mart{\'{\i}}n-Pintado}, J., {Mart{\'{\i}}n}, S., {Mauersberger}, R.:
  {Organic Chemistry in the Dark Clouds L1448 and L183: A Unique Grain Mantle
  Composition}.
\newblock \apjl \textbf{655}, L37--L40 (2007).
\newblock \doi{10.1086/511677}

\bibitem{2012MNRAS.420.2603R}
{Riaz}, B., {Honda}, M., {Campins}, H., {Micela}, G., {Guarcello}, M.G.,
  {Gledhill}, T., {Hough}, J., {Mart{\'{\i}}n}, E.L.: {The radial distribution
  of dust species in young brown dwarf discs}.
\newblock \mnras \textbf{420}, 2603--2624 (2012).
\newblock \doi{10.1111/j.1365-2966.2011.20233.x}

\bibitem{2012A&A...539L...6R}
{Ricci}, L., {Testi}, L., {Maddison}, S.T., {Wilner}, D.J.: {Fomalhaut debris
  disk emission at 7 millimeters: constraints on the collisional models of
  planetesimals}.
\newblock \aap \textbf{539}, L6 (2012).
\newblock \doi{10.1051/0004-6361/201118524}

\bibitem{2003SSRv..106...87R}
{Robert}, F.: {The D/H Ratio in Chondrites}.
\newblock \ssr \textbf{106}, 87--101 (2003).
\newblock \doi{10.1023/A:1024629402715}

\bibitem{2006M&PSA..41.5259R}
{Robert}, F., {Derenne}, S.: {The Molecular Structure and Isotopic Compositions
  of the Insoluble Organic Matter in Chondrites}.
\newblock Meteoritics and Planetary Science Supplement \textbf{41}, 5259 (2006)

\bibitem{2003ApJ...591L..41R}
{Roberts}, H., {Herbst}, E., {Millar}, T.J.: {Enhanced Deuterium Fractionation
  in Dense Interstellar Cores Resulting from Multiply Deuterated
  H$^{+}$$_{3}$}.
\newblock \apjl \textbf{591}, L41--L44 (2003).
\newblock \doi{10.1086/376962}

\bibitem{2007MNRAS.382..733R}
{Roberts}, J.F., {Rawlings}, J.M.C., {Viti}, S., {Williams}, D.A.: {Desorption
  from interstellar ices}.
\newblock \mnras \textbf{382}, 733--742 (2007).
\newblock \doi{10.1111/j.1365-2966.2007.12402.x}

\bibitem{2006ApJS..167..256R}
{Robitaille}, T.P., {Whitney}, B.A., {Indebetouw}, R., {Wood}, K., {Denzmore},
  P.: {Interpreting Spectral Energy Distributions from Young Stellar Objects.
  I. A Grid of 200,000 YSO Model SEDs}.
\newblock \apjs \textbf{167}, 256--285 (2006).
\newblock \doi{10.1086/508424}

\bibitem{2005ApJ...621L.133R}
{Rodr{\'{\i}}guez}, L.F., {Loinard}, L., {D'Alessio}, P., {Wilner}, D.J., {Ho},
  P.T.P.: {IRAS 16293-2422B: A Compact, Possibly Isolated Protoplanetary Disk
  in a Class 0 Object}.
\newblock \apjl \textbf{621}, L133--L136 (2005).
\newblock \doi{10.1086/429223}

\bibitem{2011JChPh.134h4504R}
{Romanzin}, C., {Ioppolo}, S., {Cuppen}, H.M., {van Dishoeck}, E.F.,
  {Linnartz}, H.: {Water formation by surface O3 hydrogenation}.
\newblock \jcp \textbf{134}(8), 084,504 (2011).
\newblock \doi{10.1063/1.3532087}

\bibitem{2007ApJ...663.1174S}
{Sakai}, N., {Ikeda}, M., {Morita}, M., {Sakai}, T., {Takano}, S., {Osamura},
  Y., {Yamamoto}, S.: {Production Pathways of CCS and CCCS Inferred from Their
  $^{13}$C Isotopic Species}.
\newblock \apj \textbf{663}, 1174--1179 (2007).
\newblock \doi{10.1086/518595}

\bibitem{2008ApJ...672..371S}
{Sakai}, N., {Sakai}, T., {Hirota}, T., {Yamamoto}, S.: {Abundant Carbon-Chain
  Molecules toward the Low-Mass Protostar IRAS 04368+2557 in L1527}.
\newblock \apj \textbf{672}, 371--381 (2008).
\newblock \doi{10.1086/523635}

\bibitem{2010A&A...512A..31S}
{Sakai}, N., {Saruwatari}, O., {Sakai}, T., {Takano}, S., {Yamamoto}, S.:
  {Abundance anomaly of the $^{13}$C species of CCH}.
\newblock \aap \textbf{512}, A31 (2010).
\newblock \doi{10.1051/0004-6361/200913098}

\bibitem{2010ApJ...718L..49S}
{Sakai}, N., {Shiino}, T., {Hirota}, T., {Sakai}, T., {Yamamoto}, S.: {Long
  Carbon-chain Molecules and Their Anions in the Starless Core, Lupus-1A}.
\newblock \apjl \textbf{718}, L49--L52 (2010).
\newblock \doi{10.1088/2041-8205/718/2/L49}

\bibitem{2008ApJ...676L..49S}
{Salyk}, C., {Pontoppidan}, K.M., {Blake}, G.A., {Lahuis}, F., {van Dishoeck},
  E.F., {Evans} II, N.J.: {H$_{2}$O and OH Gas in the Terrestrial
  Planet-forming Zones of Protoplanetary Disks}.
\newblock \apjl \textbf{676}, L49--L52 (2008).
\newblock \doi{10.1086/586894}

\bibitem{2011ApJ...731..130S}
{Salyk}, C., {Pontoppidan}, K.M., {Blake}, G.A., {Najita}, J.R., {Carr}, J.S.:
  {A Spitzer Survey of Mid-infrared Molecular Emission from Protoplanetary
  Disks. II. Correlations and Local Thermal Equilibrium Models}.
\newblock \apj \textbf{731}, 130 (2011).
\newblock \doi{10.1088/0004-637X/731/2/130}

\bibitem{2012A&A...538A..45S}
{Santangelo}, G., {Nisini}, B., {Giannini}, T., {Antoniucci}, S., {Vasta}, M.,
  {Codella}, C., {Lorenzani}, A., {Tafalla}, M., {Liseau}, R., {van Dishoeck},
  E.F., {Kristensen}, L.E.: {The Herschel HIFI water line survey in the
  low-mass proto-stellar outflow L1448}.
\newblock \aap \textbf{538}, A45 (2012).
\newblock \doi{10.1051/0004-6361/201118113}

\bibitem{2009ApJS..182..477S}
{Sargent}, B.A., {Forrest}, W.J., {Tayrien}, C., {McClure}, M.K., {Watson},
  D.M., {Sloan}, G.C., {Li}, A., {Manoj}, P., {Bohac}, C.J., {Furlan}, E.,
  {Kim}, K.H., {Green}, J.D.: {Dust Processing and Grain Growth in
  Protoplanetary Disks in the Taurus-Auriga Star-Forming Region}.
\newblock \apjs \textbf{182}, 477--508 (2009).
\newblock \doi{10.1088/0067-0049/182/2/477}

\bibitem{2007DPS....39.4910S}
{Schaller}, E.L., {Brown}, M.E.: {Volatile Loss and Retention on Kuiper Belt
  Objects}.
\newblock In: AAS/Division for Planetary Sciences Meeting Abstracts \#39,
  \emph{Bulletin of the American Astronomical Society}, vol.~39, p. 511 (2007)

\bibitem{2002A&A...390.1001S}
{Sch{\"o}ier}, F.L., {J{\o}rgensen}, J.K., {van Dishoeck}, E.F., {Blake}, G.A.:
  {Does IRAS 16293-2422 have a hot core? Chemical inventory and abundance
  changes in its protostellar environment}.
\newblock \aap \textbf{390}, 1001--1021 (2002).
\newblock \doi{10.1051/0004-6361:20020756}

\bibitem{2005A&A...432..369S}
{Sch{\"o}ier}, F.L., {van der Tak}, F.F.S., {van Dishoeck}, E.F., {Black},
  J.H.: {An atomic and molecular database for analysis of submillimetre line
  observations}.
\newblock \aap \textbf{432}, 369--379 (2005).
\newblock \doi{10.1051/0004-6361:20041729}

\bibitem{2002Natur.416...73S}
{Schopf}, J.W., {Kudryavtsev}, A.B., {Agresti}, D.G., {Wdowiak}, T.J., {Czaja},
  A.D.: {Laser-Raman imagery of Earth's earliest fossils}.
\newblock \nat \textbf{416}, 73--76 (2002)

\bibitem{2012arXiv1208.3095S}
{Schr{\"a}pler}, R., {Blum}, J., {Seizinger}, A., {Kley}, W.: {The physics of
  protoplanetesimal dust agglomerates. VII The low-velocity collision behavior
  of large dust agglomerates}.
\newblock ArXiv e-prints  (2012)

\bibitem{2011IAUS..280..114S}
{Semenov}, D.A.: {Chemical Evolution of a Protoplanetary Disk}.
\newblock In: IAU Symposium, \emph{IAU Symposium}, vol. 280, pp. 114--126
  (2011).
\newblock \doi{10.1017/S1743921311024914}

\bibitem{2004A&A...415..203S}
{Shen}, C.J., {Greenberg}, J.M., {Schutte}, W.A., {van Dishoeck}, E.F.: {Cosmic
  ray induced explosive chemical desorption in dense clouds}.
\newblock \aap \textbf{415}, 203--215 (2004).
\newblock \doi{10.1051/0004-6361:20031669}

\bibitem{2008ApJ...683..255S}
{Shimajiri}, Y., {Takahashi}, S., {Takakuwa}, S., {Saito}, M., {Kawabe}, R.:
  {Millimeter- and Submillimeter-Wave Observations of the OMC-2/3 Region. II.
  Observational Evidence for Outflow-triggered Star Formation in the OMC-2 FIR
  3/4 Region}.
\newblock \apj \textbf{683}, 255--266 (2008).
\newblock \doi{10.1086/588629}

\bibitem{1977ApJ...214..488S}
{Shu}, F.H.: {Self-similar collapse of isothermal spheres and star formation}.
\newblock \apj \textbf{214}, 488--497 (1977).
\newblock \doi{10.1086/155274}

\bibitem{1987ARA&A..25...23S}
{Shu}, F.H., {Adams}, F.C., {Lizano}, S.: {Star formation in molecular clouds -
  Observation and theory}.
\newblock \araa \textbf{25}, 23--81 (1987).
\newblock \doi{10.1146/annurev.aa.25.090187.000323}

\bibitem{2012A&A...543A.155S}
{Sicilia}, D., {Ioppolo}, S., {Vindigni}, T., {Baratta}, G.A., {Palumbo}, M.E.:
  {Nitrogen oxides and carbon chain oxides formed after ion irradiation of
  CO:N$_{2}$ ice mixtures}.
\newblock \aap \textbf{543}, A155 (2012).
\newblock \doi{10.1051/0004-6361/201219390}

\bibitem{2012A&A...543A..25S}
{Siebenmorgen}, R., {Heymann}, F.: {Polycyclic aromatic hydrocarbons in
  protoplanetary disks: emission and X-ray destruction}.
\newblock \aap \textbf{543}, A25 (2012).
\newblock \doi{10.1051/0004-6361/201219039}

\bibitem{2010A&A...511A...6S}
{Siebenmorgen}, R., {Kr{\"u}gel}, E.: {The destruction and survival of
  polycyclic aromatic hydrocarbons in the disks of T Tauri stars}.
\newblock \aap \textbf{511}, A6 (2010).
\newblock \doi{10.1051/0004-6361/200912035}

\bibitem{2003LPI....34.1677S}
{Skrzypczak}, A., {Binet}, L., {Gourier}, D., {Derenne}, S., {Robert}, F.: {On
  the Controversial Biogenicity of the Organic Matter in the Oldest Archean
  Cherts: Can Electron Paramagnetic Resonance Provide Clues?}
\newblock In: S.~{Mackwell}, E.~{Stansbery} (eds.) Lunar and Planetary
  Institute Science Conference Abstracts, \emph{Lunar and Planetary Institute
  Science Conference Abstracts}, vol.~34, p. 1677 (2003)

\bibitem{2006ApJ...647..412S}
{Snyder}, L.E., {Hollis}, J.M., {Jewell}, P.R., {Lovas}, F.J., {Remijan}, A.:
  {Confirmation of Interstellar Methylcyanodiacetylene (CH$_{3}$C$_{5}$N)}.
\newblock \apj \textbf{647}, 412--417 (2006).
\newblock \doi{10.1086/505323}

\bibitem{1978ppim.book.....S}
{Spitzer}, L.: {Physical processes in the interstellar medium} (1978)

\bibitem{2005A&A...440..949S}
{St{\"a}uber}, P., {Doty}, S.D., {van Dishoeck}, E.F., {Benz}, A.O.: {X-ray
  chemistry in the envelopes around young stellar objects}.
\newblock \aap \textbf{440}, 949--966 (2005).
\newblock \doi{10.1051/0004-6361:20052889}

\bibitem{2010A&A...518L.129S}
{Sturm}, B., {Bouwman}, J., {Henning}, T., {Evans}, N.J., {Acke}, B.,
  {Mulders}, G.D., {Waters}, L.B.F.M., {van Dishoeck}, E.F., {Meeus}, G.,
  {Green}, J.D., {Augereau}, J.C., {Olofsson}, J., {Salyk}, C., {Najita}, J.,
  {Herczeg}, G.J., {van Kempen}, T.A., {Kristensen}, L.E., {Dominik}, C.,
  {Carr}, J.S., {Waelkens}, C., {Bergin}, E., {Blake}, G.A., {Brown}, J.M.,
  {Chen}, J.H., {Cieza}, L., {Dunham}, M.M., {Glassgold}, A., {G{\"u}del}, M.,
  {Harvey}, P.M., {Hogerheijde}, M.R., {Jaffe}, D., {J{\o}rgensen}, J.K.,
  {Kim}, H.J., {Knez}, C., {Lacy}, J.H., {Lee}, J.E., {Maret}, S., {Meijerink},
  R., {Mer{\'{\i}}n}, B., {Mundy}, L., {Pontoppidan}, K.M., {Visser}, R.,
  {Y{\i}ld{\i}z}, U.A.: {First results of the Herschel key program ``Dust, Ice
  and Gas In Time'' (DIGIT): Dust and gas spectroscopy of HD 100546}.
\newblock \aap \textbf{518}, L129 (2010).
\newblock \doi{10.1051/0004-6361/201014674}

\bibitem{2004A&A...416..191T}
{Tafalla}, M., {Myers}, P.C., {Caselli}, P., {Walmsley}, C.M.: {On the internal
  structure of starless cores. I. Physical conditions and the distribution of
  CO, CS, N$_{2}$H$^{+}$, and NH$_{3}$ in L1498 and L1517B}.
\newblock \aap \textbf{416}, 191--212 (2004).
\newblock \doi{10.1051/0004-6361:20031704}

\bibitem{2006A&A...455..577T}
{Tafalla}, M., {Santiago-Garc{\'{\i}}a}, J., {Myers}, P.C., {Caselli}, P.,
  {Walmsley}, C.M., {Crapsi}, A.: {On the internal structure of starless cores.
  II. A molecular survey of L1498 and L1517B}.
\newblock \aap \textbf{455}, 577--593 (2006).
\newblock \doi{10.1051/0004-6361:20065311}

\bibitem{2012ApJ...748L...3T}
{Taquet}, V., {Ceccarelli}, C., {Kahane}, C.: {Formaldehyde and Methanol
  Deuteration in Protostars: Fossils from a Past Fast High-density Pre-collapse
  Phase}.
\newblock \apjl \textbf{748}, L3 (2012).
\newblock \doi{10.1088/2041-8205/748/1/L3}

\bibitem{2012A&A...538A..42T}
{Taquet}, V., {Ceccarelli}, C., {Kahane}, C.: {Multilayer modeling of porous
  grain surface chemistry. I. The GRAINOBLE model}.
\newblock \aap \textbf{538}, A42 (2012).
\newblock \doi{10.1051/0004-6361/201117802}

\bibitem{2012dA&A...submitted}
{Taquet}, V., {Lopez-Sepulcre}, A., {Ceccarelli}, C., {Kahane}, C.: {Arcsecond
  resolution observations of deuterated water towards low-mass protostar}.
\newblock \aap \textbf{in print}, A42 (2012).
\newblock \doi{10.1051/0004-6361/201117812}

\bibitem{2012A&A...submitted}
{Taquet}, V., {Peters}, P., {Kahane}, C., {Ceccarelli}, C., {Lopez-Sepulcre},
  A., {Toubin}, C., {Duflot}, D., {Wiesenfeld}, L.: {Modelling of deuterated
  water ice formation}.
\newblock \aap \textbf{in print}, A42 (2012).
\newblock \doi{10.1051/0004-6361/201117802}

\bibitem{2007A&A...464...55T}
{Tatulli}, E., {Isella}, A., {Natta}, A., {Testi}, L., {Marconi}, A., {Malbet},
  F., {Stee}, P., {Petrov}, R.G., {Millour}, F., {Chelli}, A., {Duvert}, G.,
  {Antonelli}, P., {Beckmann}, U., {Bresson}, Y., {Dugu{\'e}}, M., {Gennari},
  S., {Gl{\"u}ck}, L., {Kern}, P., {Lagarde}, S., {Le Coarer}, E., {Lisi}, F.,
  {Perraut}, K., {Puget}, P., {Rantakyr{\"o}}, F., {Robbe-Dubois}, S.,
  {Roussel}, A., {Weigelt}, G., {Zins}, G., {Accardo}, M., {Acke}, B., {Agabi},
  K., {Altariba}, E., {Arezki}, B., {Aristidi}, E., {Baffa}, C., {Behrend}, J.,
  {Bl{\"o}cker}, T., {Bonhomme}, S., {Busoni}, S., {Cassaing}, F., {Clausse},
  J.M., {Colin}, J., {Connot}, C., {Delboulb{\'e}}, A., {Domiciano de Souza},
  A., {Driebe}, T., {Feautrier}, P., {Ferruzzi}, D., {Forveille}, T., {Fossat},
  E., {Foy}, R., {Fraix-Burnet}, D., {Gallardo}, A., {Giani}, E., {Gil}, C.,
  {Glentzlin}, A., {Heiden}, M., {Heininger}, M., {Hernandez Utrera}, O.,
  {Hofmann}, K.H., {Kamm}, D., {Kiekebusch}, M., {Kraus}, S., {Le Contel}, D.,
  {Le Contel}, J.M., {Lesourd}, T., {Lopez}, B., {Lopez}, M., {Magnard}, Y.,
  {Mars}, G., {Martinot-Lagarde}, G., {Mathias}, P., {M{\`e}ge}, P., {Monin},
  J.L., {Mouillet}, D., {Mourard}, D., {Nussbaum}, E., {Ohnaka}, K., {Pacheco},
  J., {Perrier}, C., {Rabbia}, Y., {Rebattu}, S., {Reynaud}, F., {Richichi},
  A., {Robini}, A., {Sacchettini}, M., {Schertl}, D., {Sch{\"o}ller}, M.,
  {Solscheid}, W., {Spang}, A., {Stefanini}, P., {Tallon}, M., {Tallon-Bosc},
  I., {Tasso}, D., {Vakili}, F., {von der L{\"u}he}, O., {Valtier}, J.C.,
  {Vannier}, M., {Ventura}, N.: {Constraining the wind launching region in
  Herbig Ae stars: AMBER/VLTI spectroscopy of HD 104237}.
\newblock \aap \textbf{464}, 55--58 (2007).
\newblock \doi{10.1051/0004-6361:20065719}

\bibitem{2007ApJ...667..303T}
{Terada}, H., {Tokunaga}, A.T., {Kobayashi}, N., {Takato}, N., {Hayano}, Y.,
  {Takami}, H.: {Detection of Water Ice in Edge-on Protoplanetary Disks: HK
  Tauri B and HV Tauri C}.
\newblock \apj \textbf{667}, 303--307 (2007).
\newblock \doi{10.1086/520951}

\bibitem{2003A&A...403..323T}
{Testi}, L., {Natta}, A., {Shepherd}, D.S., {Wilner}, D.J.: {Large grains in
  the disk of CQ Tau}.
\newblock \aap \textbf{403}, 323--328 (2003).
\newblock \doi{10.1051/0004-6361:20030362}

\bibitem{2010A&A...518L.125T}
{Thi}, W.F., {Mathews}, G., {M{\'e}nard}, F., {Woitke}, P., {Meeus}, G.,
  {Riviere-Marichalar}, P., {Pinte}, C., {Howard}, C.D., {Roberge}, A.,
  {Sandell}, G., {Pascucci}, I., {Riaz}, B., {Grady}, C.A., {Dent}, W.R.F.,
  {Kamp}, I., {Duch{\^e}ne}, G., {Augereau}, J.C., {Pantin}, E.,
  {Vandenbussche}, B., {Tilling}, I., {Williams}, J.P., {Eiroa}, C., {Barrado},
  D., {Alacid}, J.M., {Andrews}, S., {Ardila}, D.R., {Aresu}, G., {Brittain},
  S., {Ciardi}, D.R., {Danchi}, W., {Fedele}, D., {de Gregorio-Monsalvo}, I.,
  {Heras}, A., {Huelamo}, N., {Krivov}, A., {Lebreton}, J., {Liseau}, R.,
  {Martin-Zaidi}, C., {Mendigut{\'{\i}}a}, I., {Montesinos}, B., {Mora}, A.,
  {Morales-Calderon}, M., {Nomura}, H., {Phillips}, N., {Podio}, L., {Poelman},
  D.R., {Ramsay}, S., {Rice}, K., {Solano}, E., {Walker}, H., {White}, G.J.,
  {Wright}, G.: {Herschel-PACS observation of the 10 Myr old T Tauri disk TW
  Hya. Constraining the disk gas mass}.
\newblock \aap \textbf{518}, L125 (2010).
\newblock \doi{10.1051/0004-6361/201014578}

\bibitem{2004A&A...425..955T}
{Thi}, W.F., {van Zadelhoff}, G.J., {van Dishoeck}, E.F.: {Organic molecules in
  protoplanetary disks around T Tauri and Herbig Ae stars}.
\newblock \aap \textbf{425}, 955--972 (2004).
\newblock \doi{10.1051/0004-6361:200400026}

\bibitem{2010MNRAS.407..232T}
{Thi}, W.F., {Woitke}, P., {Kamp}, I.: {Warm non-equilibrium gas phase
  chemistry as a possible origin of high HDO/H$_{2}$O ratios in hot and dense
  gases: application to inner protoplanetary discs}.
\newblock \mnras \textbf{407}, 232--246 (2010).
\newblock \doi{10.1111/j.1365-2966.2009.16162.x}

\bibitem{1993GeCoA..57.1551T}
{Thomas}, K.L., {Blanford}, G.E., {Keller}, L.P., {Klock}, W., {McKay}, D.S.:
  {Carbon abundance and silicate mineralogy of anhydrous interplanetary dust
  particles}.
\newblock \gca \textbf{57}, 1551--1566 (1993).
\newblock \doi{10.1016/0016-7037(93)90012-L}

\bibitem{1983A&A...119..177T}
{Tielens}, A.G.G.M.: {Surface chemistry of deuterated molecules}.
\newblock \aap \textbf{119}, 177--184 (1983)

\bibitem{2005pcim.book.....T}
{Tielens}, A.G.G.M.: {The Physics and Chemistry of the Interstellar Medium}
  (2005)

\bibitem{2008ApJ...680..457T}
{Troland}, T.H., {Crutcher}, R.M.: {Magnetic Fields in Dark Cloud Cores:
  Arecibo OH Zeeman Observations}.
\newblock \apj \textbf{680}, 457--465 (2008).
\newblock \doi{10.1086/587546}

\bibitem{2009A&A...506.1243T}
{Troscompt}, N., {Faure}, A., {Maret}, S., {Ceccarelli}, C., {Hily-Blant}, P.,
  {Wiesenfeld}, L.: {Constraining the ortho-to-para ratio of H$_{2}$ with
  anomalous H\_2CO absorption}.
\newblock \aap \textbf{506}, 1243--1247 (2009).
\newblock \doi{10.1051/0004-6361/200912770}

\bibitem{2005A&A...439..195V}
{van der Tak}, F.F.S., {Caselli}, P., {Ceccarelli}, C.: {Line profiles of
  molecular ions toward the pre-stellar core LDN 1544}.
\newblock \aap \textbf{439}, 195--203 (2005).
\newblock \doi{10.1051/0004-6361:20052792}

\bibitem{2002A&A...388L..53V}
{van der Tak}, F.F.S., {Schilke}, P., {M{\"u}ller}, H.S.P., {Lis}, D.C.,
  {Phillips}, T.G., {Gerin}, M., {Roueff}, E.: {Triply deuterated ammonia in
  NGC 1333}.
\newblock \aap \textbf{388}, L53--L56 (2002).
\newblock \doi{10.1051/0004-6361:20020647}

\bibitem{2011PASP..123..138V}
{van Dishoeck}, E.F., {Kristensen}, L.E., {Benz}, A.O., {Bergin}, E.A.,
  {Caselli}, P., {Cernicharo}, J., {Herpin}, F., {Hogerheijde}, M.R.,
  {Johnstone}, D., {Liseau}, R., {Nisini}, B., {Shipman}, R., {Tafalla}, M.,
  {van der Tak}, F., {Wyrowski}, F., {Aikawa}, Y., {Bachiller}, R., {Baudry},
  A., {Benedettini}, M., {Bjerkeli}, P., {Blake}, G.A., {Bontemps}, S.,
  {Braine}, J., {Brinch}, C., {Bruderer}, S., {Chavarr{\'{\i}}a}, L.,
  {Codella}, C., {Daniel}, F., {de Graauw}, T., {Deul}, E., {di Giorgio}, A.M.,
  {Dominik}, C., {Doty}, S.D., {Dubernet}, M.L., {Encrenaz}, P.,
  {Feuchtgruber}, H., {Fich}, M., {Frieswijk}, W., {Fuente}, A., {Giannini},
  T., {Goicoechea}, J.R., {Helmich}, F.P., {Herczeg}, G.J., {Jacq}, T.,
  {J{\o}rgensen}, J.K., {Karska}, A., {Kaufman}, M.J., {Keto}, E., {Larsson},
  B., {Lefloch}, B., {Lis}, D., {Marseille}, M., {McCoey}, C., {Melnick}, G.,
  {Neufeld}, D., {Olberg}, M., {Pagani}, L., {Pani{\'c}}, O., {Parise}, B.,
  {Pearson}, J.C., {Plume}, R., {Risacher}, C., {Salter}, D.,
  {Santiago-Garc{\'{\i}}a}, J., {Saraceno}, P., {St{\"a}uber}, P., {van
  Kempen}, T.A., {Visser}, R., {Viti}, S., {Walmsley}, M., {Wampfler}, S.F.,
  {Y{\i}ld{\i}z}, U.A.: {Water in Star-forming Regions with the Herschel Space
  Observatory (WISH). I. Overview of Key Program and First Results}.
\newblock \pasp \textbf{123}, 138--170 (2011).
\newblock \doi{10.1086/658676}

\bibitem{2003A&A...400L...1V}
{van Dishoeck}, E.F., {Thi}, W.F., {van Zadelhoff}, G.J.: {Detection of
  DCO$^{+}$ in a circumstellar disk}.
\newblock \aap \textbf{400}, L1--L4 (2003).
\newblock \doi{10.1051/0004-6361:20030091}

\bibitem{2009A&A...507.1425V}
{van Kempen}, T.A., {van Dishoeck}, E.F., {G{\"u}sten}, R., {Kristensen}, L.E.,
  {Schilke}, P., {Hogerheijde}, M.R., {Boland}, W., {Menten}, K.M., {Wyrowski},
  F.: {APEX-CHAMP$^{+}$ high-J CO observations of low-mass young stellar
  objects. II. Distribution and origin of warm molecular gas}.
\newblock \aap \textbf{507}, 1425--1442 (2009).
\newblock \doi{10.1051/0004-6361/200912507}

\bibitem{2001A&A...377..566V}
{van Zadelhoff}, G.J., {van Dishoeck}, E.F., {Thi}, W.F., {Blake}, G.A.:
  {Submillimeter lines from circumstellar disks around pre-main sequence
  stars}.
\newblock \aap \textbf{377}, 566--580 (2001).
\newblock \doi{10.1051/0004-6361:20011137}

\bibitem{2012A&A...537A..98V}
{Vasta}, M., {Codella}, C., {Lorenzani}, A., {Santangelo}, G., {Nisini}, B.,
  {Giannini}, T., {Tafalla}, M., {Liseau}, R., {van Dishoeck}, E.F.,
  {Kristensen}, L.: {Water emission from the chemically rich outflow L1157}.
\newblock \aap \textbf{537}, A98 (2012).
\newblock \doi{10.1051/0004-6361/201118201}

\bibitem{2010A&A...521L..31V}
{Vastel}, C., {Ceccarelli}, C., {Caux}, E., {Coutens}, A., {Cernicharo}, J.,
  {Bottinelli}, S., {Demyk}, K., {Faure}, A., {Wiesenfeld}, L., {Scribano}, Y.,
  {Bacmann}, A., {Hily-Blant}, P., {Maret}, S., {Walters}, A., {Bergin}, E.A.,
  {Blake}, G.A., {Castets}, A., {Crimier}, N., {Dominik}, C., {Encrenaz}, P.,
  {G{\'e}rin}, M., {Hennebelle}, P., {Kahane}, C., {Klotz}, A., {Melnick}, G.,
  {Pagani}, L., {Parise}, B., {Schilke}, P., {Wakelam}, V., {Baudry}, A.,
  {Bell}, T., {Benedettini}, M., {Boogert}, A., {Cabrit}, S., {Caselli}, P.,
  {Codella}, C., {Comito}, C., {Falgarone}, E., {Fuente}, A., {Goldsmith},
  P.F., {Helmich}, F., {Henning}, T., {Herbst}, E., {Jacq}, T., {Kama}, M.,
  {Langer}, W., {Lefloch}, B., {Lis}, D., {Lord}, S., {Lorenzani}, A.,
  {Neufeld}, D., {Nisini}, B., {Pacheco}, S., {Pearson}, J., {Phillips}, T.,
  {Salez}, M., {Saraceno}, P., {Schuster}, K., {Tielens}, X., {van der Tak},
  F., {van der Wiel}, M.H.D., {Viti}, S., {Wyrowski}, F., {Yorke}, H., {Cais},
  P., {Krieg}, J.M., {Olberg}, M., {Ravera}, L.: {Ortho-to-para ratio of
  interstellar heavy water}.
\newblock \aap \textbf{521}, L31 (2010).
\newblock \doi{10.1051/0004-6361/201015101}

\bibitem{2003ApJ...593L..97V}
{Vastel}, C., {Phillips}, T.G., {Ceccarelli}, C., {Pearson}, J.: {First
  Detection of Doubly Deuterated Hydrogen Sulfide}.
\newblock \apjl \textbf{593}, L97--L100 (2003).
\newblock \doi{10.1086/378261}

\bibitem{2004ApJ...606L.127V}
{Vastel}, C., {Phillips}, T.G., {Yoshida}, H.: {Detection of D$_{2}$H$^{+}$ in
  the Dense Interstellar Medium}.
\newblock \apjl \textbf{606}, L127--L130 (2004).
\newblock \doi{10.1086/421265}

\bibitem{2008ApJ...672..629V}
{Vasyunin}, A.I., {Semenov}, D., {Henning}, T., {Wakelam}, V., {Herbst}, E.,
  {Sobolev}, A.M.: {Chemistry in Protoplanetary Disks: A Sensitivity Analysis}.
\newblock \apj \textbf{672}, 629--641 (2008).
\newblock \doi{10.1086/523887}

\bibitem{2011ApJ...727...76V}
{Vasyunin}, A.I., {Wiebe}, D.S., {Birnstiel}, T., {Zhukovska}, S., {Henning},
  T., {Dullemond}, C.P.: {Impact of Grain Evolution on the Chemical Structure
  of Protoplanetary Disks}.
\newblock \apj \textbf{727}, 76 (2011).
\newblock \doi{10.1088/0004-637X/727/2/76}

\bibitem{2009Sci...325..985V}
{Villeneuve}, J., {Chaussidon}, M., {Libourel}, G.: {Homogeneous Distribution
  of $^{26}$Al in the Solar System from the Mg Isotopic Composition of
  Chondrules}.
\newblock Science \textbf{325}, 985-- (2009).
\newblock \doi{10.1126/science.1173907}

\bibitem{2011A&A...534A.132V}
{Visser}, R., {Doty}, S.D., {van Dishoeck}, E.F.: {The chemical history of
  molecules in circumstellar disks. II. Gas-phase species}.
\newblock \aap \textbf{534}, A132 (2011).
\newblock \doi{10.1051/0004-6361/201117249}

\bibitem{2007A&A...466..229V}
{Visser}, R., {Geers}, V.C., {Dullemond}, C.P., {Augereau}, J.C.,
  {Pontoppidan}, K.M., {van Dishoeck}, E.F.: {PAH chemistry and IR emission
  from circumstellar disks}.
\newblock \aap \textbf{466}, 229--241 (2007).
\newblock \doi{10.1051/0004-6361:20066829}

\bibitem{2012A&A...537A..55V}
{Visser}, R., {Kristensen}, L.E., {Bruderer}, S., {van Dishoeck}, E.F.,
  {Herczeg}, G.J., {Brinch}, C., {Doty}, S.D., {Harsono}, D., {Wolfire}, M.G.:
  {Modelling Herschel observations of hot molecular gas emission from embedded
  low-mass protostars}.
\newblock \aap \textbf{537}, A55 (2012).
\newblock \doi{10.1051/0004-6361/201117109}

\bibitem{2009A&A...495..881V}
{Visser}, R., {van Dishoeck}, E.F., {Doty}, S.D., {Dullemond}, C.P.: {The
  chemical history of molecules in circumstellar disks. I. Ices}.
\newblock \aap \textbf{495}, 881--897 (2009).
\newblock \doi{10.1051/0004-6361/200810846}

\bibitem{2004MNRAS.354.1141V}
{Viti}, S., {Collings}, M.P., {Dever}, J.W., {McCoustra}, M.R.S., {Williams},
  D.A.: {Evaporation of ices near massive stars: models based on laboratory
  temperature programmed desorption data}.
\newblock \mnras \textbf{354}, 1141--1145 (2004).
\newblock \doi{10.1111/j.1365-2966.2004.08273.x}

\bibitem{2011ApJ...729..146V}
{Vorobyov}, E.I.: {Embedded Protostellar Disks Around (Sub-)Solar Stars. II.
  Disk Masses, Sizes, Densities, Temperatures, and the Planet Formation
  Perspective}.
\newblock \apj \textbf{729}, 146 (2011).
\newblock \doi{10.1088/0004-637X/729/2/146}

\bibitem{2008ApJ...680..371W}
{Wakelam}, V., {Herbst}, E.: {Polycyclic Aromatic Hydrocarbons in Dense Cloud
  Chemistry}.
\newblock \apj \textbf{680}, 371--383 (2008).
\newblock \doi{10.1086/587734}

\bibitem{2012ApJS..199...21W}
{Wakelam}, V., {Herbst}, E., {Loison}, J.C., {Smith}, I.W.M., {Chandrasekaran},
  V., {Pavone}, B., {Adams}, N.G., {Bacchus-Montabonel}, M.C., {Bergeat}, A.,
  {B{\'e}roff}, K., {Bierbaum}, V.M., {Chabot}, M., {Dalgarno}, A., {van
  Dishoeck}, E.F., {Faure}, A., {Geppert}, W.D., {Gerlich}, D., {Galli}, D.,
  {H{\'e}brard}, E., {Hersant}, F., {Hickson}, K.M., {Honvault}, P.,
  {Klippenstein}, S.J., {Le Picard}, S., {Nyman}, G., {Pernot}, P.,
  {Schlemmer}, S., {Selsis}, F., {Sims}, I.R., {Talbi}, D., {Tennyson}, J.,
  {Troe}, J., {Wester}, R., {Wiesenfeld}, L.: {A KInetic Database for
  Astrochemistry (KIDA)}.
\newblock \apjs \textbf{199}, 21 (2012).
\newblock \doi{10.1088/0067-0049/199/1/21}

\bibitem{2006A&A...451..551W}
{Wakelam}, V., {Herbst}, E., {Selsis}, F.: {The effect of uncertainties on
  chemical models of dark clouds}.
\newblock \aap \textbf{451}, 551--562 (2006).
\newblock \doi{10.1051/0004-6361:20054682}

\bibitem{1984A&A...134L..11W}
{Walmsley}, C.M., {Jewell}, P.R., {Snyder}, L.E., {Winnewisser}, G.: {Detection
  of interstellar methyldiacetylene (CH3C4H) in the dark dust cloud TMC 1}.
\newblock \aap \textbf{134}, L11--L14 (1984)

\bibitem{2012ApJ...747..114W}
{Walsh}, C., {Nomura}, H., {Millar}, T.J., {Aikawa}, Y.: {Chemical Processes in
  Protoplanetary Disks. II. On the Importance of Photochemistry and X-Ray
  Ionization}.
\newblock \apj \textbf{747}, 114 (2012).
\newblock \doi{10.1088/0004-637X/747/2/114}

\bibitem{1999MNRAS.305..143W}
{Ward-Thompson}, D., {Motte}, F., {Andre}, P.: {The initial conditions of
  isolated star formation - III. Millimetre continuum mapping of pre-stellar
  cores}.
\newblock \mnras \textbf{305}, 143--150 (1999).
\newblock \doi{10.1046/j.1365-8711.1999.02412.x}

\bibitem{1994MNRAS.268..276W}
{Ward-Thompson}, D., {Scott}, P.F., {Hills}, R.E., {Andre}, P.: {A
  Submillimetre Continuum Survey of Pre Protostellar Cores}.
\newblock \mnras \textbf{268}, 276 (1994)

\bibitem{2004Ap&SS.292..317W}
{Wardle}, M.: {Star Formation and the Hall Effect}.
\newblock \apss \textbf{292}, 317--323 (2004).
\newblock \doi{10.1023/B:ASTR.0000045033.80068.1f}

\bibitem{2002ApJ...571L.173W}
{Watanabe}, N., {Kouchi}, A.: {Efficient Formation of Formaldehyde and Methanol
  by the Addition of Hydrogen Atoms to CO in H$_{2}$O-CO Ice at 10 K}.
\newblock \apjl \textbf{571}, L173--L176 (2002).
\newblock \doi{10.1086/341412}

\bibitem{1974ApJ...188...35W}
{Watson}, W.D.: {Ion-Molecule Reactions, Molecule Formation, and
  Hydrogen-Isotope Exchange in Dense Interstellar Clouds}.
\newblock \apj \textbf{188}, 35--42 (1974).
\newblock \doi{10.1086/152681}

\bibitem{1977MNRAS.180...57W}
{Weidenschilling}, S.J.: {Aerodynamics of solid bodies in the solar nebula}.
\newblock \mnras \textbf{180}, 57--70 (1977)

\bibitem{2010ApJ...710.1009W}
{Whittet}, D.C.B.: {Oxygen Depletion in the Interstellar Medium: Implications
  for Grain Models and the Distribution of Elemental Oxygen}.
\newblock \apj \textbf{710}, 1009--1016 (2010).
\newblock \doi{10.1088/0004-637X/710/2/1009}

\bibitem{2011ApJ...742...28W}
{Whittet}, D.C.B., {Cook}, A.M., {Herbst}, E., {Chiar}, J.E., {Shenoy}, S.S.:
  {Observational Constraints on Methanol Production in Interstellar and
  Preplanetary Ices}.
\newblock \apj \textbf{742}, 28 (2011).
\newblock \doi{10.1088/0004-637X/742/1/28}

\bibitem{1991A&ARv...2..167W}
{Whittet}, D.C.B., {Duley}, W.W.: {Carbon monoxide frosts in the interstellar
  medium}.
\newblock \aapr \textbf{2}, 167--189 (1991).
\newblock \doi{10.1007/BF00872766}

\bibitem{2012A&A...544A.146W}
{Wienen}, M., {Wyrowski}, F., {Schuller}, F., {Menten}, K.M., {Walmsley}, C.M.,
  {Bronfman}, L., {Motte}, F.: {Ammonia from cold high-mass clumps discovered
  in the inner Galactic disk by the ATLASGAL survey}.
\newblock \aap \textbf{544}, A146 (2012).
\newblock \doi{10.1051/0004-6361/201118107}

\bibitem{2006ApJ...644.1202W}
{Willacy}, K., {Langer}, W., {Allen}, M., {Bryden}, G.: {Turbulence-driven
  Diffusion in Protoplanetary Disks: Chemical Effects in the Outer Regions}.
\newblock \apj \textbf{644}, 1202--1213 (2006).
\newblock \doi{10.1086/503702}

\bibitem{2000ApJ...544..903W}
{Willacy}, K., {Langer}, W.D.: {The Importance of Photoprocessing in
  Protoplanetary Disks}.
\newblock \apj \textbf{544}, 903--920 (2000).
\newblock \doi{10.1086/317236}

\bibitem{1998ApJ...507L.171W}
{Willacy}, K., {Langer}, W.D., {Velusamy}, T.: {Dust Emission and Molecular
  Depletion in L1498}.
\newblock \apjl \textbf{507}, L171--L175 (1998).
\newblock \doi{10.1086/311695}

\bibitem{1998MNRAS.298..562W}
{Willacy}, K., {Millar}, T.J.: {Desorption processes and the deuterium
  fractionation in molecular clouds}.
\newblock \mnras \textbf{298}, 562--568 (1998).
\newblock \doi{10.1046/j.1365-8711.1998.01648.x}

\bibitem{2011ARA&A..49...67W}
{Williams}, J.P., {Cieza}, L.A.: {Protoplanetary Disks and Their Evolution}.
\newblock \araa \textbf{49}, 67--117 (2011).
\newblock \doi{10.1146/annurev-astro-081710-102548}

\bibitem{2005ApJ...626L.109W}
{Wilner}, D.J., {D'Alessio}, P., {Calvet}, N., {Claussen}, M.J., {Hartmann},
  L.: {Toward Planetesimals in the Disk around TW Hydrae: 3.5 Centimeter Dust
  Emission}.
\newblock \apjl \textbf{626}, L109--L112 (2005).
\newblock \doi{10.1086/431757}

\bibitem{2012A&A...544L..16W}
{Windmark}, F., {Birnstiel}, T., {Ormel}, C.W., {Dullemond}, C.P.: {Breaking
  through: The effects of a velocity distribution on barriers to dust growth}.
\newblock \aap \textbf{544}, L16 (2012).
\newblock \doi{10.1051/0004-6361/201220004}

\bibitem{2012ApJ...757L..11W}
{Wirstr{\"o}m}, E.S., {Charnley}, S.B., {Cordiner}, M.A., {Milam}, S.N.:
  {Isotopic Anomalies in Primitive Solar System Matter: Spin-state-dependent
  Fractionation of Nitrogen and Deuterium in Interstellar Clouds}.
\newblock \apjl \textbf{757}, L11 (2012).
\newblock \doi{10.1088/2041-8205/757/1/L11}

\bibitem{2010MNRAS.405L..26W}
{Woitke}, P., {Pinte}, C., {Tilling}, I., {M{\'e}nard}, F., {Kamp}, I., {Thi},
  W.F., {Duch{\^e}ne}, G., {Augereau}, J.C.: {Continuum and line modelling of
  discs around young stars - I. 300000 disc models for HERSCHEL/GASPS}.
\newblock \mnras \textbf{405}, L26--L30 (2010).
\newblock \doi{10.1111/j.1745-3933.2010.00852.x}

\bibitem{2002ApJ...567.1183W}
{Wood}, K., {Lada}, C.J., {Bjorkman}, J.E., {Kenyon}, S.J., {Whitney}, B.,
  {Wolff}, M.J.: {Infrared Signatures of Protoplanetary Disk Evolution}.
\newblock \apj \textbf{567}, 1183--1191 (2002).
\newblock \doi{10.1086/338662}

\bibitem{2007A&A...466.1197W}
{Woodall}, J., {Ag{\'u}ndez}, M., {Markwick-Kemper}, A.J., {Millar}, T.J.: {The
  UMIST database for astrochemistry 2006}.
\newblock \aap \textbf{466}, 1197--1204 (2007).
\newblock \doi{10.1051/0004-6361:20064981}

\bibitem{1999ApJ...517.1034W}
{Wooden}, D.H., {Harker}, D.E., {Woodward}, C.E., {Butner}, H.M., {Koike}, C.,
  {Witteborn}, F.C., {McMurtry}, C.W.: {Silicate Mineralogy of the Dust in the
  Inner Coma of Comet C/1995 01 (Hale-Bopp) Pre- and Postperihelion}.
\newblock \apj \textbf{517}, 1034--1058 (1999).
\newblock \doi{10.1086/307206}

\bibitem{2004ApJ...612L..77W}
{Wooden}, D.H., {Woodward}, C.E., {Harker}, D.E.: {Discovery of Crystalline
  Silicates in Comet C/2001 Q4 (NEAT)}.
\newblock \apjl \textbf{612}, L77--L80 (2004).
\newblock \doi{10.1086/424593}

\bibitem{1989ApJ...337..858W}
{Wootten}, A.: {The Duplicity of IRAS 16293-2422: A Protobinary Star?}
\newblock \apj \textbf{337}, 858 (1989).
\newblock \doi{10.1086/167156}

\bibitem{1979ApJ...234..876W}
{Wootten}, A., {Snell}, R., {Glassgold}, A.E.: {The determination of electron
  abundances in interstellar clouds}.
\newblock \apj \textbf{234}, 876--880 (1979).
\newblock \doi{10.1086/157569}

\bibitem{2008A&A...487..237W}
{Wouterloot}, J.G.A., {Henkel}, C., {Brand}, J., {Davis}, G.R.: {Galactic
  interstellar $^{18}$O/$\{$\^{}17$\}$O ratios - a radial gradient?}
\newblock \aap \textbf{487}, 237--246 (2008).
\newblock \doi{10.1051/0004-6361:20078156}

\bibitem{2008ARA&A..46..339W}
{Wyatt}, M.C.: {Evolution of Debris Disks}.
\newblock \araa \textbf{46}, 339--383 (2008).
\newblock \doi{10.1146/annurev.astro.45.051806.110525}

\bibitem{1995ApJ...440..674X}
{Xie}, T., {Allen}, M., {Langer}, W.D.: {Turbulent Diffusion and Its Effects on
  the Chemistry of Molecular Clouds}.
\newblock \apj \textbf{440}, 674 (1995).
\newblock \doi{10.1086/175305}

\bibitem{2011PASJ...63L..37Y}
{Yamaguchi}, T., {Takano}, S., {Sakai}, N., {Takeshi Sakai} Sheng-Yuan, T.L.,
  {Su}, Y.N., {Hirano}, N., {Takakuwa}, S., {Aikawa}, Y., {Nomura}, H.,
  {Yamamoto}, S.: {Detection of Phosphorus Nitride in the Lynds 1157 B1 Shocked
  Region}.
\newblock \pasj \textbf{63}, L37--L41 (2011)

\bibitem{2012M&PS...47...99Y}
{Yang}, L., {Ciesla}, F.J.: {The effects of disk building on the distributions
  of refractory materials in the solar nebula}.
\newblock Meteoritics and Planetary Science \textbf{47}, 99--119 (2012).
\newblock \doi{10.1111/j.1945-5100.2011.01315.x}

\bibitem{2012A&A...542A..86Y}
{Y{\i}ld{\i}z}, U.A., {Kristensen}, L.E., {van Dishoeck}, E.F., {Belloche}, A.,
  {van Kempen}, T.A., {Hogerheijde}, M.R., {G{\"u}sten}, R., {van der Marel},
  N.: {APEX-CHAMP$^{+}$ high-J CO observations of low-mass young stellar
  objects. III. NGC 1333 IRAS 4A/4B envelope, outflow, and ultraviolet
  heating}.
\newblock \aap \textbf{542}, A86 (2012).
\newblock \doi{10.1051/0004-6361/201118368}

\bibitem{2011ApJ...729...43Y}
{Young}, E.D., {Gounelle}, M., {Smith}, R.L., {Morris}, M.R., {Pontoppidan},
  K.M.: {Astronomical Oxygen Isotopic Evidence for Supernova Enrichment of the
  Solar System Birth Environment by Propagating Star Formation}.
\newblock \apj \textbf{729}, 43 (2011).
\newblock \doi{10.1088/0004-637X/729/1/43}

\bibitem{2008M&PS...43..261Z}
{Zolensky}, M., {Nakamura-Messenger}, K., {Rietmeijer}, F., {Leroux}, H.,
  {Mikouchi}, T., {Ohsumi}, K., {Simon}, S., {Grossman}, L., {Stephan}, T.,
  {Weisberg}, M., {Velbel}, M., {Zega}, T., {Stroud}, R., {Tomeoka}, K.,
  {Ohnishi}, I., {Tomioka}, N., {Nakamura}, T., {Matrajt}, G., {Joswiak}, D.,
  {Brownlee}, D., {Langenhorst}, F., {Krot}, A., {Kearsley}, A., {Ishii}, H.,
  {Graham}, G., {Dai}, Z.R., {Chi}, M., {Bradley}, J., {Hagiya}, K.,
  {Gounelle}, M., {Bridges}, J.: {Comparing Wild 2 particles to chondrites and
  IDPs}.
\newblock Meteoritics and Planetary Science \textbf{43}, 261--272 (2008).
\newblock \doi{10.1111/j.1945-5100.2008.tb00621.x}

\bibitem{2011A&A...534A..73Z}
{Zsom}, A., {Ormel}, C.W., {Dullemond}, C.P., {Henning}, T.: {The outcome of
  protoplanetary dust growth: pebbles, boulders, or planetesimals?. III.
  Sedimentation driven coagulation inside the snowline}.
\newblock \aap \textbf{534}, A73 (2011).
\newblock \doi{10.1051/0004-6361/201116515}

\end{thebibliography}

% Non-BibTeX users please use
%\begin{thebibliography}{}
%
% and use \bibitem to create references. Consult the Instructions
% for authors for reference list style.
%
%\bibitem{RefJ}
% Format for Journal Reference
%Author, Article title, Journal, Volume, page numbers (year)
% Format for books
%\bibitem{RefB}
%Author, Book title, page numbers. Publisher, place (year)
% etc
%\end{thebibliography}

%%%CC: APPENDIX
%%% vogliamo mettere un appendix con tutti gli acronyms usati??? 
%+technical terms
%EXAMPLES
%IOM
%SOM
%aliphatic
%aromatic
%carboxylic acids
%amino acids
%hydrocarbons
%L-enantiometric excess
%chiral
%short-lived nuclides
%PDR
%WCCC
%isomer
%isotopologue
%COM
%PAH
%MHD
%IMHD
%ISRF
%amine
%nitrile
%%% potrebbe essere utile mettere un appendix con le caratteristiche
%%% dei prototipi di sorgenti menzionate nel testo: L1544,
%%% iras16293-2422, TW Hya.... con le lum, mass, sizes, distances

\end{document}